\documentclass{article}

\usepackage{amssymb}
\usepackage{latexsym}

\usepackage{url}
\usepackage{xcolor}
\definecolor{newcolor}{rgb}{.8,.349,.1}
\usepackage{multirow}
\usepackage{color}

\usepackage{float}
\usepackage[utf8]{inputenc}  
\usepackage{subfiles} 
\usepackage[square,numbers,sort&compress]{natbib}
\usepackage{amssymb}
\usepackage{mathtools}
\usepackage{gensymb}
\usepackage{authblk}
\usepackage[margin=0.8in]{geometry}
\usepackage[font=footnotesize]{subfig}
\usepackage{subfiles}
\usepackage[frenchb]{babel}

\usepackage{amsmath}
\usepackage[pdftex]{graphicx}

\usepackage{array}
\newcolumntype{M}[1]{>{\centering\arraybackslash}m{#1}}

\usepackage{hyperref}

\begin{document}

\title{Physically interpretable machine learning algorithm on multidimensional non-linear fields}%

\author[1,2]{Rem-Sophia Mouradi}
\author[1]{Cédric Goeury}
\author[2,3]{Olivier Thual}
\author[1]{Fabrice Zaoui}
\author[1,4]{Pablo Tassi}

\affil[1]{EDF R\&D, National Laboratory for Hydraulics and Environment (LNHE), 6 Quai Watier, 78400 Chatou, France}
\affil[2]{Climate, Environment, Coupling and Uncertainties research unit (CECI) at the European Center for Research and Advanced Training in Scientific Computation (CERFACS), French National Research Center (CNRS), 42 Avenue Gaspard Coriolis, 31820 Toulouse, France}
\affil[3]{Institut de M\'ecanique des Fluides de Toulouse (IMFT), Universit\'e de Toulouse, CNRS, Toulouse, France}
\affil[4]{Saint-Venant Laboratory for Hydraulics (LHSV), Chatou, France}

\date{\today}
\maketitle

\begin{abstract}
In an ever-increasing interest for Machine Learning (ML) and a favorable data development context, we here propose an original methodology for data-based prediction of two-dimensional physical fields.  Polynomial Chaos Expansion (PCE), widely used in the Uncertainty Quantification community (UQ), has long been employed as a robust representation for probabilistic input-to-output mapping. It has been recently tested in a pure ML context, and shown to be as powerful as classical ML techniques for point-wise prediction. Some advantages are inherent to the method, such as its explicitness and adaptability to small training sets, in addition to the associated probabilistic framework. Simultaneously, Dimensionality Reduction (DR) techniques are increasingly used for pattern recognition and data compression and have gained interest due to improved data quality. In this study, the interest of Proper Orthogonal Decomposition (POD) for the construction of a statistical predictive model is demonstrated. Both POD and PCE have amply proved their worth in their respective frameworks. The goal of the present paper was to combine them for a field-measurement-based forecasting. The described steps are also useful to analyze the data. Some challenging issues encountered when using multidimensional field measurements are addressed, for example when dealing with few data. The POD-PCE coupling methodology is presented, with particular focus on input data characteristics and training-set choice. A simple methodology for evaluating the importance of each physical parameter is proposed for the PCE model and extended to the POD-PCE coupling.  
\end{abstract}


\shorthandoff{:}
\section{Introduction}

Deep Learning techniques (DL \citep{LeCun2015}) and more generally Machine Learning (ML \citep{ShalevShwartz2014}), and their applications to physical problems (fluid mechanics \citep{Brunton2020}; plasma physics \citep{Parsons2017}; quantum mechanics \citep{Mills2017}, etc.) have made a promising take-off in the last few years. This has been particularly the case for fields where the measurement potential has dramatically increased (e.g. Geoscience Data \citep{Karpatne2019}). In this context, learning techniques are of interest to establish non-linear physical relationships from the data by a combination of steps, in particular using transformation functions, to capture the complexity of the system \citep{ShalevShwartz2014}. \\

In particular, multi-layer Neural Networks (NN) \citep{Schmidhuber2015} are widely used for physical applications. Their popularity comes from this complex structure, which makes them adaptable for various applications \citep{Sengupta2019,Abiodun2018}. However, some limitations prevent the use of NN for physical applications: (i) it is difficult to provide an explicit input-to-output formulation, due to the combinations of steps involved in the learning (\textit{Activation Functions}, \textit{Hidden Layers} \cite{LeCun2015}). Physical interpretation of the constructed model is therefore tedious \citep{Iten2020}; (ii) too many hyper-parameters and choices are involved, depending on the number of neurons and layers (\textit{curse of dimensionality})  \citep{Laudani2015}; (iii) no general proof for the theoretical ability of approximating arbitrary functions is available, except the \textit{Universal Approximation Theorem} and its extensions \citep{Hornik1991,Hanin2019} for particular cases. \\

\begin{figure}[H]
  \centering
    \includegraphics[trim={0.5cm 3cm 0.5cm 3.2cm},clip,scale=0.52]{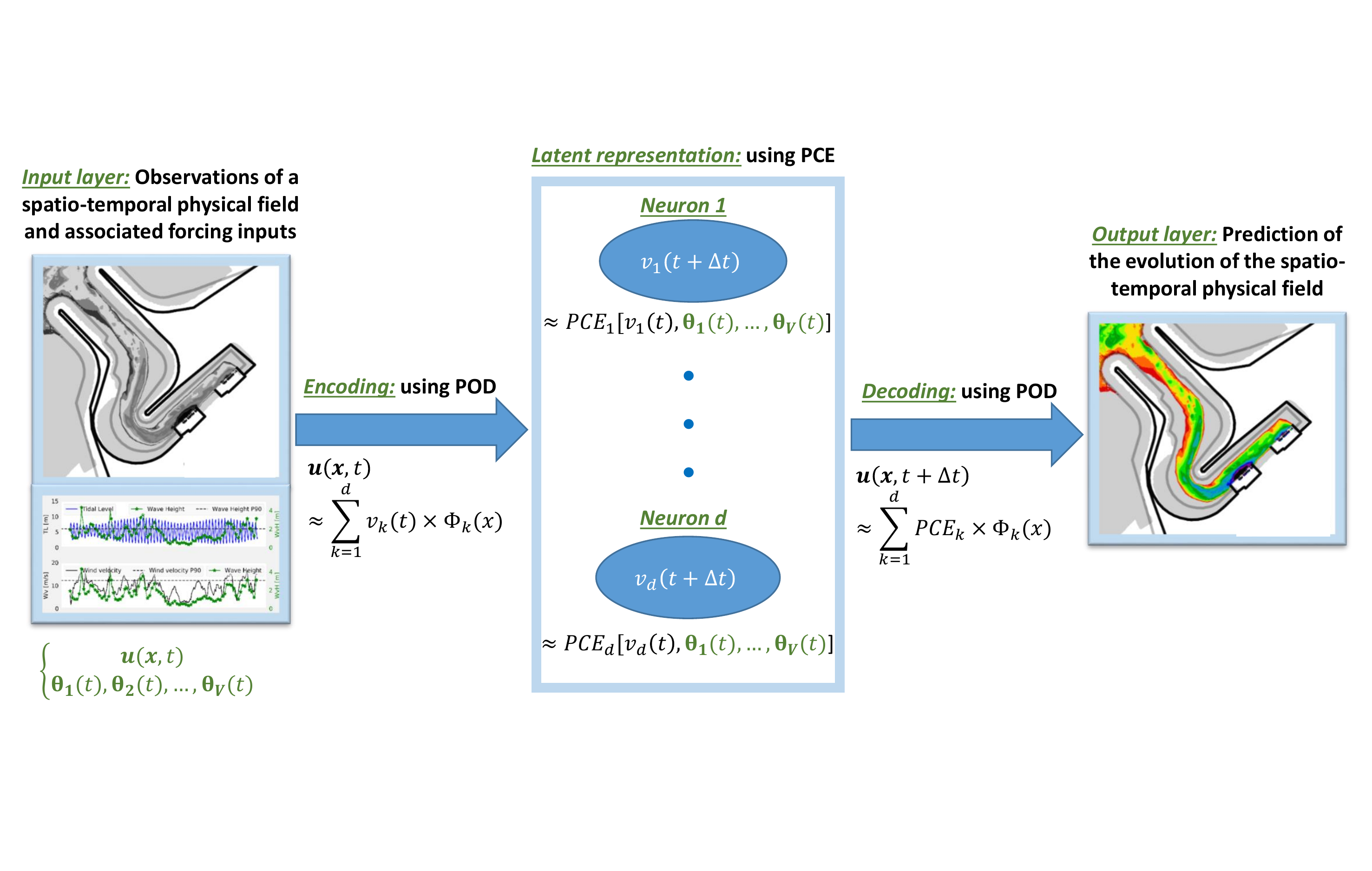}
    \caption{Representation of the POD-PCE ML approach.}
    \label{fig:POD-PCE_ML}
\end{figure}

To overcome these limitations, we propose an alternative ML method, based on a coupling between Proper Orthogonal Decomposition (POD) \citep{Cordier2008} and Polynomial Chaos Expansion (PCE) \citep{Lemaitre2001_a,Lemaitre2002_b}. This approach is proposed for the prediction of spatially-distributed physical fields that vary in time. The idea is to use POD to separate the spatial patterns from the temporal variations, that are related to the conditioning parameters using PCE. To correspond to common NN paradigms, an adequate representation of this idea is given in Figure \ref{fig:POD-PCE_ML}. In particular, POD is used for both \textit{Encoding} and \textit{Decoding} whereas PCE is used as an  \textit{Activation Function} in the \textit{Latent Representation} \cite{LeCun2015}. 

The proposed POD-PCE addresses these drawbacks of ML.\begin{itemize}
\item[$(i)$] It is explicit and simple to implement, as it consists of the association of two linear decompositions. POD is a linear separation of the spatiotemporal patterns \citep{Lumley1967}, shown to be accurate for both linear and non-linear problems \citep{Taira2017}, combining simplicity and relevance. PCE is a well-established method in Uncertainty Quantification (UQ) \citep{XiuKarniadakis2003_flow,Sudret2014}, widely used for the study of stochastic behavior in physics \citep{Tarakanov2019,Jones2013}. It is a linear polynomial expansion that allows non-linearities to be gradually added to the model by increasing the polynomial degree. The linearity and orthonormality of the POD and PCE components and the probabilistic framework of PCE make the output's statistical moments easier to study \citep{Sudret2008}, enabling straightforward physical interpretation of the model \citep{GarciaCabrejo2014}.
\item[$(ii)$] It only has two hyper-parameters: a number of POD components, and a PCE polynomial degree. Both can be chosen according to quantitative criteria \citep{Cordier2008,Blatman2011}. All other forms of parameterization (choice of the polynomial basis) can be achieved with robust physical and/or statistical arguments \citep{SoizeGhanem2004}, as assessed in the present paper. Furthermore, the orthonormality of the POD and PCE bases minimizes the number of components necessary to capture essential variations in data. Additionally, the POD modes capture more energy than any other decomposition \citep{Muller2008_phd}, PCE is known to exponentially converge with polynomial degree \citep{Lemaitre2002_b}, and the cardinality of the latter can be reduced by sparse basis selection \citep{Blatman2009}.
\item[$(iii)$] It can be considered as a universal expansion for physical field approximation: a physical field has a finite variance, which implies that it belongs to the Hilbert space of random variables with finite second order moments. There therefore exists a numerable set of orthogonal random variables, that form the basis of this Hilbert space, on which the field of interest can be expanded (strict equality, not approximation) \citep{Sudret2014}. A mathematical setting for basis construction based on input was established by \citet{SoizeGhanem2004} for the general case of dependent variables with arbitrary density, provided that the set of inputs is finite. \\
\end{itemize}

In the literature, associating regression techniques to Reduced Order Models (ROM), that include POD, is not novel \citep{Larson2003,Wang2019_ROM-NN}. The cited studies, however, focused on dimensionality reduction, whereas the explicit formulation and applicability to complex physical processes are emphasized in the present study. Secondly, coupling PCE to ROM was recently addressed \citep{Nagel2020,Lataniotis2018} and the use of PCE as ML is consistent with the work of \citet{Torre2019}, where the authors showed that PCE is as powerful as classical ML techniques. However, neither spatiotemporal fields nor physical interpretability were addressed. The data in these studies were either obtained from numerical experiments, emulated from analytical benchmark functions such as Sobol or Ishigami, or based on one-dimensional data sets \citep{Torre2019}. In contrast, the proposed POD-PCE methodology is herein assessed on two-dimensional physical fields. In particular, a toy example where synthetic data are emulated using an analytical function (groundwater perturbations due to tidal loadings \cite{Li2000}), and a real data set (high-resolution field measurements of underwater topography) are used. Although similar from a learning point of view, these two applications are characterized with differences. In particular, the toy problem is purely parametric and controllable, whereas the real data concern temporal dynamics and are of limited size. The cases are therefore complementary, in the sense that they allow demonstrating different properties of the proposed methodology. Hence, using the particularities of each case, the study consists in: i) the evaluation of the combined use of POD and PCE as ML for point-wise prediction; ii) the robustness of the methodology to noise; iii) the application to field data with the inherent challenges not encountered with numerical data (e.g. paucity); iv) a focus on model explicitness as a key condition for physical understanding and v) the influence of forcing variables study, based on a classical measure of importance (Garson weights \citep{Gevrey2003}) directly computed with the POD-PCE expansion coefficients. \\

The paper is organized as follows. Section \ref{section:theory} gives a detailed explanation of the methodology, with a proposal for physical importance measures in Subsection \ref{subsection:theory:PCE:original}. Section \ref{section:toy} deals with the assessment of the methodology on synthetic data, for both prediction and physical interpretation. In particular, the robustness of the approach to noise is evaluated in Subsection \ref{subsection:toy:noise}. The model is then deployed on field measurements in Section \ref{section:application}. The study case and data are described in \ref{subsection:application:case}. POD and PCE performances are then demonstrated independently in \ref{subsection:application:learning} with a deep physical analysis. The performance of the POD-PCE predictor is discussed in \ref{subsection:application:prediction}. A summary of the study and perspectives of the proposed methodology are presented in Section \ref{section:summary}.



\shorthandoff{:}

\section{Theoretical framework}
\label{section:theory}
In this section, the objective is to define the framework of the proposed POD-PCE Machine Learning methodology, along with physical influence indicators for the inputs. This is the object of Subsection \ref{subsection:theory:methodology}, but first, a reminder of the existing POD and PCE theoretical bases is presented in \ref{subsection:theory:POD} and \ref{subsection:theory:PCE} respectively.

\subsection{Proper Orthogonal Decomposition}
\label{subsection:theory:POD}

POD is a dimensionality reduction technique \citep{Lumley1967} that is well documented in literature \citep{Cordier2008,Taira2017}. Theoretical details and demonstrations can be found in \citep{Muller2008_phd,Couplet2005_phd}. For clarity's sake, the essential elements of POD are summarized below. \\

The goal of POD is to extract the main patterns of continuous bi-variate functions.  These patterns, when added and multiplied by appropriate coefficients, explain the dynamics of the variable of interest: a real-valued physical field.  \\

Let $\mathbf{u}:\Omega\times \mathbb{T} \rightarrow \mathbb{D}$ be a continuous function of two variables $(\mathbf{x},t) \in \Omega\times \mathbb{T}$. The following relationships and properties hold for any $\Omega\times \mathbb{T}$ and Hilbert space $\mathbb{D}$ characterized by its scalar product $(.~,.)_{\mathbb{D}}$ and induced norm $||.||_{\mathbb{D}}$. However, as is the case for a majority of physical fields, we shall consider $\Omega$ as a set of spatial coordinates (e.g. $\mathbb{R}^2$ or $\mathbb{R}^3$), $\mathbb{T}$ an event space (e.g. parameters space $\mathbb{R}^V$ with $V \in \mathbb{N}^*$, or a temporal subset $[0,T] \subseteq \mathbb{R}^+$), and $\mathbb{D}$ as a set of scalar real values or vector real values (e.g. $\mathbb{R}$ or $\mathbb{R}^2$). POD consists in an approximation of $\mathbf{u}(\mathbf{x},t)$ at a given order $d\in \mathbb{N}$ (Lumley \citep{Lumley1967}) as in Equation \ref{eq:POD:approx},
\begin{equation}
\label{eq:POD:approx}
\mathbf{u}(\mathbf{x},t) \approx \sum_{k=1}^{d} v_k(t) \boldsymbol{\phi}_k(\mathbf{x}) \ ,
\end{equation}
where $\{v_k(.)\}_{k=1}^{d} \subset \mathcal{C}(\mathbb{T},\mathbb{R})$ and $\{\boldsymbol{\phi}_k(.)\}_{k=1}^{d}  \subset \mathcal{C}(\Omega,\mathbb{D})$, with $\mathcal{C}(\mathbb{A},\mathbb{B})$ denoting the space of continuous functions defined over $\mathbb{A}$ and arriving at $\mathbb{B}$. The objective of POD is to identify $\{\boldsymbol{\phi}_k(.)\}_{k=1}^{d}$ that minimizes the distance of the approximation from the true value $\mathbf{u}(.,.)$, over the whole $\Omega\times\mathbb{T}$ domain, with an orthogonality constraint for $\{\boldsymbol{\phi}_k(.)\}_{k=1}^{d}$ using the scalar product $(.~,.)_{\mathbb{D}}$. This can be defined, in the least-squares sense, as a minimization problem. \\

The minimization problem is defined for all orders $d\in \mathbb{N}$, so that the members $\boldsymbol{\phi}_k$ are ordered according to their importance. In particular, for order 1, $\boldsymbol{\phi}_1$ is the linear generator of the sub-vector space most representative of $\mathbf{u}(\mathbf{x},t)$ in $\mathbb{D}$. For  $\mathbb{D}= Im(\mathbf{u})$, the family $\{\boldsymbol{\phi}_k(.)\}_{k=1}^{d}$ is called the POD basis of $\mathbb{D}$ of rank $d$. The solution to this problem has already been established in literature \citep{Lumley1967,Sirovich1987}. The theoretical aspects of POD and demonstrations of mathematical properties can, for example, be found in \citep{Muller2008_phd}: the POD basis of $\mathbb{D}$ of order $d$ is the orthonormal set of eigenvectors of an operator $\mathcal{R} :\mathbb{D} \rightarrow \mathbb{D}$ defined as $\mathcal{R}\boldsymbol{\phi} = \left< (\mathbf{u},\boldsymbol{\phi})_{\mathbb{D}} \times \mathbf{u} \right>_{\mathbb{T}}$, if the eigenvectors are taken in decreasing order of the corresponding eigenvalues $\{\lambda_k\}_{k=1}^d$.  \\

For this expansion, an accuracy rate, also called the Explained Variance Rate (EVR), denoted $e_d$ at rank $d$, can be calculated as in Equation \ref{eq:POD:RIC}~. EVR tends to 1 (perfect approximation) when $d \rightarrow +\infty$.
\begin{equation}
\label{eq:POD:RIC}
e_d=\dfrac{ \sum_{k \leq d} \lambda_k}{ \sum_{k=1}^{+\infty} \lambda_k} \ .
\end{equation}

In practice, for $\mathbb{D}=\mathbb{R}$, when $\mathbf{u}(.,.)$ is a discrete sample on a set of $m \in \mathbb{N}$ space coordinates $\boldsymbol{\mathcal{X}}=\{\mathbf{x}_1,\dots,\mathbf{x}_m\}$ and for $n \in \mathbb{N}$ measurement events $\mathcal{T}=\{t_1,\dots,t_n\}$ (e.g. realizations of the parameters, time coordinates, etc.), the available data set is arranged in a matrix $\mathbf{U}(\mathcal{X},\mathcal{T})=[\mathbf{u}(\mathbf{x}_i,t_j)]_{i,j} \in \mathbb{R}^{m \times n}$, called the snapshot matrix, so as to be able to work in a discrete space. The POD problem formulated in Equation \ref{eq:POD:approx} can be written in its discrete form as $\mathbf{U}(\boldsymbol{\mathcal{X}},\mathcal{T}) = \boldsymbol{\Phi}^{(d)}(\boldsymbol{\mathcal{X}}) \mathbf{V}^{(d)}(\mathcal{T})$, where $\boldsymbol{\Phi}^{(d)}(\boldsymbol{\mathcal{X}}) \coloneqq[\boldsymbol{\phi}_j(\mathbf{x}_i)]_{i,j} \in \mathbb{R}^{m \times d}$ and $ \mathbf{V}^{(d)}(\mathcal{T}) \coloneqq [v_i(t_j)]_{i,j} \in \mathbb{R}^{d \times n}$. The problem can therefore be viewed as if working with a new function $\mathbf{U}(\boldsymbol{\mathcal{X}},.)=[\mathbf{u}(\mathbf{x}_i,~.)]_{i\in\{1,\dots,m\}}: \mathcal{T} \rightarrow \mathbb{D}=\mathbb{R}^M$. Then, the average over $\mathbb{T}$ can be defined as the statistical mean over the subset $\mathcal{T}$, and the scalar product $(.~,.)_{\mathbb{D}}$ as the canonical product over $\mathbb{R}^m$. The POD operator $\mathcal{R}$ can be written as in Equation \ref{eq:POD:R},
\begin{equation}
\label{eq:POD:R}
\mathcal{R} \boldsymbol{\phi}(\boldsymbol{\mathcal{X}}) = \frac{1}{n} \sum_{j=1}^{n} \mathbf{U}(\boldsymbol{\mathcal{X}},t_j)^{T} \boldsymbol{\Phi}(\boldsymbol{\mathcal{X}}) \mathbf{U}(\boldsymbol{\mathcal{X}},t_j) = \frac{1}{n} \mathbf{U}(\boldsymbol{\mathcal{X}},\mathcal{T})\mathbf{U}(\boldsymbol{\mathcal{X}},\mathcal{T})^{T}\boldsymbol{\Phi}(\boldsymbol{\mathcal{X}}) \ ,
\end{equation}
where $\mathbf{U}(\mathcal{X},t_j)=[\mathbf{u}(\mathbf{x}_i,~.)]_{i\in\{1,\dots,m\}}$ is the column number $j$ of the matrix $\mathbf{U}(\boldsymbol{\mathcal{X}},\mathcal{T})$ (i.e realization $t_j$ of the measurement over $\boldsymbol{\mathcal{X}}$), and $\boldsymbol{\Phi}(\mathcal{X})=[\boldsymbol{\phi}(\mathbf{x}_i)]_{i\in\{1,\dots,m\}}$. As finding the POD basis is equivalent to identifying the orthonormal set of eigenvectors of the operator $\mathcal{R}$, then for this discrete representation the problem becomes equivalent to solving the eigen problem of the matrix $\mathbf{R} \coloneqq \frac{1}{n}\mathbf{U}(\boldsymbol{\mathcal{X}},\mathcal{T})\mathbf{U}(\boldsymbol{\mathcal{X}},\mathcal{T})^{T}$, called the covariance matrix. A number $d\in\mathbb{N}$ of eigen vectors $\boldsymbol{\Phi}(\boldsymbol{\mathcal{X}})$ are identified and stored in the columns of the matrix $\boldsymbol{\Phi}^{(d)}(\boldsymbol{\mathcal{X}})$. For the eigenvalues of the covariance matrix $\mathbf{R}$ denoted $\{\lambda_k\}_{k=1}^d$, the expansion in Equation \ref{eq:POD:approx} can also be written as in Equation \ref{eq:POD:KLT}, where $\{\boldsymbol{\phi}_k(.)\}_{k=1}^{d}$ together with $\{a_k(.)\}_{k=1}^{d}$ are bi-orthonormal, and $v_k(.)=a_k(.) \sqrt{n \times \lambda_k}$. 
\begin{equation}
\label{eq:POD:KLT}
\mathbf{u}(\mathbf{x},t) \approx \sum_{k=1}^{d} a_k(t) \sqrt{n \times \lambda_k} \boldsymbol{\phi}_k(\mathbf{x}) \ .
\end{equation}
By defining the matrix $\mathbf{A}^{(d)}(\mathcal{T}) \coloneqq [a_i(t_j)]_{i,j} \in \mathbb{R}^{d \times n}$ and the operator $\mathbf{D}^{(d)}(\lambda_1, ..., \lambda_d)$ corresponding to the diagonal matrix of elements $\lambda_i$, we have $\mathbf{U}(\boldsymbol{\mathcal{X}},\mathcal{T}) = \boldsymbol{\Phi}^{(d)}(\boldsymbol{\mathcal{X}}) \mathbf{D}^{(d)}(\sqrt{n \times \lambda_1}, ..., \sqrt{n \times \lambda_d})\mathbf{A}^{(d)}(\mathcal{T})$. Therefore the transposed form is $\mathbf{U}(\boldsymbol{\mathcal{X}},\mathcal{T})^T = \mathbf{A}^{(d)}(\mathcal{T})^T  \mathbf{D}^{(d)}(\sqrt{n \times \lambda_1}, ..., \sqrt{n \times \lambda_d}) \boldsymbol{\Phi}^{(d)}(\boldsymbol{\mathcal{X}})^T$. Thanks to the orthonormality of $\{a_k(.)\}_{k=1}^{d}$, the covariance matrix reads $\mathbf{R}= \frac{1}{n}\boldsymbol{\Phi}^{(d)}(\boldsymbol{\mathcal{X}}) \mathbf{D}^{(d)}(n \times \lambda_1, ..., n \times \lambda_d) \boldsymbol{\Phi}^{(d)}(\boldsymbol{\mathcal{X}})^T = \boldsymbol{\Phi}^{(d)}(\boldsymbol{\mathcal{X}}) \mathbf{D}^{(d)}(\lambda_1, ..., \lambda_d) \boldsymbol{\Phi}^{(d)}(\boldsymbol{\mathcal{X}})^T$. \\

When $n << m$, it is more computationally efficient to solve the eigenproblem of $\mathbf{R}^T$ instead of the eigenproblem of $\mathbf{R}$ as highlighted by \citet{Sirovich1987}~. This is often the case when a limited number of occurrences is measured for a two-dimensional physical field, as is the case encountered for the application described in Section \ref{section:application}. \\

When an order $d<<min(m,n)$ corresponds to a high EVR as defined in Equation \ref{eq:POD:RIC}, we speak of dimensionality reduction, because the data are projected in a sub-space that is of much smaller dimension than $\mathbb{R}^{m \times n}$. When diverse enough records are available for the variable under study, we may consider that $\{\boldsymbol{\phi}_k(\boldsymbol{\mathcal{X}})\}_{k=1}^{d}=\{[\boldsymbol{\phi}_k(\mathbf{x}_i)]_{i\in\{1,\dots,m\}}\}_{k=1}^{d}$, i.e. the resulting POD basis, is a generator of all possible states. Predicting the associated expansion coefficients $\{a_k(t)\}_{k=1}^{d}$ for a given event $t$ would therefore be enough to predict the whole state. Hence, we propose to use the POD as a basis extractor. This would first enable study of the dynamics of the variable of interest and eventually extraction of physical information, as shown in the applications Sections \ref{section:application} and  \ref{section:toy}. Then, the basis can be used as a generator for the prediction of diverse states. This implies predicting $\{a_k(t)\}_{k=1}^{d}$, for which we propose to use Polynomial Chaos Expansion (PCE), as described in the following Section \ref{subsection:theory:PCE}. 

\subsection{Polynomial Chaos Expansion}
\label{subsection:theory:PCE}
A reminder of the theoretical base of PCE is presented in Subsection \ref{subsection:theory:PCE:existing}. Theoretical details, demonstrations and interesting references can be found in \citep{Sudret2008,XiuKarniadakis2003_flow}. Then, a simple indicator is proposed in Subsection \ref{subsection:theory:PCE:original} for the analysis of the variables influence on the output value. The latter is later generalized for POD-PCE in Section \ref{subsection:theory:methodology}. 

\subsubsection{Learning}
\label{subsection:theory:PCE:existing}
The idea behind Polynomial Chaos Expansion (PCE) is to formulate an explicit model that links a variable of interest (output) to conditioning parameters (inputs), both in a probability space. This enables the propagation path of probabilistic information (uncertainties, occurrence frequencies) to be mapped from the input to the output space. The variable of interest, $\mathbf{Y}$, and the input parameters $\boldsymbol{\Theta} = (\theta_1, \theta_2, ..., \theta_V)$ are therefore considered random variables, characterized by a given Probability Density Function (PDF) denoted $f_{\boldsymbol{\Theta}}$. It should be kept in mind that the outputs of our problem are the POD expansion coefficients $\mathbf{Y}=[a_k(t)]_{k\in\{1,\dots,d\}}$, and that the inputs correspond to physical forcings, as described later in Section \ref{subsection:theory:methodology}. The objective is to derive the variations of the POD coefficients as the outcome of the forcings. Let us now recall some fundamentals of the mathematical probabilistic framework, taking the example of a one dimensional real-valued variable. The definitions can be easily extended to $\mathbb{R}^{M}$. \\

Let $(\Omega,F,\mathbb{P})$ be a probability space, where $\Omega$ is the event space (space of all the possible events $\omega$) equipped with $\sigma$-algebra $F$ (some events of $\Omega$) and its probability measure $\mathbb{P}$ (likelihood of a given event occurrence). A random variable defines an application $Y(\omega): \Omega \rightarrow D_Y \subseteq \mathbb{R}$, with realizations denoted by $y \in D_Y$. The PDF of $Y$ is a function $f_Y: D_Y \rightarrow \mathbb{R}$ that verifies $\mathbb{P}(Y \in E \subseteq D_Y) = \int_E f_Y(y) dy$. The \textit{$k^{th}$ moments} of $Y$ are defined as $\mathbb{E}[Y^k] \coloneqq \int_{D_Y} y^kf_Y(y)dy$, the first being the expectation denoted $\mathbb{E}[Y]$. In the same manner, we define the \textit{$k^{th}$ central moments} of $Y$ as $\mathbb{E}[(Y-\mathbb{E}[Y])^k]$, the first being $0$ and the second the variance of $Y$ denoted by $\mathbb{V}[Y]$.  The covariance of two random variables is defined as $cov(X,Y)=\mathbb{E}[(X-\mathbb{E}[X])(Y-\mathbb{E}[Y])]$ and a resulting property is $\mathbb{V}[Y] = cov(Y,Y)$. \\

Returning to the PCE construction, inputs $\boldsymbol{\Theta} = (\theta_1, \theta_2, ..., \theta_V)$ are considered to live in the space of real random variables with finite second moments (and finite variances). This space is denoted by $\mathcal{L}^2_{\mathbb{R}}(\Omega,F,\mathbb{P};\mathbb{R})$ and is a Hilbert space equipped with the inner product $(\theta_1,\theta_2)_{\mathcal{L}^2_{\mathbb{R}}} \coloneqq \mathbb{E}[\theta_1\theta_2] = \int_{\Omega}\theta_1(\omega)\theta_2(\omega)d\mathbb{P}(\omega)$ and its induced norm $||\theta_1||_{\mathcal{L}^2_{\mathbb{R}}} \coloneqq \sqrt{\mathbb{E}[\theta_1^2]}$. The PCE objective is to map the output space from the input space with a model $\mathcal{M}$ as in Equation \ref{eq:PCE:metamodel}:
\begin{equation}
\label{eq:PCE:metamodel}
\begin{matrix}
Y & = & \mathcal{M}(\mathbf{\boldsymbol{\Theta}}) =  \sum_{\mathcal{I} \subseteq \{1, ..., V \}} \mathcal{M}_\mathcal{I}(\theta_\mathcal{I}) \\
& = & \mathcal{M}_0 + \sum_{i=1}^{V} \mathcal{M}_i(\theta_i) + \sum_{1 \leq i < j \leq V}  \mathcal{M}_{i,j}(\theta_i,\theta_j) + ... +  \mathcal{M}_{1,...,V}(\theta_1, \theta_2, ..., \theta_V)  \ ,
\end{matrix}
\end{equation}
where $\mathcal{M}_0$ is the expectation of $Y$ and $\mathcal{M}_{\mathcal{I} \subseteq \{1,...,V\}}$ represents the common contribution of the variables $\mathcal{I} \subseteq \{1,...,V\}$ to the variation in $Y$. For the PCE model, these contributions have a polynomial form. We shall define, for each input variable $\theta_i$, an orthonormal univariate polynomial basis $\left\{ \xi_{ \beta}^{(i)}(.), \beta \in [|0,p|] \right\}$ where $p \in \mathbb{N}$ is a chosen polynomial degree and $\xi^{(i)}_{ \beta}(.)$ is of degree $\beta$. The orthonormality is defined with respect to the inner product $(.,~.)_{\mathcal{L}^2_{\mathbb{R}}}$. If we introduce the multi-index notation $\boldsymbol{\alpha} = (\alpha_1, ..., \alpha_V) \in \mathbb{N}^V$ so that $|\boldsymbol{\alpha}| =\sum_{i=1}^{V}\alpha_i $, we can define a multivariate basis $\left\{\zeta_{\boldsymbol{\alpha}}^{\boldsymbol{\Theta}}(.), |\boldsymbol{\alpha}| \in [|0,p|]\right\}$ as $\zeta_{\boldsymbol{\alpha}}^{\boldsymbol{\Theta}}(\theta_1, \theta_2, ..., \theta_V) \coloneqq \prod_{i=1}^{V}\xi_{\alpha_i}^{(i)}(\theta_i)$. Therefore, the model in Equation \ref{eq:PCE:metamodel} can be written as:
\begin{equation}
  \label{eq:PCE:polynomial}
Y = \mathcal{M}(\boldsymbol{\Theta}) = \sum_{|\boldsymbol{\alpha}| \leq P} c_{\boldsymbol{\alpha}} \zeta_{\boldsymbol{\alpha}}^{\boldsymbol{\Theta}}(\theta_1, \theta_2, ..., \theta_V) \ ,
\end{equation}
where $c_{\boldsymbol{\alpha}} \in \mathbb{R}$ are deterministic coefficients that can be estimated thanks to different methods. It can be formulated as a minimization problem, and regularization methods can be used when dealing with small data sets. In the present study, we used the Least Angle Regression Stagewise method (LARS) in order to construct an adaptive sparse PCE. It is an iterative procedure, consisting on an improved version of forward selection. The algorithm begins by finding the polynomial pattern, here denoted $\zeta_i$ for simplicity, that is the most correlated to the output. The latter is linearly approximated by $\epsilon_i\zeta_i$, where $\epsilon_i \in \mathbb{R}$. Coefficient $\epsilon_i$ is not set to its maximal value, but increased starting from 0, until another pattern $\zeta_j$ is found to be as correlated to $Y - \epsilon_i\zeta_i$, and so on. In this approach, a collection of possible PCE, ordered by sparsity, is provided and an optimum can be chosen with an accuracy estimate. It was performed in this study using corrected leave-one-out error. The reader can refer to the work of \citet{Blatman2011} for further details on LARS and more generally on sparse constructions.  \\

The choice of the basis is crucial and is directly related to the choice of input variable marginals, via the inner product  $(.,.)_{\mathcal{L}^2_{\mathbb{R}}}$. Chaos polynomials were first introduced in \citep{Wiener1938} for input variables characterized by Gaussian distributions. The orthonormal basis with respect to this marginal is the Hermite polynomials family. Later, other Askey scheme hypergeometric polynomial families were associated to some well-known parametric distributions \citep{XiuKarniadakis2002}. For example, the Legendre family is orthonormal with respect to the Uniform marginals. This is called $gPC$ (generalized Polynomial Chaos) when variables of different PDFs are used as inputs. In practice however, the input distributions of physical variables can be different from usual parametric marginals. In such cases, the marginals can be inferred by empirical methods such as the Kernel Smoother (see \citep{Hastie2009} for theoretical elements). In this case, an orthonormal polynomial basis with respect to arbitrary marginals can be built with a Gram-Schmidt orthonormalization process as in \citep{Witteveen2006} or via the Stieltjes three-term recurrence procedure as in \citep{WanKarniadakis2006}.   \\

To highlight the importance of the marginals and choice of polynomial basis for the learning process, several configurations are attempted in Section \ref{section:application}. Different input sets and distributions (Gaussian, Uniform, inferred by Kernel Smoothing) were tested. The influence of the polynomial basis on the ML is investigated in Section \ref{subsubsection:application:learning:PCE}. 

\subsubsection{Physical importance measures}
\label{subsection:theory:PCE:original}
\begin{figure}[H]
  \centering
    \includegraphics[trim={1cm 10.8cm 12cm 1.cm},clip,scale=0.52]{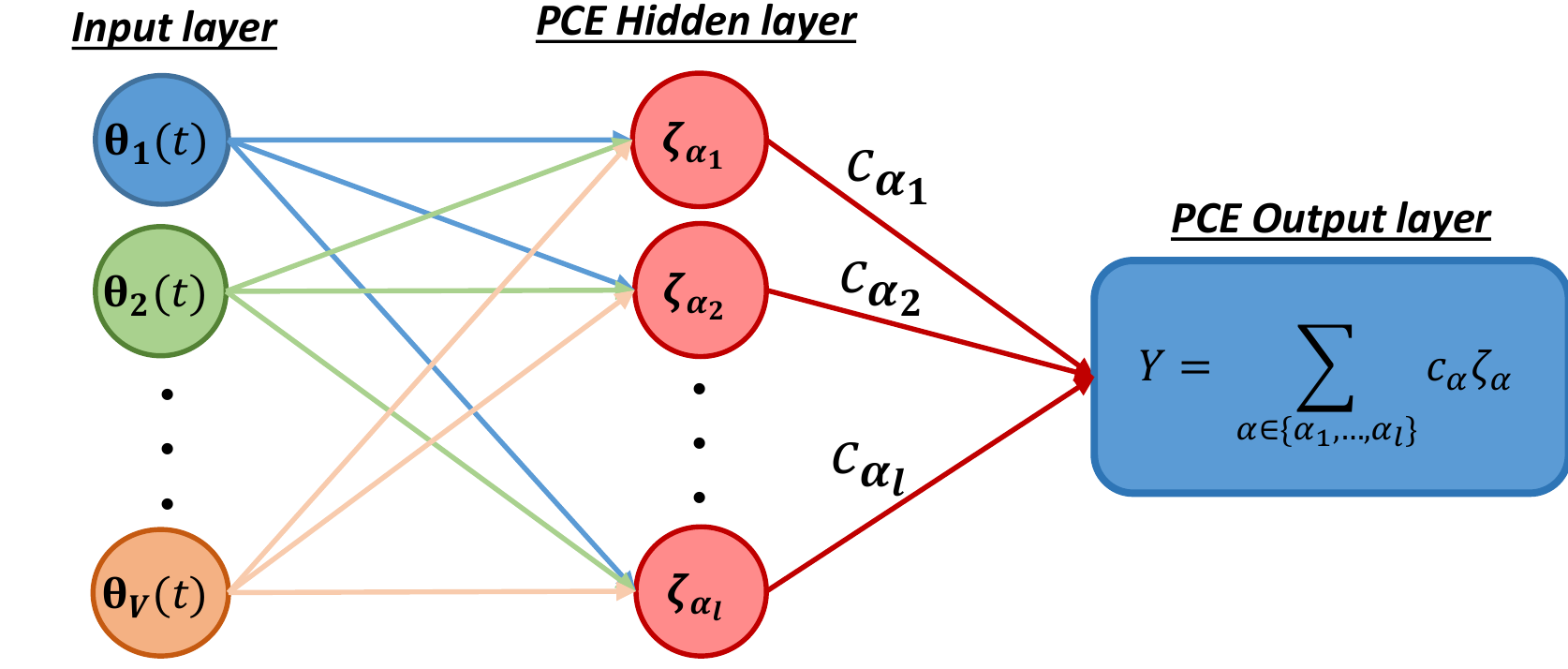}
    \caption{Representation of the PCE learning in the NN paradigm.}
    \label{fig:PCE_NN}
\end{figure}

Once the PCE construction is achieved, a physical interpretation can be performed. It is notable that classical NN indicators can be used \citep{Gevrey2003}. The PCE can be represented in the Feedforward NN paradigm as in Figure \ref{fig:PCE_NN}.  Such networks are classically composed, in addition to the input and output layers, of successive \textit{hidden layers}. Each hidden layer is composed of \textit{neurons} that transform the variables of the previous layer (outputs of the previous \textit{neurons}) into a new set of variables. This is done by combining a linear transformation, giving different \textit{weights} to the previous \textit{neurons}, and a transformation function, called \textit{Activation Function} (AF). This succession of layers is called the \textit{latent representation}. For a number of hidden layers $L \geq 1$, the NN can be formally written as $\mathbf{Y} \approx f_{out}(\mathbf{A}_L~f_L(\dots \mathbf{A}_2~f_2(\mathbf{A}_1~f_{1}(\mathbf{A}_{in}~\boldsymbol{\Theta}))))$ where $\{\mathbf{A}_k\}_{k\in[|1,L|]}$ and $\{f_k\}_{k\in[|1,L|]}$ are the hidden layer weight matrices and AFs, $\mathbf{A}_{in}$ is the input-to-hidden connection matrix and $f_{out}$ is the final hidden-output transformation \citep{ShalevShwartz2014}. \\

The PCE-based NN represented in Figure \ref{fig:PCE_NN} is a single layer feedforward, composed of $l \in \mathbb{N}$ neurons, that can be written as $\mathbf{Y} \approx f_{out}(\mathbf{A}_1~f_1(\mathbf{A}_{in}\boldsymbol{\Theta}))$. The first matrix $\mathbf{A}_{in}$ is the input-to-hidden connection matrix of dimension $V \times V$, that links the input layer to the PCE hidden layer containing the multivariate polynomials $\left\{\zeta_{\boldsymbol{\alpha}}^{\boldsymbol{\Theta}}, \boldsymbol{\alpha} \in \{\boldsymbol{\alpha}_1, ...,\boldsymbol{\alpha}_l\}\right\}$, where $V$ is the number of inputs and the multivariate indexes $\{\boldsymbol{\alpha}_1, ..., \boldsymbol{\alpha}_l\}$ are conditioned by the chosen polynomial degree $p$ such as $\forall i \in [|1,l|] ~ 0 \leq |\boldsymbol{\alpha}_i| \leq p$, and by the number of selected features if a sparse polynomial is constructed, as in the present case using LARS \citep{Blatman2009}. Matrix $\mathbf{A}_{in}$ represents the contributions of the $V$ variables to the multivariate polynomials $\left\{\zeta_{\boldsymbol{\alpha}}^{\boldsymbol{\Theta}}, \boldsymbol{\alpha} \in \{\boldsymbol{\alpha}_1, ...,\boldsymbol{\alpha}_l\}\right\}$. It is a diagonal matrix such that $[\boldsymbol{A}_{in}]_{j, j \in [|1,V|]^2}$ is 0 if $\forall i \in [|1,l|] ~ (\alpha_i)_j=0$ and 1 if not. The first multi-dimensional AF $f_1$ is a vector of multivariate functions that transforms the set of selected inputs corresponding to $[\boldsymbol{A}_{in}]_{i,i \in [|1,V|]^2}=1$ to the multivariate polynomials of the chosen basis (Hermite, Legendre, etc.) by tensor product over the univariate basis. The hidden layer weight matrix $\mathbf{A}_1$ gives different weights to the constructed polynomial features. It is a diagonal matrix composed of the PCE expansion coefficients such as $[\mathbf{A}_1]_{i,j \in [|1,l|]^2}=[c_{\boldsymbol{\alpha}_i}]_{i \in [|1,l|]}$. \\

The final hidden-output transformation $f_{out}$ is a summation. Figure \ref{fig:PCE_NN} can also be presented differently: another hidden layer can be added to the PCE latent representation as $\mathbf{Y} \approx f_{out}(\mathbf{A}_2~f_2(\mathbf{A}_1~f_1(\mathbf{A}_{in}~\boldsymbol{\Theta}))$. The first AF $f_1$ would represent a transformation of each input variable to a list of monomials of degrees 1 to $p$ (here, $\mathbf{A}_{in}$ is identity). The second AF $f_2$ therefore represents the tensor product that transforms the different monomials to multivariate features, with $\mathbf{A}_{1}$ appropriately filled with zeros and ones, and   $[\mathbf{A}_2]_{i,j \in [|1,l|]^2}=[c_{\boldsymbol{\alpha}_i}]_{i \in [|1,l|]}$.

To capture the importance of each feature, the Garson relative Weights (GW) defined in Equation \ref{eq:PCE:weights} are a classical measure to quantify the relative importance of each neuron of the last hidden layer, and therefore of each polynomial pattern, for the output value \citep{Gevrey2003,Tsang2017}.
\begin{equation}
    \label{eq:PCE:weights}
    w_{\zeta_{\boldsymbol{\alpha}}^{\boldsymbol{\Theta}}} = \dfrac{|c_{\mathcal{\boldsymbol{\alpha}}}|}{\sum_{0 \leq \boldsymbol{\beta} \leq 1} |c_{\mathcal{\boldsymbol{\beta}}}|}  \ .
\end{equation}
This measure can be used to understand the importance given by the NN algorithm to the variables and their possible interactions, especially when using feature selection algorithms as LARS: "feature interactions [...]  are created at hidden units with nonlinear activation functions,  and the influences of the interactions  are  propagated  layer-by-layer  to  the  final  output"  \citep{Tsang2017}. In the particular case of a polynomial expansion, the interpretation is straightforward, the importance of each variable alone corresponds to its monomials, and the importance of its interactions with other variables corresponds to the multivariate polynomials in which it is involved. \\

For the particular case of the orthonormal basis provided by PCE, the GW defined in \ref{eq:PCE:weights} can be interpreted in terms of Pearson's correlations between output $Y$ and the polynomial basis elements $\zeta_{\boldsymbol{\alpha}}^{\boldsymbol{\Theta}}$ denoted $\rho_{Y,~\zeta_{\boldsymbol{\alpha}}^{\boldsymbol{\Theta}}}$, with $\alpha \neq (0, ..., 0)$. Indeed, Pearson's correlations $\rho_{Y,~\zeta_{\boldsymbol{\alpha}}^{\boldsymbol{\Theta}}}$ are defined as in Equation \ref{eq:PCE:pearson},
\begin{equation}
\label{eq:PCE:pearson}
\rho_{Y,~\zeta_{\boldsymbol{\alpha}}^{\boldsymbol{\Theta}}} = \dfrac{\mathbb{E}\left[(Y-\mathbb{E}(Y))(\zeta_{\boldsymbol{\alpha}}^{\boldsymbol{\Theta}}-\mathbb{E}(\zeta_{\boldsymbol{\alpha}}^{\boldsymbol{\Theta}}))\right]}{\sqrt{\mathbb{V}(Y)\mathbb{V}(\zeta_{\boldsymbol{\alpha}}^{\boldsymbol{\Theta}})}} =  \dfrac{c_{\boldsymbol{\alpha}}}{\sqrt{\sum_{1 \leq |\boldsymbol{\beta}| \leq p} c_{\boldsymbol{\beta}} ^2}} \ ,
\end{equation}

thanks to the orthonormality of the basis with respect to the scalar product $(.~,.)_{\mathcal{L}^2_{\mathbb{R}}}$ that guarantees: \begin{itemize}
\item[$\bullet$] $\mathbb{E} \left[\zeta_{\boldsymbol{\alpha}}^{\boldsymbol{\Theta}}\right] = \left(\zeta_{\boldsymbol{\alpha}}^{\boldsymbol{\Theta}},\zeta_{\boldsymbol{\beta=(0,...,0)}}^{\boldsymbol{\Theta}}=1\right)_{\mathcal{L}^2_{\mathbb{R}}} = 0$;

\item[$\bullet$] $\mathbb{E}\left[Y\right] = \left(\sum_{0 \leq |\boldsymbol{\beta}| \leq p} c_{\boldsymbol{\beta}} \zeta_{\boldsymbol{\beta}}^{\boldsymbol{\Theta}} ~,\zeta_{\boldsymbol{\beta=(0,...,0)}}^{\boldsymbol{\Theta}}\right)_{\mathcal{L}^2_{\mathbb{R}}} = c_{\beta=(0,...,0)}$;
\item[$\bullet$] $\mathbb{E} \left[Y, \zeta_{\boldsymbol{\alpha}}^{\boldsymbol{\Theta}}\right] = \left(\sum_{0 \leq |\boldsymbol{\beta}| \leq p} c_{\boldsymbol{\beta}} \zeta_{\boldsymbol{\beta}}^{\boldsymbol{\Theta}} ~, \zeta_{\boldsymbol{\alpha}}^{\boldsymbol{\Theta}} \right)_{\mathcal{L}^2_{\mathbb{R}}}=c_{\boldsymbol{\alpha}}$ ;

\item[$\bullet$] $\mathbb{V} \left[\zeta_{\boldsymbol{\alpha}}^{\boldsymbol{\Theta}}\right] = \mathbb{E} \left[\left(\zeta_{\boldsymbol{\alpha}}^{\boldsymbol{\Theta}} - \mathbb{E} \left[ \zeta_{\boldsymbol{\alpha}}^{\boldsymbol{\Theta}} \right]\right)^2\right] = \mathbb{E} \left[\left(\zeta_{\boldsymbol{\alpha}}^{\boldsymbol{\Theta}}\right)^2\right] = ||\zeta_{\boldsymbol{\alpha}}^{\boldsymbol{\Theta}} ||_{\mathcal{L}^2_{\mathbb{R}}}^2 = 1$  ;

\item[$\bullet$] $\mathbb{V} \left[Y\right] = \mathbb{E} \left[\left(Y - \mathbb{E} \left[ Y \right]\right)^2\right] = \left(\sum_{1 \leq |\boldsymbol{\beta}| \leq p} c_{\boldsymbol{\beta}} \zeta_{\boldsymbol{\beta}}^{\boldsymbol{\Theta}} ~, \sum_{1 \leq |\boldsymbol{\beta}| \leq p} c_{\boldsymbol{\beta}} \zeta_{\boldsymbol{\beta}}^{\boldsymbol{\Theta}} \right)_{\mathcal{L}^2_{\mathbb{R}}} =  \sum_{1 \leq |\boldsymbol{\beta}| \leq p} c_{\boldsymbol{\beta}} ^2 $ . \\

\end{itemize}

Therefore, the weights $w_{\zeta_{\boldsymbol{\alpha}}^{\boldsymbol{\Theta}}}$ can also be computed as $|\rho_{Y,~\zeta_{\boldsymbol{\alpha}}^{\boldsymbol{\Theta}}}|/\sum_{1 \leq |\boldsymbol{\beta}| \leq p} |\rho_{Y,~\zeta_{\boldsymbol{\beta}}^{\boldsymbol{\Theta}}}|$. This means that they measure the relative importance of the basis element in the expansion of the output, in terms of linear correlation, regardless of the sign of the latter. These "relative Pearson's correlations" can be seen as a physical contribution since the PCE model is strictly linear. The sum of the GW $w_{\zeta_{\boldsymbol{\alpha}}^{\boldsymbol{\Theta}}}$ for all the polynomial features equals 1. This means that they allow $\{\zeta_{\boldsymbol{\alpha}}\}_{|\boldsymbol{\alpha}| \leq p}$ to be ranked in terms of relative contribution to the output $Y$. The contributions can be analyzed either for each polynomial pattern separately, or for a single variable $\theta_i$ by adding all the polynomial shares related to this variable alone, or by adding all the polynomial shares related to this variable and its interactions (Sobol' indices analogy \citep{Sudret2008}). 

\subsection{POD-PCE based predictor}
\label{subsection:theory:methodology}

POD and PCE were introduced separately in Subsections \ref{subsection:theory:POD} and  \ref{subsection:theory:PCE} respectively. We are now fully equipped with the adequate theoretical basis and mathematical notations, to present the POD-PCE ML methodology for a data-based model learning of a multidimensional physical field. In this Subsection, we will first summarize the proposed approach, then the formal details of the coupling will be given with the definition of adequate accuracy measures. Finally the previously discussed importance measures will be generalized for the POD-PCE physical study. \\

The proposed POD-PCE ML consists of steps, in a learning and a prediction phase, summed up as follows:
\begin{itemize}
  \item[$\bullet$] Learning phase: \begin{itemize}
\item[$\ast$] POD basis construction: given a set of measurements $\mathbf{U}(\mathcal{X},\mathcal{T})=[\mathbf{u}(\mathbf{x}_i,t_j)]_{i,j} \in \mathbb{R}^{m \times n}$ (snapshot matrix), construct a spatial POD basis accordingly. Variable $t_j$ can represent time in case of temporal dynamics, or more generally an occurrence of $\mathbf{U}(\mathcal{X},.)$. Then, in general, $\mathcal{T}$ would be an event space;
\item[$\ast$] PCE learning: construct PCE models that map each POD coefficient, obtained in the previous step along with the spatial basis, to a set of inputs. In the particular case of temporal dynamics, previous values of the physical field, represented by previous POD coefficients, can be part of the learning inputs. For example, one could use an initial field value $\mathbf{U}(\boldsymbol{\mathcal{X}},t_j)$ to learn a future field $\mathbf{U}(\boldsymbol{\mathcal{X}},t_{j+1})$ from a set of physical parameters that condition the evolution over $[t_j,t_{j+1}]$. The latter can consist in time series of physical variables, representative statistics of the latter, physical constants, etc. and can be denoted $\boldsymbol{\Theta}(t_j \rightarrow t_{j+1})$;
  \end{itemize}
  
  \item[$\bullet$] Prediction phase: \begin{itemize}
  \item[$\ast$] Given a new realization of the inputs, predict the new POD coefficients using the learned PCE models, then reconstruct the new estimate $\mathbf{U}(\boldsymbol{\mathcal{X}}, t_k)$ on the POD basis. As previously explained, an initial value to the physical field, denoted  $\mathbf{U}(\boldsymbol{\mathcal{X}}, t_{k-1})$, may be part of the inputs for temporal dynamics. In particular, its reduced form, consisting in temporal POD coefficients, is used. In this case, an additional step is needed: $\mathbf{U}(\boldsymbol{\mathcal{X}}, t_{k-1})$ is projected on the constructed POD basis in order to retrieve the values of associated POD coefficients which are then used as PCE inputs. \\
  \end{itemize}
\end{itemize}

The learning and prediction set-ups are more complex to establish for temporal evolution problems, because the field information at previous times are required. Therefore, for the sake of clarity, the steps are explicitly developed in the following Subsection \ref{subsubsection:theory:methodology:ML}. The accuracy of the methodology is later demonstrated on both a parametric toy problem in Section \ref{section:toy}, and a field measurements-based temporal problem in Section \ref{section:application}. These two can be considered as complementary applications, and demonstrate that the POD-PCE ML can be applied in different learning set-ups of multi-dimensional physical fields. Similarities in the treatment of both problems can be noticed, but their particularities are also used to demonstrate different properties of the POD-PCE learning, that are shortly described at the beginning of each section. 

\subsubsection{Machine learning methodology}
\label{subsubsection:theory:methodology:ML}
Here, the formal hypothesis behind the POD-PCE ML reasoning and its mathematical formulation are discussed. Let $\mathbf{U}(\boldsymbol{\mathcal{X}},.)=[\mathbf{u}(\mathbf{x}_i,~.)]_{i\in\{1,\dots,m\}}$ be a field of interest defined on a set of $m \in \mathbb{N}$ space coordinates $\boldsymbol{\mathcal{X}}=\{\mathbf{x}_1,\dots,\mathbf{x}_m\}$. Let $\boldsymbol{\Theta(.)} = (\theta_1(.), \theta_2(.), ..., \theta_V(.))$ be a vector of the inputs supposed to condition the evolution of $\mathbf{U}(\boldsymbol{\mathcal{X}},.)$ over time. The dynamic model, denoted $\mathcal{H}$, that gives an estimation of a future state $\mathbf{U}(\boldsymbol{\mathcal{X}},t_{j+1})$ from a past state $\mathbf{U}(\boldsymbol{\mathcal{X}},t_{j})$ and an estimation of $\boldsymbol{\Theta}(t_{j} \rightarrow t_{j+1})$ over the time interval $[t_{j},t_{j+1}]$, where $t_{j}<t_{j+1} \in \mathbb{R}^+$, is formulated as in Equation \ref{eq:prediction:model}~.
\begin{equation}
  \mathbf{U}(\boldsymbol{\mathcal{X}},t_{j+1}) \approx \mathcal{H} \left[ \mathbf{U}(\boldsymbol{\mathcal{X}},t_{j}),t_{j+1}-t_{j},\boldsymbol{\Theta}(t_{j} \rightarrow t_{j+1})\right] \ .
  \label{eq:prediction:model}
\end{equation}

If the field of interest has been recorded over a set of past times $\mathcal{T}=\{t_1,\dots,t_n\} \subset \mathbb{R}^+$, where $t_j<t_{j+1}$, a POD basis can be constructed as in Section \ref{subsection:theory:POD}, consisting of  $d\in\mathbb{N}$ vectors of dimension $m$ stored in a matrix as $\boldsymbol{\Phi}^{(d)}(\boldsymbol{\mathcal{X}})=(\boldsymbol{\Phi}^{(d)}_1(\boldsymbol{\mathcal{X}}),\dots,\boldsymbol{\Phi}^{(d)}_d(\boldsymbol{\mathcal{X}})) \in \mathbb{R}^{m\times n}$, and can be seen as a generator of all possible states if enough records are available. If so, any future state $\mathbf{U}(\boldsymbol{\mathcal{X}},t_{j})$ can be expanded on this POD basis and the associated temporal coefficients are simply the weights of $\mathbf{U}(\boldsymbol{\mathcal{X}},t_{j})$ on the POD basis. They are therefore obtained using the canonical scalar product over $\mathbb{R}^m$, as in Equation \ref{eq:theory:methodology:PODprojection}.

\begin{equation}
  \label{eq:theory:methodology:PODprojection}  
  \begin{matrix} \mathbf{U}(\boldsymbol{\mathcal{X}},t_{j})
    & \approx & \sum_{k=1}^{d} a_k(t_j) \sqrt{n \times \lambda_k} \boldsymbol{\Phi}_k^{(d)}(\boldsymbol{\mathcal{X}}) \\
    & \approx & \sum_{k=1}^{d} (\mathbf{U}(\boldsymbol{\mathcal{X}},t_{j}), \boldsymbol{\Phi}_k^{(d)}(\boldsymbol{\mathcal{X}}))_{\mathbb{R}^m}\boldsymbol{\Phi}_k^{(d)}(\boldsymbol{\mathcal{X}})  \\
    & \approx & \sum_{k=1}^{d} \mathbf{U}(\boldsymbol{\mathcal{X}},t_{j})^T \boldsymbol{\Phi}_k^{(d)}(\boldsymbol{\mathcal{X}})\boldsymbol{\Phi}_k^{(d)}(\boldsymbol{\mathcal{X}}) \ .
    \end{matrix}
\end{equation}
  
Hence, the variable part of $\mathbf{U}(\boldsymbol{\mathcal{X}},t_{j})$ is fully expressed in the temporal coefficients $a_k(t_j)$. The field of interest $\mathbf{U}(\boldsymbol{\mathcal{X}},t_{j})$ can be either a field measurement, a laboratory or a numerical experiment. In any-case, it can be considered as being generated by a random process "in the sense that nature happens without consideration of what could be the best realizations for the learning algorithm" \citep{ShalevShwartz2014}. Therefore, the coefficients $a_k(t_j)$ can also be seen as the $j^{th}$ realization of a random variable $A_k$. We can therefore construct a PCE approximation $\mathcal{H}_k$ that maps  $A_k$ from its input space. The latter is taken as a collection of random variables, composed from the set $(A_1,...,A_d)$ at a previous time, the duration of the dynamic, and the input variables $\boldsymbol{\Theta}(t_{j} \rightarrow t_{j+1})$.  This is formulated as a classical dynamic model in  Equation \ref{eq:prediction:minimodel}~.
\begin{equation}
  a_k(t_{j+1}) \approx \mathcal{H}_k \left[a_1(t_{j}),\dots,a_d(t_{j}),t_{j+1}-t_j,\boldsymbol{\Theta}(t_{j} \rightarrow t_{j+1})\right] \ .
  \label{eq:prediction:minimodel}
\end{equation}

The model $\mathcal{H}$ in  Equation \ref{eq:prediction:model} is approximated as in Equation \ref{eq:prediction:fullmodel}~.
\begin{equation}
  \mathcal{H} \left[ \mathbf{U}(\boldsymbol{\mathcal{X}},t_{j}),t_{j+1}-t_{j},\boldsymbol{\Theta}(t_{j} \rightarrow t_{j+1})\right]  \approx \sum_{k=1}^{d} \mathcal{H}_k \left[a_1(t_{j}),\dots,a_d(t_{j}),t_{j+1}-t_j,\boldsymbol{\Theta}(t_{j} \rightarrow t_{j+1})\right] \sqrt{n \times \lambda_k} \boldsymbol{\Phi}_k^{(d)}(\boldsymbol{\mathcal{X}}) \ .
  \label{eq:prediction:fullmodel}
  \end{equation}
  
Some limitations to the introduced formulations in Equations \ref{eq:prediction:minimodel} and \ref{eq:prediction:fullmodel} can be highlighted. A first limitation concerns discontinuities that can be met in physical fields. This can occur either in the complete spatial field $\mathbf{U}(.,.)$, in its reduced version represented by the POD coefficients $a_k(.)$, or in the inputs $\boldsymbol{\Theta}$. In the first case, the classical linear approximations as POD may be inefficient \cite{Taddei2020}. One solution developed by \citet{Taddei2020}, called RePOD (Registration POD), consists in a parametric transformation of the interest discontinuous field into a smoother one for linear transformations. In the second case, where discontinuity happens in the POD temporal coefficients, this would impact the learning with PCE. Innovative solutions were identified to apply PCE when the output's space is characterized with rapid variations or discontinuities, for instance near a critical point in the inputs space. As an example, a method called adaptive Multi-Element PCE was developped for Legendre polynomials in \cite{WanKarniadakis2005} and extended to arbitrary measures in \cite{WanKarniadakis2006}. The inputs space is decomposed to a union of subsets, and the output variable is locally expanded on each subset. The final solution is then a combination of PCE sub-problems. In the last discontinuity case that concerns the inputs $\boldsymbol{\Theta}$, the previous splitting techniques can also be used. For example, the sub-intervals in the inputs space can be constructed in such way to avoid the discontinuity. PCE sub-problems would therefore be treated as usual. \\

A second limitation concerns the choice of input variables for regression models, and is an ongoing research question in statistics \citep{Guyon2003}. As a practical illustration, the dynamical problem written in Equation \ref{eq:prediction:minimodel} can incorporate additional inputs, for example the information at previous times $t_{j-1}$, $t_{j-2}$, etc. However, when a large set of inputs can be used and only a small set of realizations is available for learning, a well-posedness problem occurs. One solution consists in transforming the large set of inputs to a reduced version, for example with the help of PCA \citep{Jolliffe2011} for DR. This approach was not studied here and will be the topic of a future study. However, different input configurations will be evaluated, to investigate the influence of variable selection on the proposed learning. For example, the hypothesis of dependence between the random variables $(A_1,\hdots,A_d)$ could be relaxed. This would imply writing the approximation in Equation \ref{eq:prediction:minimodel} in a relaxed form as $\mathcal{H}_k\left[a_k(t_{j}),t_{j+1}-t_j,\boldsymbol{\Theta}(t_{j} \rightarrow t_{j+1})\right] $. In that case a simpler model $\mathcal{H}$, under the strong independence assumption, can be formulated as in Equation \ref{eq:prediction:fullmodelRelaxed}.
\begin{equation}
  \mathcal{H} \left[ \mathbf{U}(\boldsymbol{\mathcal{X}},t_{j}),t_{j+1}-t_{j},\boldsymbol{\Theta}(t_{j} \rightarrow t_{j+1})\right]  \approx \sum_{k=1}^{d} \mathcal{H}_k \left[a_k(t_{j}),t_{j+1}-t_j, \boldsymbol{\Theta}(t_{j} \rightarrow t_{j+1})\right] \sqrt{n \times \lambda_k}\boldsymbol{\Phi}_k^{(d)}(\boldsymbol{\mathcal{X}}) \ .
  \label{eq:prediction:fullmodelRelaxed}
  \end{equation}

Both alternatives are tested in Section \ref{section:application}. To investigate the influence of input selection on learning accuracy, a quantitative evaluation of the hypothesis is needed. More generally, whether for the above-mentioned simplifications or for the approximated form of the model in general, accuracy estimators are needed. These are presented below.  

\subsubsection{Accuracy tests for the approximation}
There are two determining parts in the POD-PCE learning process. Firstly, the PCE learning $\mathcal{H}_k(.)$ of each mode $A_k$ should be as accurate as possible. Secondly, the reconstructed field $\sum_{k=1}^{d} \mathcal{H}_k(.) \sqrt{n \times \lambda_k}\boldsymbol{\Phi}_k^{(d)}(\boldsymbol{\mathcal{X}})$ for a given rank $d$ should be as close to the real field $\mathbf{U}(\boldsymbol{\mathcal{X}})$ as possible. \\

The distance between each mode and its PCE approximate can be evaluated using the \textit{generalization error}, denoted $\delta(A_k,\mathcal{H}_k)$ and defined as in Equation \ref{eq:theory:methodology:generalizationError}.
\begin{equation}
  \label{eq:theory:methodology:generalizationError}
  \delta(A_k,\mathcal{H}_k) = \mathbb{E} \left[ (A_k - \mathcal{H}_k(.))^2 \right] \ .
  \end{equation}

For the model defined in Equation \ref{eq:prediction:fullmodelRelaxed}~, this error can be estimated, on a set of paired realizations $(a_k(t_1),\dots,a_j(t_n))$ and $(\boldsymbol{\Theta}(t_0 \rightarrow t_1),\dots,\boldsymbol{\Theta}(t_{n-1} \rightarrow t_n))$ ,  as in Equation \ref{eq:empiricalError} as explained by \citet{Blatman2009}~. This approximated version of the \textit{generalization error} is called the \textit{empirical error}. 
\begin{equation}
  \label{eq:empiricalError}
  \delta(A_k,\mathcal{H}_k) \approx \delta_{emp}(A_k,\mathcal{H}_k) \coloneqq \dfrac{1}{n} \sum_{j=1}^n \left( a_k(t_j) - \mathcal{H}_k\left[a_k(t_{j-1}),t_{j+1}-t_{j},\boldsymbol{\Theta}(t_{j-1} \rightarrow t_{j})\right] \right)^2\ .
  \end{equation}

Its relative estimate denoted $\epsilon_{emp}(A_k,\mathcal{H}_k)$ can be defined as in Equation \ref{eq:relativeEmpiricalError}~.
\begin{equation}
  \epsilon_{emp}(A_k,\mathcal{H}_k) \coloneqq \dfrac{\delta_{emp}(A_k,\mathcal{H}_k)}{\mathbb{V}[A_k]} \ . 
  \label{eq:relativeEmpiricalError}
\end{equation}

Once the PCE learnings can be trusted, the distance at time $t_j$ between the true state $\mathbf{U}(\boldsymbol{\mathcal{X}},t_j)$ and the POD-PCE approximation $\mathcal{H} \left[ \mathbf{U}(\boldsymbol{\mathcal{X}},t_{j}),t_{j+1}-t_{j},\boldsymbol{\Theta}(t_{j} \rightarrow t_{j+1})\right]$ can be defined. It might be estimated using the relative Root Mean Squared Error (relative RMSE), denoted $r[\mathbf{U},\mathcal{H}](t_j)$ and calculated as in Equation \ref{eq:relativeRMSE}~,  where $\mathbf{h}(\mathbf{x}_i,t_j)$ refers to the value of the POD-PCE approximation at coordinate $\mathbf{x}_i$ and time $t_j$. 
\begin{equation}
  \label{eq:relativeRMSE}
    r[\mathbf{U},\mathcal{H}](t_j) \coloneqq  \sqrt{ \dfrac{\sum_{i=1}^m \left[\mathbf{u}(\mathbf{x}_i,t_j)-\mathcal{\mathbf{h}}(\mathbf{x}_i,t_j)\right]^2}{\sum_{i=1}^m \left[\mathbf{u}(\mathbf{x}_i,t_j)\right]^2} \ .}
\end{equation}

A mean value of the relative RMSE is calculated over a set of realizations corresponding to a set of times $\mathcal{T}=\{t_1,\dots,t_n\}$. It is denoted $r[\mathbf{U},\mathcal{H}]^{(\mathcal{T})}$ and estimated as in Equation \ref{eq:timeAveragedRelativeRMSE}.
\begin{equation}
  \label{eq:timeAveragedRelativeRMSE}
    r[\mathbf{U},\mathcal{H}]^{(\mathcal{T})} \coloneqq \dfrac{1}{n} \sum_{j=1}^n r[\mathbf{U},\mathcal{H}](t_j)   \ .
\end{equation}

Once the accuracies of the PCE learnings and the POD-PCE coupling have been evaluated, a final model, which will be the most accurate one, can be chosen. This model would, for our ML set-up, be the best representation of the dependence structure between inputs and outputs. It is used to shed light on the underlying physical relationships. Therefore the inputs are ranked in terms of physical influence, using an appropriate ranking indicator, presented in the following Subsection. 

\subsubsection{Physical influence of inputs based on the POD-PCE model}
The GW influence measures presented for the PCE models in Subsection \ref{subsection:theory:PCE} are here extended for the POD-PCE coupling. These indicators are adequate for the analysis of each PCE model $\mathcal{H}_k$, i.e. for interpreting the contribution of the inputs to each random variable $A_k$ separately. However, calculating the contributions to each $A_k$ independently precludes putting them in perspective according to the importance of $A_k$ in the final reconstructed model $\mathcal{H}$ that approximates $\mathbf{U}(\boldsymbol{\mathcal{X}},.)$. Hence, adapted indicators should be calculated. \\

Let $\mathbf{U}(\boldsymbol{\mathcal{X}},.)$ be the random spatiotemporal field approximated by the POD-PCE ML, for prediction from time $t_j$ to time $t_{j+1}$ and let $\mathcal{H}_k$ be the PCE approximation at degree $p^{(k)}$ that maps the random POD temporal coefficient $A_k$ from a set of input variables, using the expansion on the multivariate polynomial basis $\left\{ \zeta_{\boldsymbol{\alpha}}^{(k)}(.) \right\}_{|\boldsymbol{\alpha}| \leq p^{(k)}}$. The POD-PCE model formulated in Equation \ref{eq:prediction:fullmodel} is written as in Equation \ref{eq:prediction:fullmodelPCE}: 
\begin{equation}
\label{eq:prediction:fullmodelPCE}
   \mathbf{U}(\boldsymbol{\mathcal{X}},.) \approx \sum_{k=1}^{d} A_k \sqrt{n \times \lambda_k} \boldsymbol{\Phi}_k^{(d)}(\mathbf{\boldsymbol{\mathcal{X}}}) \approx \sum_{k=1}^{d} \left( \sum_{|\boldsymbol{\alpha}| \leq p^{(k)}} c_{\boldsymbol{\alpha}}^{(k)}  \zeta_{\boldsymbol{\alpha}}^{(k)}(.) \right)  \sqrt{n \times \lambda_k} \boldsymbol{\Phi}_k^{(d)}(\mathbf{\boldsymbol{\mathcal{X}}}) \ .
  \end{equation}
  
\begin{figure}[H]
  \centering
    \includegraphics[trim={0cm 2.3cm 0cm 0.5cm},clip,scale=0.52]{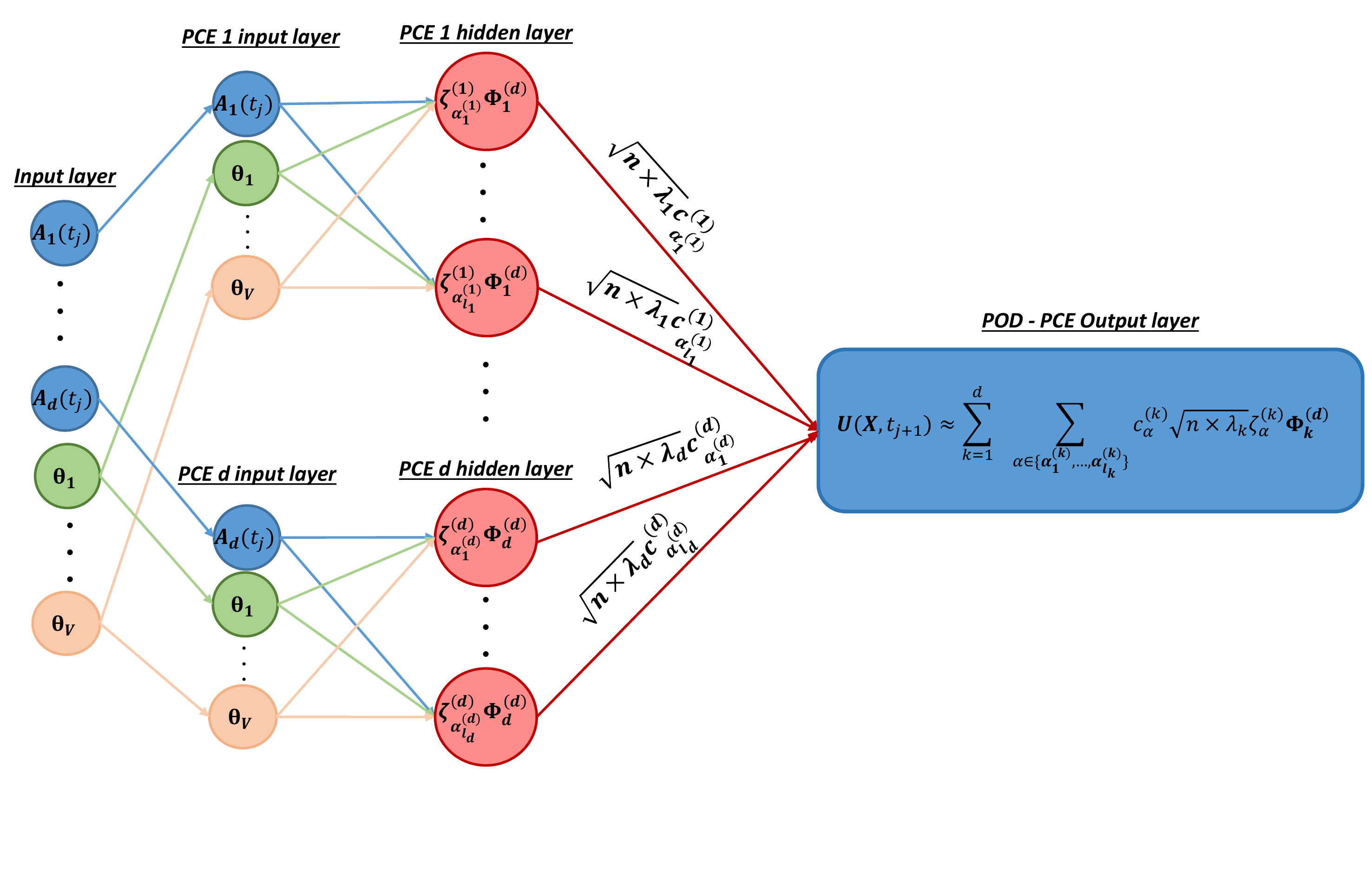}
    \caption{Representation of the POD-PCE ML approach in the NN paradigm.}
    \label{fig:POD_PCE_NN}
\end{figure}

Thanks to its linearity, the POD-PCE ML can be represented as a single-layered NN, as shown in Figure \ref{fig:POD_PCE_NN}.

Therefore, a new indicator, \textit{Generalized Garson Weights} (GGW), denoted $W_{\zeta_{\boldsymbol{\alpha}}^{(k)}}$, is computed and simply re-evaluated from the PCE Garson weights (GW), here denoted $w_{\zeta_{\boldsymbol{\alpha}}^{(k)}}$, as in Equation \ref{eq:PODPCE:generalizedWeights}.
\begin{equation}
  \label{eq:PODPCE:generalizedWeights}
	\begin{matrix}
  W_{\zeta_{\boldsymbol{\alpha}}^{(k)}} & \coloneqq & \dfrac{|c_{\boldsymbol{\alpha}}^{(k)}|\sqrt{n \times \lambda_k}}{  \sum_{e=1}^d  \sum_{|\boldsymbol{\beta}| \leq p^{(e)}} \left(|c_{\boldsymbol{\beta}}^{(e)}| \sqrt{n \times \lambda_e}\right) } \\
	& = &  \dfrac{ \left(\sum_{|\boldsymbol{\beta}| \leq p^{(k)}} |c_{\boldsymbol{\beta}}^{(k)}|\right) w_{\zeta_{\boldsymbol{\alpha}}^{(k)}} \sqrt{\lambda_k}}{  \sum_{e=1}^d  \sum_{|\boldsymbol{\beta}| \leq p^{(e)}} \left(|c_{\boldsymbol{\beta}}^{(e)}| \sqrt{\lambda_e}\right) } = \left(\dfrac{ \sum_{|\boldsymbol{\beta}| \leq p^{(k)}} \left( |c_{\boldsymbol{\beta}}^{(k)}| \sqrt{\lambda_k} \right) }{  \sum_{e=1}^d  \sum_{|\boldsymbol{\beta}| \leq p^{(e)}} \left(|c_{\boldsymbol{\beta}}^{(e)}| \sqrt{\lambda_e}\right) } \right) w_{\zeta_{\boldsymbol{\alpha}}^{(k)}} \ .
	\end{matrix}
\end{equation}

These GGW indicators show that the contribution of the polynomials $\{\zeta_{\boldsymbol{\alpha}}^{(k)}\}_{|\boldsymbol{\alpha}| \leq p^{(k)}}$ of $A_k$ are enhanced with the eigenvalue  $\lambda_k$, which is directly linked to the importance of the POD mode $\boldsymbol{\Phi}_k^{(d)}(\boldsymbol{\mathcal{X}})$ (EVR in Equation $\ref{eq:POD:RIC}$). An analogy can be drawn with the generalized sensitivity indices for a reduced order model \citep{Lamboni2011}. The $\sum_{k=1}^d \sum_{|\boldsymbol{\alpha}| \leq p^{(k)}} W_{\zeta_{\underline{\alpha}}^{(k)}}=1$ property holds. This means that the indices allow $\{\{\zeta_{\boldsymbol{\alpha}}^{(k)}\}_{|\boldsymbol{\alpha}| \leq p^{(k)}}\}_{k\in\{1,\dots,d\}}$ to be ranked altogether in terms of contribution to output $\mathbf{U}$. The influences can be analyzed following the indications of Section \ref{subsection:theory:PCE:original}.


\shorthandoff{:}


\shorthandoff{:}

\section{Application on a parametric toy problem}
\label{section:toy}
The theoretical framework of the proposed POD-PCE learning was presented in the previous Section \ref{section:theory}, including the detailed coupling formulation, accuracy estimators and physical influence measures in Subsection \ref{subsection:theory:methodology}. In the latter, it was highlighted that there is a slight difference in the learning and prediction steps between temporal problems and parametric problems. In this section, the POD-PCE ML is applied to a parametric toy problem, for which the analytical solution is introduced in Subsection \ref{subsection:toy:analytical}. The problem is simple and controllable, and allows demonstrating the learning performance, the consistency of physical interpretations in comparison with the analytical information, and the robustness of the learning to noise in the data. Subsection \ref{subsection:toy:learning} therefore deals with the application of the POD-PCE methodology for physical analysis and prediction, while in Subsection \ref{subsection:toy:noise}, robustness to different noise levels is investigated. 

\subsection{Problem description}
\label{subsection:toy:analytical}
The chosen toy problem deals with the representation of groundwater flow in a confined aquifer. Such a flow can be complex to describe and is generally represented using the depth-averaged groundwater flow equations \cite{Li2000}. Analytical solutions for these equations can be found for particular configurations. For example, a solution was identified by \citet{Li2000} in case of a semi-infinite coastal aquifer subject to oscillating boundary conditions, resulting from oceanic and estuarine tidal loadings. The solution is given for the particular case where the estuary and coastline are perpendicular. The oceanic BC (along coastline) is taken as a single and spatially uniform tidal harmonic constituent $Acos(\omega t)$, where $A$ and $\omega$ are the tidal amplitude and pulsation respectively. The corresponding BC along the estuary is a non-uniform tidal loading $Aexp(-\kappa_{er}x)cos(\omega t-\kappa_{ei}x)$, where $\kappa_{er}$ and $\kappa_{ei}$ are the estuary's tidal damping coefficient and tidal wave number respectively, that represent changes in the amplitude and phase along the estuary. \\

This forcing results with fluctuations in the \textit{water table}, that is defined as the level separating the water and saturated ground from the remaining upper unsaturated ground. The fluctuations, denoted $f$, can be calculated using the analytical solution defined in \cite{Li2000} as in Equation \ref{eq:toy:solution}.
\begin{equation}
\left\{ \begin{matrix}
\label{eq:toy:solution}
f(x,y,t) & = & f_0(x,t) + f_1(x,y,t) \\
f_0(x,t) & = & Aexp\left(-\sqrt{\frac{\omega}{2D}}x\right)cos\left(\omega t-\sqrt{\frac{\omega}{2D}}x\right) \\
f_1(x,y,t) & = & A \times Re\left\{ \int_0^t \left[g(k_1,x) - g(k_1,-x) - g(k_2,x) + g(k_2,-x)\right]dt_0\right\}
\end{matrix} \right.
\end{equation}
where constant $D$ is the diffusivity of the aquifer \cite{Li2000}, $t$ is the time variable, and $(x,y)$ are the cartesian cross- and long-shore coordinates, corresponding to the distance from ocean and estuary respectively. The operator $Re\{z\}$ denotes the real part of complex $z$. Coefficients $k_1$ and $k_2$ are defined as $k_1\coloneqq -(\kappa_{er} + \kappa_{ei}i)$ and $k_2 \coloneqq -( \sqrt{\frac{\omega}{2D}} +  \sqrt{\frac{\omega}{2D}}i) $, where $i=\sqrt{-1}$. Function $f$ is defined in equation \ref{eq:toy:function}, where $erf(z) \coloneqq \frac{2}{\sqrt{\pi}}\int_0^z e^{-t^2}dt$ is the Gauss error function.
\begin{equation}
\label{eq:toy:function}
\begin{matrix}
g(\psi,\xi) & = & \frac{y}{4\sqrt{\pi}[D(t-t_0)]^{3/2}} \times exp\left(\psi^2D(t-t_0) + i\omega t_0 + \psi\xi - \frac{y^2}{4D(t-t_0)}\right) \\ 
& & \times \left[ 1 + erf\left(\dfrac{2\psi D(t-t_0) + \xi}{2\sqrt{D(t-t_0)}}\right) \right]
\end{matrix}
\end{equation}

\begin{figure}[H]
  \centering
  \subfloat[][Amplitude (m)]{\includegraphics[trim={0cm 1.5cm 0.5cm 3.2cm},clip,scale=0.35]{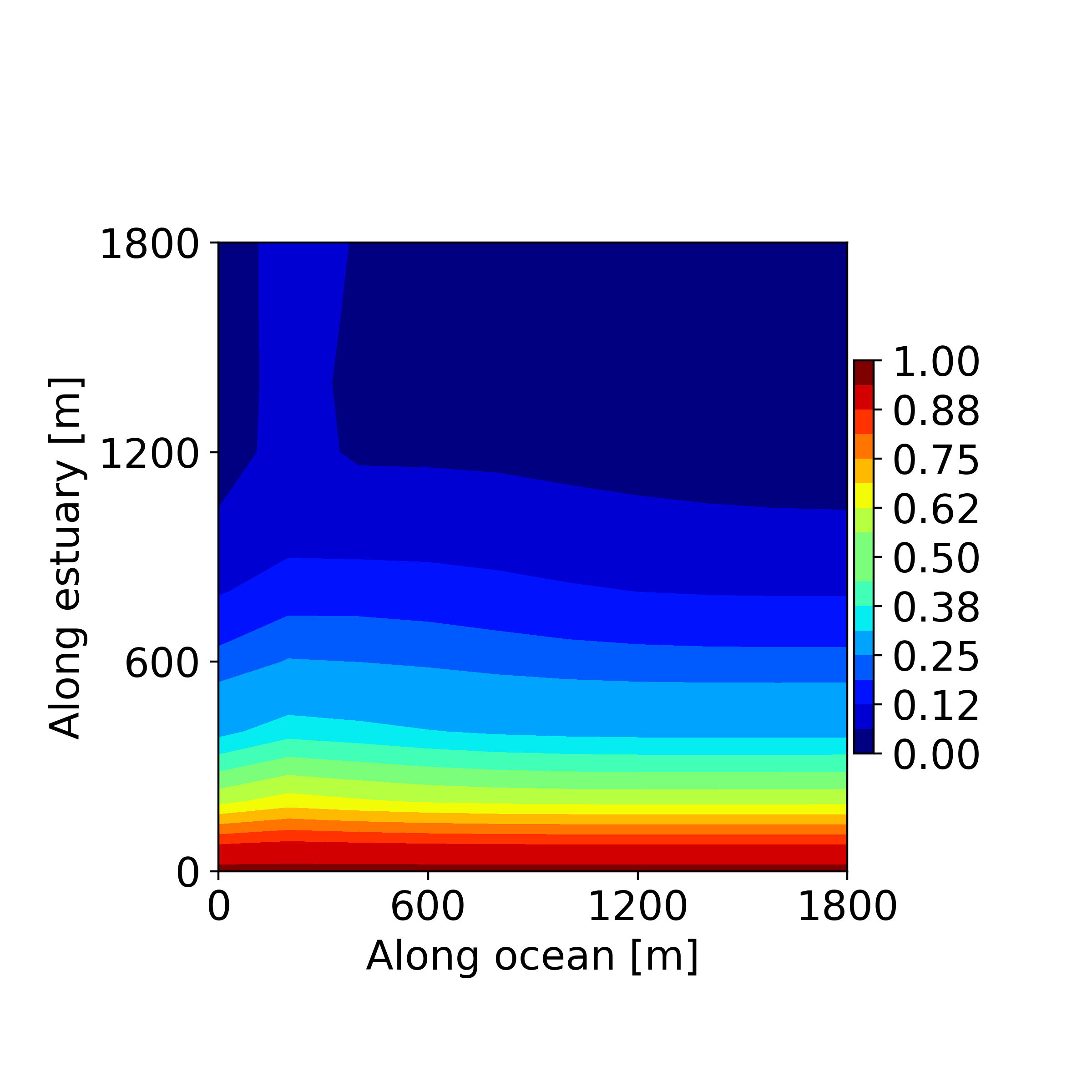}}
  \subfloat[][log-Amplitude]{\includegraphics[trim={0cm 1.5cm 0.5cm 3.2cm},clip,scale=0.35]{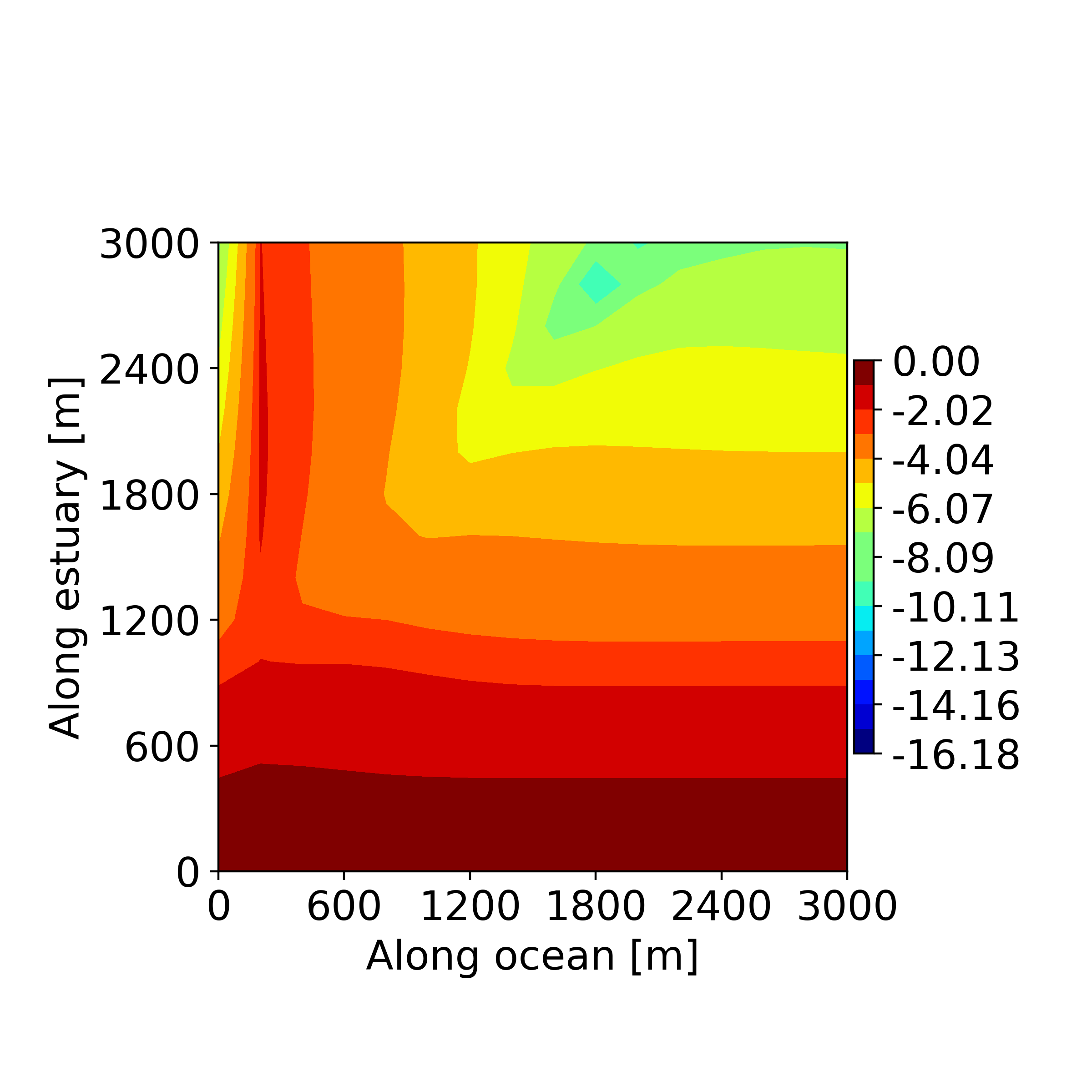}}
  \subfloat[][Phase]{\includegraphics[trim={0cm 1.5cm 0.5cm 3.2cm},clip,scale=0.35]{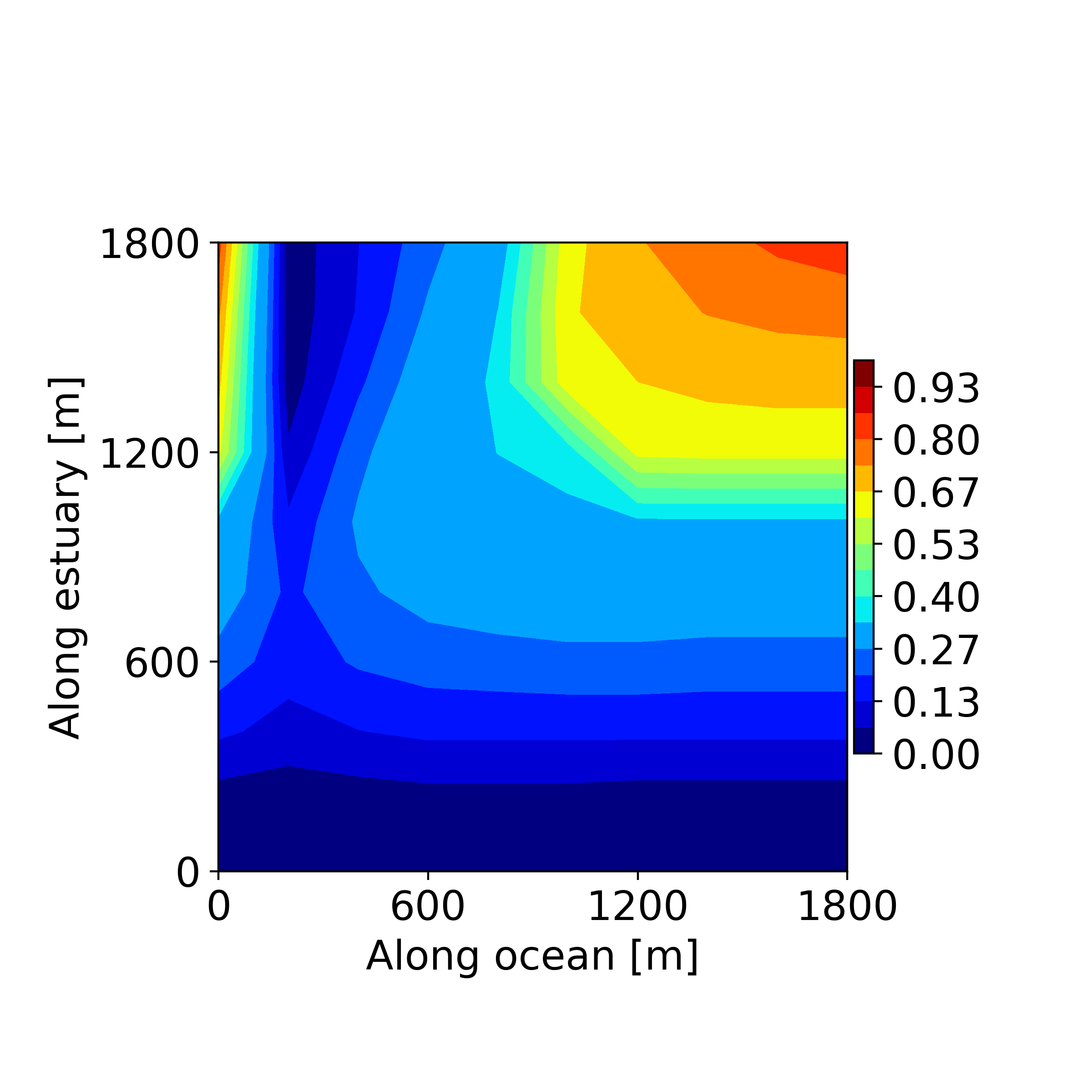}}
  \caption{Amplitude and phase of the water table fluctuation, using the parameters proposed by  \cite{Li2000} in Table 1.}
  \label{fig:toy:analytical_amplitude_phase}
\end{figure}

This solution is complex and non-linear due to the presence of an interaction zone where the effects of the ocean and estuary are coupled. This results with complex fluctuation patterns that can extend to several square kilometers \cite{Li2000}, depending on the aquifer configuration. For example, the diurnal tide configuration proposed by \cite{Li2000} in Table 1 is used for illustration. The amplitude of the fluctuation calculated at each $(x,y)$ location as $[\max_{t}(f(x,y,t))-\min_{t}(f(x,y,t))]/2$ over $t\in[0,T]$, and the phase calculated at each $(x,y)$ location as the time lag, relative to $T$, between the time series $f(x,y,t)$ and $f(0,0,t)$ over $t\in[0,T]$, are shown in Figure \ref{fig:toy:analytical_amplitude_phase}. 
The amplitude is decreasing through the aquifer (Figures \ref{fig:toy:analytical_amplitude_phase}-a and  \ref{fig:toy:analytical_amplitude_phase}-b), and a time lag is noticed in the tidal propagation (Figure \ref{fig:toy:analytical_amplitude_phase}-c). Both the amplitude damping and time lag are increasing through the aquifer and along the estuary. It can therefore be interesting to see if the POD-PCE methodology succeeds in recovering and explaining such patterns, in particular by learning their dependency to the tidal, estuary and aquifer parameters, from a statistical sample of the solution. \\

In order to apply the POD-PCE methodology on the aquifer case, a statistical sample of the solution and corresponding input sample of parameters are needed. In the presented study, the tidal period $T$ (and therefore pulsation $\omega$) is fixed to the diurnal configuration of \citet{Li2000}, whereas an ensemble of realizations is generated for the remaining control parameters $(A,k_{er},k_{ei},D)$. For this, Gaussian PDFs are used with mean values corresponding to the setting used by \citet{Li2000}, and a variation coefficient (standard deviation divided by mean) of $20\%$. This value corresponds to the average variation coefficient associated to optimal fitting of groundwater flow parameters performed by \cite{Wagner1992} on several cases. Indeed, the maximum variation coefficient was between $12\%$ and $28\%$ depending on the case. A random sample of size $n=200$ is produced using the Gaussian PDFs (Monte Carlo), and each realization of the inputs denoted $\boldsymbol{\Theta}_{j \in \{1,...,n\}}$ is associated to a realization of the output by calculating $f(x,y,t)_{j \in \{1,...,n\}}$. 

\subsection{POD-PCE learning}
\label{subsection:toy:learning}

The methodology is applied on the perturbation amplitude in the aquifer. The objective is to understand how the perturbation propagates from the boundaries, for different tidal, aquifer and estuary characteristics. The perturbations are calculated over a tidal period on a cartesian spatial grid composed of $m \in \mathbb{N}$ points, denoted $(x,y)_{i \in \{1,...,m\}}$. The spatial discretization step is $200~m$ in both directions, and the temporal step is $1$ hour. The amplitude of the perturbation, denoted $a'$, is then locally computed on each point of the grid. It depends on both the spatial location in the aquifer and the simulation parameters $\boldsymbol{\Theta}$. The solutions can be stored in a snapshot matrix as $\boldsymbol{A}'(\boldsymbol{\mathcal{X}},\mathcal{T}) = \left[  a'\left( (x,y)_i, \boldsymbol{\Theta}_j \right) \right]_{i,j} \in \mathbb{R}^{m \times n}$, where  $\boldsymbol{\mathcal{X}}$ designates the spatial coordinates space and $\boldsymbol{\mathcal{T}}$ designates the parameters space. The snapshot matrix is then POD-processed as explained in Section \ref{subsection:theory:POD}. Therefore, at each spatial coordinate $(x,y)$, each realization of the amplitude associated to a given parameterization $\boldsymbol{\Theta}$ can be approximated as $a'(x,y, \boldsymbol{\Theta}) \approx \sum_{k=1}^{d} a_k(\boldsymbol{\Theta}) \phi_k(x,y)$, where $d \in \mathbb{N}$ is a chosen POD approximation rank. \\

\begin{figure}[H]
  \centering
  \subfloat[][Mode  $\Phi_1(x,y)$]{\includegraphics[trim={0cm 0cm 3.37cm 1.7cm},clip,scale=0.35]{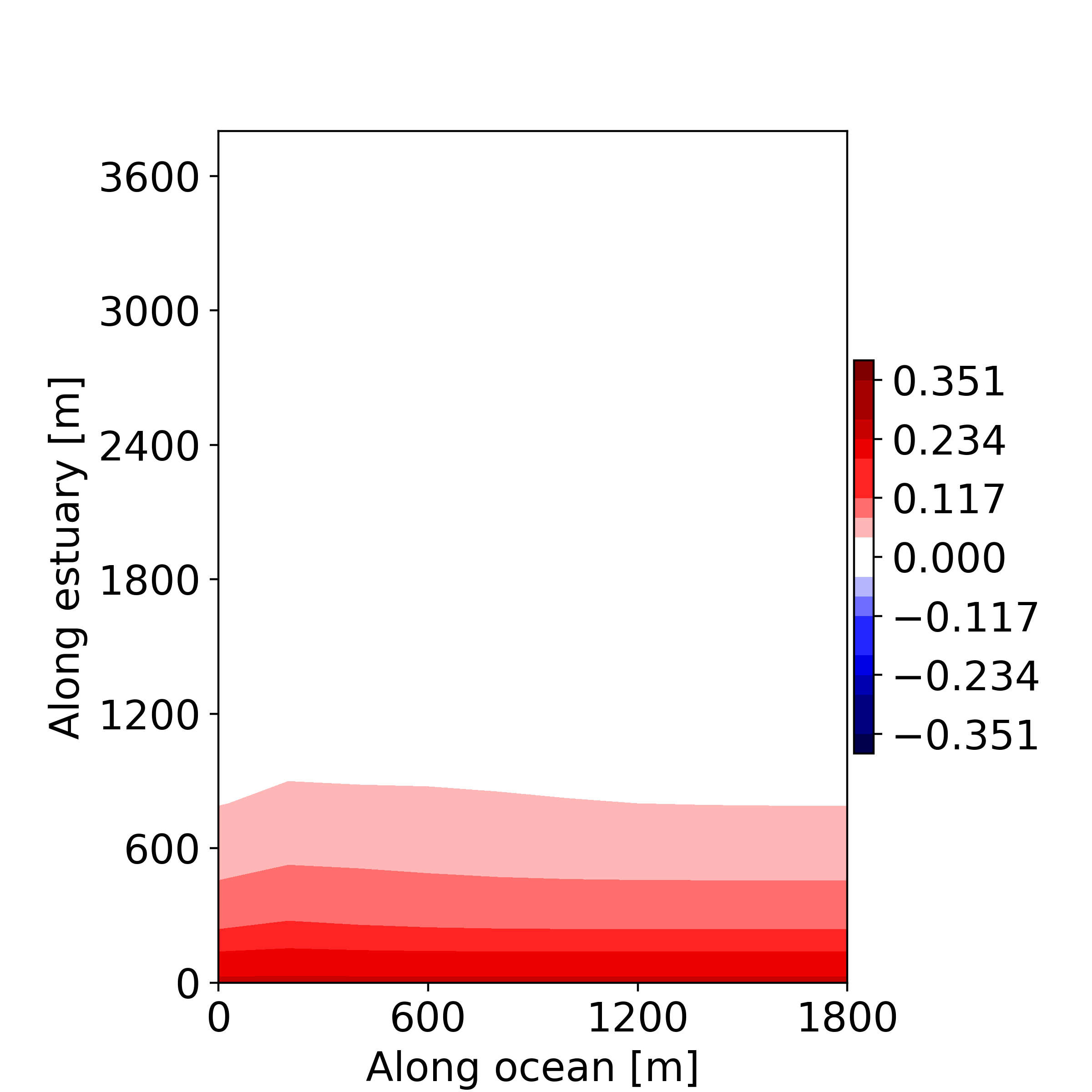}}
  \subfloat[][Mode $\Phi_2(x,y)$]{\hspace{0.3cm} \includegraphics[trim={2.8cm 0cm 3.37cm 1.7cm},clip,scale=0.35]{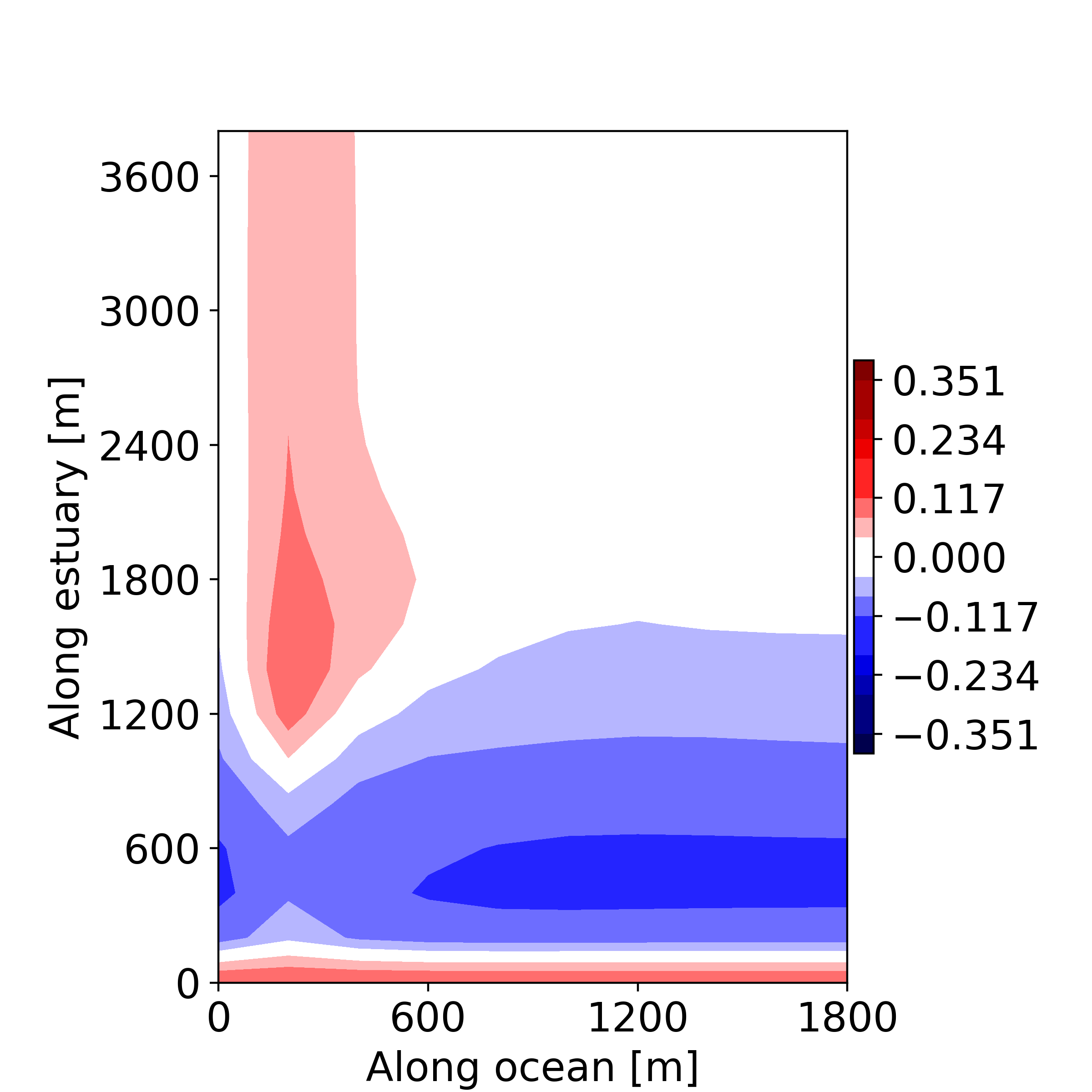}}
  \subfloat[][Mode $\Phi_3(x,y)$]{\hspace{0.3cm} \includegraphics[trim={2.8cm 0cm 3.37cm 1.7cm},clip,scale=0.35]{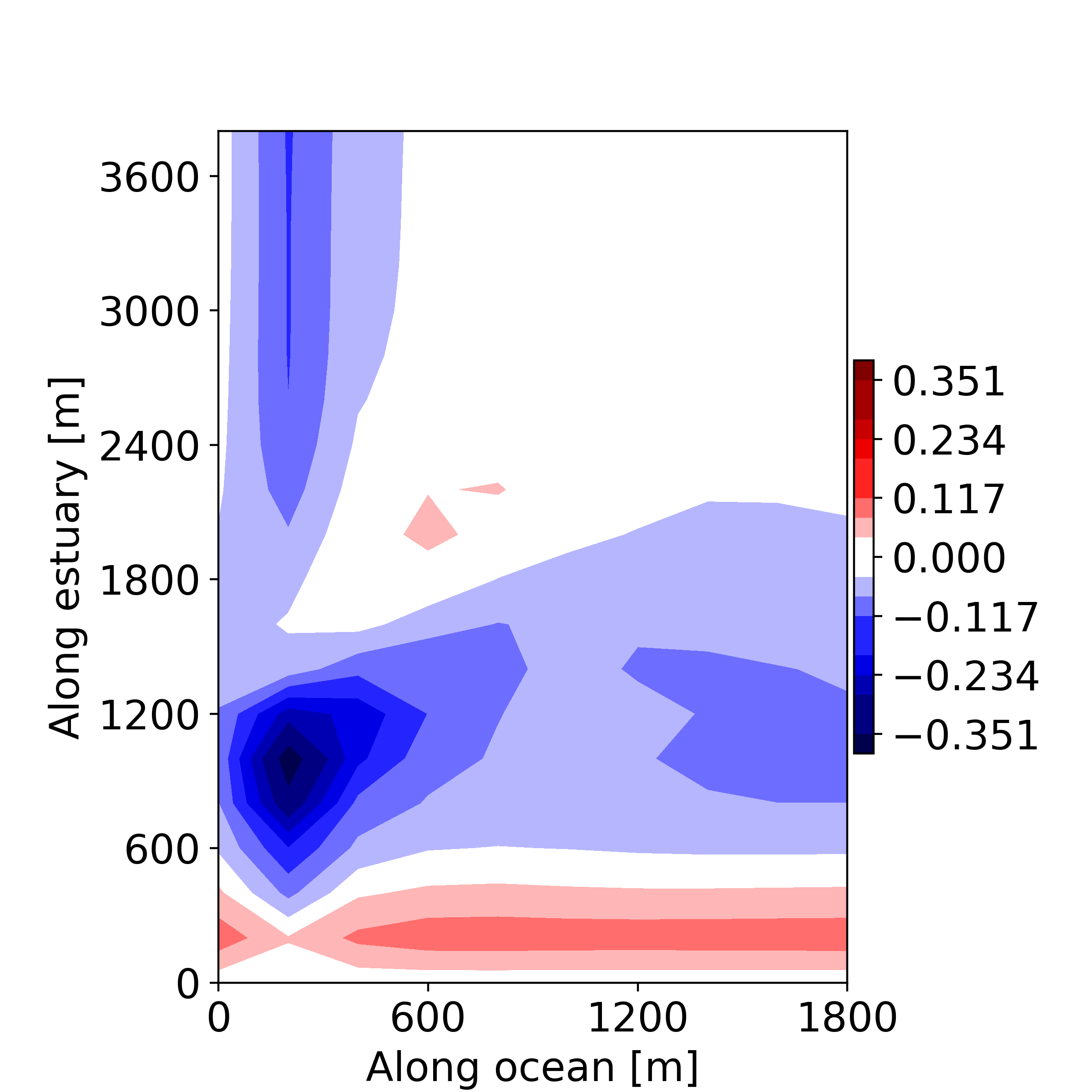}}
  \subfloat[][Mode $\Phi_3(x,y)$]{\hspace{0.3cm} \includegraphics[trim={2.8cm 0cm 0.8cm 1.7cm},clip,scale=0.35]{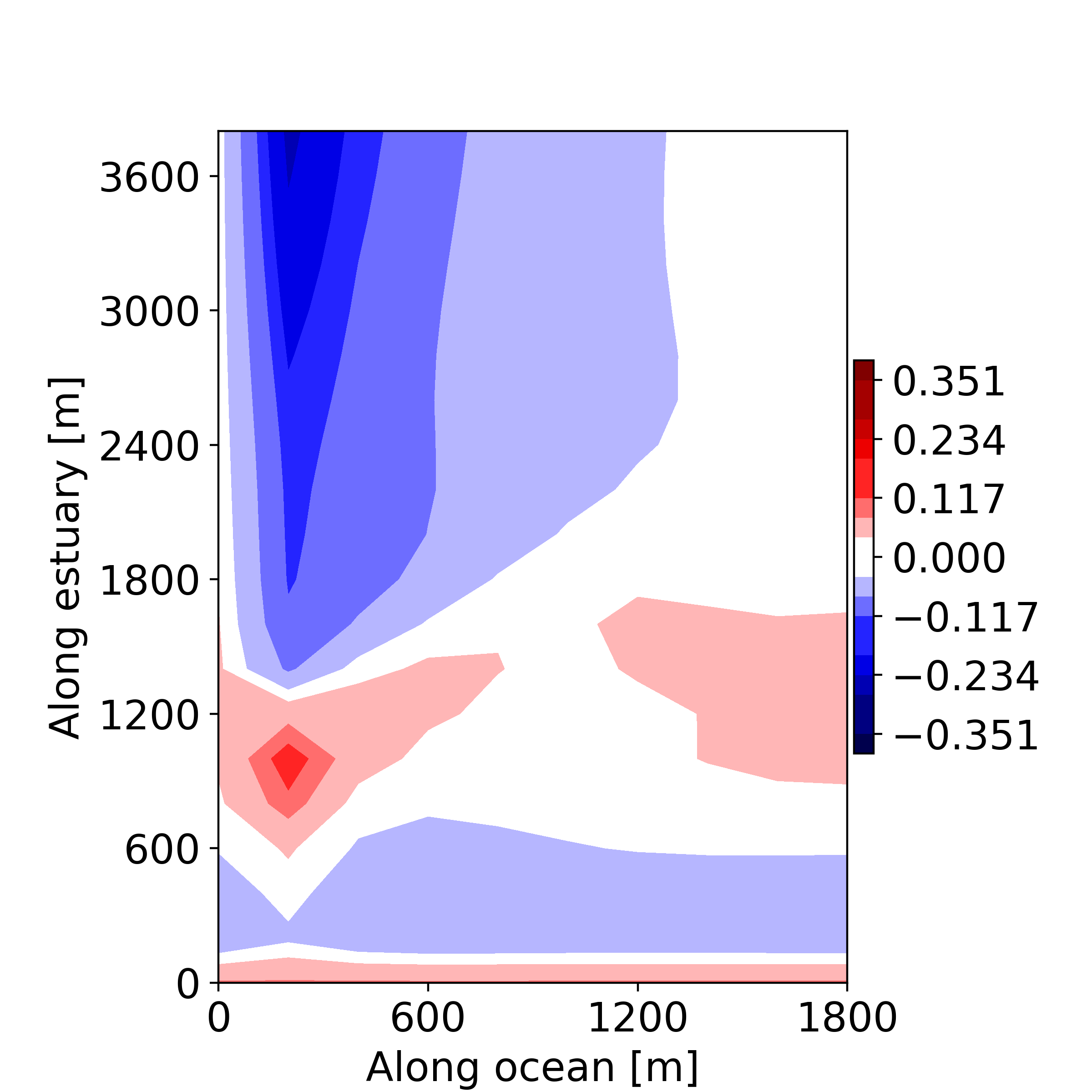}}
  \caption{The first four spatial patterns of the POD applied to aquifer toy problem.}
  \label{fig:toy:POD:spatial_amplitude}
\end{figure}

The EVR defined in Equation \ref{eq:POD:RIC} is calculated for each POD approximation rank. More than $99\%$ of the variance is already captured by the first mode, and the problem is therefore highly reducible. The spatial components of the first four POD modes are plotted in Figure \ref{fig:toy:POD:spatial_amplitude}. The first mode shows a gradual damping of the amplitude in the cross-shore direction. Its spatial values are all positive and the corresponding POD coefficient is strictly positive as well.
 As the coefficient directly multiplies the spatial mode, it plays, at the same time, the role of a magnitude enhancer and a gradient intensification. Indeed, the higher the coefficient, the higher the amplitude at the ocean boundary, and the higher the difference between the latter and the aquifer amplitudes. The second spatial mode plays a regulation role, through a succession of positive and negative spatial values in the cross-shore direction. The corresponding POD coefficients are also either positive or negative. When positive, they enhance the amplitude gradient in the cross-shore direction, and the opposite occurs when they are negative. Added to that is a variation in the longitudinal direction, from the estuary onward. The third modes looks similar to the second with added spatial details, whereas the fourth mode puts more emphasis on the damping in the longitudinal direction, from the estuary onward. \\
 
\begin{figure}[H]
  \centering
  \includegraphics[trim={0cm 0.5cm 0cm 0.2cm},clip,scale=0.7]{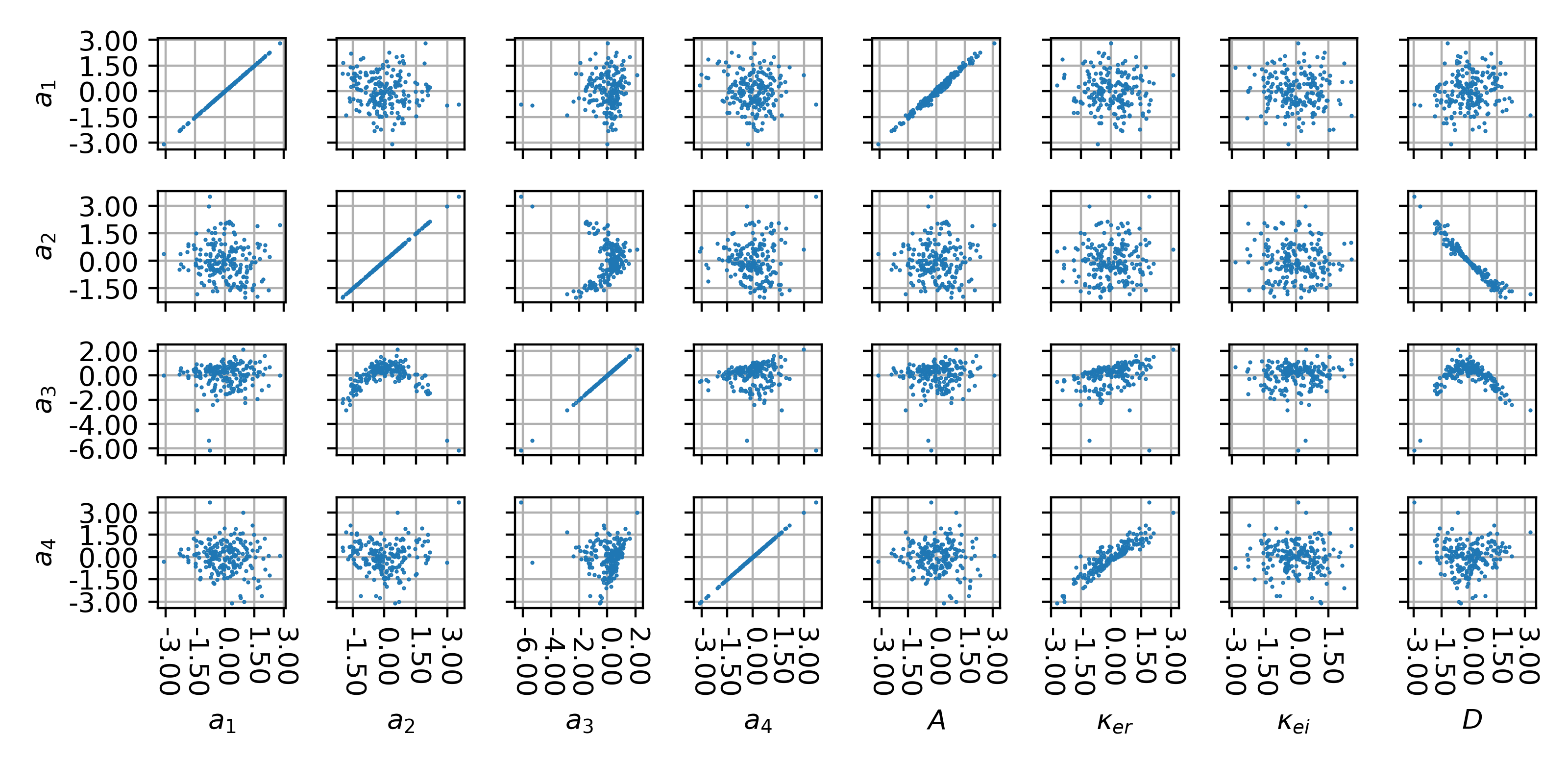}
    \caption{Scatter plot of the first four POD coefficients and control parameters of the aquifer. The variables are centered and reduced.}
    \label{fig:toy:POD:scatter_amplitude}
\end{figure}

A scatter plot can be used to understand the dependencies between the modes and parameters, as in Figure \ref{fig:toy:POD:scatter_amplitude}, and confirms the previous interpretations. Namely, a clear linear dependency between Mode 1 and the amplitude $A$ is noticed. The relation of Mode 1 to damping, that is rather related to diffusivity $D$ and estuary coefficient $\kappa_{er}$, is however not visible, although a dispersion of the mode around the linear tendency is noticed. This dispersion may be related to $D$ or $\kappa_{er}$, even in smaller proportions, or to possible interactions, later clarified using PCE. The dependency of Modes 2 and 3 to the diffusivity $D$ is also obvious, and the shapes indicate existing non-linearities. Mode 4 is highly dependent on the estuary amplitude damping coefficient $\kappa_{er}$, and no obvious dependency to the wave number in the estuary $\kappa_{ei}$, whatsoever, is noticed.
 
The dependencies that may explain the dispersion of the clouds around their main shapes need to be investigated. Hence, PCE models (theory in Section \ref{subsection:theory:PCE}) can be used to detect additional physical relationships. They are learned from the data for each POD coefficient as $a_i = \mathcal{H}_i(\boldsymbol{\Theta})$, using Hermite polynomials (orthonormal basis with regards to the used Gaussian marginals). The statistical set is separated to a learning set of size $150$ and a prediction set of size $50$. The PCE polynomial degree is optimized for each mode separately. Degrees from 1 to 7 were tested, and the associated relative empirical errors on the training and prediction sets, respectively denoted $\epsilon_T$ and $\epsilon_P$, were calculated as in Equation \ref{eq:relativeEmpiricalError}~. The PCE degree that minimized the training and prediction errors for each model was chosen. \\

The optimal PCE fitting for the first four modes shows good point-wise evaluations on the learning and prediction data-sets. PCE performs better for the modes of higher variance percentages. The smaller the variance rate, the higher the errors. Consequently, in this particular case, PCE succeeds in constructing causal models for the first four modes, but stops at an average evaluation (constant) for modes of higher ranks (smaller variance). For illustration, the prediction relative empirical errors $\epsilon_P$ of the first four modes are $6 \times 10^{-5}\%$, $4\times 10^{-3}\%$, $9\times 10^{-2}\%$ and $3 \times 10^{-1}\%$ respectively. At least one order of magnitude of precision is lost at each rank. The relative prediction residuals between the POD coefficient and their PCE estimation are also calculated for each sample member. For the first four modes, $90\%$ of the absolute relative residuals are lower than $1\times 10^{-2}\%$, $6.9\%$, $5.65 \%$ and $37.9 \%$ respectively.\\

The good performance of PCE encourages its use for POD-PCE prediction, as well as for physical interpretation. Firstly, in order to choose the adequate number of POD modes for the full model, the errors generated by the various steps of the algorithm (POD, PCE and coupling) are analyzed. To do so, the mean relative RMSE (averaged over the prediction set, as in Equation \ref{eq:timeAveragedRelativeRMSE}) was calculated for each step and for each approximation rank $d$, as follows:
\begin{itemize}
\item[$\bullet$] \textit{Reduction error:} distance between the POD approximation $\sum_{k=1}^{d} a_k\boldsymbol{\Phi}_k(x,y)$ and the corresponding amplitudes two-dimensional field $\mathbf{a}'(x,y,\boldsymbol{\Theta})$;
\item[$\bullet$] \textit{Learning error}: distance between the POD approximation $\sum_{k=1}^{d} a_k\boldsymbol{\Phi}_k(x,y)$ and the prediction using the POD-PCE coupling  formulated as $\sum_{k=1}^{d} \mathcal{H}_k(\boldsymbol{\Theta})\boldsymbol{\Phi}_k(x,y)$~;
  \item[$\bullet$] \textit{Prediction error:} the resulting final error between the prediction using POD-PCE coupling and the corresponding amplitudes two-dimensional field $\mathbf{a}'(x,y,\boldsymbol{\Theta})$. \\
  \end{itemize}
  
\begin{figure}[H]
  \centering
  \includegraphics[trim={0cm 0cm 0cm 0cm},clip,scale=0.35]{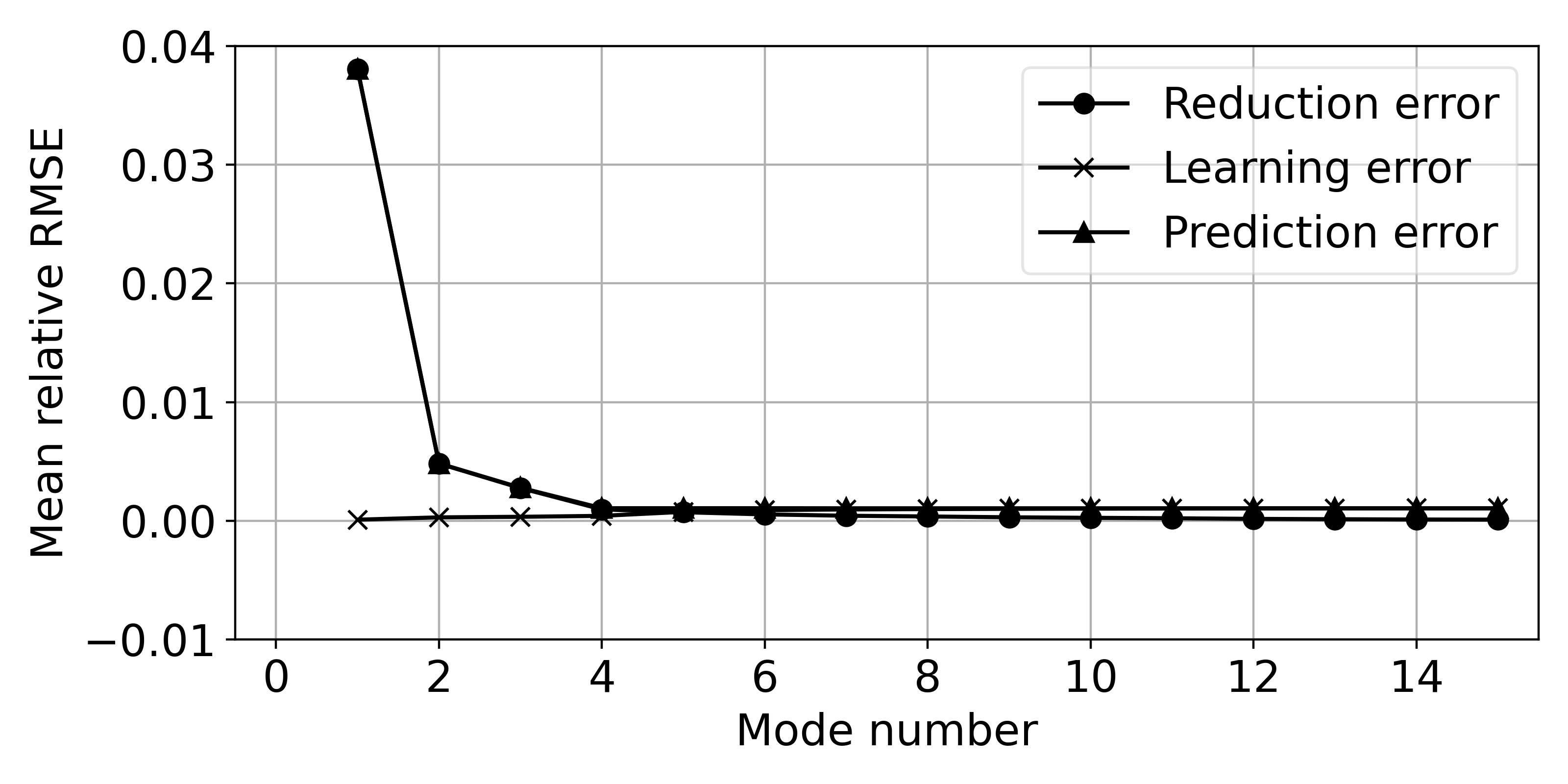}
    \caption{Mean relative RMSE generated at different steps of the POD-PCE ML applied to the aquifer case, with different approximation ranks.}
    \label{fig:toy:Prediction:errors_amplitude}
\end{figure}

The results are shown in Figure \ref{fig:toy:Prediction:errors_amplitude}~. Reduction error decreased from $3.8\%$ at rank 1, to $9.5 \times 10^{-2} \%$ at rank 4. The decrease is exponential, with a stabilization starting from rank 4. The learning error increased from $8.8 \times 10^{-3} \%$ at rank 1 to $4.1 \times 10^{-2} \%$ at rank 4, which is related to the increase in the PCE prediction error of the POD modes coefficients. In fact, a prediction of rank $d+1$ has an additional temporal coefficient that is predicted as compared to rank $d$. It is therefore natural that the distance between the approximation $\sum_{k=1}^d a_k(\boldsymbol{\Theta}) \phi_k (x,y)$ and its prediction $\sum_{k=1}^d \mathcal{H}_k(\boldsymbol{\Theta}) \phi_k (x,y)$ increased with increasing rank. The learning error order of magnitude keeps however low. Hence, the prediction error trend is almost identical to reduction error, decreasing from $3.8\%$ at rank 1 (identical to reduction error) to $1 \times 10^{-1} \%$ at rank 4, where it stabilizes. It is the balance of, on the one hand, the increase in accuracy by adding POD modes and, on the other hand, the increase in forecasting error with increasing number of POD coefficients to be predicted. Hence, a 4-Modes POD-PCE model was selected for prediction.  

\begin{figure}[H]
  \centering
  \subfloat[][Analytical]{\includegraphics[trim={0.7cm 0.1cm 1.2cm 1.8cm},clip,scale=0.35]{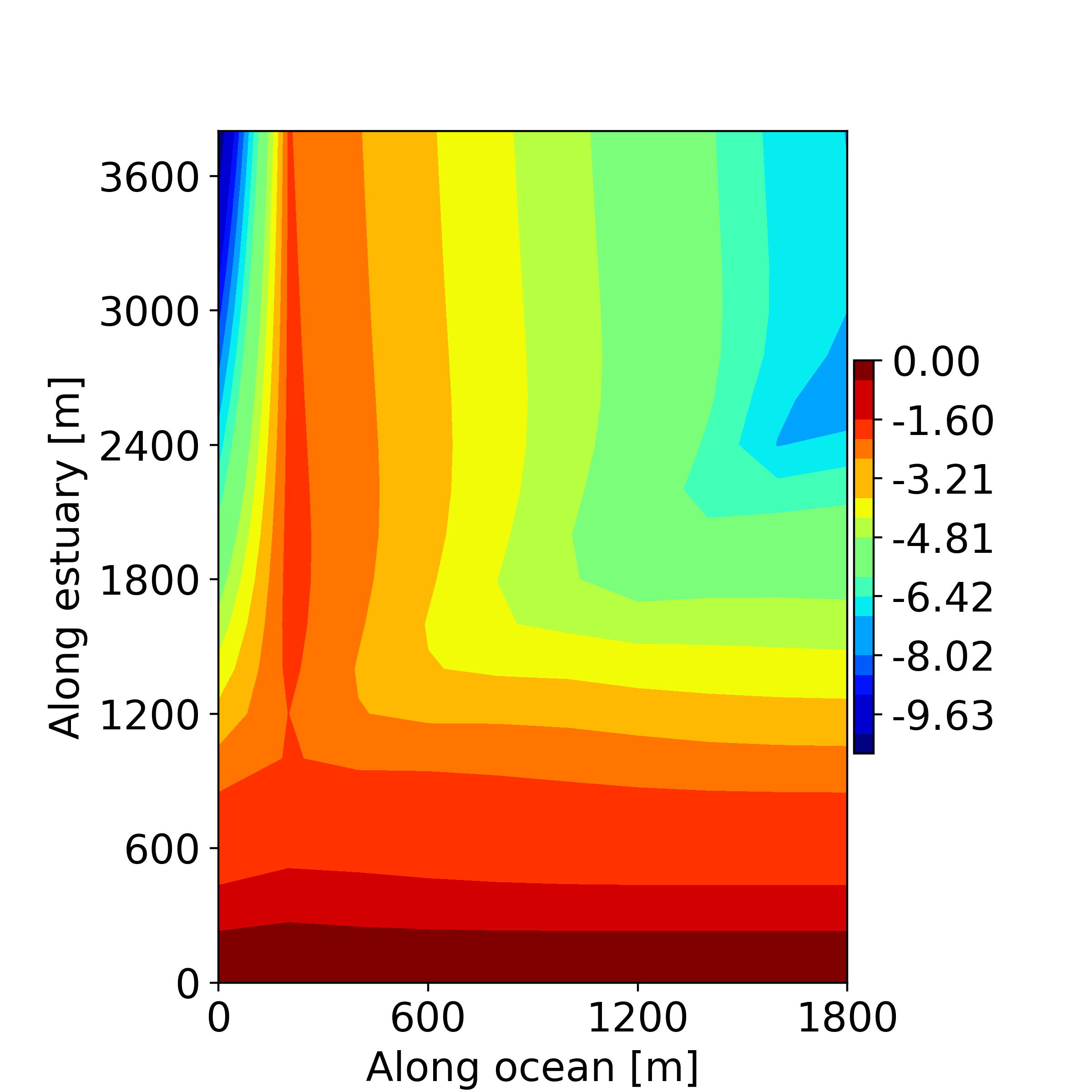}}
  \subfloat[][POD-PCE]{\hspace{0.3cm} \includegraphics[trim={0.7cm 0.1cm 1.2cm 1.8cm},clip,scale=0.35]{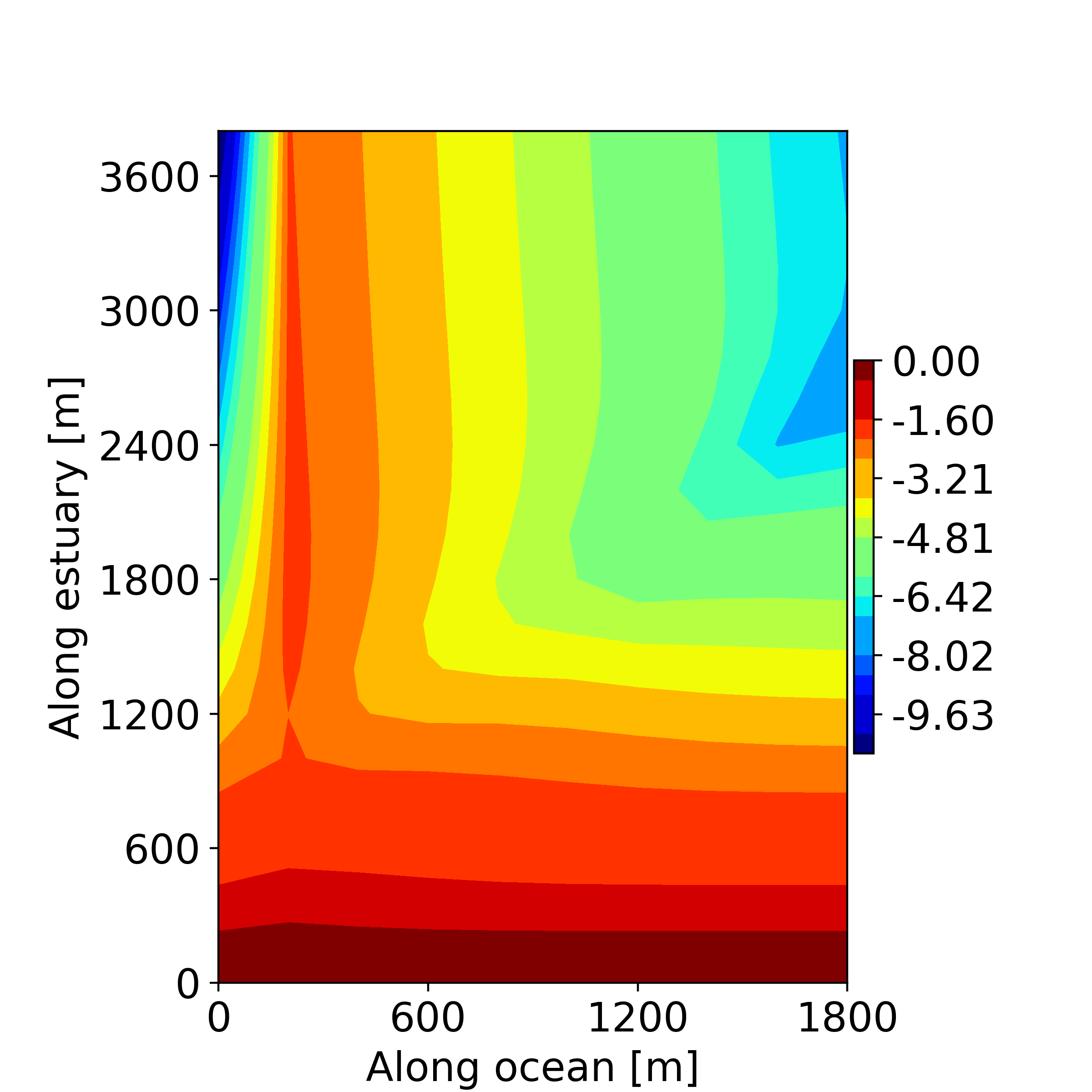}}
  \subfloat[][Absolute residual]{\hspace{0.3cm} \includegraphics[trim={0.7cm 0.1cm 0.8cm 1.8cm},clip,scale=0.35]{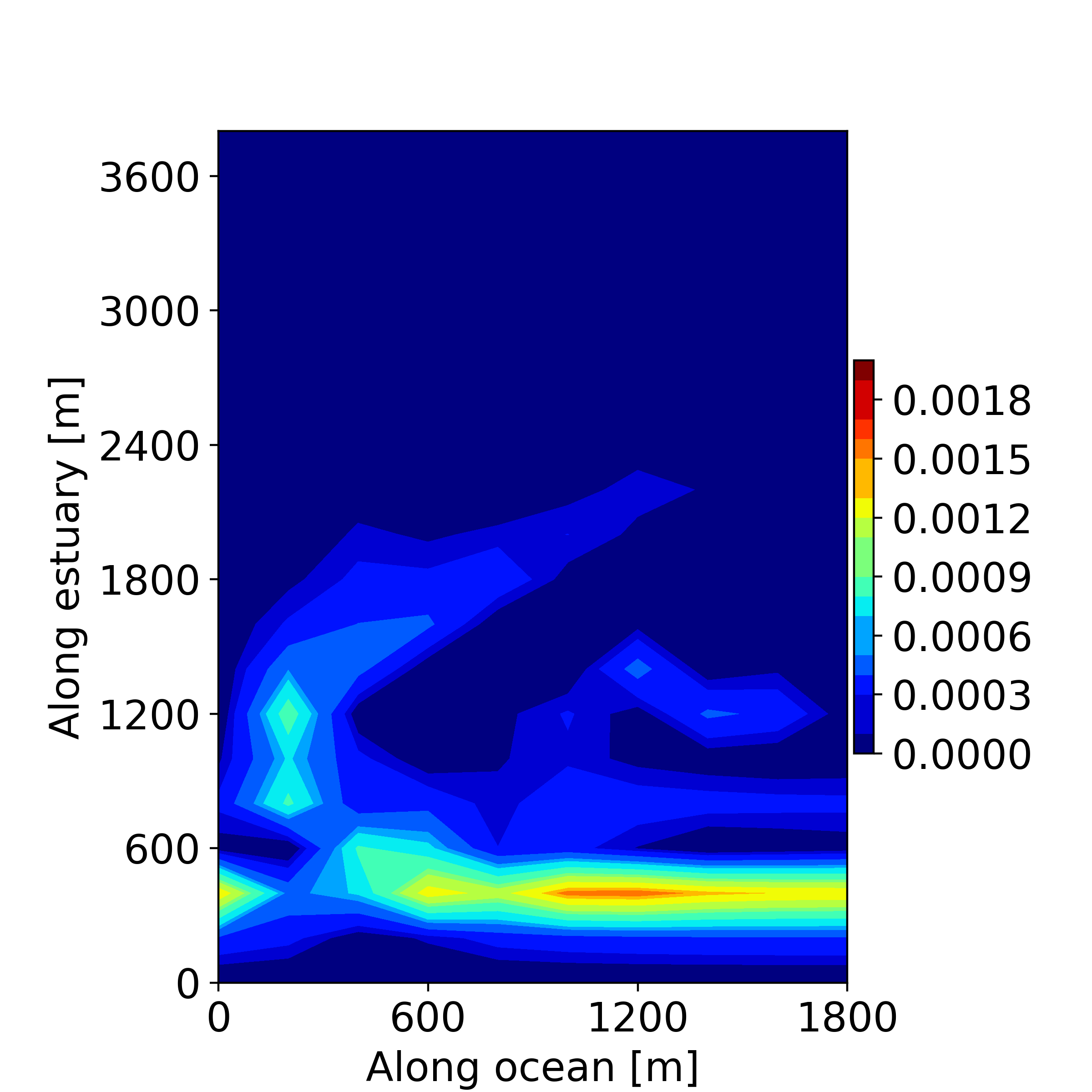}}
  \caption{Analytical solution vs. POD-PCE prediction of the aquifer's log-amplitude in meters, and resulting absolute residual of the amplitude.} 
  \label{fig:toy:PODPCE:amplitude}
\end{figure}

An example of prediction is shown in Figure \ref{fig:toy:PODPCE:amplitude}. The model gives good qualitative estimation of the two-dimensional amplitude distribution along the estuary and through the aquifer. Slight differences may be noticed however between the analytical solution and POD-PCE prediction. Namely, the absolute residuals can go up to $0.002~m$, but this occurs, for example in Figure \ref{fig:toy:PODPCE:amplitude}-c, in a zone where the amplitude is $0.3~m$, which represents a local error of $0.7\%$.
The fitted PCE models for the first four POD modes were used to rank and analyze the physical contributions. To do this, the \textit{Garson Weights} (GW) and \textit{Generalized Garson Weights} (GGW), respectively presented in Sections \ref{subsection:theory:PCE} and \ref{subsection:theory:methodology}, were calculated for each polynomial term. The indicators values are shown in \hyperref[Appendix:A]{Appendix A}, Table \ref{table:toy:sensitivity:generalizedWeights_amplitude}. The GW results confirm the great dependency of Mode 1 on the tidal amplitude $A$ ($85\%$), and a dispersion mainly caused by the diffusivity $D$ ($11\%$). Mode $2$ and Mode $3$ are principally dependent on the diffusivity $D$, and the noticed dispersion is related to the interaction of the diffusivity $D$ with the tidal amplitude $A$ in Mode 2  ($15 \%$), whereas it is explained by the tidal damping in the estuary $\kappa_{er}$ for Mode 3  ($19 \%$). The main variation of Mode 4 is captured linearly around $\kappa_{er}$. The GGW show that the most important parameter is the tidal amplitude, with a total of $80\%$ of influence (without interactions), followed by the diffusivity (more than $15\%$ without interactions). The main dynamics (a total of $93\%$) are explained by first degree monomials. This means that the non-linearities are represented in the spatial POD patterns. Simple interactions expressed by second degree polynomials are then added to complete the dynamics, as well as other non-linear contributions, for example second to third degree monomials of the diffusivity $D$. \\

The same strategy is adopted to learn the phase, or time lag between $p(x,y,t)$ and $p(0,0,t)$, in \hyperref[Appendix:A]{Appendix A}. The model is optimal when 3 POD modes are used, stabilizing around a $6\%$ prediction error. It gives a good mapping of the two-dimensional time lag distribution in the aquifer, even though more important differences between the analytical solution and the prediction are noticed compared to the amplitude prediction. The error can locally go up to $25\%$, but the global performance of the model remains satisfying. Lastly, calculation of GW and GGW indicators shows that the phase problem involves higher polynomial degrees, and higher orders of interaction. Additionally, the wave number in the estuary $\kappa_{ei}$, which did not appear as an influencing parameter for the amplitude distribution, is necessary for the phase representation.  

\subsection{Robustness to noise}
\label{subsection:toy:noise}
In this section, three noise levels ($1\%$, $10\%$ and $20\%$) are added to the data in order to evaluate the POD-PCE methodology. Perturbations are directly added to the $2D$ amplitude distributions in the aquifer. For each realization of the latter, the local value at a given location is perturbed using a zero-mean Gaussian PDF (white noise), with a standard deviation calculated as the average local value over the ensemble, multiplied by the chosen noise level percentage. \\

\begin{figure}[H]
  \centering
  \subfloat[][EVR]{\includegraphics[trim={0cm 0cm 0cm 0.2cm},clip,scale=0.4]{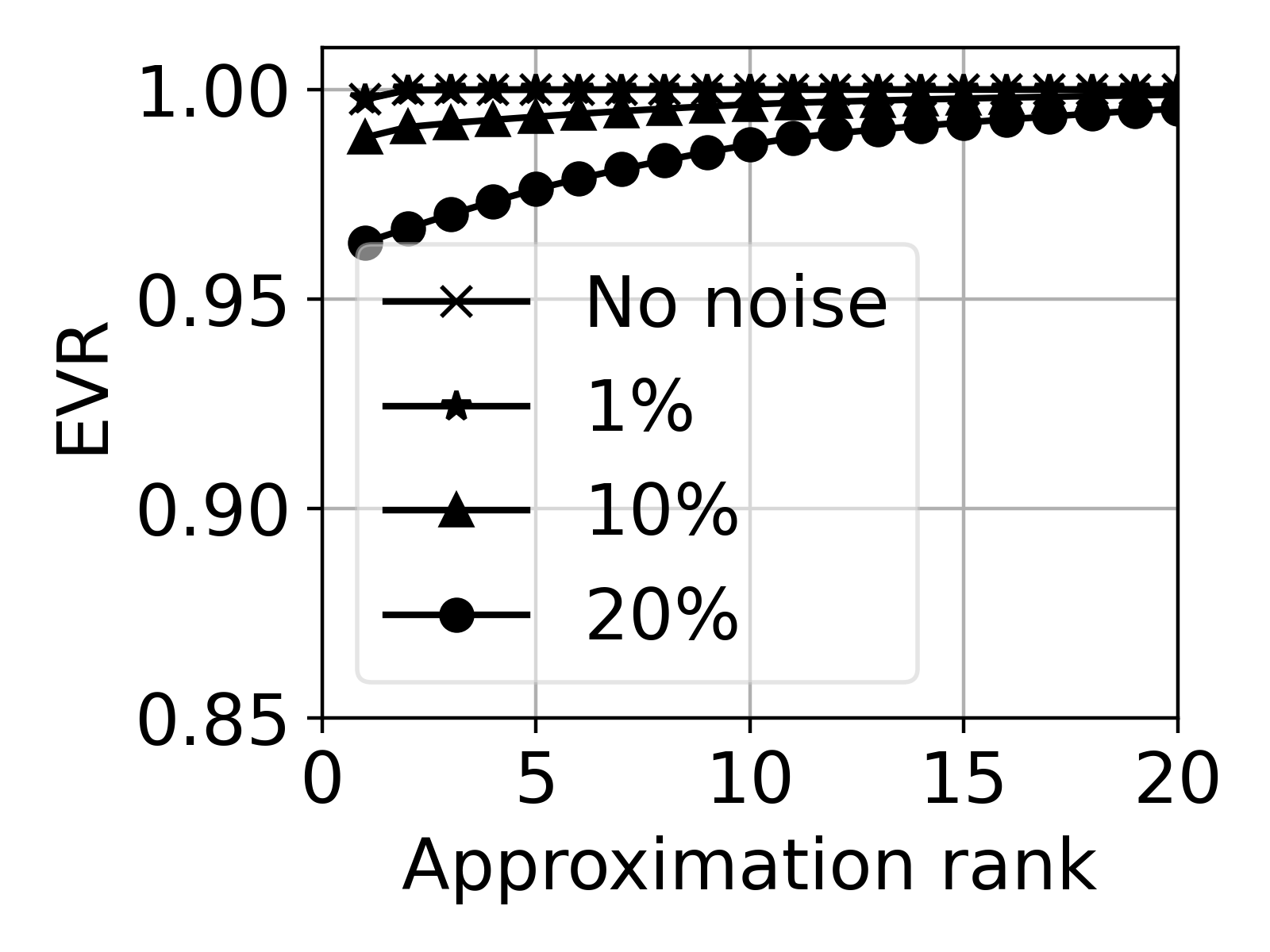}}
  \subfloat[][Reduction RMSE]{\includegraphics[trim={0cm 0cm 0cm 0.2cm},clip,scale=0.4]{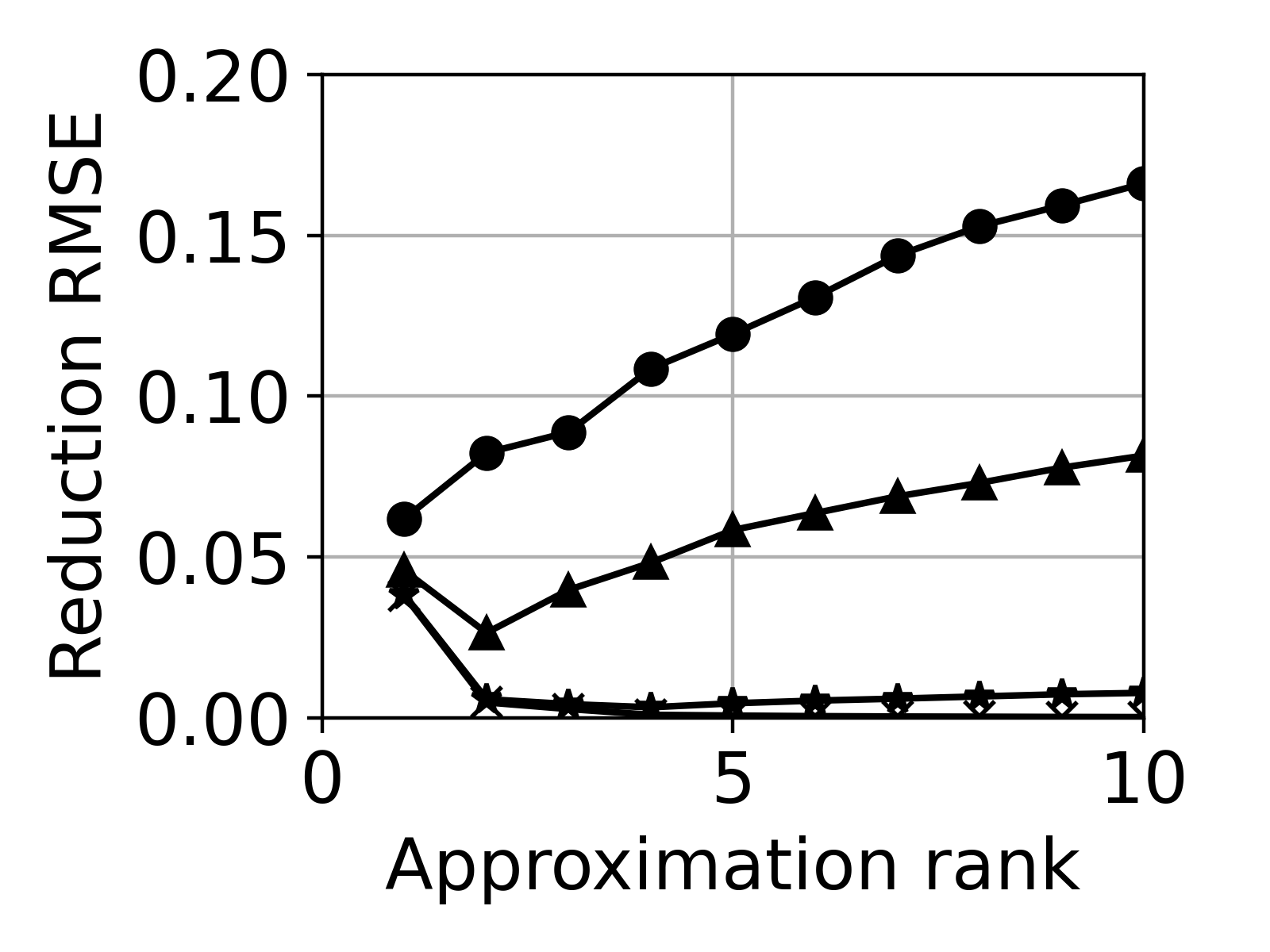}}
  \subfloat[][Learning RMSE]{\includegraphics[trim={0cm 0cm 0cm 0.2cm},clip,scale=0.4]{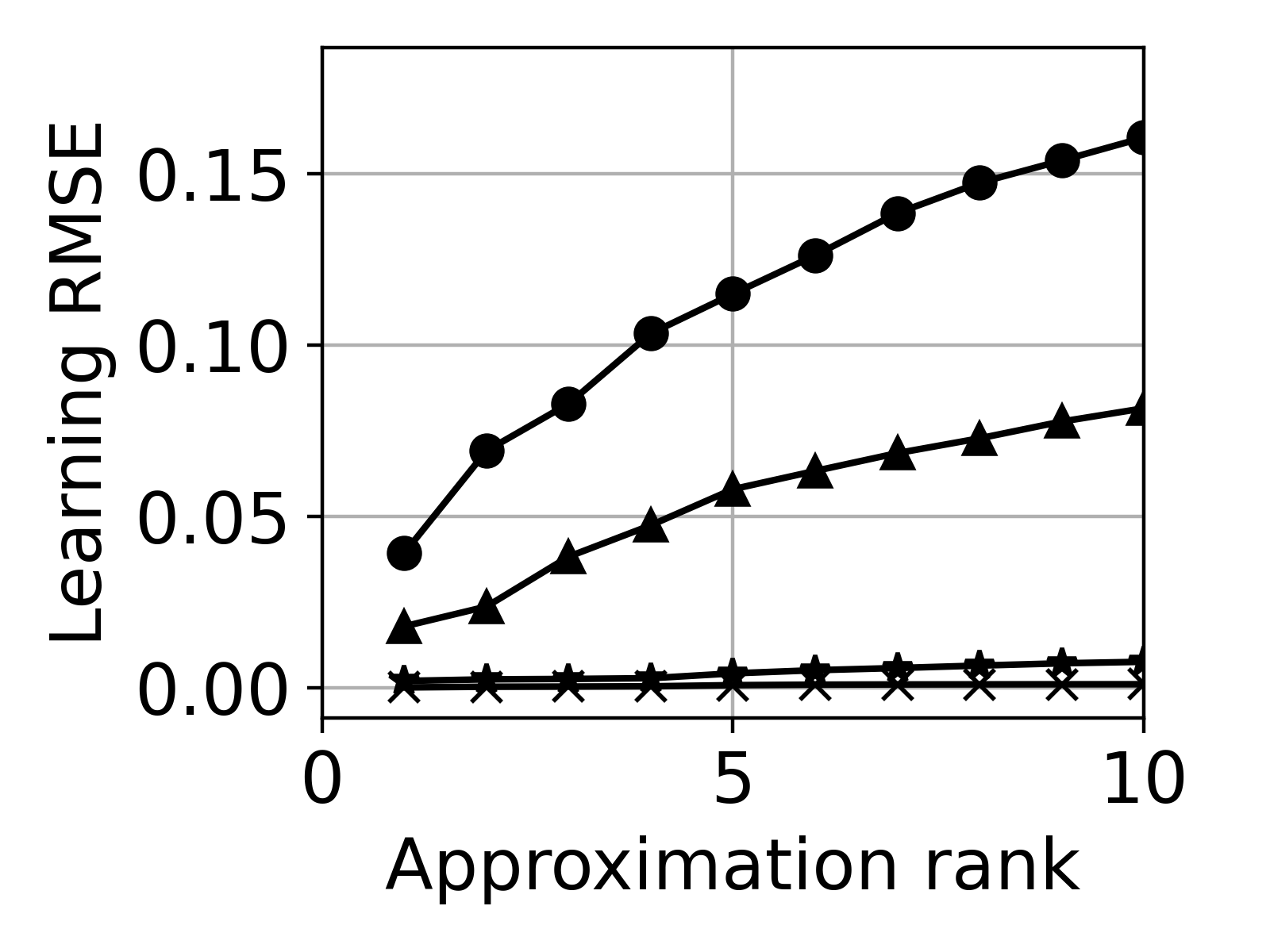}}
  \subfloat[][Prediction RMSE]{\includegraphics[trim={0cm 0cm 0cm 0.2cm},clip,scale=0.4]{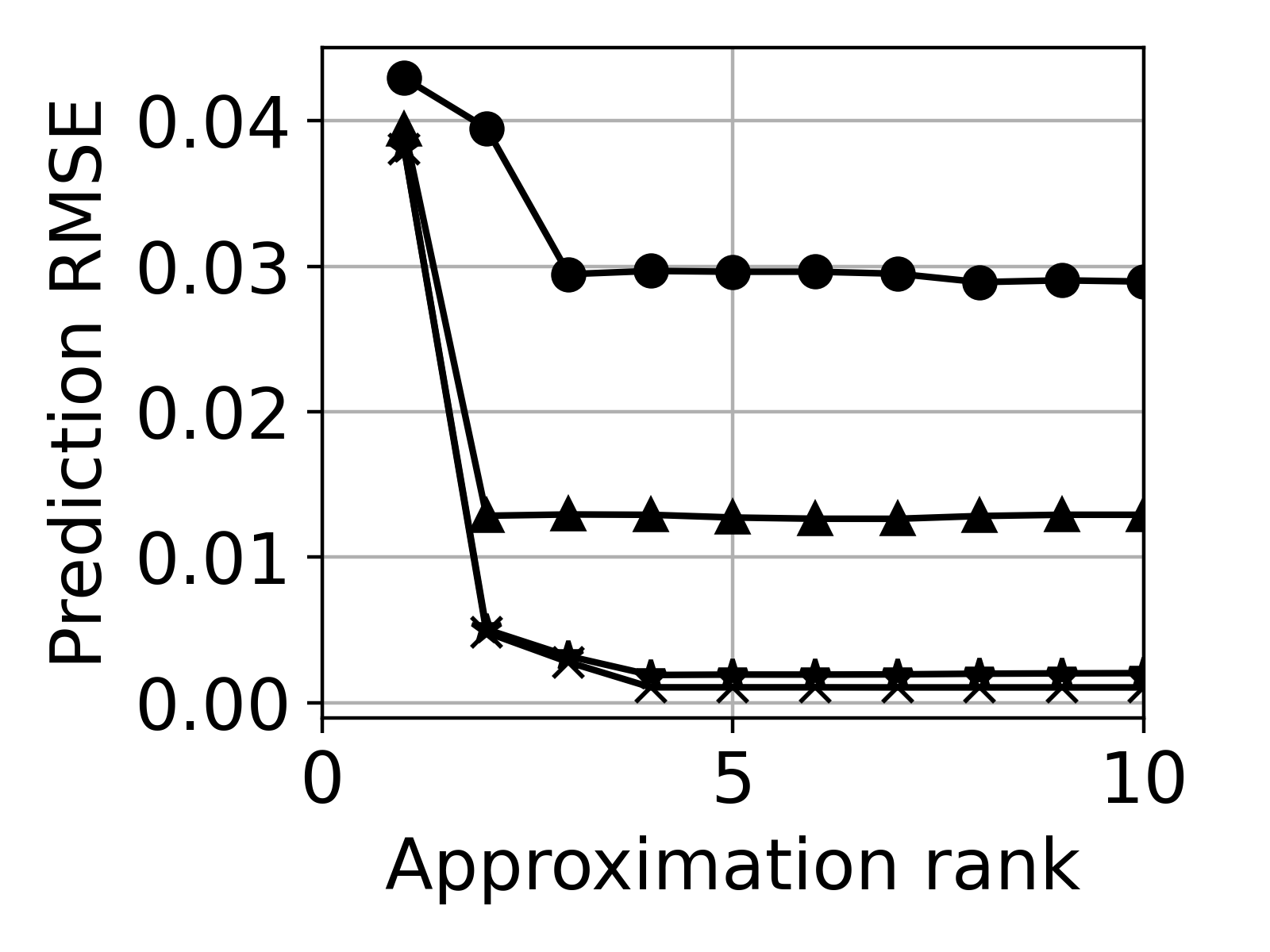}}
    \caption{EVR and POD-PCE steps RMSE with different noise levels added to the aquifer toy problem.}
    \label{fig:toy:POD:noise_comparison}
\end{figure}

\begin{figure}[H]
  \centering
  \subfloat[][Modes 1 and 2]{\includegraphics[trim={0cm 1cm 0cm 0.36cm},clip,scale=0.8]{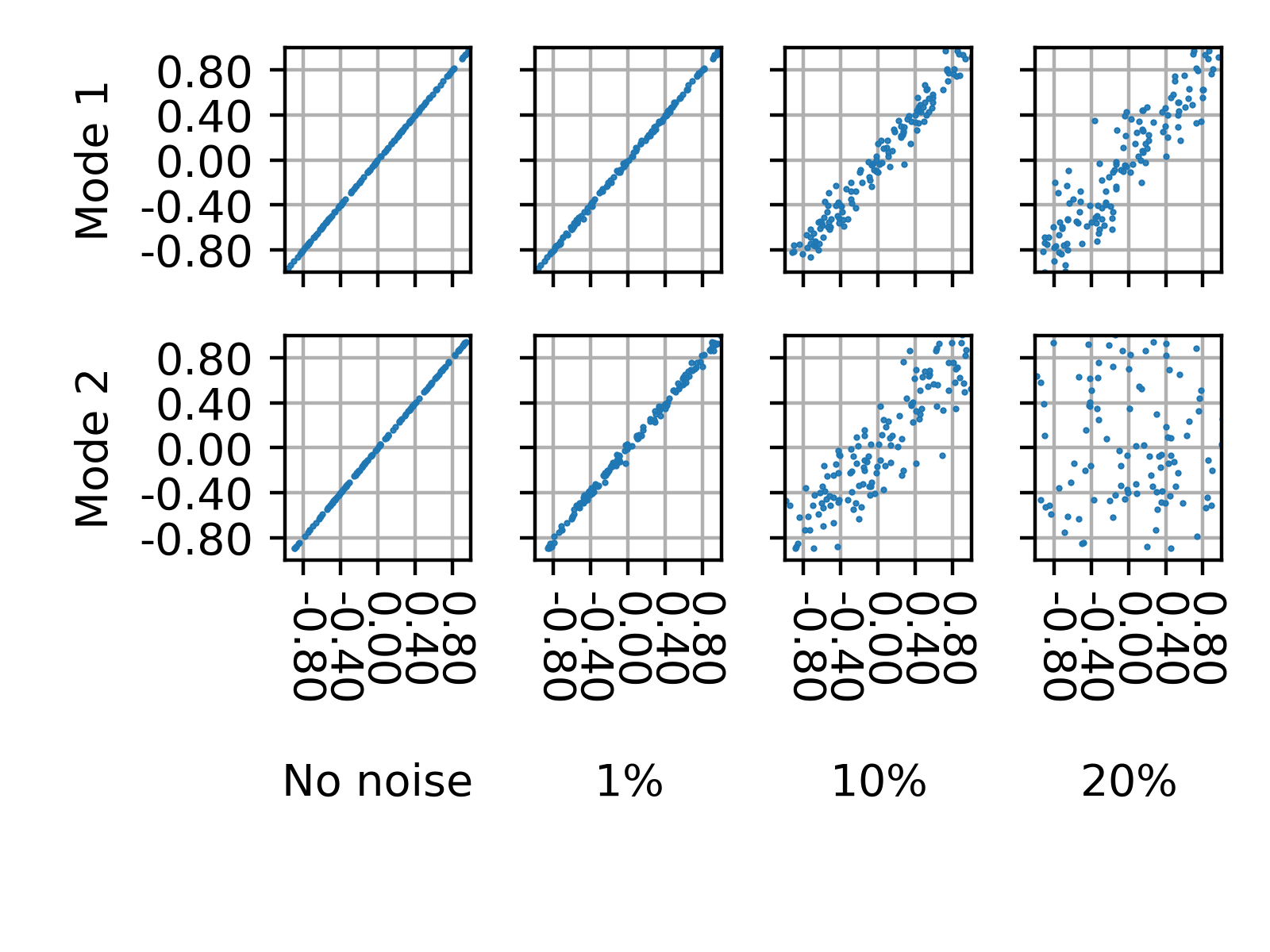}}
	\subfloat[][Modes 3 and 4]{\includegraphics[trim={0cm 1cm 0cm 0.36cm},clip,scale=0.8]{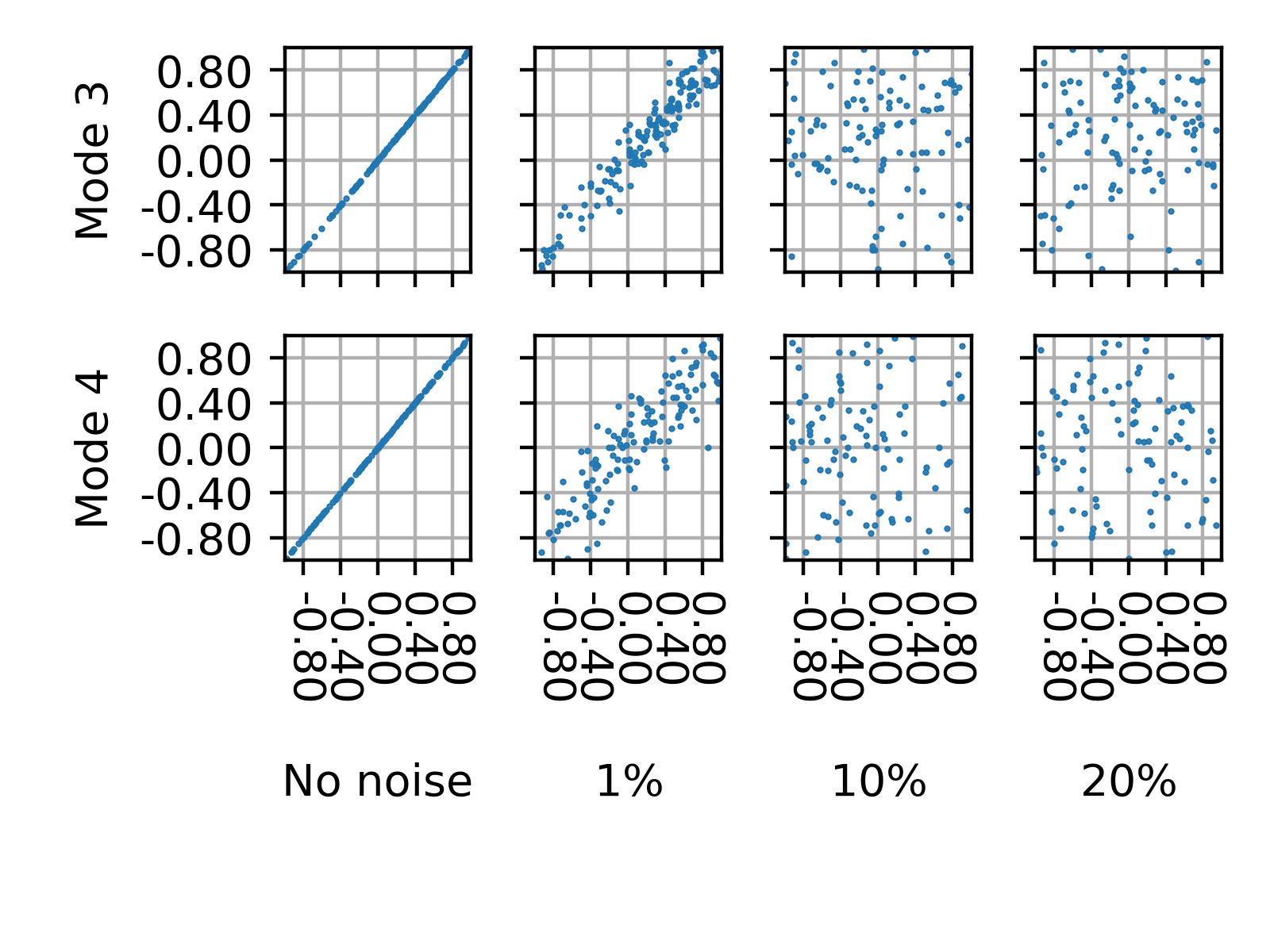}}
    \caption{Original vs. noisy POD modes resulting from $2D$ perturbations. The plotted data are centered and reduced.}
    \label{fig:toy:POD:scatter_noise}
\end{figure}

The EVR are compared for the different noises in Figure \ref{fig:toy:POD:noise_comparison}-a. The represented variance is smaller with the same approximation rank for the noisiest data. This is natural because the variance of the random Gaussian noise is added, and the highest the noise level, the more it is statistically important. Hence, the POD modes are either impacted with random dispersion, or are totally random. A scatter plot, where the modes at different noise levels are plotted against the original modes without noise, is shown in Figure \ref{fig:toy:POD:scatter_noise}, where dispersion is clearly visible.

The perturbed data are then used to evaluate the POD-PCE methodology. The reduction, learning and prediction average RMSE are calculated for each approximation rank as previously described in Section \ref{subsection:toy:learning}. The results are shown in Figures \ref{fig:toy:POD:noise_comparison}-b, \ref{fig:toy:POD:noise_comparison}-c and \ref{fig:toy:POD:noise_comparison}-d respectively.  A difference in the calculation should be noted however: for the reduction and prediction steps, the RMSE are evaluated between the original field (without noise), and the approximation resulting from the noisy field. For reduction for example, if the original field is denoted $\mathbf{a}'(x,y,\boldsymbol{\Theta})$ and the noisy field is denoted $\mathbf{b}'(x,y,\boldsymbol{\Theta})$, the POD approximation $\sum_{k=1}^{d} a_k\boldsymbol{\Phi}_k(x,y)$ is deduced from realizations of $\mathbf{b}'(x,y,\boldsymbol{\Theta})$, but the RMSE is calculated between $\sum_{k=1}^{d}a_k\boldsymbol{\Phi}_k(x,y)$ and the original field $\mathbf{a}'(x,y,\boldsymbol{\Theta})$ that represents the "truth". \\

Firstly, it can be noticed in Figure \ref{fig:toy:POD:noise_comparison}-b that the higher the noise, the more difficult the reduction. This is coherent with the previous EVR analysis. Additionally, while reduction error decreases with the mode number up to $1\%$ of noise, it may increase with the mode number for higher noise levels. Indeed, when noise perturbs the modes, adding them to the approximation may move the resulting field away from the "truth" (original field without noise). Secondly, Figure \ref{fig:toy:POD:noise_comparison}-c shows that learning is more difficult with noisy data. In fact, if the higher rank modes are purely random, then it is impossible for PCE to provide a causal model from the inputs. If a given mode contains a physical information and a random perturbation at the same time (dispersion in Figure \ref{fig:toy:POD:scatter_noise}), PCE may succeed in capturing the physical dependencies, and is shown to be robust up to $30\%$ of noise in \cite{Torre2019}. However, in both cases, the PCE expansion does not represent pure randomness and the learning error naturally increases with the noise level. 
Lastly, for the prediction RMSE (Figure \ref{fig:toy:POD:noise_comparison}-d), the $1\%$ and $10\%$ noisy data perform similar to unperturbed ones, where prediction error follows the same decreasing trend as reduction error, then stabilizes to a minimal value where learning is not interesting anymore. Conversely, for the $20\%$ noise level, prediction error decreases from Mode 1 to 3 although reduction error increases. This can be explained by the fact that PCE succeeds in detecting the physical patterns (explainable with inputs), while eliminating the noise from the approximation. Adding the PCE models to the POD-PCE prediction, contribute in the constitution of realistic physical fields (prediction error decrease), while directly adding the noisy POD modes moves the approximation away from the "truth" (reduction error increase). \\ 

As a conclusion, even with a maximum of $20\%$ of added noise, a POD-PCE model of rank 1 that could be considered as the simplest approximation, does not exceed an average of $5\%$ RMSE compared to the "truth" for the amplitude prediction. The most optimal model in this case decreases to $3 \%$ of RMSE. 

\subsection{Summary of the POD-PCE ML performances on a toy problem}
\label{subsection:toy:summary}

The proposed POD-PCE ML approach was applied to this toy problem for different motivations. Firstly, the coherence of physical interpretation needed to be confronted to reality for validation, which is here possible due to the availability of analytical solution. Secondly, demonstrating the proposed ML capacity on a parametric problem is complementary with the application to a temporal problem, as in Section \ref{section:application}, and allows at the same time clarifying the learning steps on a simpler case. Lastly, assessing the robustness of the methodology to noise was a capital question to investigate, and here made possible by adding artificial perturbations of different levels to the data. \\

Firstly, the POD spatial patterns were interpretable, and the associated coefficients show dependencies to the control parameters. The physical analysis was completed using PCE and inputs ranking indicators (GW and GGW). For example, it is noticed that while the wave number in the estuary, denoted $\kappa_{ei}$, has no influence on the amplitude distribution, it controls however the time-lag in the aquifer (see \hyperref[Appendix:A]{Appendix A}). This completely makes sense regarding the analytical formula. Next, the errors at the different POD-PCE algorithm steps were analyzed. While a gain in accuracy is established by increasing the POD modes number in the reduction phase, the learning error using PCE is inversely increased with model complexity. However, only a small number of modes is necessary for an accurate prediction, as the average RMSE for the optimal 4-Modes model is around $0.1\%$, reaching a local maximum of $0.7\%$ for the amplitude. Lastly, robustness to noise was tested using different perturbation levels. The prediction's average RMSE settles around $3\%$ even for a noise level of $20\%$. PCE with LARS is here of particular interest, as it allows incorporating physical dependencies in the model, while ignoring random perturbations. This assures that the method is trustworthy for an application to a purely measurement-based set-up, as in the following Section \ref{section:application}.


\shorthandoff{:}

\section{Application to a measurement based temporal problem}
\label{section:application}
The POD-PCE ML properties are now investigated on a temporal problem. In particular, the approach is tested on field measurements, introduced with an industrial study case and inherent challenges. As is the case in many measurement based problems, noise can occur due to device errors, and the problem is characterized with data paucity. The noise problematic was treated in Section \ref{section:toy} on the parametric problem. However, the data paucity was not an issue, and supplementary tests are here necessary. They consist in the evaluation of learning choices (selected inputs, marginals, polynomial basis) and thorough analysis of the statistical convergence at different learning stages. The latter is of capital importance to demonstrate the trustworthiness of the physical analysis, as no analytical model is available to confront the conclusions. In particular, convergence as a function of the training set size, with associated confidence intervals, is shown in Subsection \ref{subsubsection:application:learning:POD} for POD and in Subsection \ref{subsubsection:application:learning:PCE} for PCE. Additionally, different learning set-ups, with different marginals choices and input configurations are compared in Subsection \ref{subsubsection:application:learning:PCE}. Lastly, random selection of the training members allows presenting probability distributions of the GW indicators in Subsection \ref{subsubsection:application:learning:PCE}. This helps demonstrating the robustness of physical interpretation to the learning set selection. \\

The physics, data and industrial context are described in Subsection \ref{subsection:application:case}. Subsection \ref{subsection:application:learning} deals with application of the POD-PCE learning phase to the data and assessment of accuracy and robustness with respect to the numerical choices (data set, inputs, marginals and polynomial basis). Finally, the prediction phase using POD-PCE is dealt with in Subsection \ref{subsection:application:prediction}, and the ability of the proposed ML to predict mean quantities and spatial details is demonstrated.

\subsection{Study case}
\label{subsection:application:case}
Sedimentation processes in nearshore areas can be responsible for the excessive sediment deposition commonly observed in cooling water intakes in power plants. As a result, the carrying capacity of the water intake can be drastically reduced, by decreasing its effective area of transport~\citep{Sruthi2017}. Cooling water intakes usually incorporate jetties, of which the angle with the shoreline and position relative to the direction of the net longshore sediment transport influence the amount of sediments diverted into the channel inlet by waves and tidal currents. Jetties also reduces littoral drift, resulting in localized sediment accretion against the shore-normal structure due to the longshore sediment transport being trapped by the jetty~\citep{Dean2004}. In addition, a return current is prone to develop, in the form of a swirling vortex at the end of the structure, and can induce sediment deposition in the vicinity of the channel entrance, consequently affecting the amount of sediments delivered into the cooling water intake~\citep{Costa2015}. Consequently, effective water intake management involves frequent dredging, with high operational costs and usually hindered by a tight schedule. It is then necessary to assess intake sedimentation under different natural forcing and plant operation scenarios in order to optimize dredging operations to help mitigate the potentially adverse impact of the waves, tidal currents, and meteorological forcing combined with plant functioning. 

\subsubsection*{Site characteristics}
\begin{figure}[H]
  \centering
    \includegraphics[trim={0cm 0cm 0cm 0cm},clip,scale=0.11]{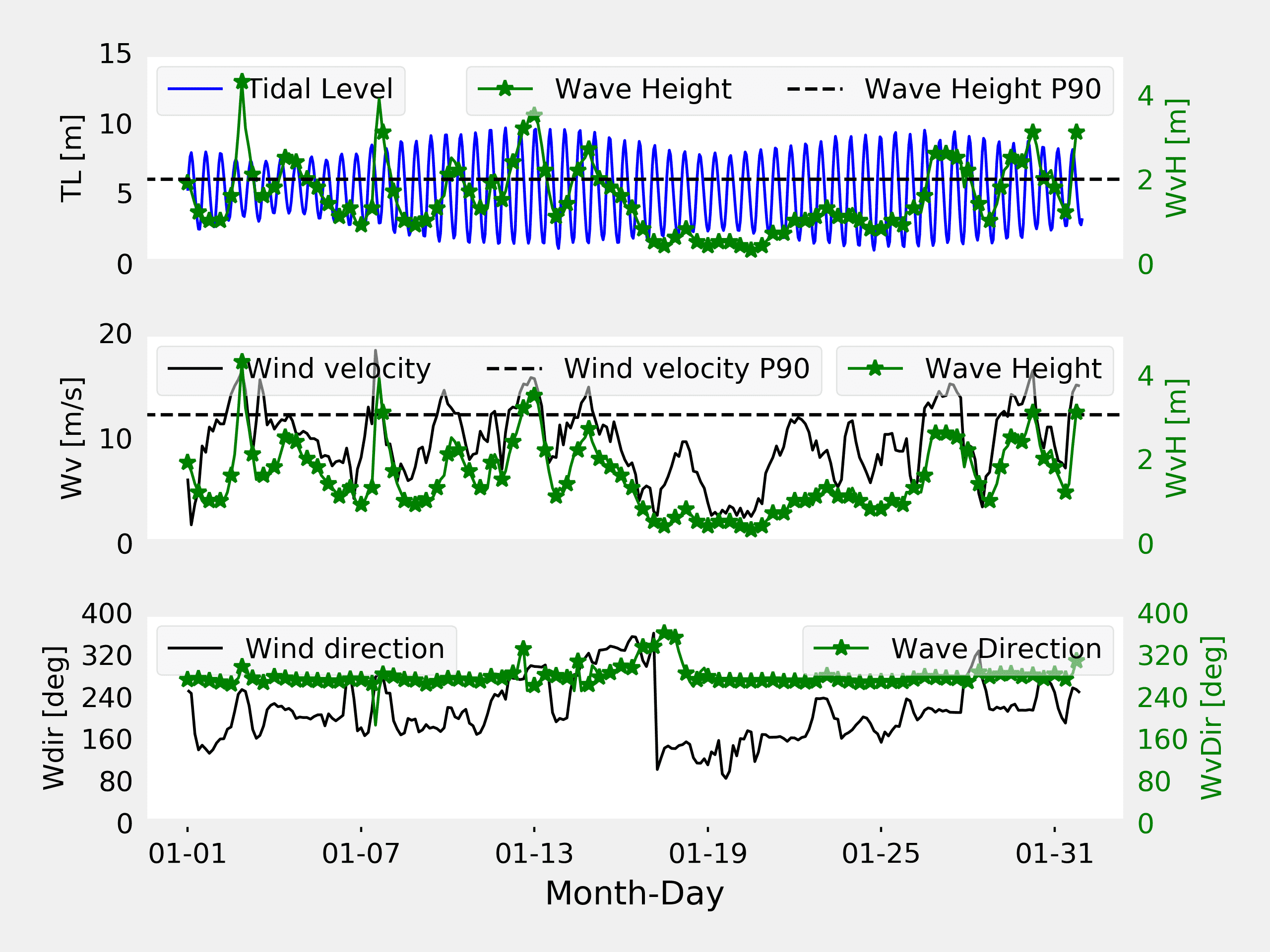}
    \caption{Measurements of Tidal Level ($TL$), Wind velocity ($Wv$), Wind direction ($Wdir$), Wave Height ($WvH$) and Wave Direction ($WvD$) on January 2016. (P: Percentile).}
    \label{fig:case:monthData}
\end{figure}

The study site is located on the eastern English Channel coast in northern France. Tide in the study zone is classified as mega-tidal and is dominated by semi-diurnal circulation, with low-tide water depth of $10-15$ m, and a mean tidal range of approximately 8.5 m, reaching 10 m during the spring tide~\citep{LeBot2010}. Hydrodynamics are influenced by asymmetrical current velocities, with flood and ebb currents in the E-NE and W-SW directions, respectively. Current velocity at $2.2~m$ above seabed vary from $0.70$ to $0.98$ m/s, depending on flood/ebb phase, respectively~\citep{Michel2017}. Wave activity in this open exposed environment is moderate, with significant annual and decennial wave height of 3.8 m and 4.7 m, respectively, with maximum values of $4.2-5.8$ m, averaged period of $7-9$ s and a predominant W direction. Orbital velocities measured during the spring-tide period ranges between 0.5-1.3 m/s. An example of tidal levels, wind direction and velocity and wave height and direction in January 2016 is shown in Figure~\ref{fig:case:monthData}~. In the study area, bed sediment varies from medium to fine silted sands, with a morphology characterized by the presence of mega-rides parallel to the coast. In this zone, rock occupies less than 4\% of bed surface~\citep{Costa2015}. \\

\subsubsection*{Data}
Hydrodynamic and meteorological information comprise wave and wind variables, provided by the VAG prediction model of the sea state \citep{Guillaume1987}, using retrospective 3-hourly simulations between 2009 and 2018. Tidal water levels were obtained from the SHOM-REFMAR tide gauge station located in the vicinity of the study zone, with hourly survey frequency~\citep{Refmar}. Bathymetric measurements were available from Single-Beam Echo Sounding on 39 cross-sectional profiles of intake measured at 25~m intervals, collected fortnightly between 2005 and 2018. Mean profiles were 100~m long with 0.5~m spatial resolution of bathymetric data. Additional information such as the daily coolant flow rates, and channel dredging volumes and frequency, were provided by the plant operator.\\

The available measurements of the forcings did not have the same frequencies. One solution to homogenize frequencies consists in reducing the measured data to representative statistics over the sedimentation interval $\Delta t \approx 15 \ days$ separating two bed elevations measurements. Hence, the following statistics were used: \begin{itemize}
\item[$\bullet$] \textit{Tidal level indicators:} average low tide ($TLmean$), minimum low tide ($TLmin$), maximum tidal range ($TLrange$) and standard deviation ($TLstd$);
\item[$\bullet$] \textit{Wind indicators:} average wind velocity ($Wmean$) and average direction weighted by velocity ($Wdir$);
\item[$\bullet$] \textit{Wave indicators:} average wave height ($WvH$), standard deviation ($Wvstd$), average wave period and average wave direction weighted by height (\textit{resp.} $Wvper$ and $Wvdir$), average wave height exceeding the $90^{th}$ percentile (arbitrary storm indicator, $Wv2m$) and percentage occurence ($Wv2m\%$);
\item[$\bullet$] \textit{Operational indicators:} average pumping flowrate (Qmean); time lapse since last dredging ($Dp$), and last dredged volume ($Dv$). \\
\end{itemize}

These statistical indicators were calculated for each sedimentation interval, and may be characterized with correlations. For example, a positive correlation was noted between mean low tide $TLmean$ and wave parameters $Wvper$ and $WvH$. Mean wave periods $Wvper$ and mean wave heights $WvH$ were also positively correlated. \\

For the learning part, the data overlapped only over a limited period. A maximum of $60$ measurements could therefore be used, with up to $15$ forcing variables. Obviously, this "small data" configuration is a considerable handicap for the dimension of the problem, especially given that the variable of interest is a two-dimensional bathymetric field. However, permanent intake monitoring ensures that the data set will always grow and can be used to update the learning. This limitation shall not prevent testing the accuracy of the methodology on small sets such as are often encountered in physical applications, as attempted below, where learning and prediction using POD-PCE is applied to the described data. For the learning algorithm, input variables are needed, corresponding to the reduced statistical indicators described above, and denoted $(\theta_1,...,\theta_V)$, where $V$ is the supposed dimension of the problem.


\shorthandoff{:}

\subsection{Measurement-based learning of a physical field using POD and PCE}
\label{subsection:application:learning}
This section concerns learning the spatio-temporal bathymetric field using POD and PCE independently. The POD modes are extracted in Subsection \ref{subsubsection:application:learning:POD} and the temporal patterns are learned from the forcing parameters using PCE in Subsection \ref{subsubsection:application:learning:PCE}. Throughout this investigation, particular attention is given to the convergence of the learning and to its robustness with respect to the numerical choices. Trusted POD-PCE learning is immediately used for physical interpretation and the most important physical insights are summarized in Subsection \ref{subsubsection:application:learning:summary}.

\subsubsection{Physical analysis and data reduction using POD}
\label{subsubsection:application:learning:POD}
\begin{figure}[H]
  \centering
  \vspace{-0.4cm}
  \includegraphics[trim={0cm 0.2cm 0cm 0.5cm},clip,scale=0.09]{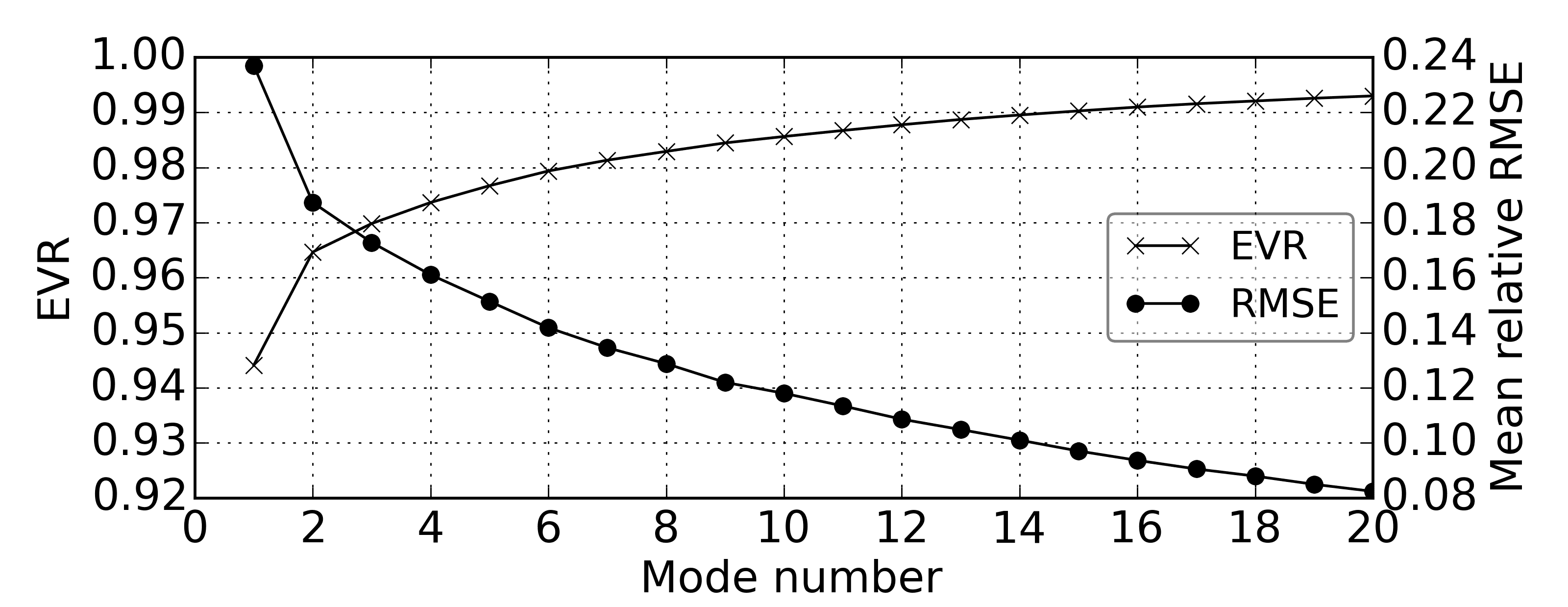}
    \caption{Evolution of the EVR and mean relative RMSE with mode number for the POD applied to the intake bathymetries.}
    \label{fig:POD:SR:RIC}
\end{figure}

\begin{figure}[H]
  \centering
  \vspace{-0.5cm}
    \subfloat[][Coefficients $a_1(t)$ and $a_2(t)$]{\includegraphics[trim={0cm 0cm 0cm 1.5cm},clip,scale=0.07]{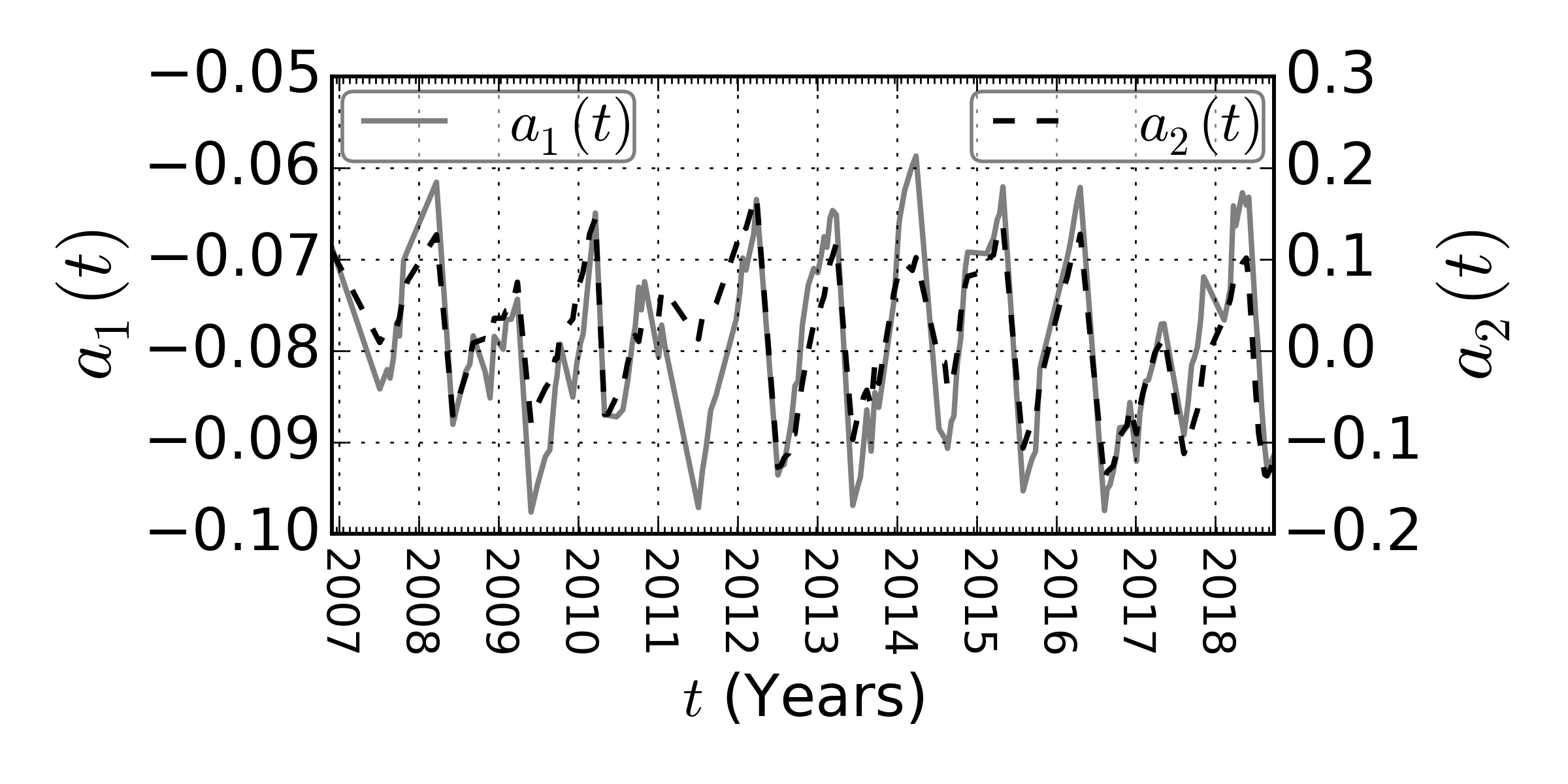}}
    \subfloat[][Coefficients $a_3(t)$ and $a_3(t)$]{\includegraphics[trim={0cm 0cm 0cm 1.5cm},clip,scale=0.07]{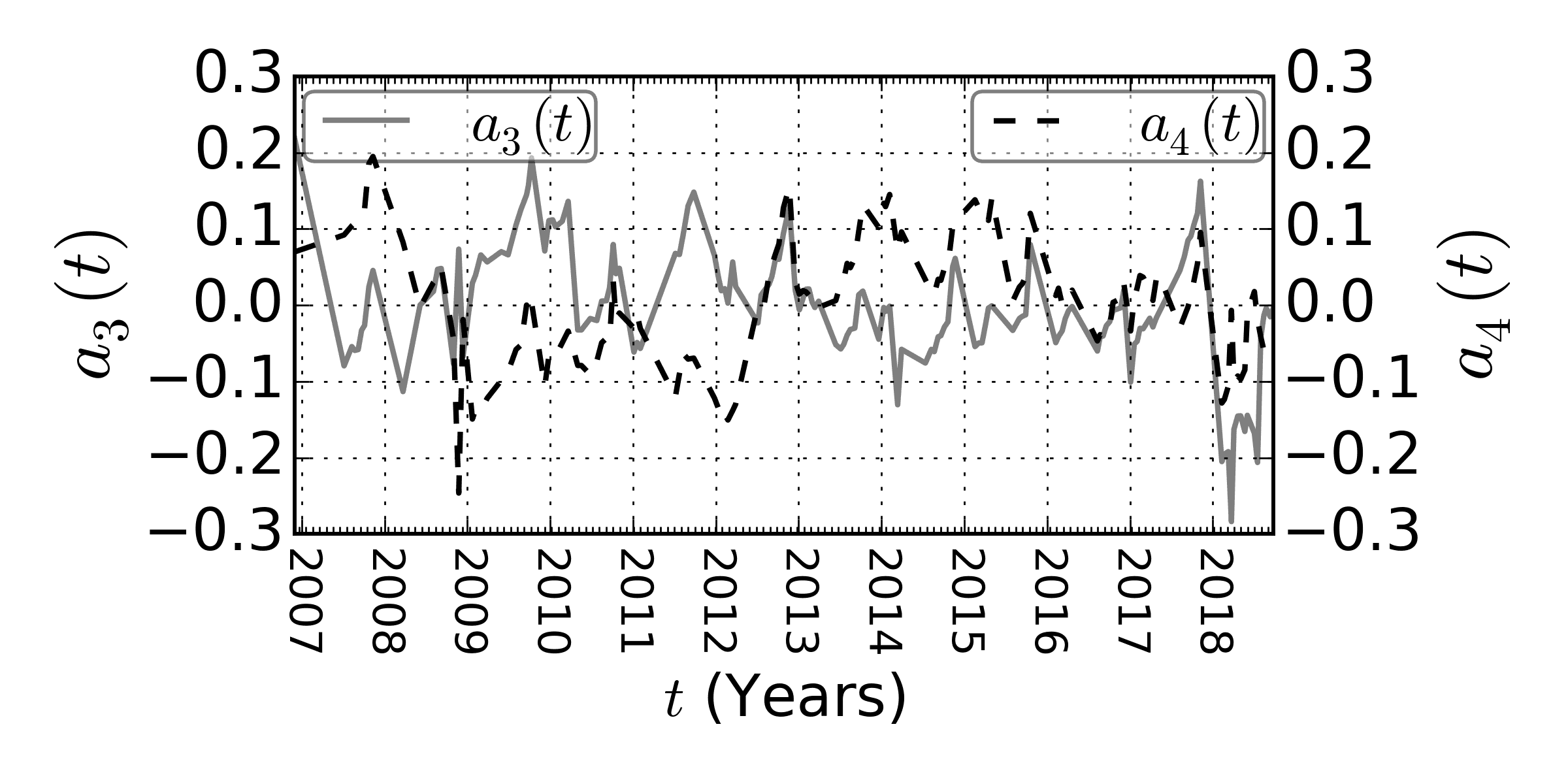}}
    \caption{The first four temporal coefficients of the POD applied to intake bathymetries.}
    \label{fig:POD:SR:temporal}
\end{figure}

First, POD was applied on the bathymetry measurements. The aim was to identify morphodynamic patterns so as to better understand the sediment deposition inside the channel, and to characterize variations in depositions with the external forcing variables. After setting aside poor-quality measurements (e.g.~incomplete bathymetries), a total $n=156$ realizations were used. The bathymetry points were sonar boat measurements on $m_p=39$ cross-sections inside the intake. Linear interpolation was performed on $m_i=100$ fixed points for each profile, in order to express all measurements on the same grid, giving a total $m=m_i \times m_p = 3,900$ spatial points. The interpolated realizations were then stored in a snapshot matrix $\boldsymbol{Z}(\boldsymbol{\mathcal{X}},\mathcal{T}) = [z(x_i,t_j)]_{i,j} \in \mathbb{R}^{m \times n}$ and POD-processed as explained in Section \ref{subsection:theory:POD}. The EVR defined in Equation \ref{eq:POD:RIC} and the mean relative RMSE between the POD approximation and the complete measurement (averaged over the realization set as in Equation \ref{eq:timeAveragedRelativeRMSE}~) were calculated for each POD approximation rank and are plotted in Figure \ref{fig:POD:SR:RIC}.
The first pattern represents over 94\% of the variance, and explains most of the variation in dynamics. The variance percentage reached 99\% at rank 14, where the mean error was slightly over 10\%, decreasing to 8\% at rank 20. Dimensionality reduction is therefore a realistic option for this specific dynamic problem. 
\begin{figure}[H]
  \centering
  \vspace{-1cm}  
    \subfloat[][Mode $\Phi_1(x)$]{\includegraphics[trim={20cm 10cm 20cm 10cm},clip,scale=0.04]{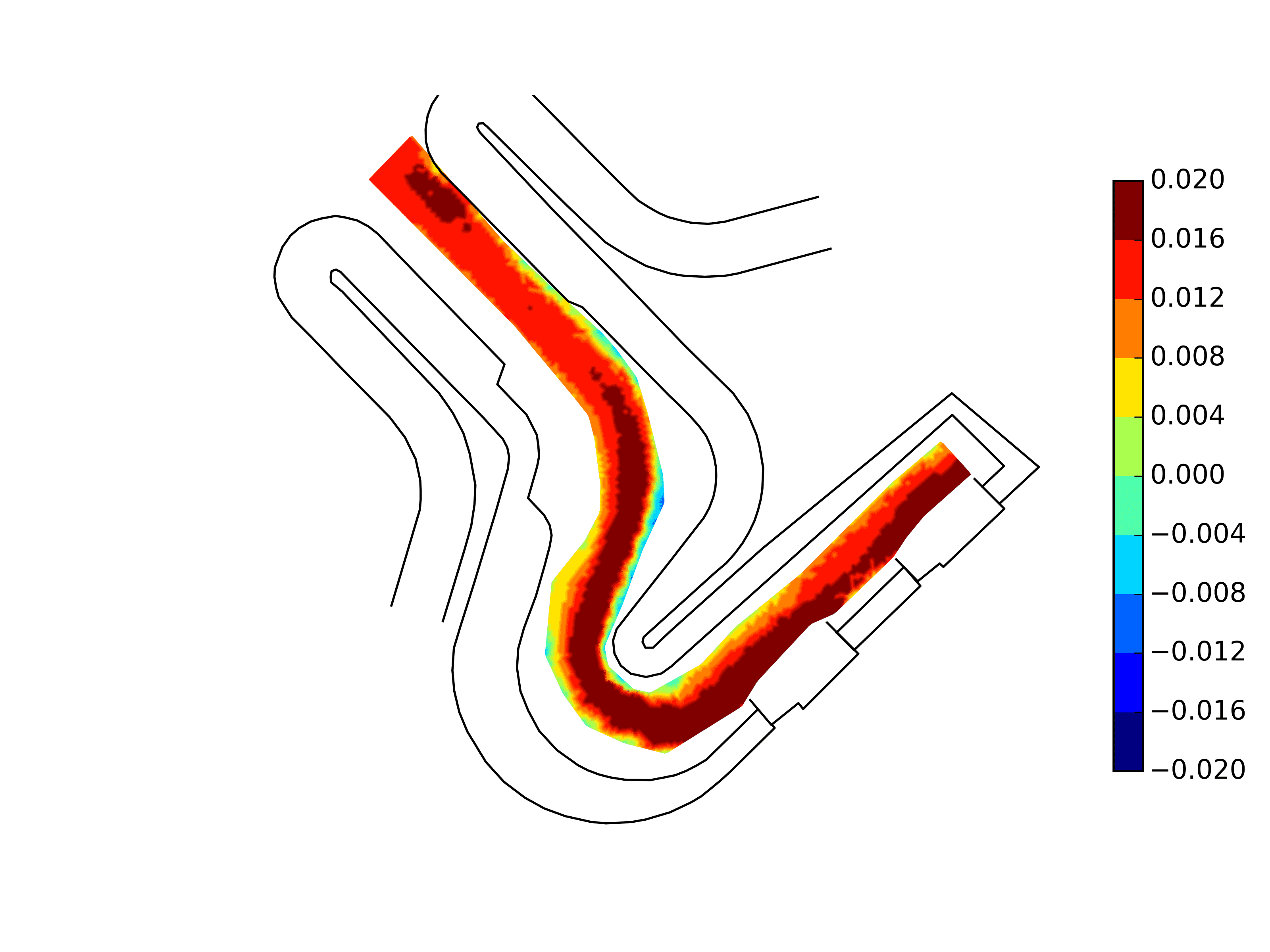}} 
    \subfloat[][Mode $\Phi_2(x)$]{\includegraphics[trim={20cm 10cm 20cm 10cm},clip,scale=0.04]{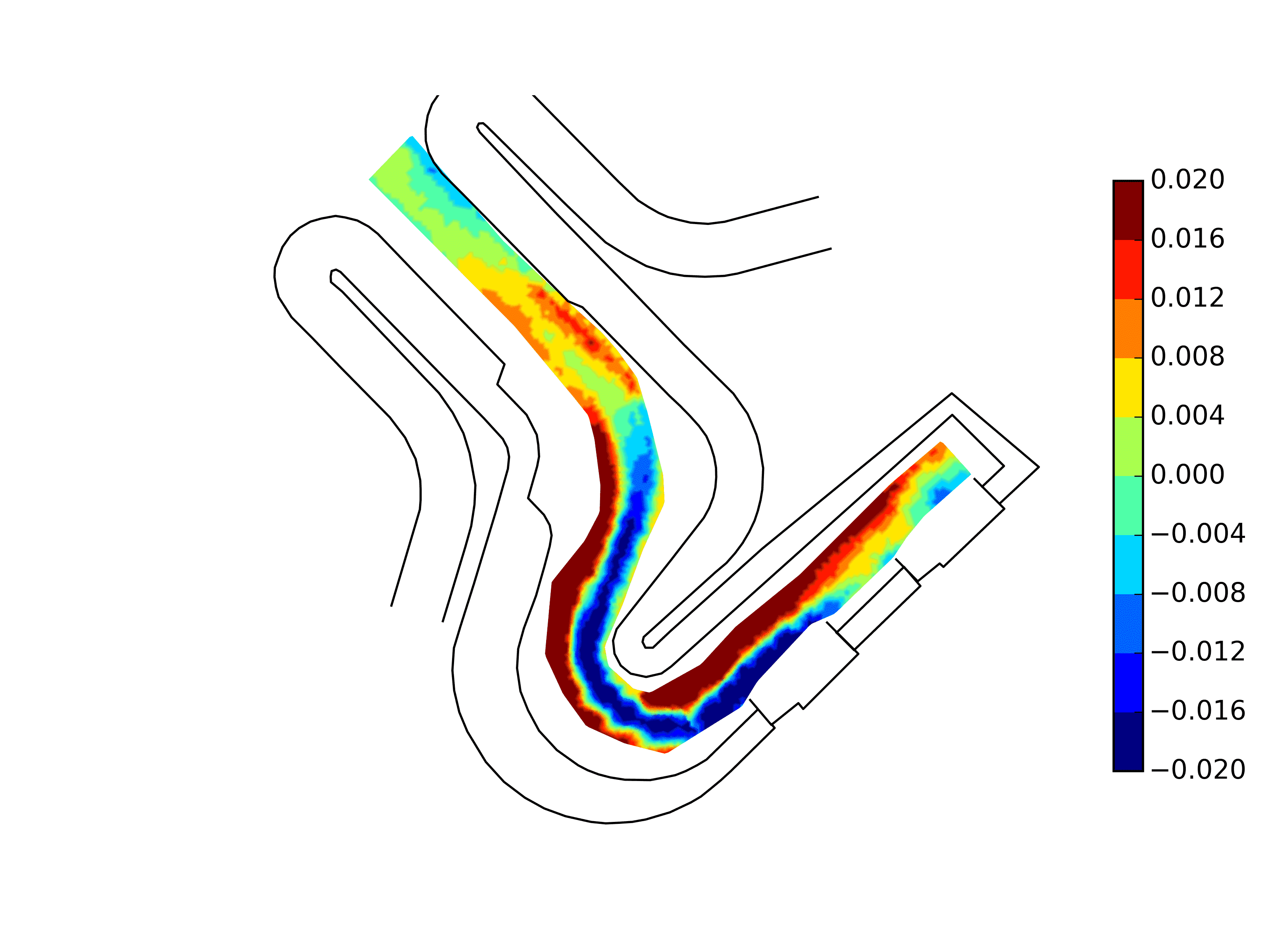}}
    \subfloat[][Mode $\Phi_3(x)$]{\includegraphics[trim={20cm 10cm 20cm 10cm},clip,scale=0.04]{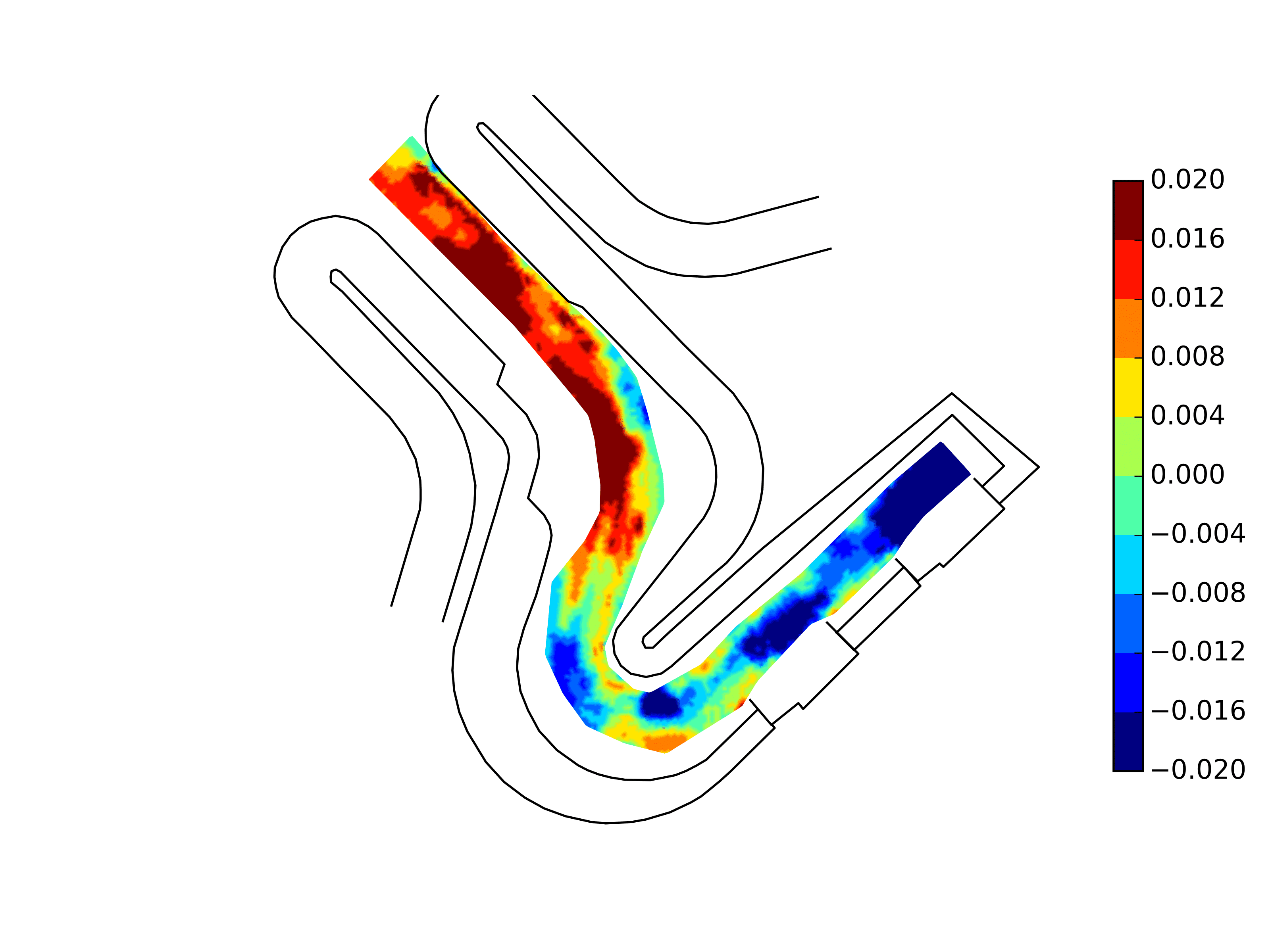}}
    \subfloat[][Mode $\Phi_4(x)$]{\includegraphics[trim={20cm 10cm 20cm 10cm},clip,scale=0.04]{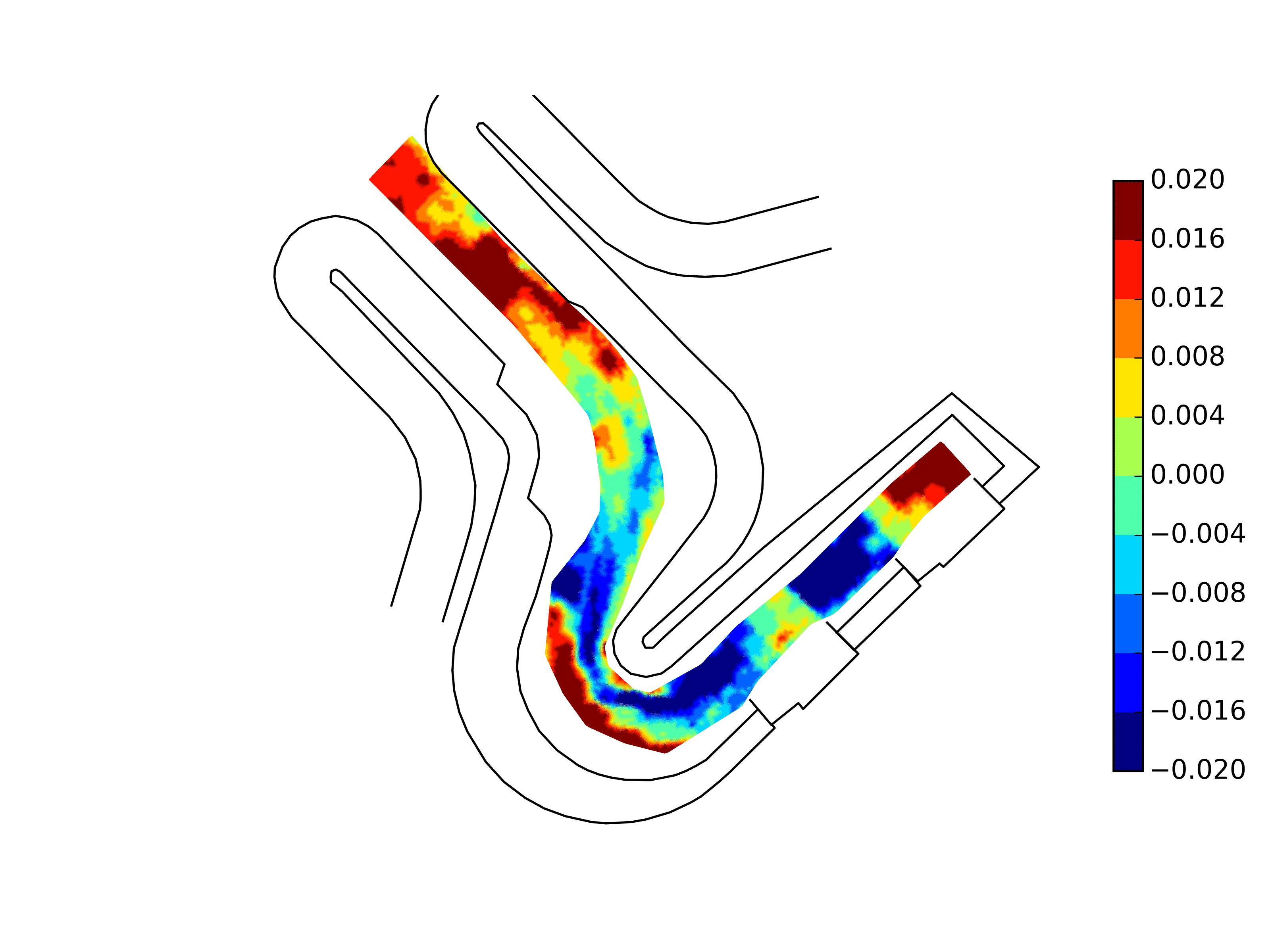}}
    \subfloat[]{\includegraphics[trim={93cm 10cm 0cm 10cm},clip,scale=0.05]{figs/PODMode_4.png}}
    \caption{The first four spatial patterns of the POD applied to intake bathymetries.}
    \label{fig:POD:SR:spatial}
\end{figure}
This encouraged the learning and prediction attempts undertaken in Subsections \ref{subsubsection:application:learning:PCE} and \ref{subsection:application:prediction} respectively. The spatial and temporal components of the first four POD modes corresponding to an EVR higher than $97\%$ are respectively plotted in Figures \ref{fig:POD:SR:spatial} and \ref{fig:POD:SR:temporal}. The first spatial pattern (Figure \ref{fig:POD:SR:spatial}-a) represents the channel's slope. Its temporal coefficient (Figure \ref{fig:POD:SR:temporal}-a) shows regularity in time that is almost periodicity. When it increased, overall sediment deposition in the channel increased, because the difference between the upstream and the downstream bed elevations, and therefore the slope, diminished. The  sediment deposition in the channel might be related to the increasing sediment supply caused by the external forcing influence. Decrease always corresponded to a dredging episode. The apparent periodicity is therefore not natural or seasonal but due to periodicity of operational intervention: sediment deposition in the channel is tolerated up to a certain level and then dredging is always undertaken at a certain point, which corresponds to the maximum of the temporal coefficient. 
The second pattern (Figure \ref{fig:POD:SR:spatial}-b) acts as a geometric distribution function of the sediment deposition. In general, when the first temporal coefficient was maximal, the second coefficient was positive, meaning that the sedimentation mainly occurred in the middle of the first portion of the channel (upstream), on the right bank of the bend and on the left bank in front of the pumps. This spatial distribution can be associated to the internal flow characterized by a velocity distribution inside the channel. In fact, the sediments settle where velocity is the lowest, which is probably the case where the banks appear. The computed sediment deposition and erosional patterns are analogous to those commonly observed in meandering rivers \citep{Janocko2013}. The third pattern (Figure \ref{fig:POD:SR:spatial}-c) shows sediment deposition concentrated in the first portion of the intake, and the fourth pattern (Figure \ref{fig:POD:SR:spatial}-c) emphasizes sediment dynamics, particularly in the downstream part of the channel. This behavior is statistical proof and quantification of finer sediment supply. The finer sediment fraction was transported in suspension and deposited at the end of the intake channel. The temporal coefficients associated with the third and fourth mode (Figure \ref{fig:POD:SR:temporal}-b) were less regular than those of the first and second mode, and seemed to follow a more stochastic dynamic. The peaks may represent unusual sediment supply, probably linked to extreme events (e.g. storms).
\begin{figure}[H]
  \centering
  \vspace{-0.5cm}
    \subfloat[][Mode 1]{\includegraphics[trim={0cm 0cm 0cm 0cm},clip,scale=0.07]{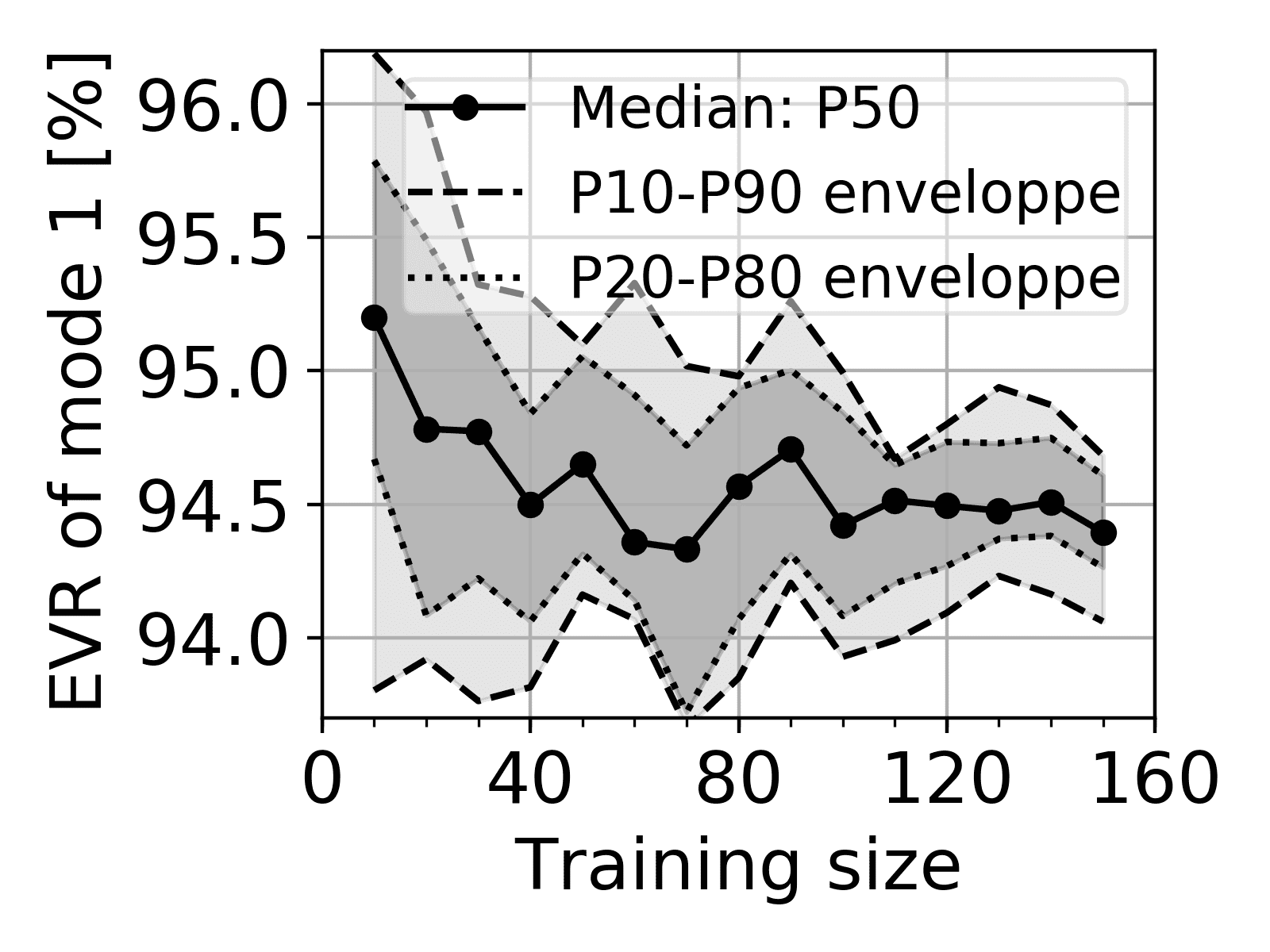}}
    \subfloat[][Mode 2]{\includegraphics[trim={0cm 0cm 0cm 0cm},clip,scale=0.07]{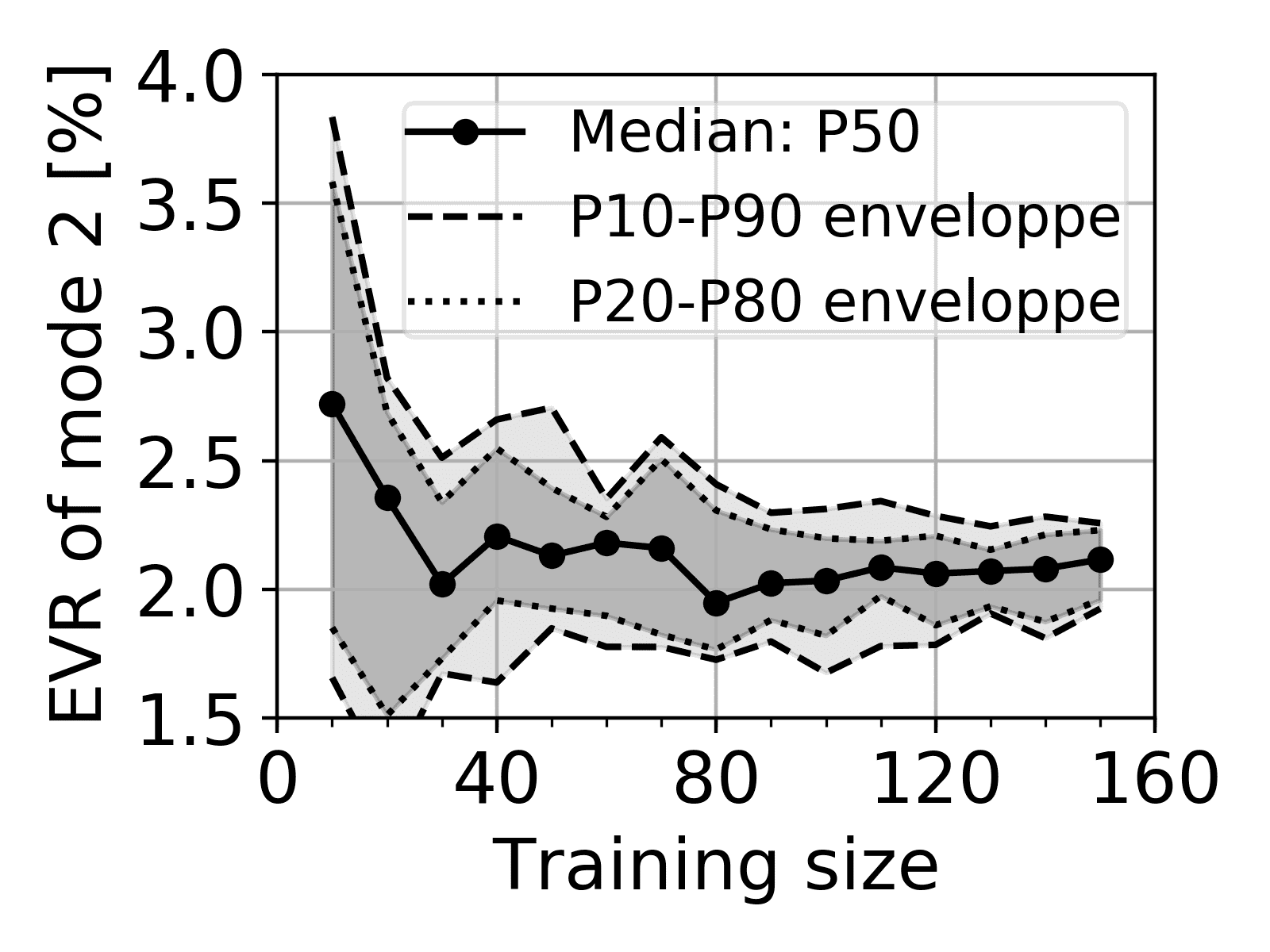}}
    \subfloat[][Mode 3]{\includegraphics[trim={0cm 0cm 0cm 0cm},clip,scale=0.07]{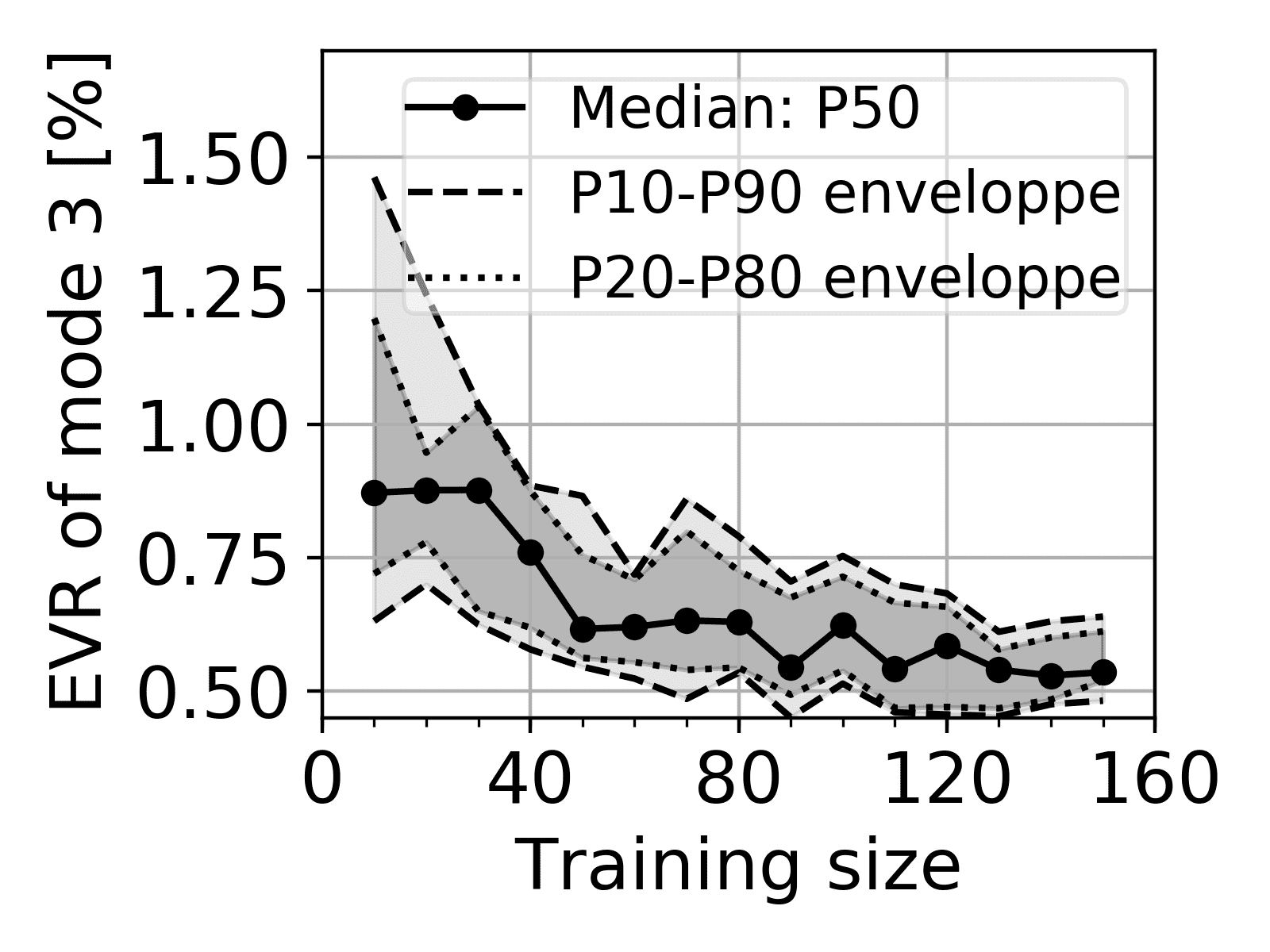}}
    \subfloat[][Mode 4]{\includegraphics[trim={0cm 0cm 0cm 0cm},clip,scale=0.07]{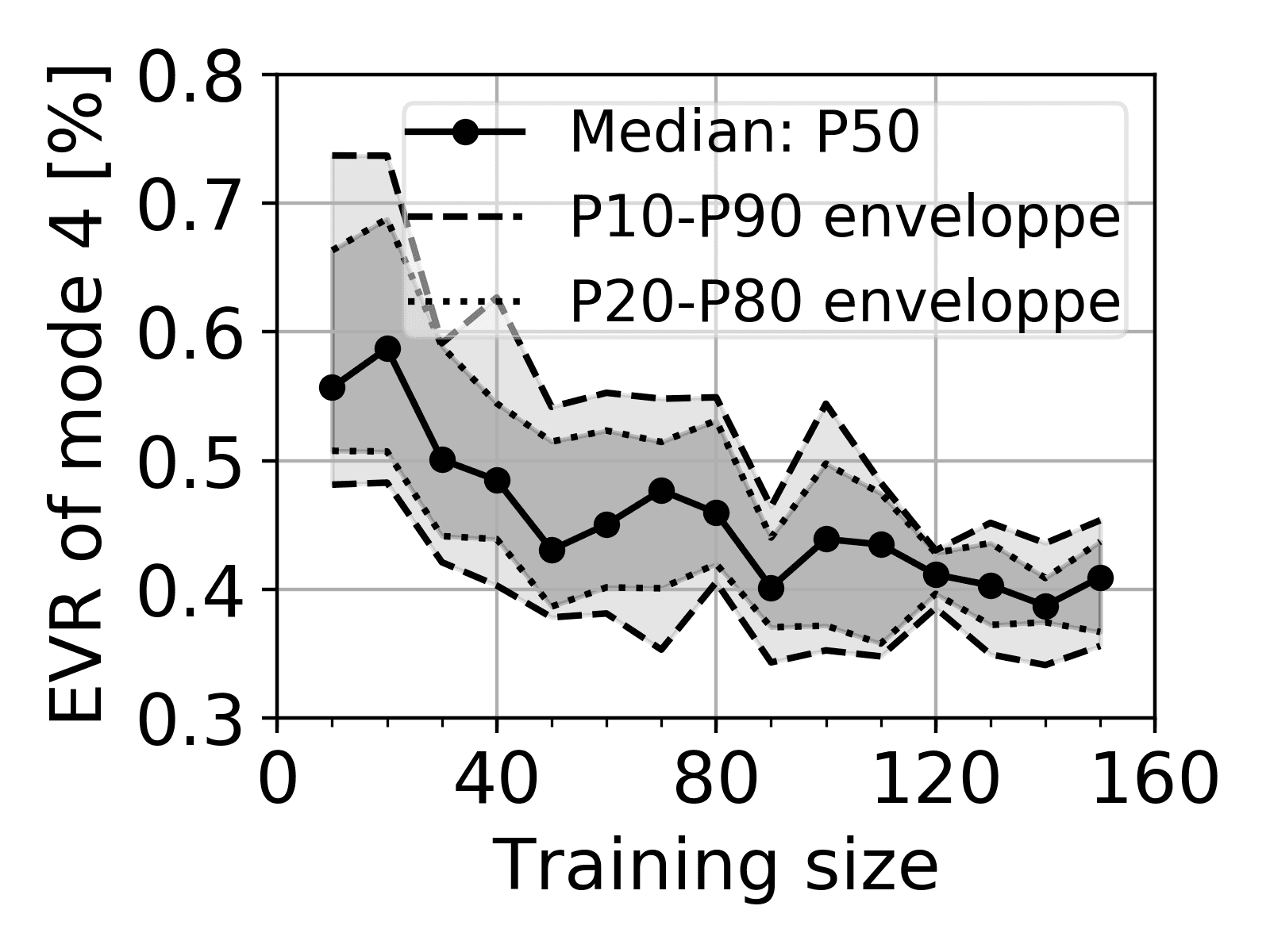}}
    \caption{EVR convergence of the first four bathymetry POD modes, using a bootstrap of size 20. Plots show median values and confidence intervals (P: Percentile).}
    \label{fig:POD:convergence:eV}
\end{figure}
To check the robustness of the statistical conclusions deduced from POD, convergence analysis is necessary. This was performed on the EVR values associated with the first four patterns, using bootstrap analysis \citep{Efron1986}. The results are shown in Figure \ref{fig:POD:convergence:eV}. The convergence of the mean values and the tightening of the confidence intervals around the mean with increasing matrix size are clear for these first four modes. However, whereas the confidence intervals represent at most an error of $\pm 0.6 \%$ around the mean for the first mode, they reached respectively $\pm 12.5\%$, $\pm 25\%$ and $\pm 12.5\%$ for the second, third and fourth modes.\\

The analysis proved that the POD results could be used to pursue the learning. Firstly, a high EVR and low RMSE were associated with a small number of modes, guaranteeing optimal data reduction ($d<<min(m,n)$ as explained in Section \ref{subsection:theory:POD}). The number of POD modes to accurately represent the bathymetry can be chosen accordingly. In the present study, the configuration was $d=11$ modes (discussed in \textit{Step 3}), guaranteeing EVR $\geq 98\%$ and information loss $\leq 12\%$ (mean relative RMSE). Secondly, the EVRs were guaranteed to converge statistically at least for the first four modes, with error of $\pm 0.6\%$ around the mean for the most important mode, representing over $94\%$ of the variance. Thirdly, the deduced patterns were physically coherent. Lastly, more than a decade of evolution was used to extract the POD basis, under variable operational and environmental conditions. As long as the operating conditions of the intake remained unchanged, it can be assumed that a wide range of evolutions has been covered, except for extreme events that rarely occur and that are not specifically treated in this study \citep{Ghil2011}. Hence, the POD basis can be considered as a physically trustworthy and mathematically complete basis to understand past evolutions and to predict future ones. The learning of temporal coefficients is therefore attempted in Subsection \ref{subsubsection:application:learning:PCE}. 

\subsubsection{Learning of the POD patterns using PCE}
\label{subsubsection:application:learning:PCE}
The temporal coefficients calculated with the POD in Section \ref{subsubsection:application:learning:POD} (Figure \ref{fig:POD:SR:temporal})  were learned using PCE (theory in Section \ref{subsection:theory:PCE}). The aim was to learn the way these coefficients evolve over time, as a function of the forcing parameters presented in Section \ref{subsection:application:case}~, with the ultimate objective of field prediction as explained in Section \ref{subsection:theory:methodology} and applied in Section \ref{subsection:application:prediction}. The present section focuses strictly on the learning phase and the physical analysis of the learned model, highlighting quality of learning (robustness, convergence, etc.). \\

The investigation of learning is organized in four steps. \begin{itemize}
\item[$\bullet$ \textit{Step 1 - Sensitivity of learning to inputs and marginals:}] different configurations were tested to practically demonstrate the implications of these choices on the accuracy of fit.
\item[$\bullet$ \textit{Step 2 - Convergence and Robustness of fit:}] The best model resulting from Step 1 was studied more deeply. Its convergence and robustness with respect to the choice of training members are were analyzed.
\item[$\bullet$ \textit{Step 3 - Physical interpretation of the best learned model:}] the best model was chosen, and the most influential forcings were ranked using the \textit{Garson Weights} (GW) and \textit{Generalized Garson Weights} (GGW) presented in Sections \ref{subsection:theory:PCE} and \ref{subsection:theory:methodology} respectively. 
\item[$\bullet$ \textit{Step 4 - Robustness of the physical interpretation with respect to the learning-set members:}] the physical conclusions of the model were shown to be statistically meaningful. \\
\end{itemize}

These steps, in the above order, follow the logic of statistical model construction to build a trustworthy prediction algorithm, used in Section \ref{subsection:application:prediction}.

\paragraph{\textit{Step 1 - Sensitivity of learning to inputs and marginals}}\mbox{}\\
Input variable selection is capital, and marginals must be chosen wisely. Below, we demonstrate the influence of these choices on the performance of the learning. Different configurations were tested. \\

Before introducing the tested configurations, the training steps that are common to all configuration need to be defined. A learning data-set is classically separated into different sub-sets, corresponding to different steps of the learning algorithm. This is commonly referred to as the "Train-Validation-Test Split" \citep{ShalevShwartz2014}: a training set is used for the learning, a validation set is used to check the learning and for further calibration, and a test set is used to assess the prediction capability of the statistical model. However, the data-set used in this study was small: the bathymetry and forcings measurements shown in Subsection \ref{subsection:application:case} overlap for the 2012-2017 period only, leaving 64 sedimentation periods to study. Therefore, only a "Train-Predict" split was performed, where the prediction set played the role of both the test set and validation set. Hence, the numerical choices were calibrated on the training set, and validated on the prediction set for both statistical accuracy and physical prediction. The learning was then performed with an arbitrary choice of training-set size at $50$, which left a prediction set of $14$. The training data were chosen in chronological order (first $50$ records), to mimic the learning process in an industrial context. This arbitrary training-set choice had consequences for learning; the sensitivity of learning to training set choice was investigated (\textit{Step 2}). All the model choices presented below (choice of inputs and marginals) were assessed on this training configuration. To assure that comparison is made between models at their best performances, the PCE polynomial degree was optimized for each separately. Degrees from 1 to 7 were tested, and the associated relative empirical errors on the training and prediction sets, respectively $\epsilon_T$ and $\epsilon_P$, were calculated as in Equation \ref{eq:relativeEmpiricalError}~. The PCE degree that minimized the training and prediction errors for each model was chosen; the corresponding result is referred to as "optimal" learning. \\

Three different input configurations were used for the learning of each temporal coefficient $a_i$ as generally formulated in Equation \ref{eq:prediction:minimodel}~. \begin{itemize}
\item[$\bullet$] $\mathcal{H}_i$-model: a first simple configuration where all the statistical indicators described in Subsection \ref{subsection:application:case} were used and an independence hypothesis between the POD temporal coefficients is considered. The model is written as in Equation \ref{eq:prediction:fullmodelRelaxed}~: $a_i(t_{j+1}) \approx \mathcal{H}_i \left[a_i(t_{j}),t_{j+1}-t_j, \boldsymbol{\Theta}(t_{j} \rightarrow t_{j+1})\right]$. This is a model of dimension $17$.
\item[$\bullet$] $\mathcal{H}_i^F$-model: a more complex configuration where a "Full" $15$-mode POD approximation is considered with possible dependencies between the temporal coefficients. Of course, the choice of the basis size and the dependency structure can be optimized, but the objective here was to make a first step toward a more optimal configuration.  The model can be written as: $a_i(t_{j+1}) \approx \mathcal{H}_i^F \left[a_{1}(t_{j}), \hdots, a_{15}(t_{j}), t_{j+1}-t_j, \boldsymbol{\Theta}(t_{j} \rightarrow t_{j+1})\right]$. This is a model of dimension $31$.
\item[$\bullet$] $\mathcal{H}_i^P$-model: a smaller set of inputs, used by the operators to qualitatively evaluate sediment deposition risk, was used.  It corresponds to the six variables $TLmean$, $WvH$, $Wvper$, $Wvdir$, $Wv2m$ and $Wv2m\%$. This mimics the physical expertise that may be engaged when building a statistical model. It is written as $a_i(t_{j+1}) \approx \mathcal{H}_i^P \left[a_i(t_{j}),t_{j+1}-t_j,\boldsymbol{\Theta^P}(t_{j} \rightarrow t_{j+1})\right]$, where $\boldsymbol{\Theta^P}$ stands for the "physical". This is a model of dimension $8$. 
  \end{itemize}
\begin{figure}[H]
  \centering
  \vspace{-0.5cm}
  \subfloat[][Training error]{\includegraphics[trim={0cm 2cm 0cm 2.5cm},clip,scale=0.083]{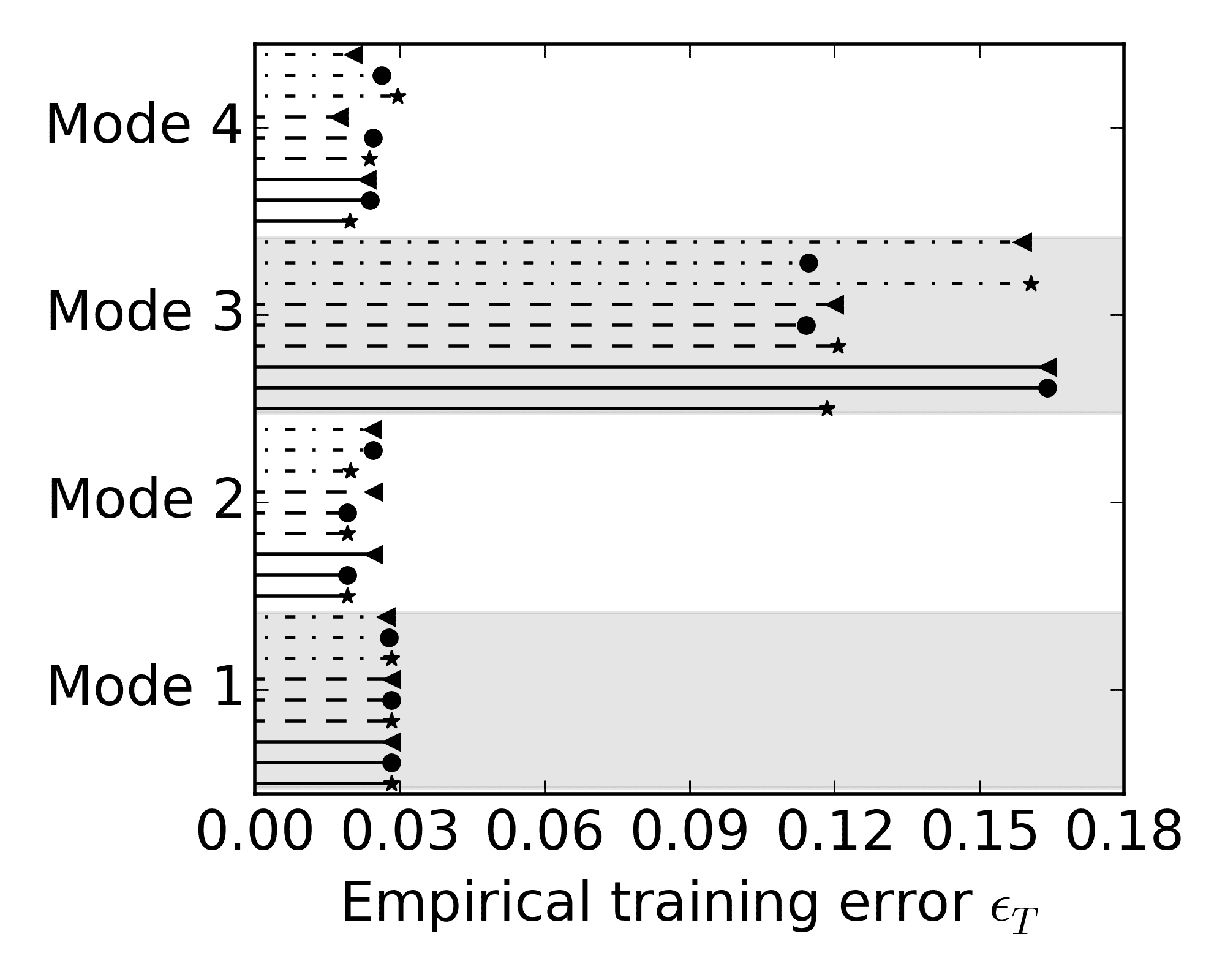}}
 \subfloat[][Prediction error]{\includegraphics[trim={12cm 2cm 2.0cm 2.5cm},clip,scale=0.083]{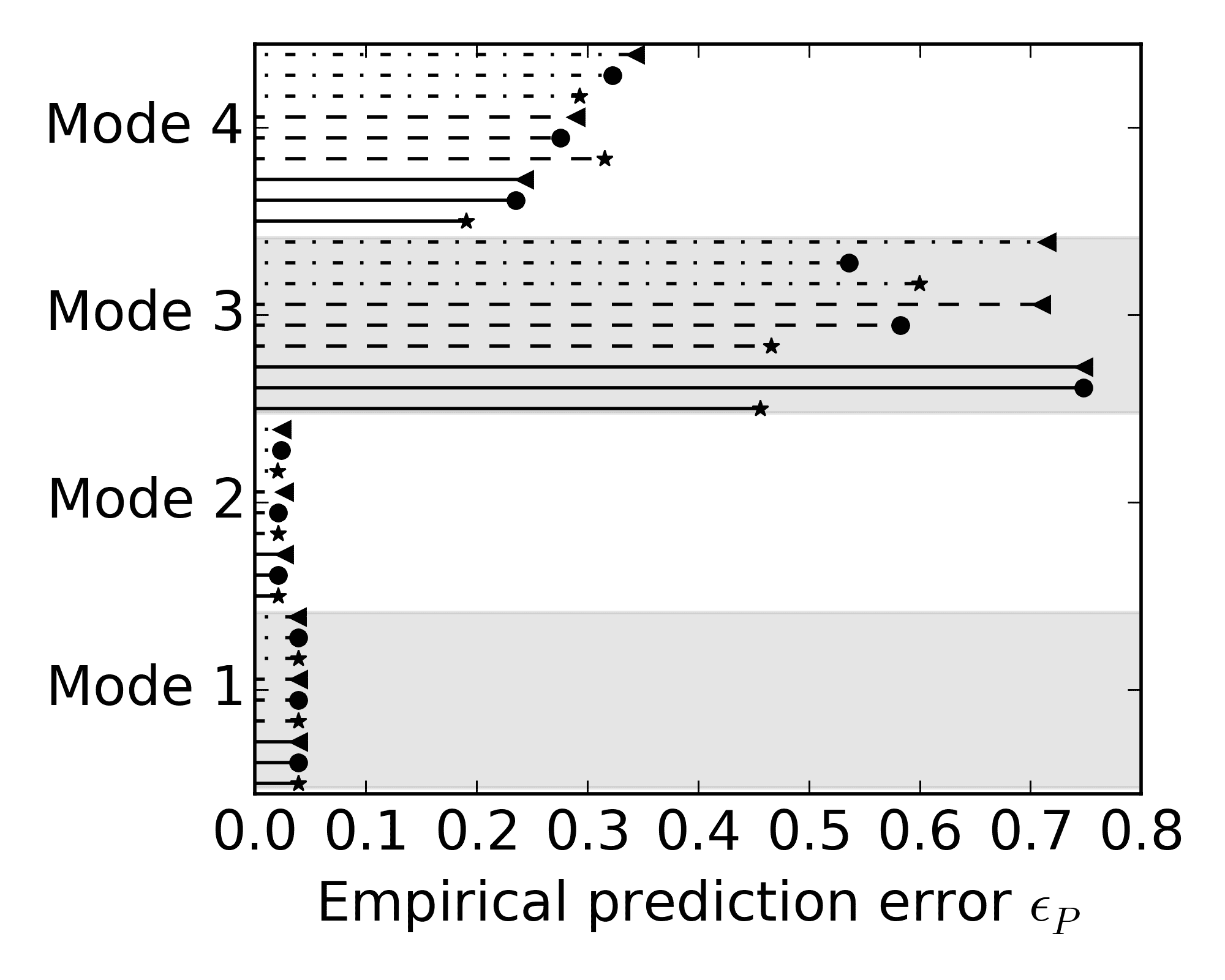}}
 \subfloat{\includegraphics[trim={60cm -8.5cm 0cm 3.6cm},clip,scale=0.07]{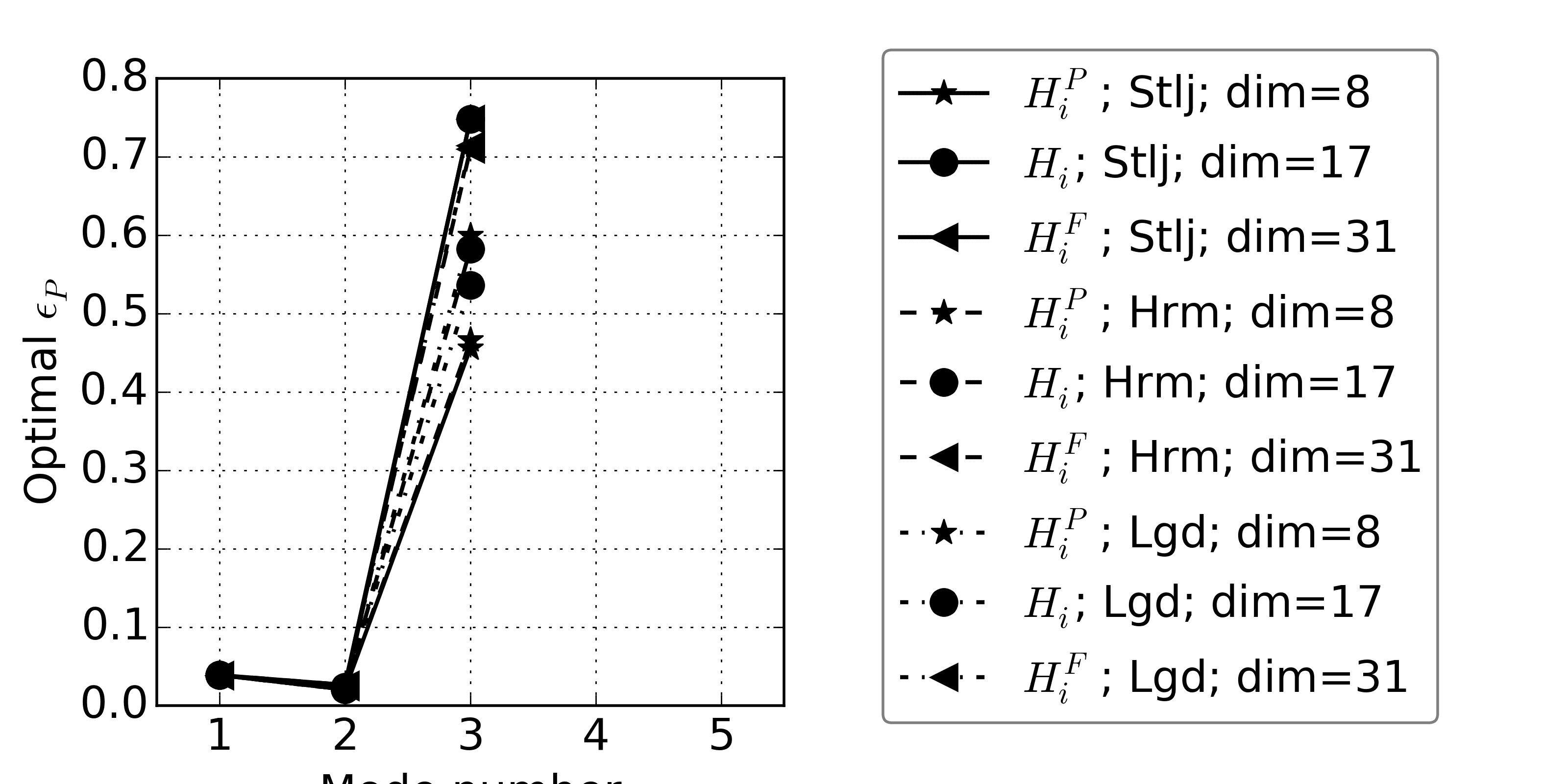}}
    \caption{The empirical training error $\epsilon_T$ and prediction error $\epsilon_P$ corresponding to the optimal fitting of models with different dimensions and marginals. The figure is organized as follows: errors are plotted for each mode vertically, separated by a gray band. Each marginal type corresponds to the same line style, and each dimension to the same marker style. The legend is shown in the order of the plots, down to top for each mode.}
    \label{fig:PCE:errorsComparison}
\end{figure}

To these variable choices were associated three choices of marginals, conditioning the choice of the PCE orthonormal polynomial (Section \ref{subsection:theory:PCE}). \begin{itemize}
\item[$\bullet$]  \textit{Lgd:} all the variables follow a Uniform PDF. The bounds of the marginal were set to the minimum and maximum chronological values $\pm 1\%$ as in \citep{Torre2019}. The associated orthonormal polynomial basis is the Legendre family.
\item[$\bullet$]  \textit{Hrm:} all the variables have Gaussian marginals characterized by the empirical mean and variance deduced from the data. The associated orthonormal polynomial basis is the Hermite family.
\item[$\bullet$]  \textit{Stlj:} the marginals were inferred from the data using Gaussian Kernel density estimates. The orthonormal polynomial basis was constructed from the knowledge of the marginal using a Stieltjes orthogonalization.  \\
\end{itemize}

The three marginal choices \textit{(Lgd, Hrm, Stlj)} were trained with the three dimension choices ($\mathcal{H}_i^P$ dim=8, $\mathcal{H}_i$ dim=17, $\mathcal{H}_i^F$ dim=31). The empirical errors of the "optimal" learnings are compared in Figure \ref{fig:PCE:errorsComparison}~. For Modes 1 and 2, training and prediction errors were almost identical for all configurations, although with a slight advantage with the smallest dimensions for all the marginal types in the learning of Mode 2. Starting from Mode 3, bigger differences emerged. At the learning step of Mode 3, models of dimension 17 and 31 were poorly fitted for the \textit{Stlj} and \textit{Lgd} configurations compared to others. At the prediction step of Mode 3, the errors of models with dimension 31 were much greater than smaller dimensions for all marginal choices. There seemed to be an overfitting of the model by selecting a larger number of inputs. The best models for Mode 3 were those of the smallest dimension, 8, with either the \textit{Stlj} or the \textit{Hrm} model. Lastly, for Mode 4, two orderings were observed for the prediction error. Firstly, for each marginal choice, prediction error increased with dimension, which confirmed the overfitting hypothesis. Secondly, error was the smallest with the \textit{Stlj} model (Kernel density), followed by the \textit{Hrm} model (Gaussian) and lastly by the \textit{Lgd} model (Uniform). Here, Uniform marginals performed worst; they were probably too different from the real data marginals and did not account for particularities in the inputs. In the parametric family, Gaussian marginals probably fitted real density better. \\

To conclude this comparison, the best marginal choice was the Kernel density estimate. The  smallest dimensions performed the best, with the \textit{Stlj} choice for the polynomial basis. The $\mathcal{H}_i^P;Stlj$ model of dimension 8 was therefore selected. However, the training was performed with an arbitrary split of the available statistical set. The sensitivity of the model to the learning set size and members is performed in the following step. 

\begin{figure}[H]
  \centering
     \vspace{-0.5cm} 
     \subfloat[][Mode 1 - Stlj]{\includegraphics[trim={0cm 1cm 0cm 1cm},clip,scale=0.07]{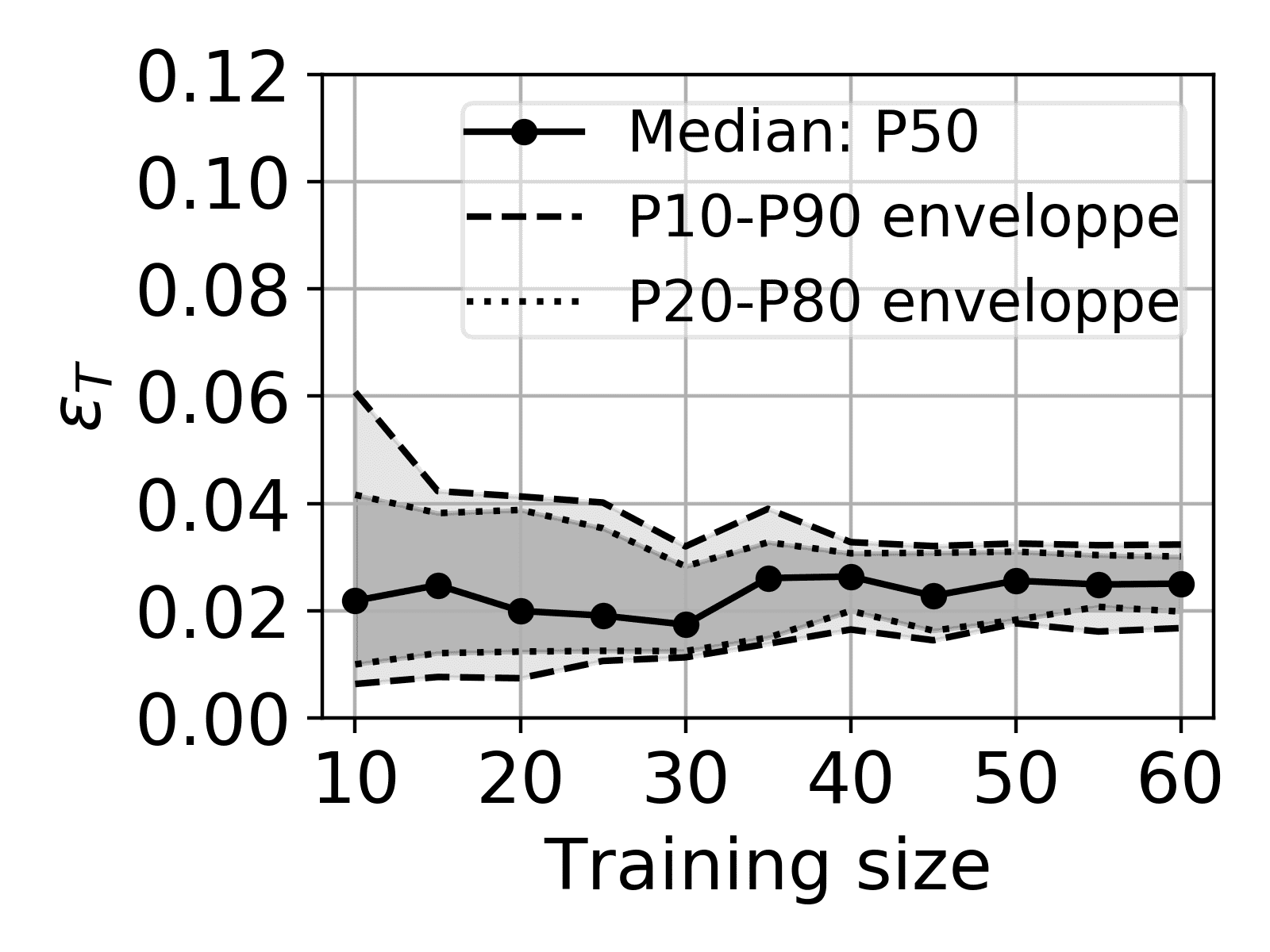}}
     \subfloat[][Mode 2 - Stlj]{\includegraphics[trim={0cm 1cm 0cm 1cm},clip,scale=0.07]{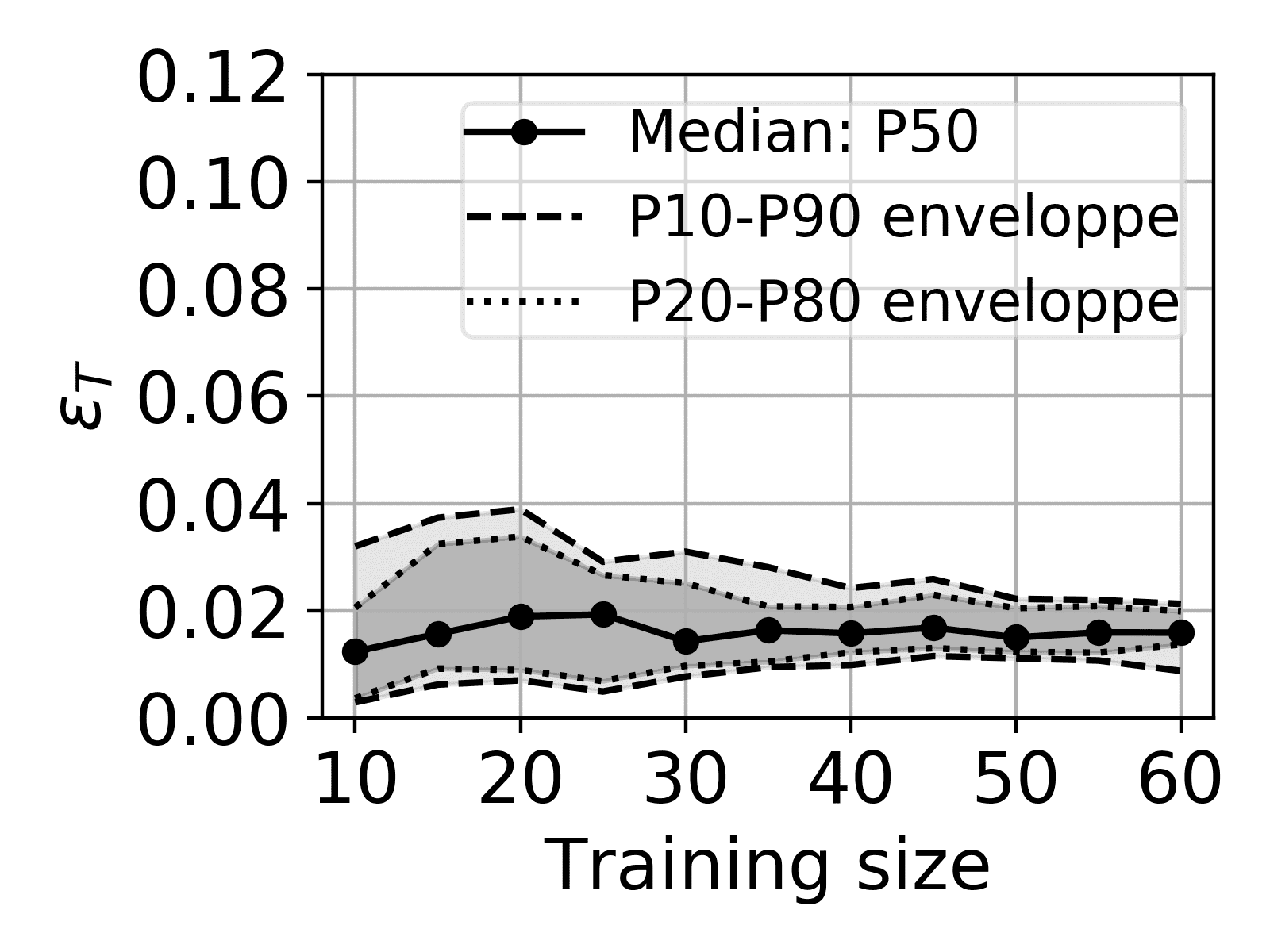}}
     \subfloat[][Mode 3 - Stlj]{\includegraphics[trim={0cm 1cm 0cm 1cm},clip,scale=0.07]{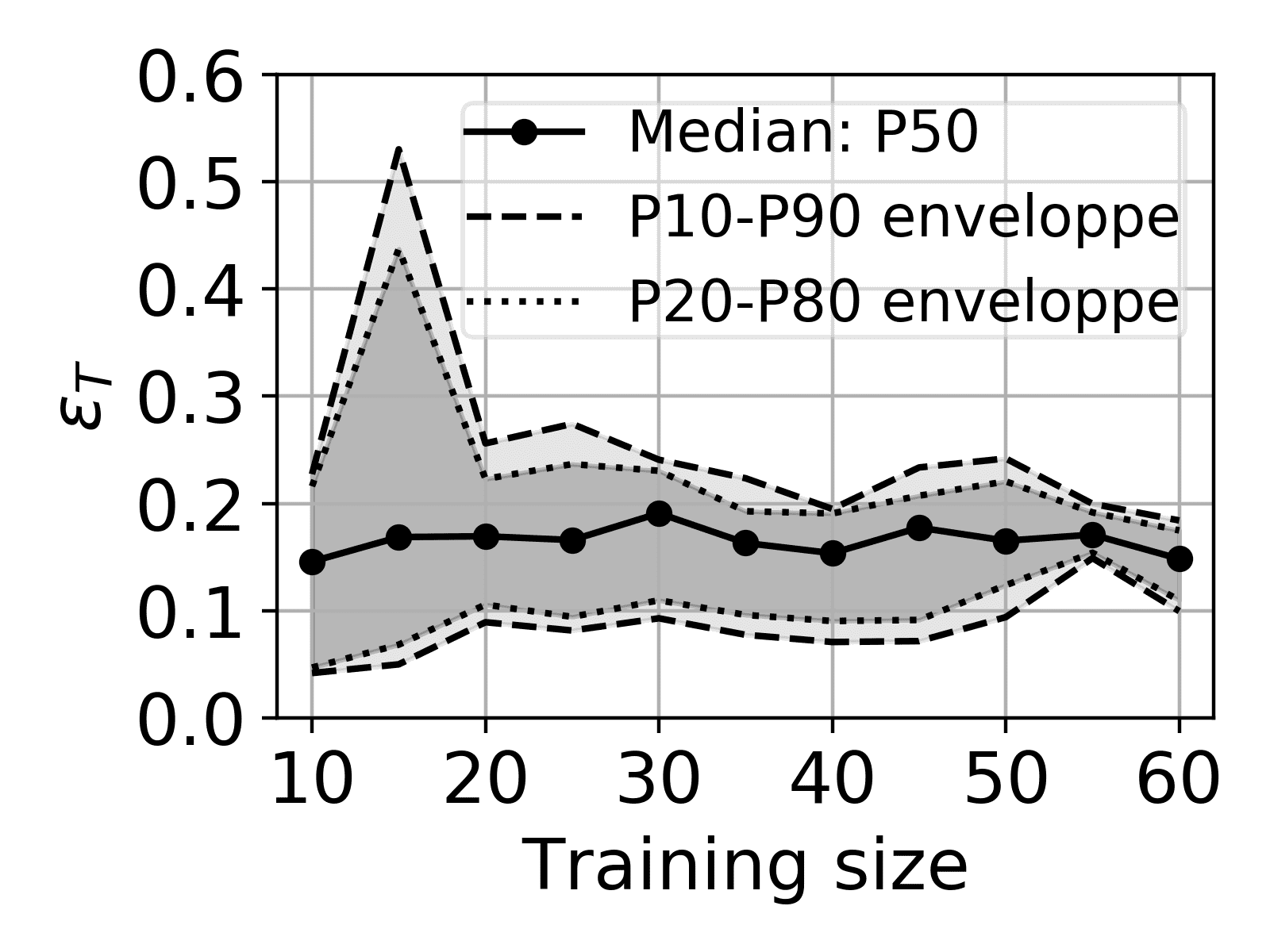}}
     \subfloat[][Mode 3 - Hrm]{\includegraphics[trim={0cm 1cm 0cm 1cm},clip,scale=0.07]{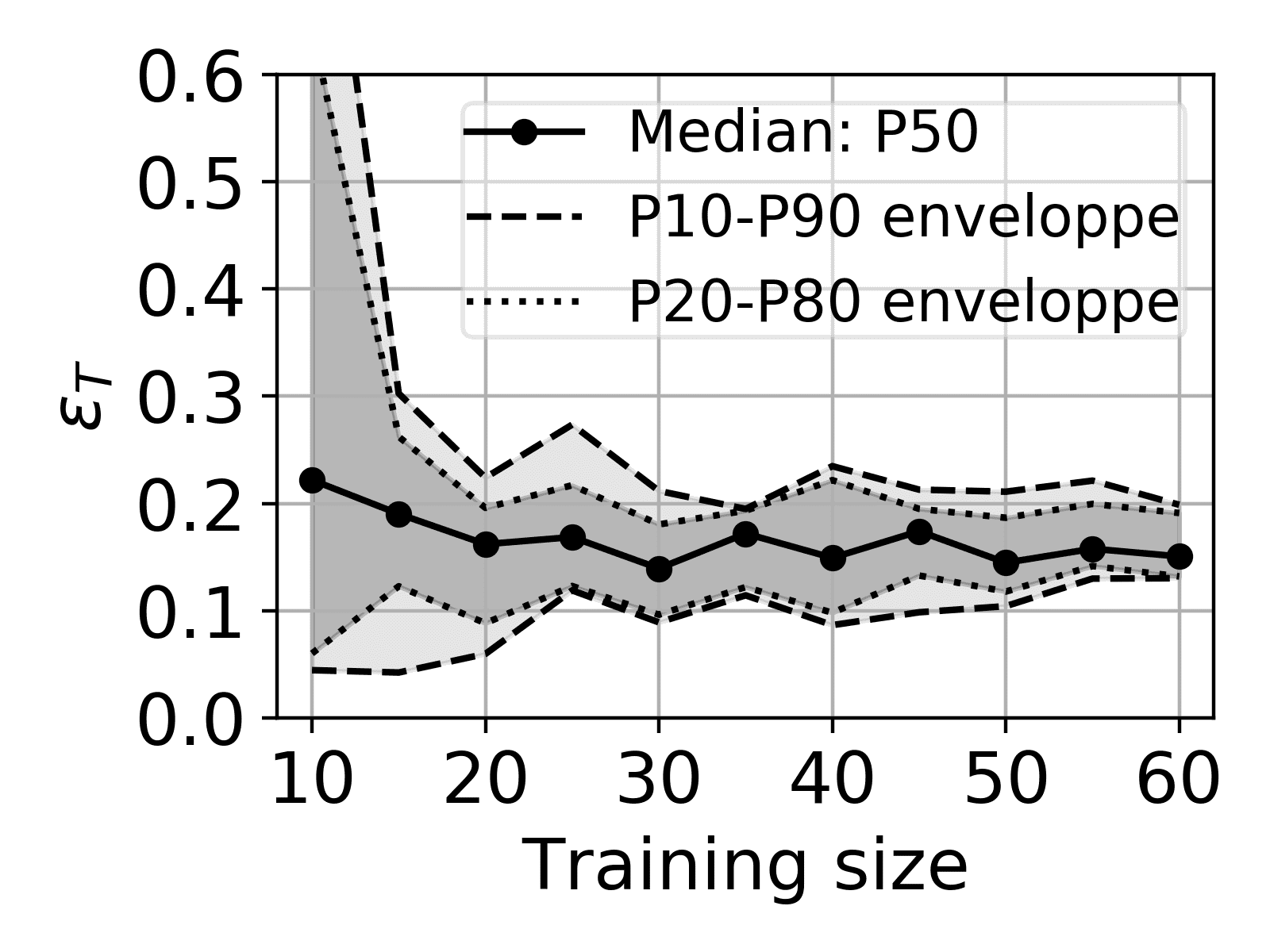}}
    \caption{Training empirical errors $\epsilon_T$ calculated for diverse training sizes with a Bootstrap of size 20. Plots show median value and confidence intervals (P: Percentile).}
    \label{fig:PCE:convergenceTraining}
\end{figure}

\begin{figure}[H]
  \centering
     \vspace{-0.5cm} 
		\subfloat[][Mode 1 - Stlj]{\includegraphics[trim={0cm 1cm 0cm 1cm},clip,scale=0.07]{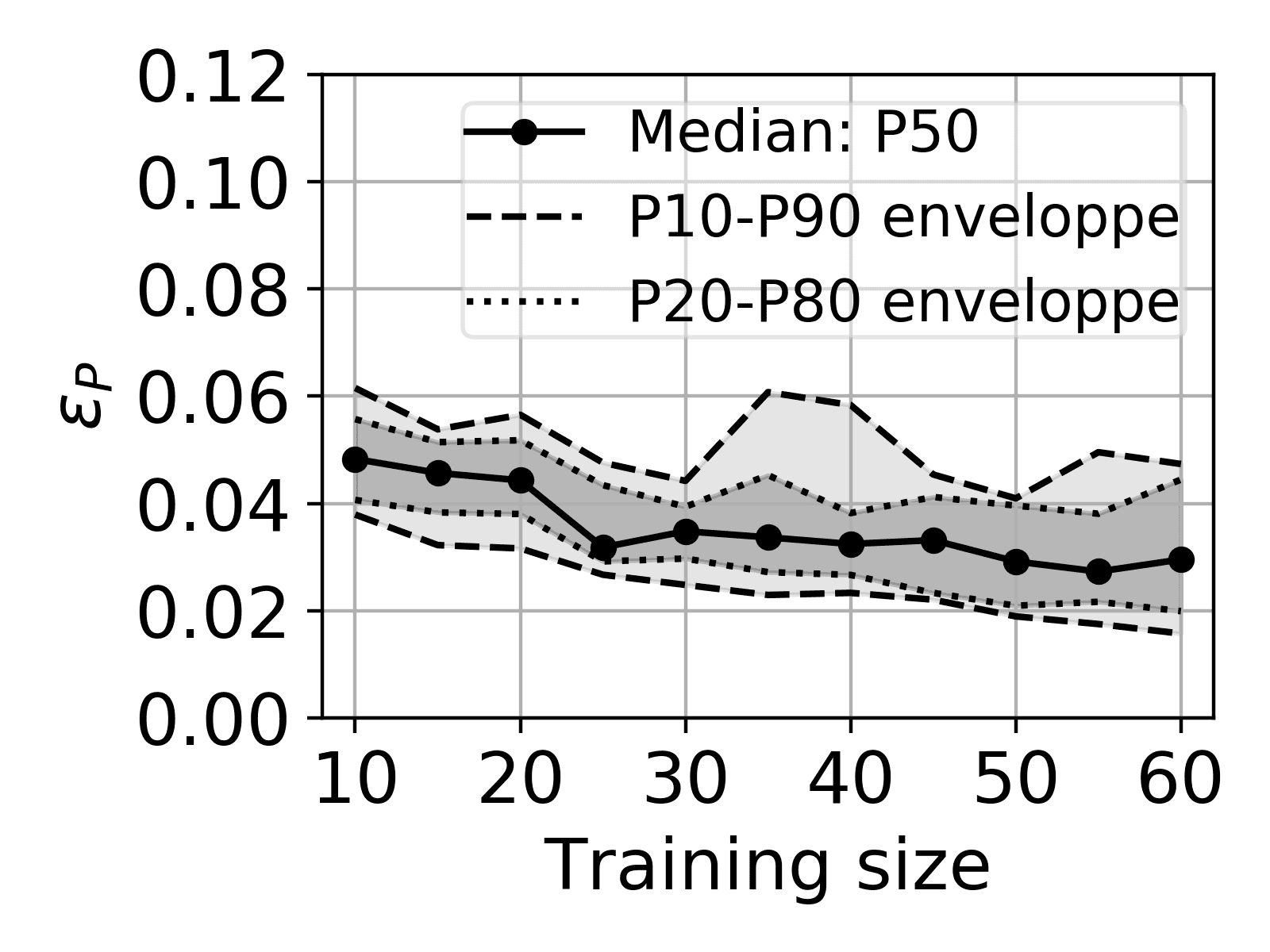}}
     \subfloat[][Mode 2 - Stlj]{\includegraphics[trim={0cm 1cm 0cm 1cm},clip,scale=0.07]{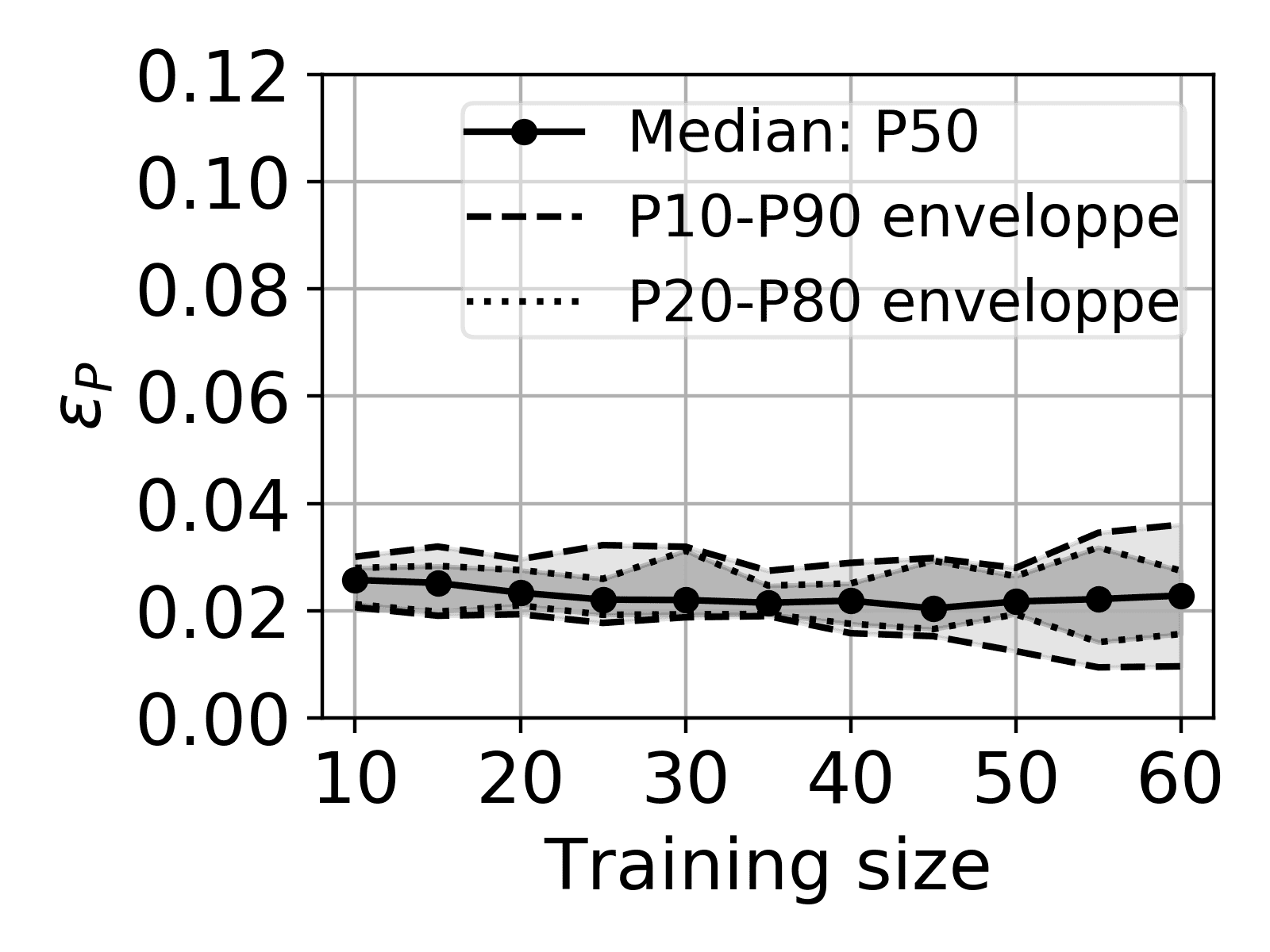}}
     \subfloat[][Mode 3 - Stlj]{\includegraphics[trim={0cm 1cm 0cm 1cm},clip,scale=0.07]{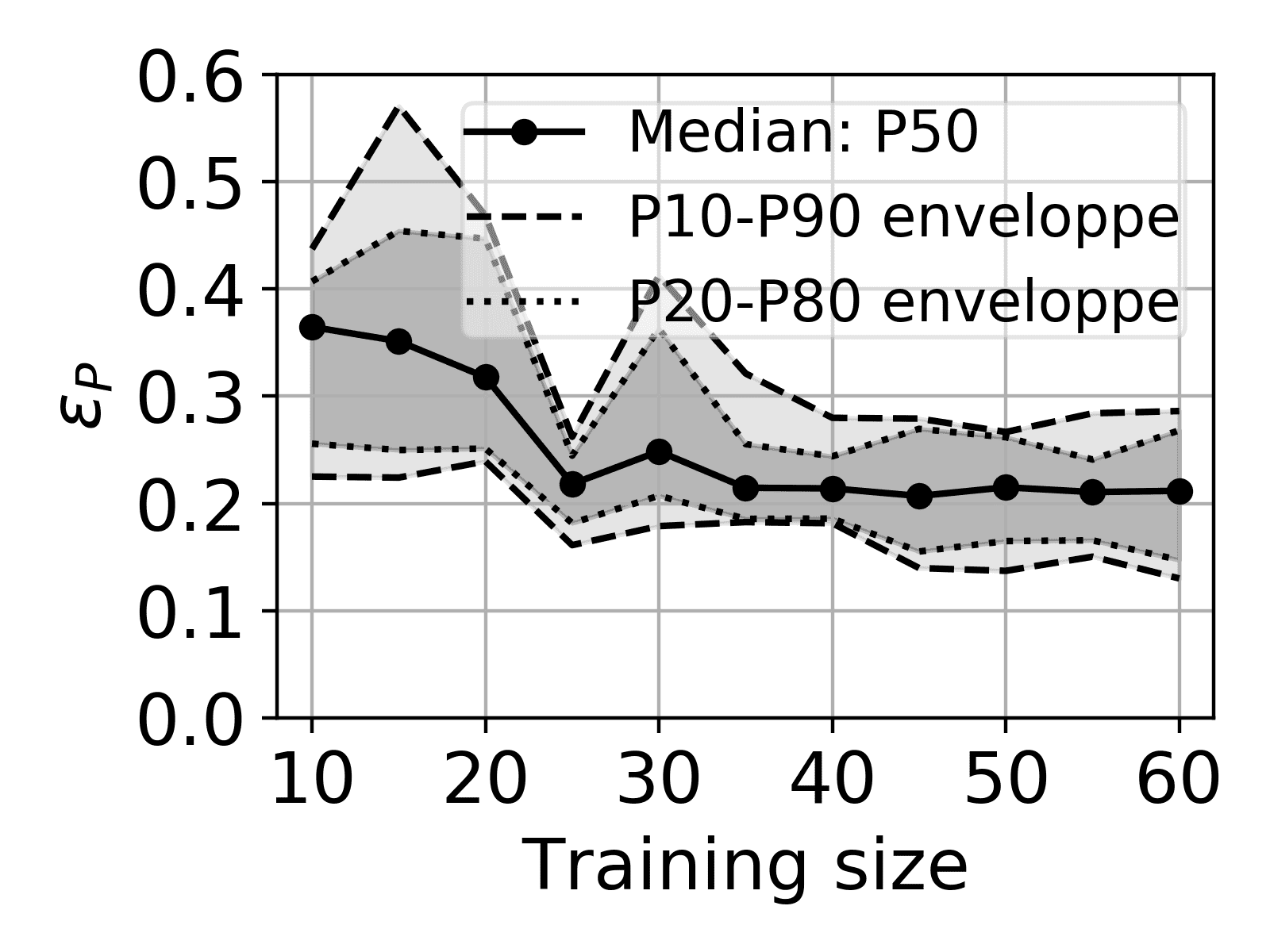}}
     \subfloat[][Mode 3 - Hrm]{\includegraphics[trim={0cm 1cm 0cm 1cm},clip,scale=0.07]{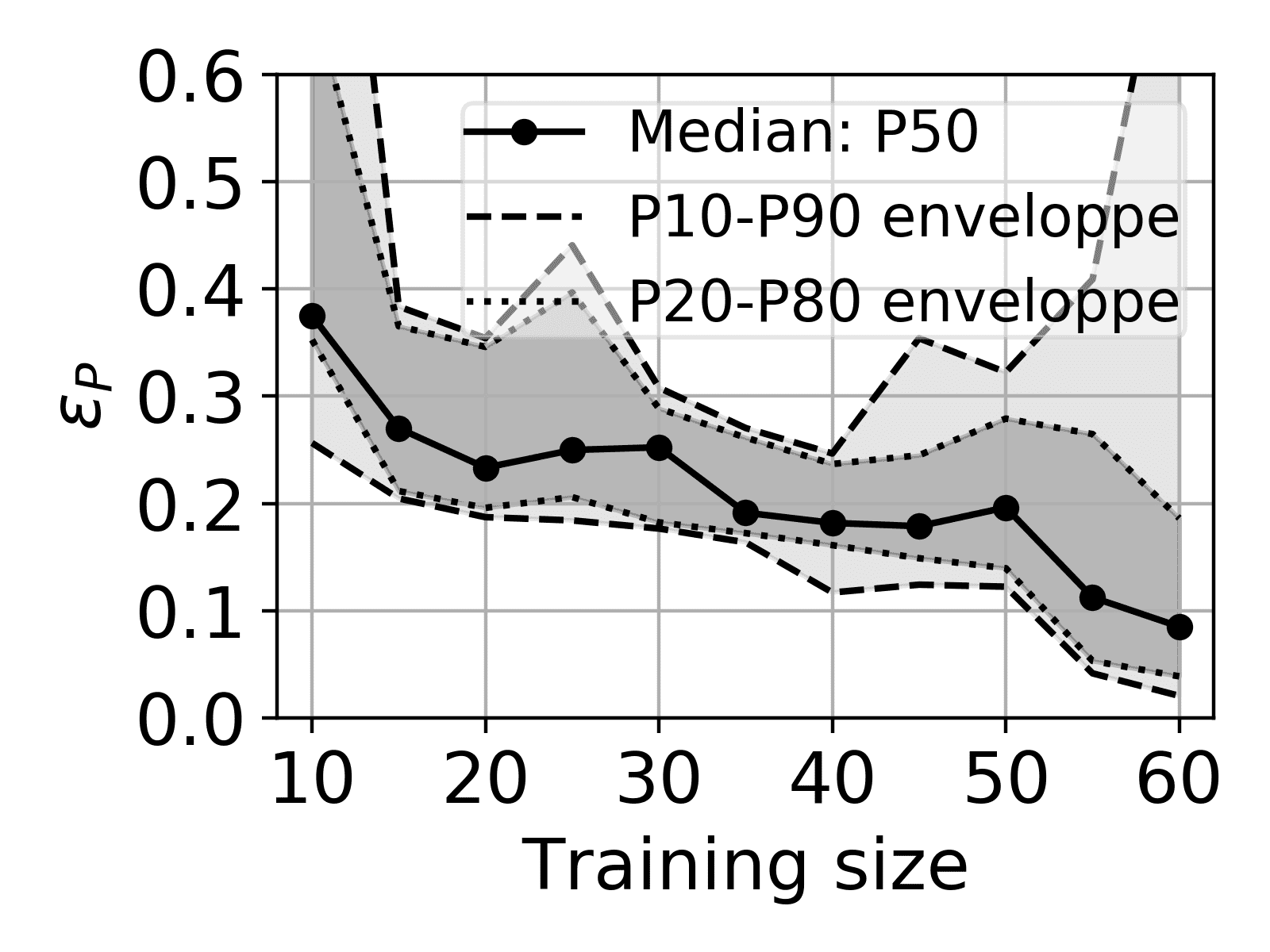}}
    \caption{The prediction empirical errors $\epsilon_P$ calculated for diverse training sizes with a Bootstrap of size 20. Plots show median value and confidence intervals (P: Percentile).}
    \label{fig:PCE:convergencePrediction}
\end{figure}

\paragraph{\textit{Step 2 - Convergence and robustness of the fit}}\mbox{}\\
Up to this point, an arbitrary number of $50$ measurements was used for training, leaving $14$ prediction points for testing purposes. In the following, the influence of training set size on the learning and prediction error is assessed. The objective is to check the robustness of the previous best model $\mathcal{H}_i^P;Stlj$ with respect to the data-set size and members. The evolution of the training and prediction empirical errors according to training set size is shown in Figure \ref{fig:PCE:convergenceTraining} and  \ref{fig:PCE:convergencePrediction} respectively. For comparison, the convergence of the $\mathcal{H}_i^P;Hrm$ model for Mode 3 is also shown for both errors. For each training set size, members were chosen randomly among the full data-set, and the remaining members were used for the prediction phase. For the estimation of the confidence intervals, bootstrap analysis was performed \citep{Efron1986}.\\

\begin{figure}[H]
  \centering
  \vspace{-0.5cm}
     \subfloat[][First mode $a_1(t)$]{\includegraphics[trim={0cm 1cm 0cm 2.5cm},clip,scale=0.08]{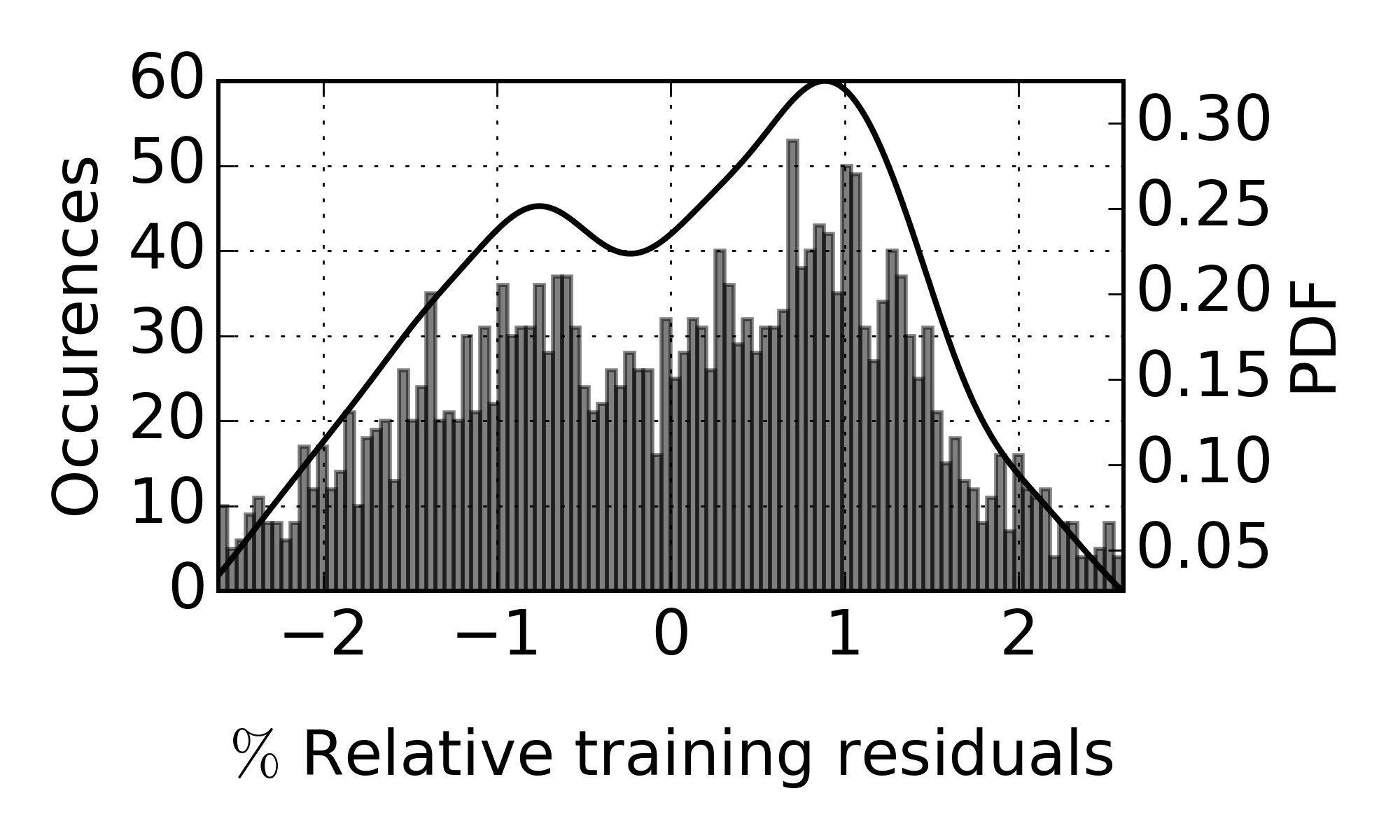}}
     \subfloat[][Second mode $a_2(t)$]{\includegraphics[trim={0cm 1cm 0cm 2.5cm},clip,scale=0.08]{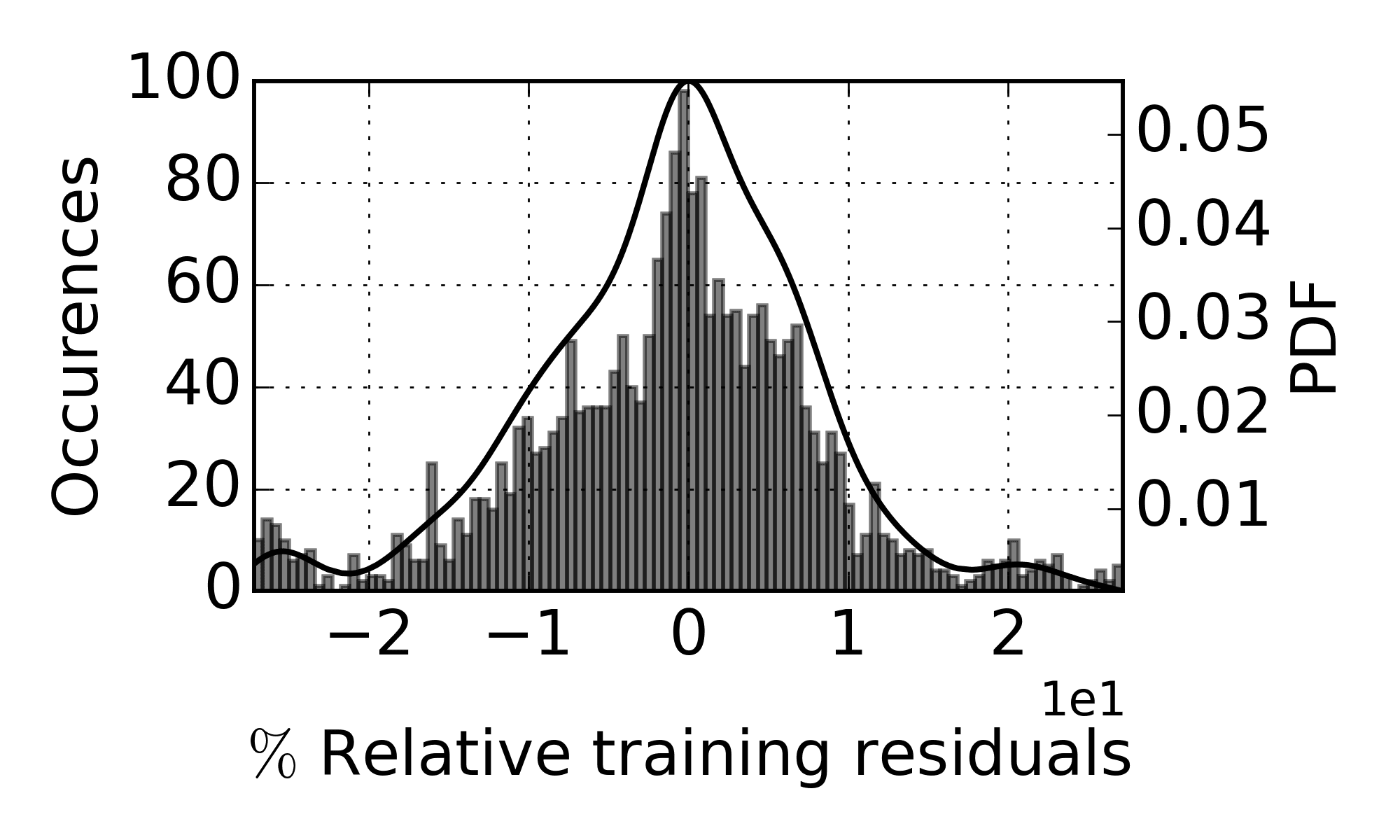}}
     \subfloat[][Third mode $a_3(t)$]{\includegraphics[trim={0cm 1cm 0cm 2.5cm},clip,scale=0.08]{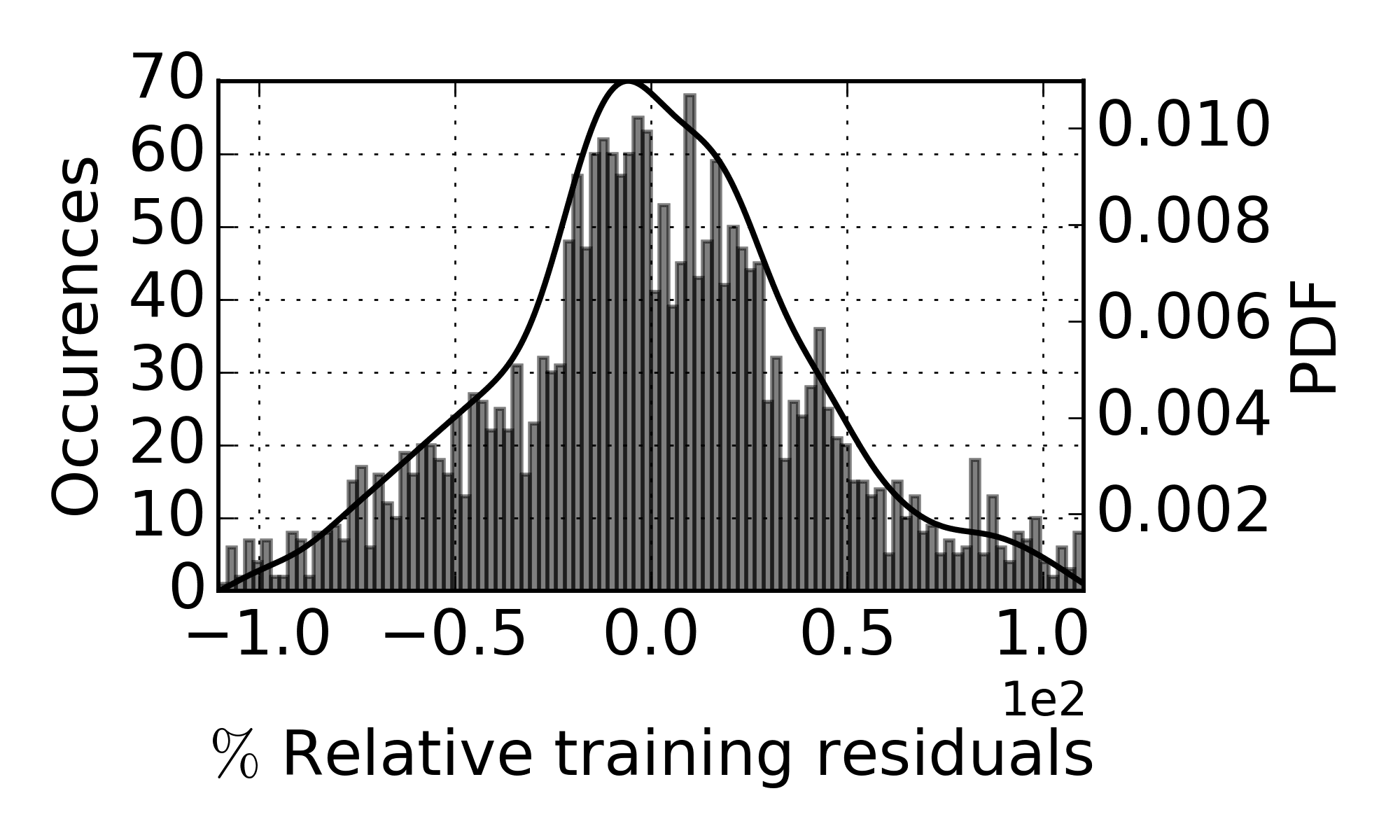}}
    \caption{The training residuals distributions of the $\mathcal{H}_i^P;Stlj$ model calculated for diverse training sizes with a Bootstrap of size 20. }
    \label{fig:PCE:training_residuals}
\end{figure}

\begin{figure}[H]
  \centering
  \vspace{-0.5cm}
     \subfloat[][First mode $a_1(t)$]{\includegraphics[trim={0cm 1cm 0cm 2.5cm},clip,scale=0.08]{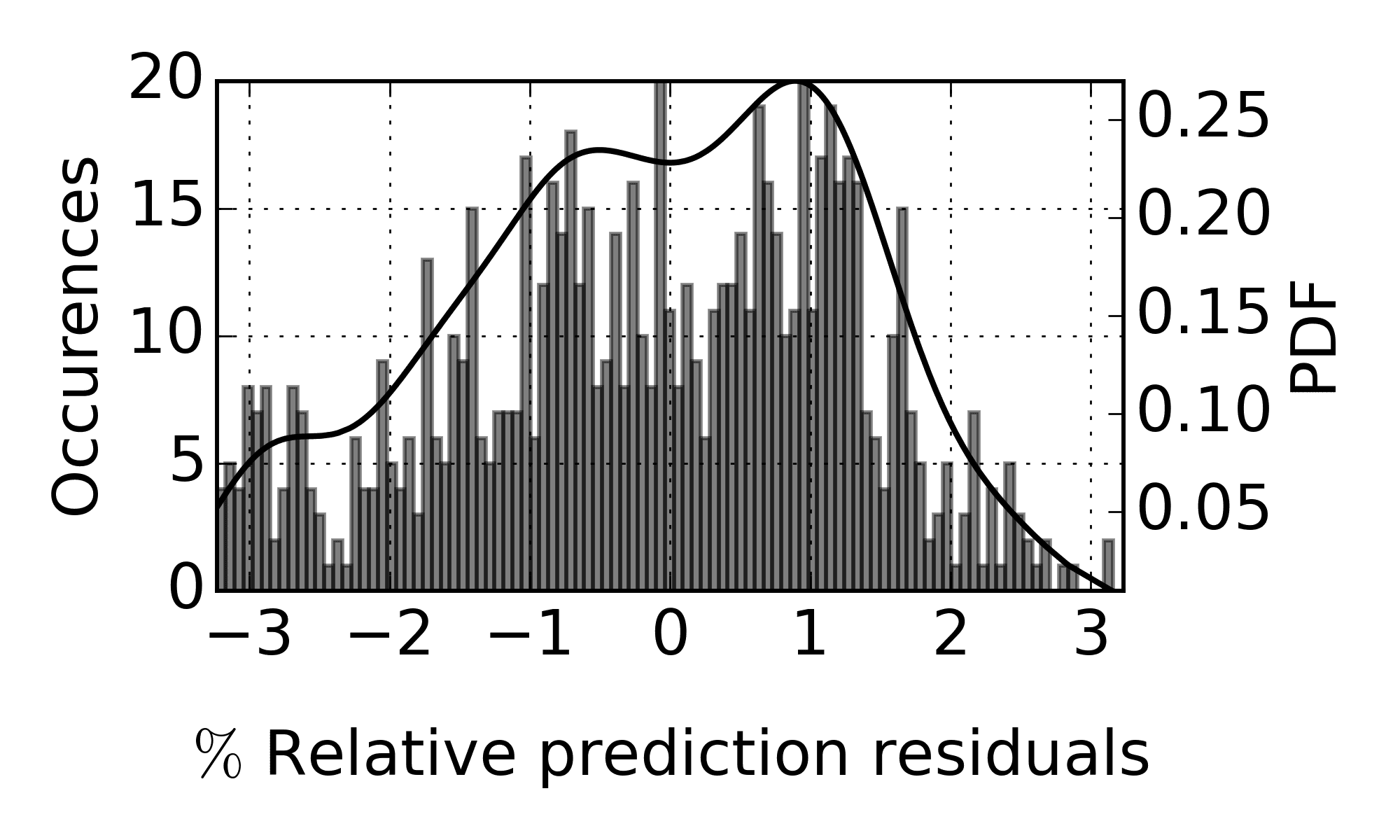}}
     \subfloat[][Second mode $a_2(t)$]{\includegraphics[trim={0cm 1cm 0cm 2.5cm},clip,scale=0.08]{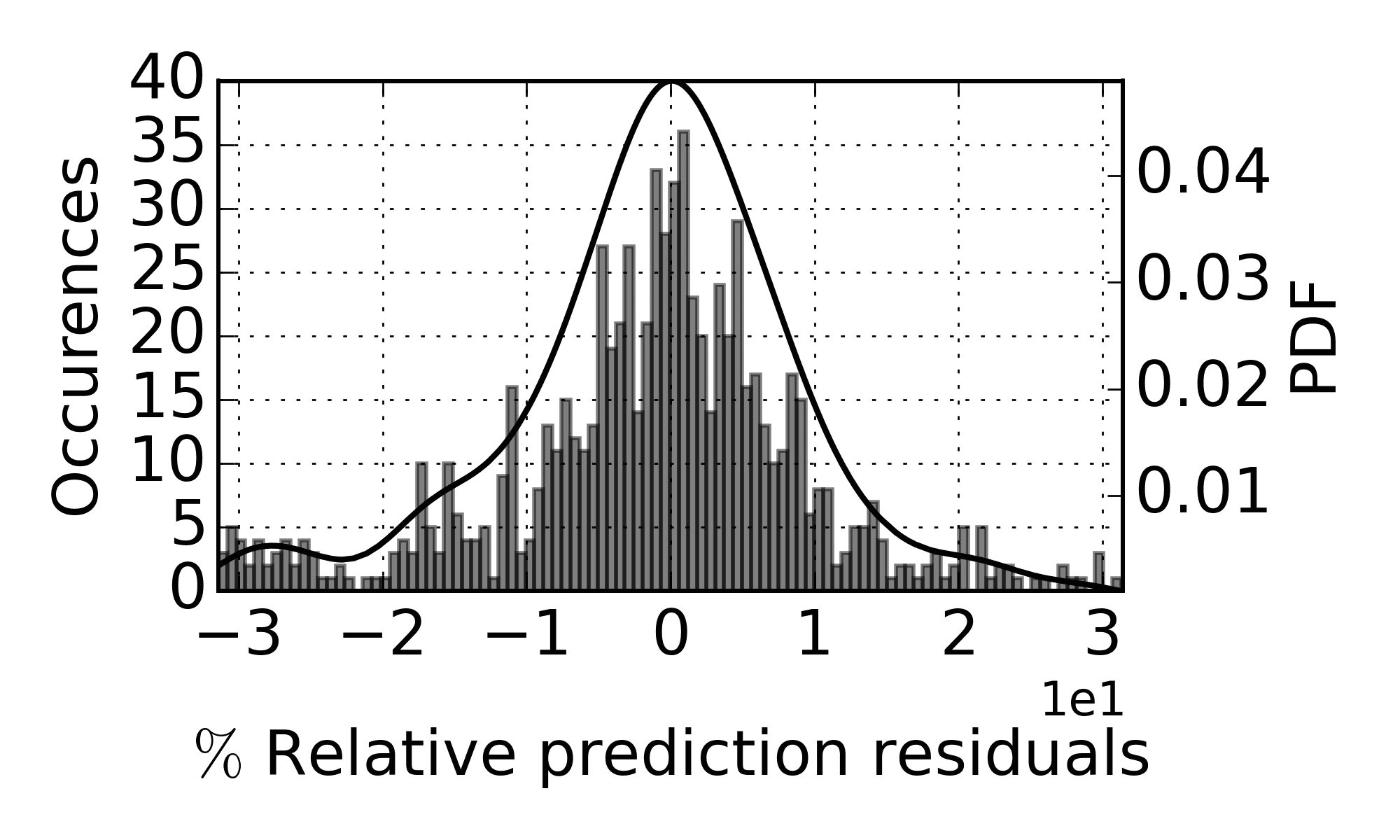}}
     \subfloat[][Third mode $a_3(t)$]{\includegraphics[trim={0cm 1cm 0cm 2.5cm},clip,scale=0.08]{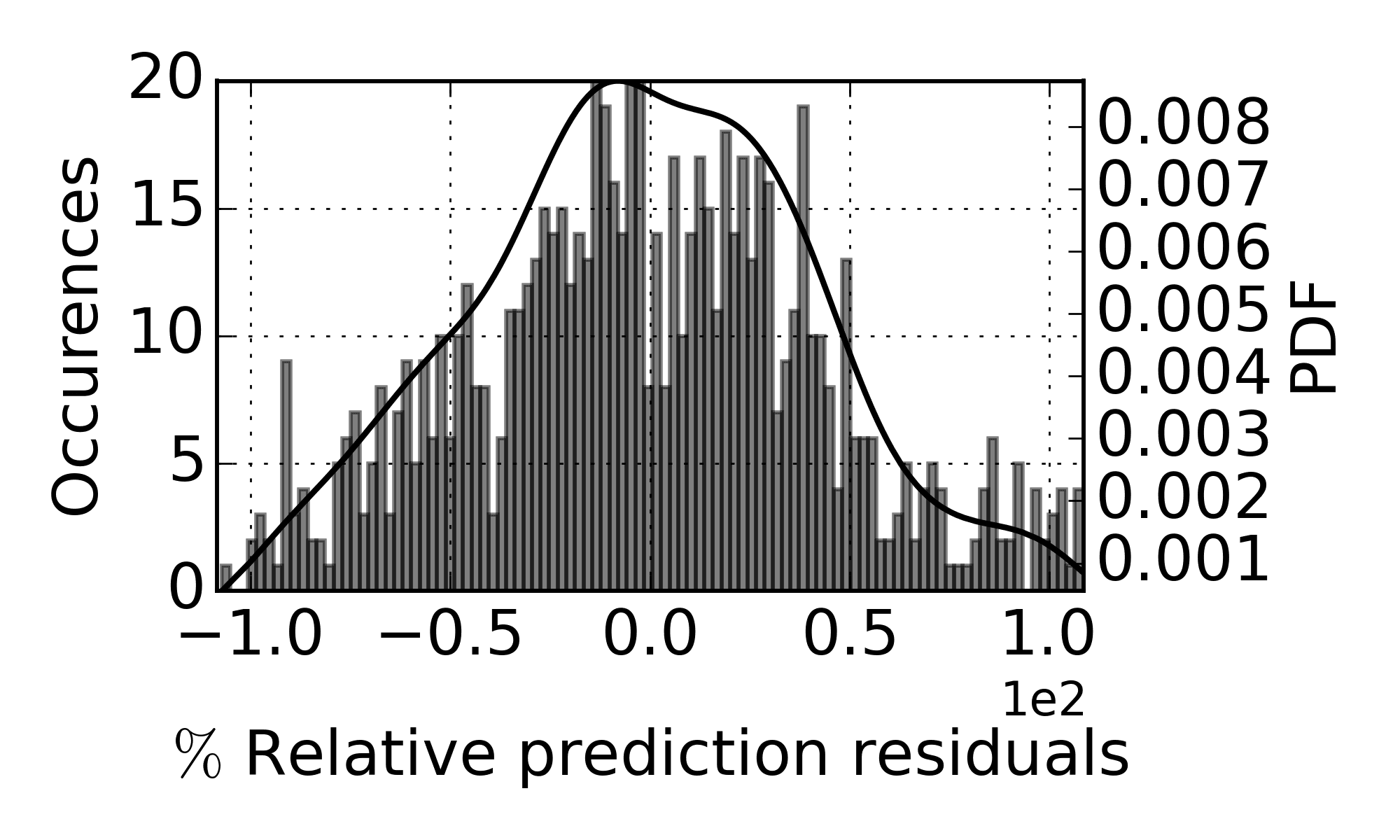}}
    \caption{The prediction residuals distributions using \textit{Stlj} model of dimension 8 calculated for diverse training sizes with a Bootstrap of size 20.}
    \label{fig:PCE:prediction_residuals}
\end{figure}

For the first two modes, the training errors in Figure \ref{fig:PCE:convergenceTraining} show a convergence of the median value and a tightening of the confidence intervals. The trainings can be considered as converging from around training size 40.
The associated median prediction errors in Figure \ref{fig:PCE:convergencePrediction} globally decreased with increasing training set size. However, although the final median values were lower for the \textit{Hrm} model, the confidence intervals were much larger than for the \textit{Stlj} model. The latter seems much more robust with respect to changes in training scenario.
The residuals distributions of the $\mathcal{H}_i^P;Stlj$ model, calculated as $a_i(.) - \mathcal{H}_i^P[.]$ on all the training sizes and Bootstraps, are shown in Figure \ref{fig:PCE:training_residuals} and Figure \ref{fig:PCE:prediction_residuals} for training and prediction, respectively.
 Only the middle $80\%$ portion of the residuals range is plotted, in order to analyze the center of the distribution; the full residuals distribution was long tailed, because the confidence intervals associated with small training set sizes were too large and produced extreme behaviors of the model. The training residuals were generally centered around zero: i.e., the models are unbiased. A slight asymmetry was, however, observed for Mode 1, which means that $a_1(.)$ was more often overestimated by $\mathcal{H}_i^P;Stlj$. Consequently, the mean elevation in the channel and the mean global sedimentation may be slightly exaggerated. These exaggerations, however, remained within a reasonable range, as most of the residuals fell within the $\pm 2\%$ interval. The training residuals of Modes 2 and 3 were perfectly centered, but percentage error dramatically increased. Most of the residuals fell within the $\pm 10\%$ interval for Mode 2, whereas they reached $\pm 50\%$ for Mode 3. However, this error concerns modes that represent at most $4\%$ of the total bathymetry variance, as more than $96\%$ of the total variance was already captured by the addition of the first two modes.

The residuals shapes (i.e. slight overestimation for Mode 1 and perfect centering for Modes 2 and 3) were maintained through the prediction phase. Furthermore, the residuals mostly fell within the ranges identified in the training phase. $\mathcal{H}_i^P;Stlj$  model behavior was stable. The prediction uncertainty could therefore be measured and trusted and the physical interpretation was consequently  robust, as discussed below in \textit{Step 3}.
 
\begin{figure}[H]
  \centering
  \vspace{-0.5cm}
  \subfloat[][PCE degrees]{\includegraphics[trim={0cm 1cm 0cm 1.9cm},clip,scale=0.086]{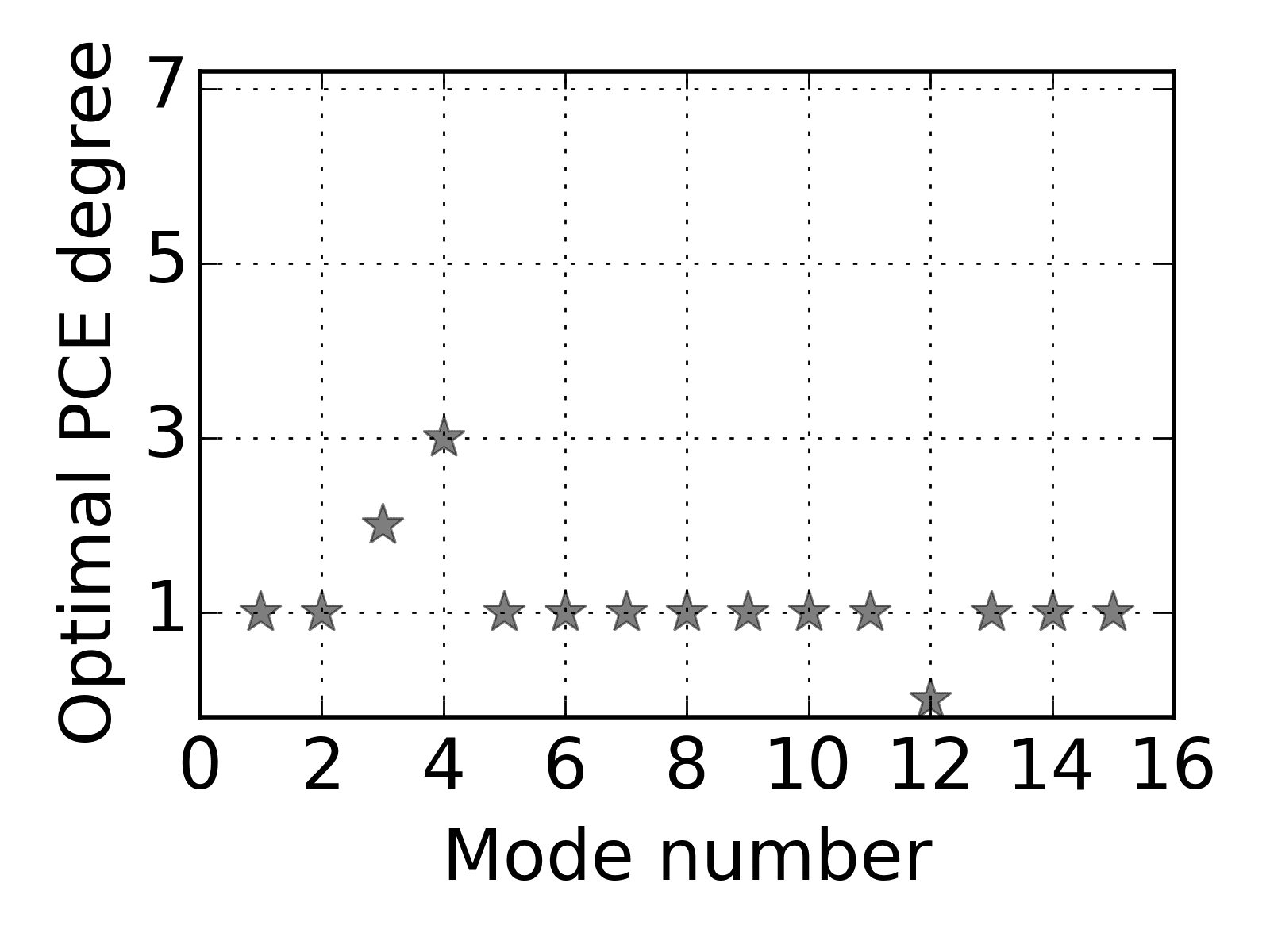}}
  \subfloat[][Training errors]{\includegraphics[trim={0cm 1cm 0cm 2.5cm},clip,scale=0.088]{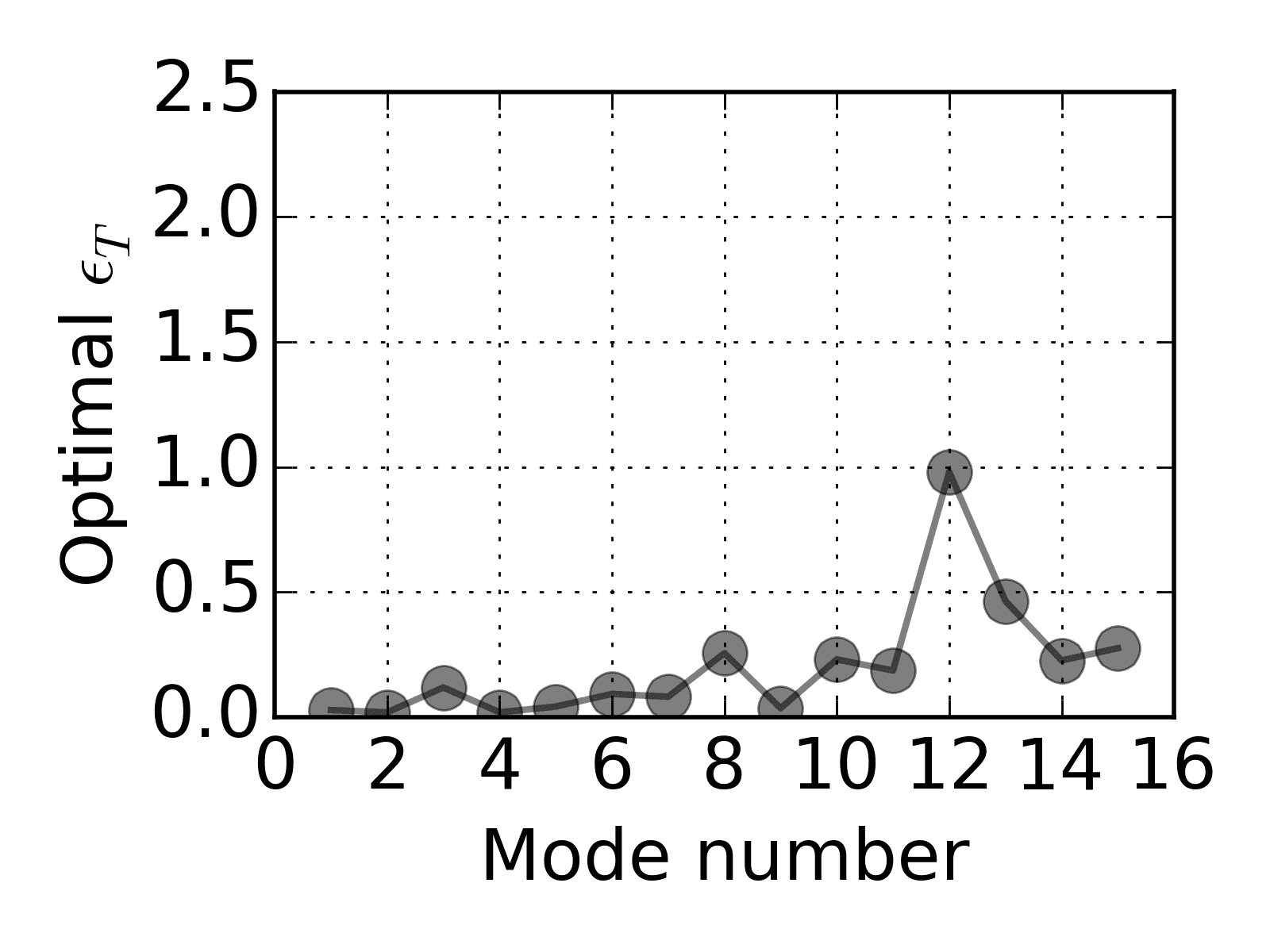}}
  \subfloat[][Prediction errors]{\includegraphics[trim={0cm 1cm 0cm 2.5cm},clip,scale=0.088]{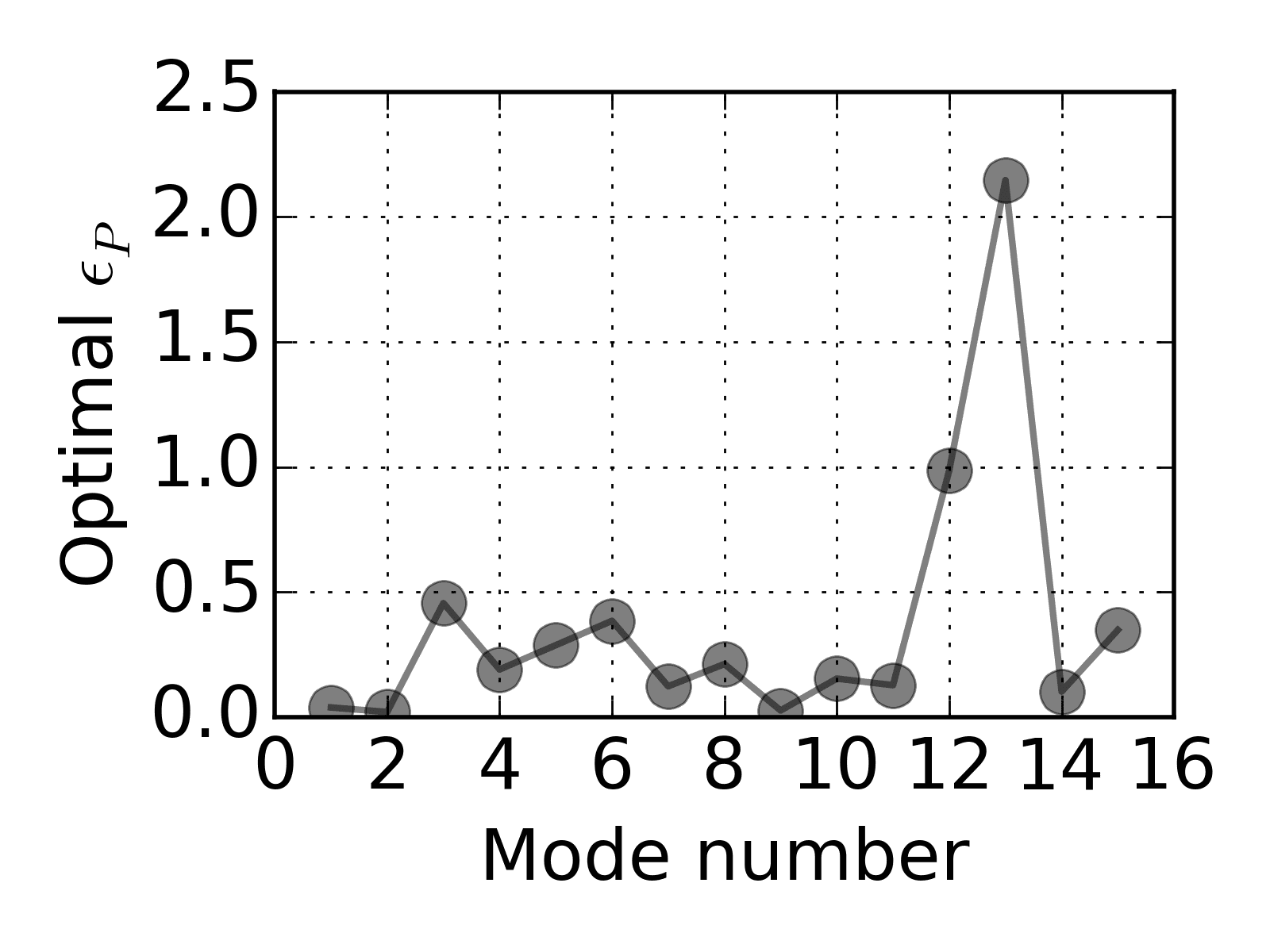}}
    \caption{Optimal PCE degrees for the $\mathcal{H}_i^P;Stlj$ model and associated empirical errors of the training ($\epsilon_T$) and the prediction ($\epsilon_P$) sets.}
    \label{fig:PCE:errors}
\end{figure}

\paragraph{\textit{Step 3 - Physical interpretation of the best learned model}}\mbox{}\\
The calibrated $\mathcal{H}_i^P;Stlj$ PCE models were considered optimal, as they showed good fit, convergence and robustness with respect to the training choices. Here, they are analyzed to deduce physical information. Firstly, the optimal polynomial degrees selected for each mode and the associated training and prediction empirical errors are shown in Figure \ref{fig:PCE:errors}. Linear models were optimal for Modes 1 and 2  (degree 1), and the associated errors were low for both the training and the prediction sets. For Modes 3 and 4, the optimal polynomial degrees increased, which implies higher-order contributions and/or higher-order interactions for the input variables. For modes of higher ranks, the models were either linear (degree 1) or approximated by a simple average value (degree 0). This means that LARS rejects polynomial terms of higher degrees because they do not significantly improve the learning \citep{Blatman2009}.

Prediction relative empirical errors in Figure \ref{fig:PCE:errors}-c (calculated as in Equation \ref{eq:relativeEmpiricalError}~) increased from Mode 3, but remained under $50\%$ up to mode 11. This must be interpreted according to the meaning of this indicator: it is a measure of the missing variations (distance between the model and reality) relative to the variance of the data. It therefore represents the amount of variance that was not captured by the PCE model (also called "the fraction of unexplained variance" \citep{Cruciani1992}). For example, for mode 12, estimated with a degree 0 PCE model, $100\%$ of the variance is not captured, which is natural because only the average value is accounted for with degree 0. For Modes 3 to 11 with error up $50\%$, this means that either the training set or the used inputs made it impossible to predict more than $50\%$ of the variance. However, as presented in \textit{Step 2}, this $50\%$ error concerned at most $4\%$ of the total bathymetry variance. Hence, the errors starting from Mode $3$ represented at most $2\%$ of missing variance. Beyond Mode 11, prediction with PCE would not be optimal, as the prediction error dramatically increases. \\ 

\begin{figure}[H]
  \centering
     \includegraphics[trim={0cm 6cm 0cm 0.5cm},clip,scale=0.15]{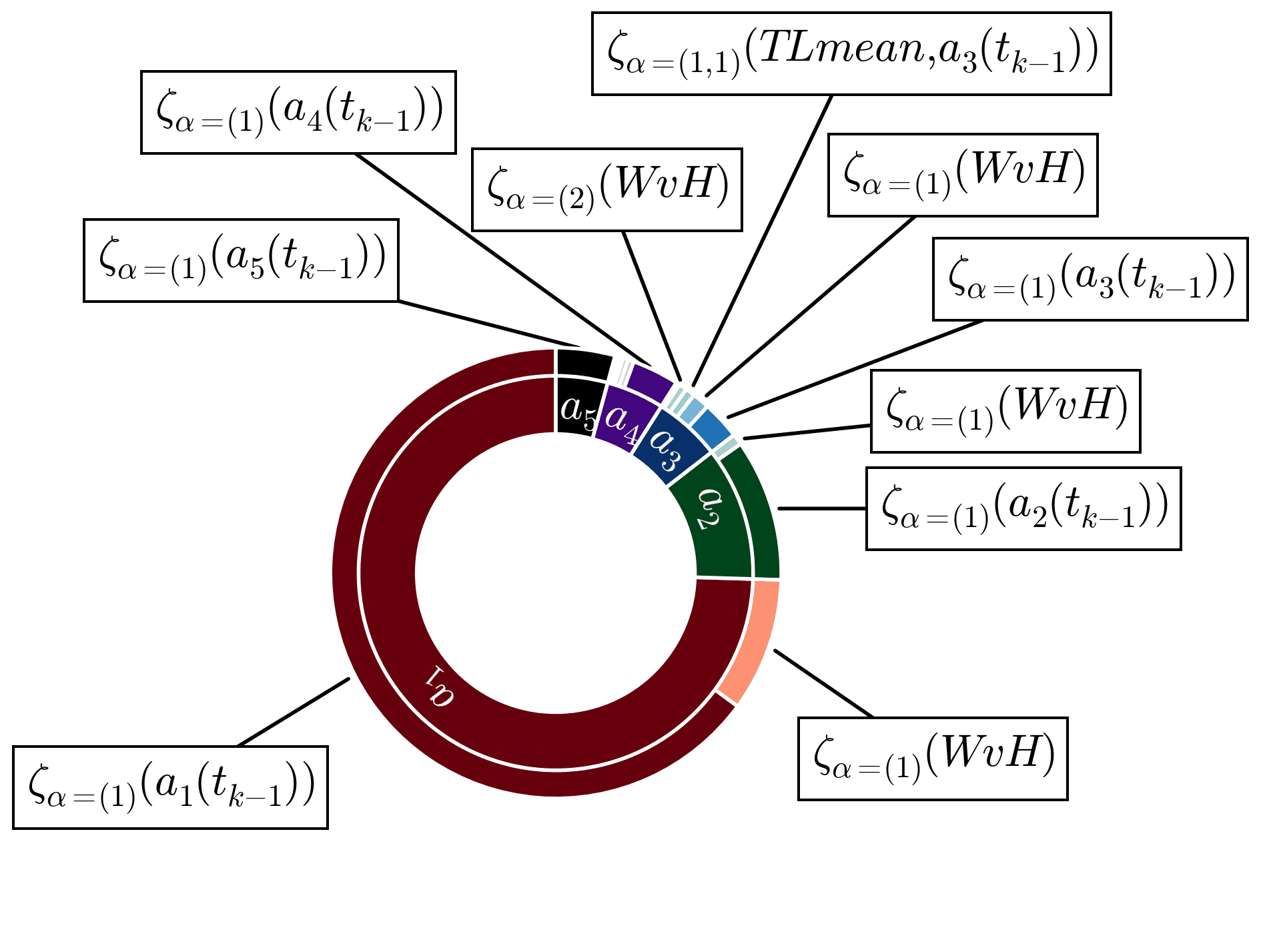}
    \caption{Piechart of the most influential parameters, using GW and GGW on the POD-PCE. The inner circle represents the share of each mode. The outer circle represents the share of each polynomial term. The polynomial terms corresponding to GGW higher than $0.5\%$ are shown.}
    \label{fig:PCE:pie_sensitivity}
\end{figure}

Secondly, PCE models were used to analyze the contribution of each forcing variable to the dynamics. For this, the \textit{Garson Weights} (GW) defined in Equation \ref{eq:PCE:weights} were used to estimate the influence of the forcings on each temporal coefficient. The global influence on the whole bathymetry field was quantified using the \textit{Generalized Garson Weights} (GGW), as in Equation \ref{eq:PODPCE:generalizedWeights}. The ranking of the modes and the impact of the inputs is represented in Figure \ref{fig:PCE:pie_sensitivity}.
 Mode 1 corresponds to a major contribution and the following modes are ranked according to their POD importance. The share of each polynomial term corresponds to the GGW in relation to the global contribution (full circle). When this share is compared to the importance of the corresponding mode, it corresponds to the GW. Lastly, the polynomial terms corresponding to more than $0.5\%$ GGW are indicated. $\zeta_{\boldsymbol{\alpha}=(.)}(.)$ corresponds to the notation introduced in Subsection \ref{subsection:theory:POD}, with the multi-index notation for $\alpha$ that represents the polynomial degree of each monomial. For example, $\zeta_{\boldsymbol{\alpha}=(\alpha_1,\alpha_2)}(\theta_1,\theta_2)$ corresponds to a polynomial of degree $\alpha_1+\alpha_2$, where $\theta_1$ contributes as a monomial of degree $\alpha_1$ and $\theta_2$ as a monomial of degree $\alpha_2$. The meaning of the variables that appear in Figure \ref{fig:PCE:pie_sensitivity} can be found in Subsection \ref{subsection:application:case}~. \\

For all the temporal coefficients $a_i(.)$, the most influential contributor by far was the value of the previous state $a_i(t_{j-1})$, in the form of a monomial of degree 1. It is followed by contributions involving the mean wave height during the sedimentation period period $WvH$ for all the modes, which makes $WvH$ the most important external forcing, figuring in the third position among all the forcings, with a contribution of $9.6\%$ through the first mode, a total of $12.6\%$ if only $WvH$ monomials are considered, and $13.3\%$ if interactions with other variables are taken into account. \\

The other forcing contributions also appeared, but with much less importance: e.g., the influence of mean low tide level $TLmean$, which took an interaction form with the previous bathymetry shape for Mode 3. Firstly, this interaction makes sense in terms of physics, as sediment deposition is conditioned by the value of bed shear stress \citep{VanRijn_2007_a}, which depends on velocity and water depth. The water depth value is exactly the tidal level minus the bed elevation value, which here appears as a multiplicative interaction between $TLmean$ and Mode 3. Second, the value of this contribution was only $0.7\%$ GGW, which is negligible when compared to the first contribution of $WvH$. The learned model gave much more importance to waves than to tides. This does not necessarily mean that tides have no influence on sediment deposition, but may simply suggest that, in the present configuration, sediment mobilization by the tide is always more or less the same, and that the forcing that makes a considerable difference is the variation in wave heights. Waves are a determining factor for sediment mobilization in coastal configurations, through the influence they have on bed shear stress \citep{VanRijn_2007_a}. Further more, a noticeable correlation between $WvH$ and $TLmean$ was noticed in the used data-set, which means that the information of low-tide levels is to a certain extent contained in the mean wave height. There is therefore a probable dependency between these variables. In case of dependencies, the iterative process used by LARS may drop a variable that is physically important because the important information is already contained in another variable, due to their dependency. \\

Lastly, it is important to note that the contribution of less frequent wave events was also present but to a much smaller extent. It is represented by a polynomial term in the form $\zeta_{\alpha=(1,1)(Wv2m,Wv2m\%)}$, where $Wv2m$ and $Wv2m\%$ are respectively mean wave height exceeding $2m$ and the associated frequency of occurrence (arbitrary storm indicator chosen in Subsection \ref{subsection:application:case}). This term appears in Modes 3 and 4 for a maximum total influence of $0.3\%$. Higher-order interactions and less frequent events are therefore represented by modes of higher rank, associated with smaller variance percentages. 

\paragraph{\textit{Step 4 - Robustness of the physical interpretation with respect to the learning-set members}}\mbox{}\\
As a last proof of the robustness of the proposed learning algorithm, specifically concerning physical interpretation, a sensitivity analysis with respect to the training set members was performed. The robustness of the calculated \textit{Garson Weights} (GW) with respect to the choice the training members was studied. This is equivalent to studying the robustness of the polynomial basis term selection as produced by LARS, and their associated multiplicative coefficients.\\

\begin{figure}[H]
  \centering
  \vspace{-0.5cm}
    \subfloat[][Polynomial term $a_k(t_{j-1})$]{\includegraphics[trim={0cm 1.5cm 0cm 2.5cm},clip,scale=0.088]{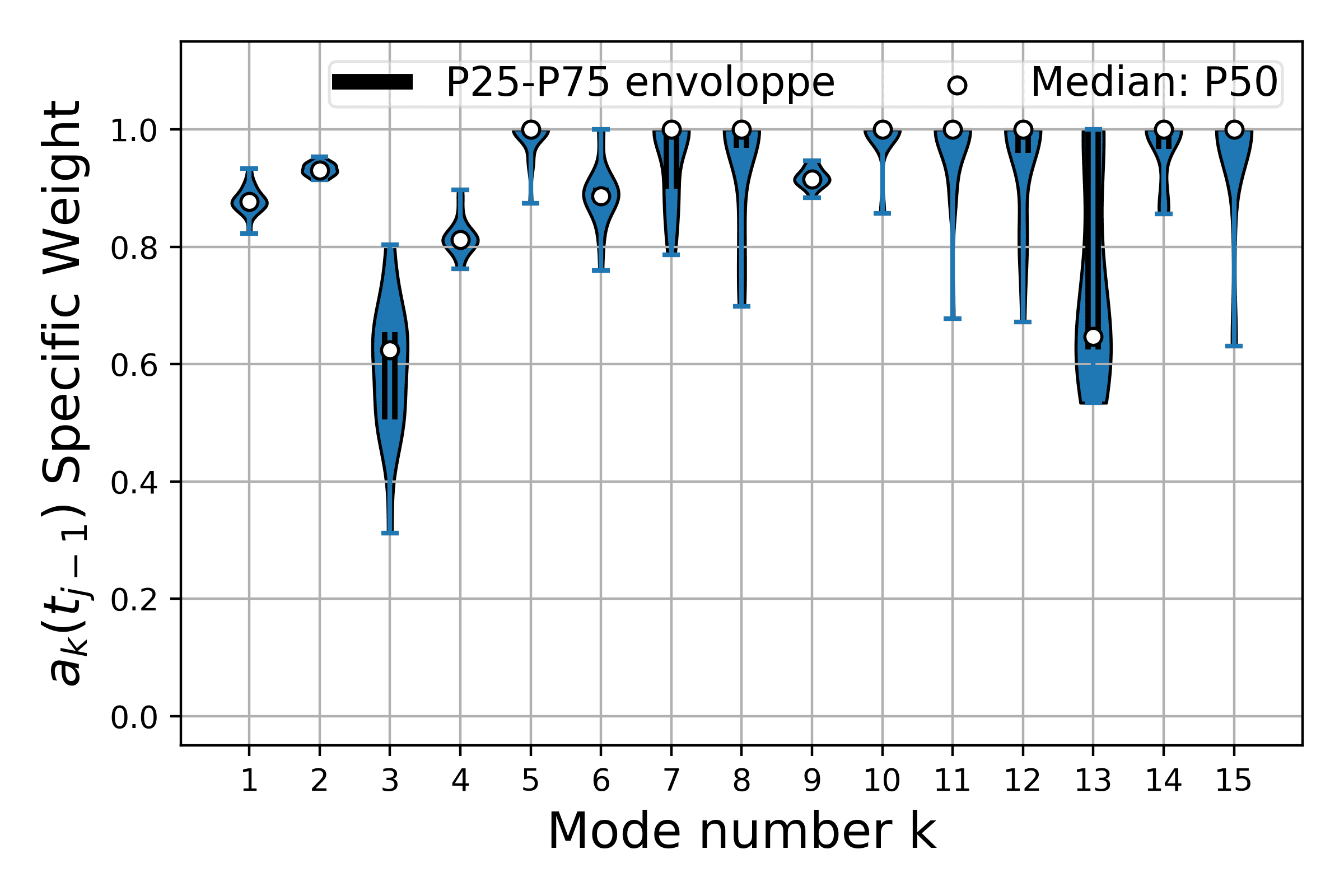}}
    \subfloat[][Polynomial term $WvH$]{\includegraphics[trim={0cm 1.5cm 0cm 2.5cm},clip,scale=0.088]{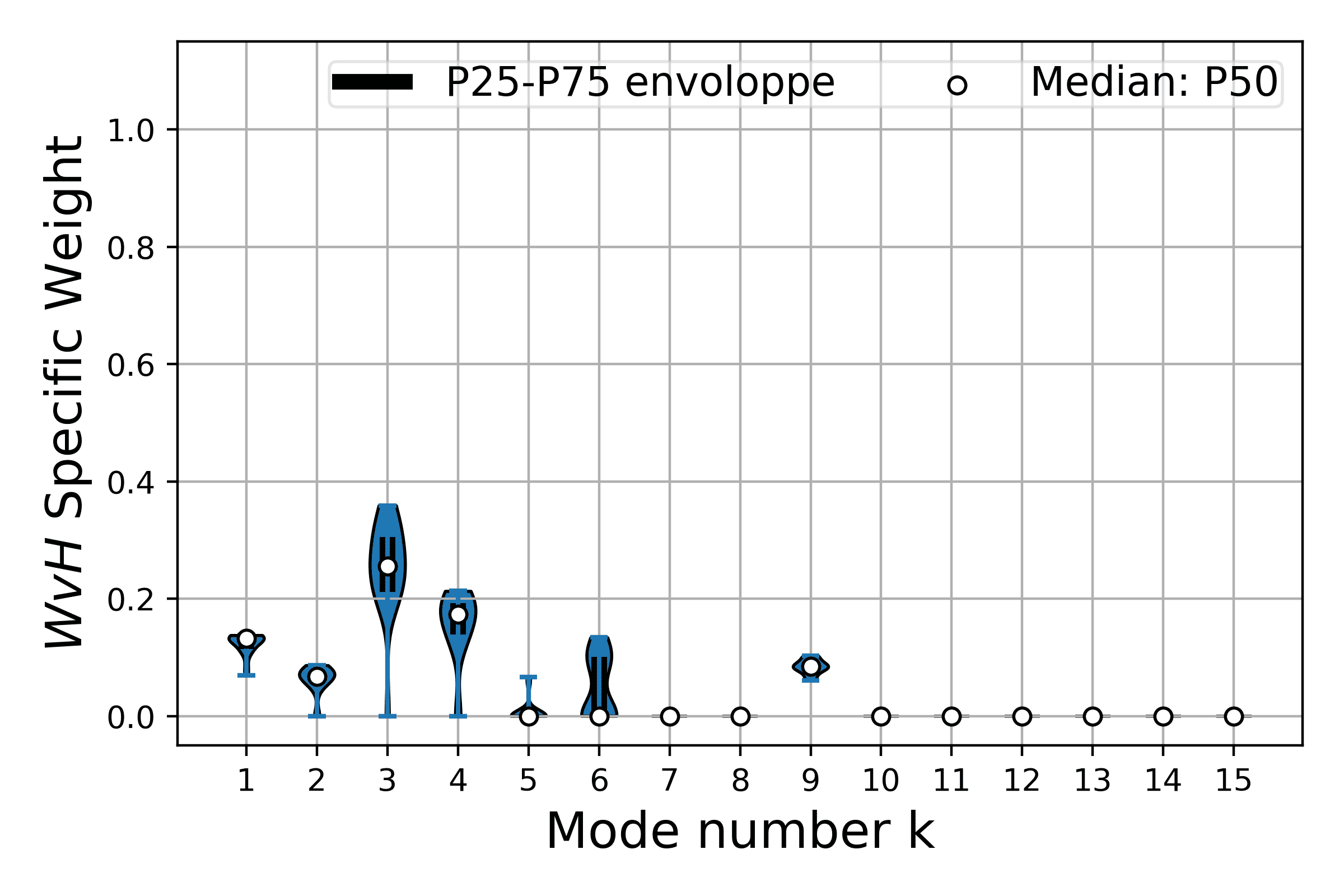}}
  \caption{Probability density functions of the GWs associated with the degree 1 monomials of variables $a_k(t_{j-1})$ and $WvH$. The training size is 50 using 20 different random picks for each size.}
    \label{fig:PCE:sensitivity:specificWeights}
\end{figure}

For this, a Bootstrap analysis was again used to construct different learning sets of size $50$, instead of choosing the first $50$ measurements. This produces a distribution of the GWs rather than a single value, for each polynomial term. The result is shown in Figure \ref{fig:PCE:sensitivity:specificWeights} for the weights of the $a_k(t_{j-1})$ and $WvH$ monomials.
For modes 1 to 4, the median weights P50 (Percentile 50) of the $a_k(t_{j-1})$ monomials, represented in Figure \ref{fig:PCE:sensitivity:specificWeights}-a, were always over $0.6$, but the variation range was strictly less than $1$, with density functions centered around the median and a small standard deviation for modes 1, 2 and 4. This means that whatever the training set, the previous state $a_k(t_{j-1})$ value was always predominant but never enough to estimate the evolution of the first four modes. A tendency (in particular linear) using the last state was not sufficient, and additional information was always needed (forcing). In parallel, Figure \ref{fig:PCE:sensitivity:specificWeights}-b shows that this information is certainly the waves, as the median values of the GW for the first four modes were between $10$ and $25\%$, corresponding to the information gap left by the previous value variable $a_k(t_{j-1})$ in the fitted PCE model. Starting from mode 5, the median values of $a_k(t_{j-1})$'s GW had greater chance of falling around 1, which means that the associated polynomial models only rely on the last recorded value of the mode for the future guess. In other terms, the constructed model consists of a linearization around the previous value (tendency capturing) and does not incorporate the correlations between the future-state and the forcing variables (causality model). This can be explained by the small variances of the higher-rank modes and the difficulty of learning the PCE models from statistics averaged over the sedimentation periods. Additionally, the P25-P75 confidence interval moved to the upper bound of the density functions.

\subsubsection{Summary of the physical insights from the learning}
\label{subsubsection:application:learning:summary}
The spatial patterns as deduced by POD express the spatial correlation in the sediment deposition from the upstream to the downstream part of the channel. The EVR reached 99\% with $d=20$ modes only, where the mean relative RMSE between the approximation and reality was slightly over 10\%. This is a statistical proof that the spatial correlations expressed in the POD patterns are explanatory of the physical dynamics over their whole range of variation (at least that observed from 2010 to 2018), with a low approximation rank. In conclusion, the dynamic problem exhibits fairly strong spatial correlations, and the solution to the problem can be expressed on a finite orthonormal basis. \\

The temporal patterns express the evolution of the sedimentation, as they multiply the spatial patterns. They were learned using PCE as a function of the previously cited inputs (previous states and forcings). The statistical model configuration (dimension and marginals) was chosen after an investigation of different options. The associated training and prediction error converged for the first three modes, and are characterized by tight confidence intervals. The residuals of the selected model were either negligible or centered around zero, demonstrating the unbiased character of the learning and prediction. The fitted models are of lower degree for the low-rank modes 1 and 2 and of higher degrees for modes 3 and 4, which are higher-rank, due to the emergence of interactions between the forcings, namely variables related to extreme behavior (storm events). The model mainly relies on the last state information, showing a strong correlation/continuity in time of the studied physics. Using GW, which measures the forcing influence for the first five modes, the action of waves was highlighted by the PCE model as a determining phenomenon. The first mode influenced the dynamic with a rate of  $64.9\%$, the previous value of the second mode with a rate of $10.2\%$ and, in third position, the mean wave height with a rate of $9.6\%$. The remaining $15.3\%$ is essentially associated with previous values of higher order modes ($10.3\%$), interactions with tides and contributions of other wave indicators. The GWs show robustness with respect to the choice of the training set members, which makes them trustworthy, at least for temporal correlation and analysis of wave influence. The main physical conclusions are that the dynamic problem is characterized by strong temporal correlations, representing more than $85\%$ of the evolution, with an external sediment source, mainly represented by the waves, representing not more than $15\%$.


\shorthandoff{:}

\subsection{Prediction of a physical field using POD-PCE coupling}
\label{subsection:application:prediction}
After performing both POD and PCE independently, the accuracy of a Machine Learning process using a POD-PCE coupling was assessed as in Section \ref{subsection:theory:methodology}. In the continuity with Section \ref{subsection:application:learning}, the first $50$ historical bathymetries were used for training and the other $14$ for forecasting. First, the impact of the size of the POD basis on the prediction process is assessed in Subsection \ref{subsubsection:application:prediction:size}. Then, the best size was determined and the average prediction behavior is analyzed in Subsection \ref{subsubsection:application:prediction:average}. The accuracy of the POD-PCE ML in predicting spatial details is assessed on cross-section examples in Subsection \ref{subsubsection:application:prediction:spatial} and a summary is given in Subsection \ref{subsubsection:application:prediction:summary}. 

\subsubsection{Influence of POD basis size}
\label{subsubsection:application:prediction:size}
\begin{figure}[H]
  \centering
  \includegraphics[trim={0cm 3cm 0cm 2.5cm},clip,scale=0.075]{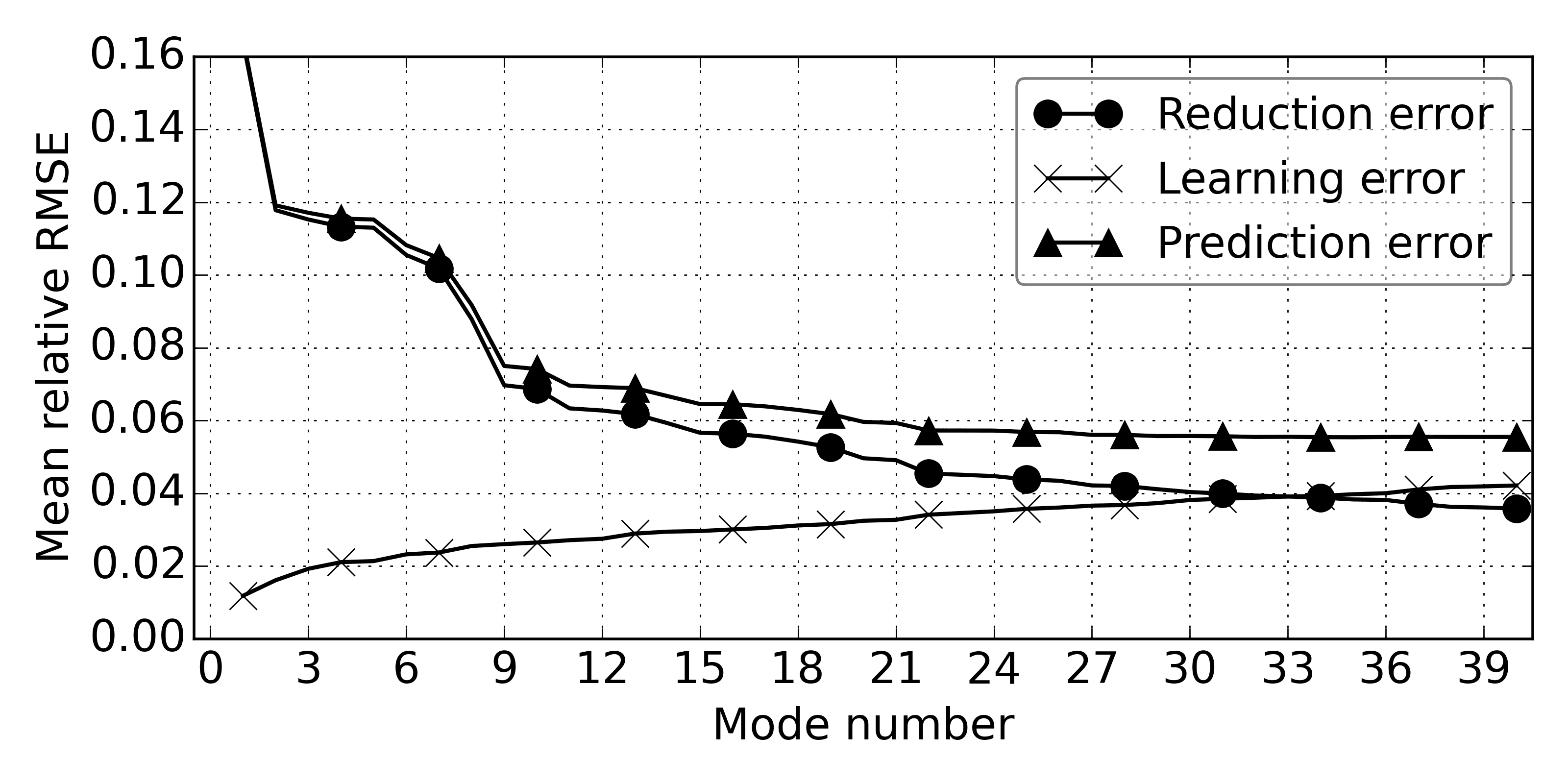}
    \caption{Mean relative RMSE generated by the reduction and the learning, and the resulting prediction errors for different approximation ranks.}
    \label{fig:Prediction:errors}
\end{figure}

In order to track the errors generated by the various steps of the algorithm (POD, PCE and coupling), the mean relative RMSE (averaged over the prediction set, as in Equation \ref{eq:timeAveragedRelativeRMSE}~) was calculated for each step (reduction, learning and prediction) and for each approximation rank $d$, as described in Section \ref{subsection:toy:learning}. The results are shown in Figure \ref{fig:Prediction:errors}~. 
Reduction error decreased from $16\%$ to $3\%$, with increasing approximation rank. The error followed a logarithmic trend, with a significant slowdown from rank 9. These errors are coherent with the errors averaged over the full set (rather than the prediction set only) in the POD results Section \ref{subsubsection:application:learning:POD} (around $8\%$). The learning error increased from $1\%$ to $5\%$ with increasing approximation rank, which is natural because the complexity of the model is increased. The learning error order of magnitude was consistent with the empirical prediction error of $4\%$ for mode 1 (as calculated in Section \ref{subsubsection:application:learning:PCE}), associated with an EVR of over $94\%$. Lastly, the prediction error decrease is the balance of, on the one hand, the increase in accuracy by adding POD modes and, on the other hand, the increase in forecasting error with increasing number of temporal coefficients to be predicted. Consequently, the prediction error decreased from 16 to 6.9\% up to rank 11, following almost the same decreasing trend as the reduction error. However, the decrease rate became slower and increasingly subdued, being overtaken by the learning errors, which dramatically increased starting from mode 12, as seen in Figure \ref{fig:PCE:errors}. Hence, a POD-PCE model of size 11 was selected for prediction.  

\subsubsection{Average performance of the chosen model}
\label{subsubsection:application:prediction:average}
\begin{figure}[H]
  \centering
  \vspace{-0.4cm}
  \includegraphics[trim={0cm 3cm 0cm 2.5cm},clip,scale=0.075]{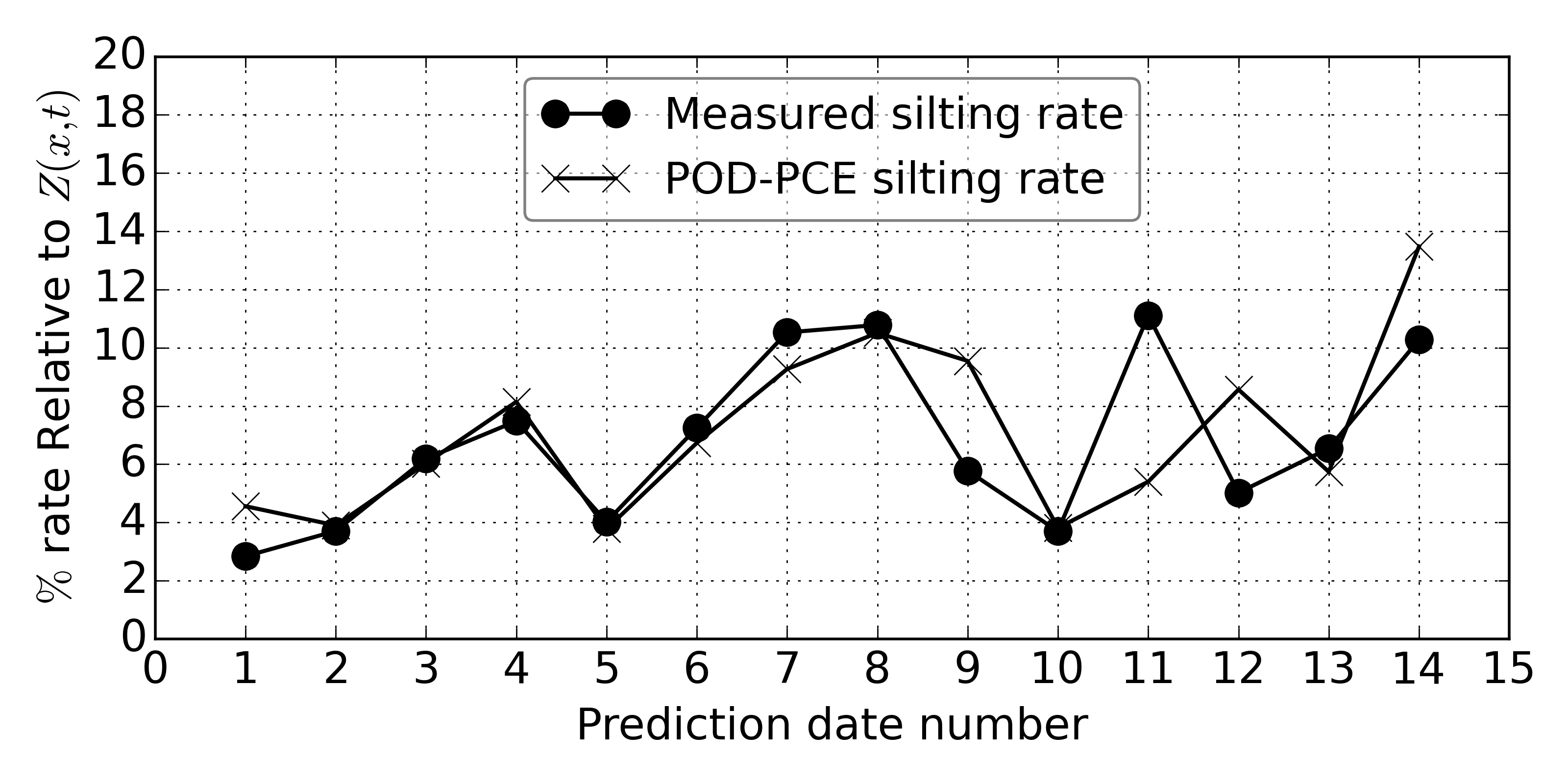}
    \caption{A comparison between the real sedimentation rates and the POD-PCE prediction of the sedimentation.}
    \label{fig:Prediction:sedimentationRates}
\end{figure}

Average sediment deposition was predicted using the POD-PCE model of rank 11, for each of the $14$ prediction dates. The average sedimentation rate, denoted $S_r$, was calculated for time $t_j$ representing the sedimentation over $[t_{j-1},t_j]$, as in Equation \ref{eq:prediction:sedimentationRate}. For operational estimation of sediment deposition, only the positive evolutions are of interest; therefore, the erosion points were discarded in calculating rate $S_r$ by cancelling negative evolutions. Indeed, $z(\mathbf{x}_i,t_j)<z(\mathbf{x}_i,t_{j-1})$ implies $\left(z(\mathbf{x}_i,t_j) - z(\mathbf{x}_i,t_{j-1})\right) = - |z(\mathbf{x}_i,t_j) - z(\mathbf{x}_i,t_{j-1})|$, and therefore a null contribution to the sedimentation rate $S_r$. Furthermore, only regions of considerable depth are of interest. Therefore only $n_p$ bathymetry points under $-1\ m$ ($\mathbf{x}_i, i \in \mathcal{N}_p$) were taken into account. The results are shown in Figure \ref{fig:Prediction:sedimentationRates}
\begin{equation}
  \label{eq:prediction:sedimentationRate}
  S_r = \frac{1}{2n_P} \sum_{i \in \mathcal{N}_p } \dfrac{\left(z(\mathbf{x}_i,t_j) - z(\mathbf{x}_i,t_{j-1})\right) + |z(\mathbf{x}_i,t_j) - z(\mathbf{x}_i,t_{j-1})| }{|z(\mathbf{x}_i,t_j)|} \qquad .
  \end{equation}

The POD-PCE prediction globally followed the real sedimentation trend, for example from Dates 1 to 8. When it was not equal to the real sedimentation rate, it was generally an overestimation, which is coherent with the asymmetry observed in the distribution of Mode 1 training and prediction residuals (Figure \ref{fig:PCE:training_residuals} in Subsection \ref{subsubsection:application:learning:PCE}). \\

For the particular Date 11 however, half of the sedimentation was missing. Investigation of the data for this particular measurement showed that the previous record, taken as input, had been made 29 days previously, which is far from the average $\Delta t \approx 15~\text{days}$; it is twice the mean interval, thus underestimating sedimentation by half. As measurement intervals were in general around the average, sedimentation time interval $\Delta t$ was not selected as a key parameter by LARS for the dynamic model, although it was given as an input and is physically significant. For the particular case of the time variable, multiplicative enhancement can be intended as a correction. However, this shows one of the limitations of statistical modeling: statistical significance can be confused with physical importance. Indeed, for the statistical conclusions to be physically significant, the measurements should be diverse enough to account for the variations in the inputs and the impact of these variations on the output. This was unfortunately not guaranteed for sedimentation measurement intervals, as they were often equal to 2 weeks. Additionally, for Dates 12 and 13, a large part of the wave measurements were missing in the sedimentation time interval. Consequently, mean wave height $WvH$ was estimated over only a small portion of the time interval. This may lead to a good prediction (Date 13) if the interval used is representative enough of the full interval, and to bad prediction (Date 12) when not, and highlights the limitations of statistical averaging.

\subsubsection{Spatial details of prediction by the chosen model}
\label{subsubsection:application:prediction:spatial}
\begin{figure}[H]
  \centering
	\vspace{-0.3cm}
  \subfloat{\includegraphics[trim={0cm 6.3cm 0cm 0cm},clip,scale=0.5]{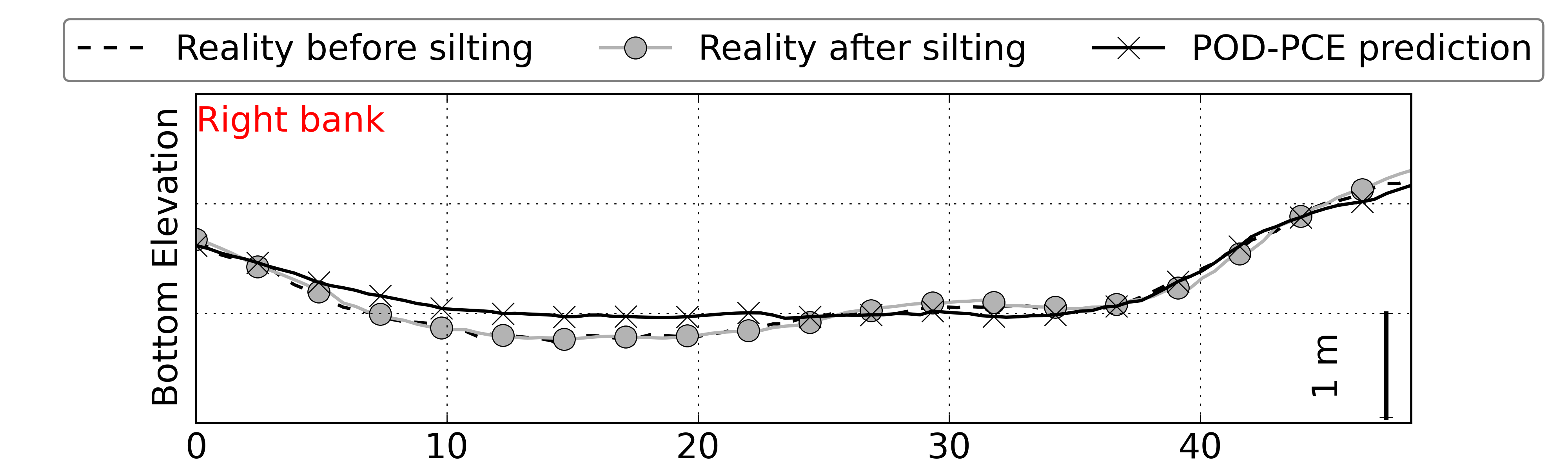}} \\
	\vspace{-0.1cm}
	\setcounter{subfigure}{0}
  \subfloat[][Entrance - Date 1]{\includegraphics[trim={0cm 1cm 0cm 2.0cm},clip,scale=0.09]{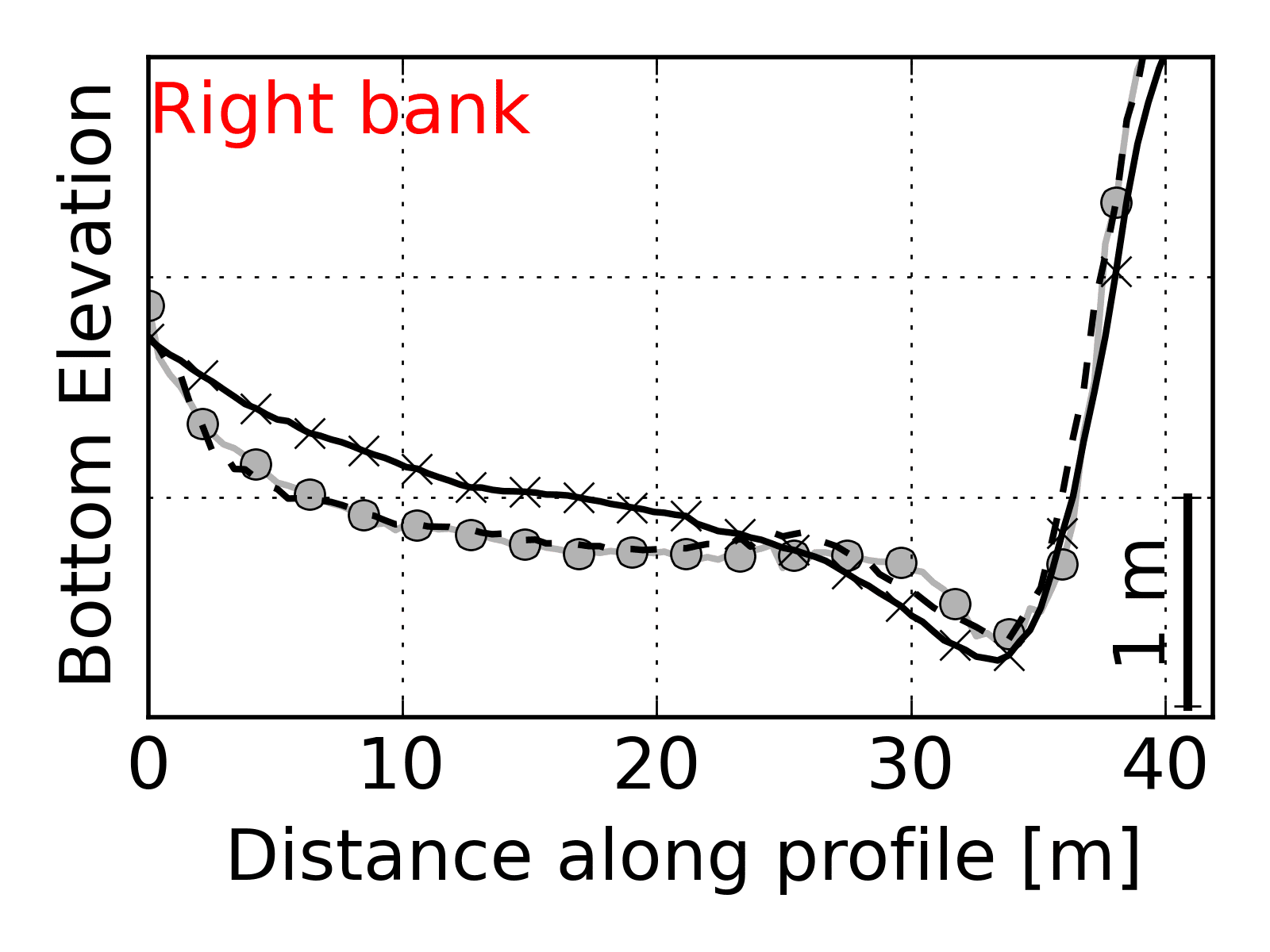}} 
  \subfloat[][Entrance - Date 3]{\includegraphics[trim={0cm 1cm 0cm 2.0cm},clip,scale=0.09]{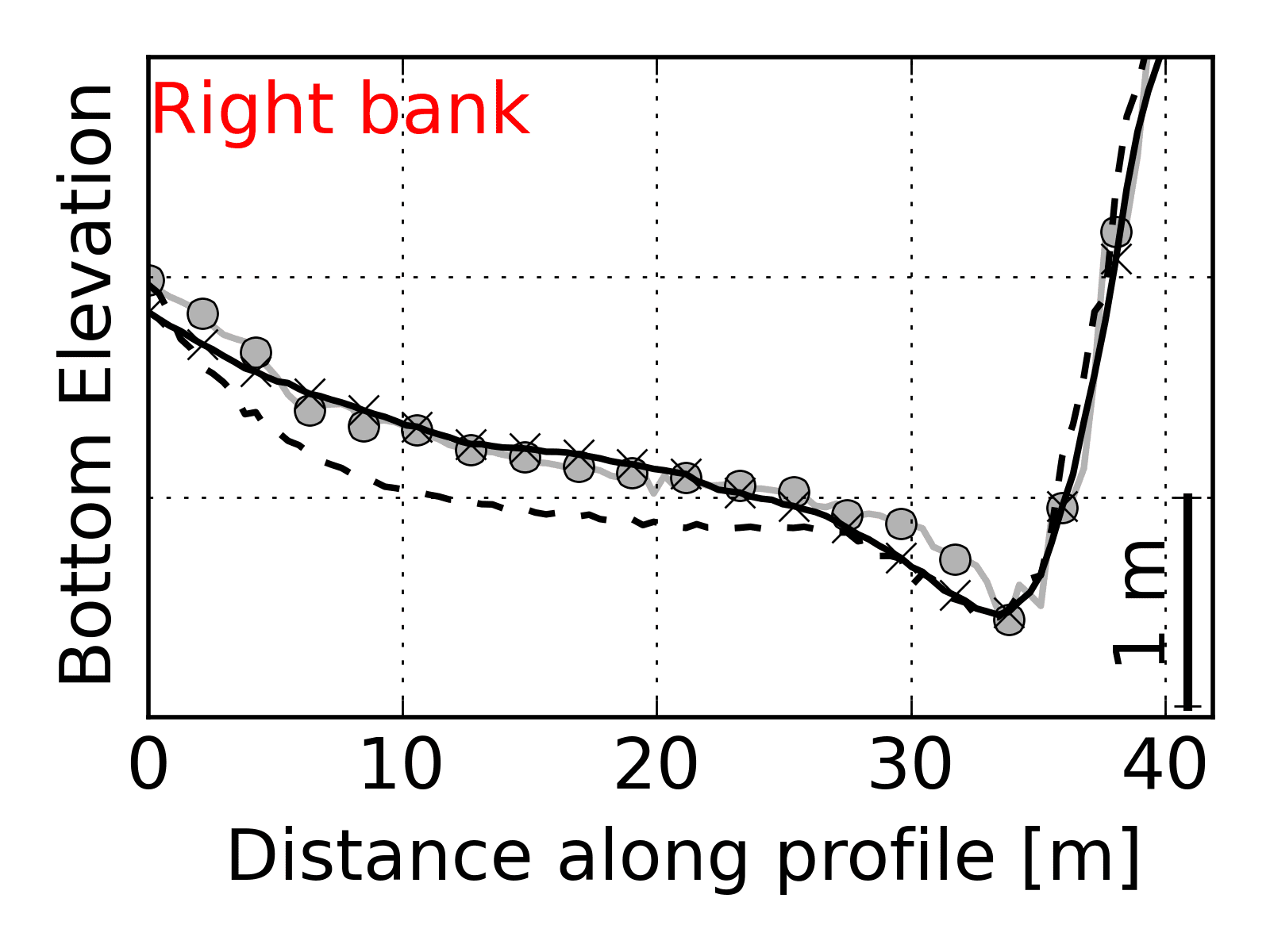}} 
  \subfloat[][Entrance - Date 7]{\includegraphics[trim={0cm 1cm 0cm 2.0cm},clip,scale=0.09]{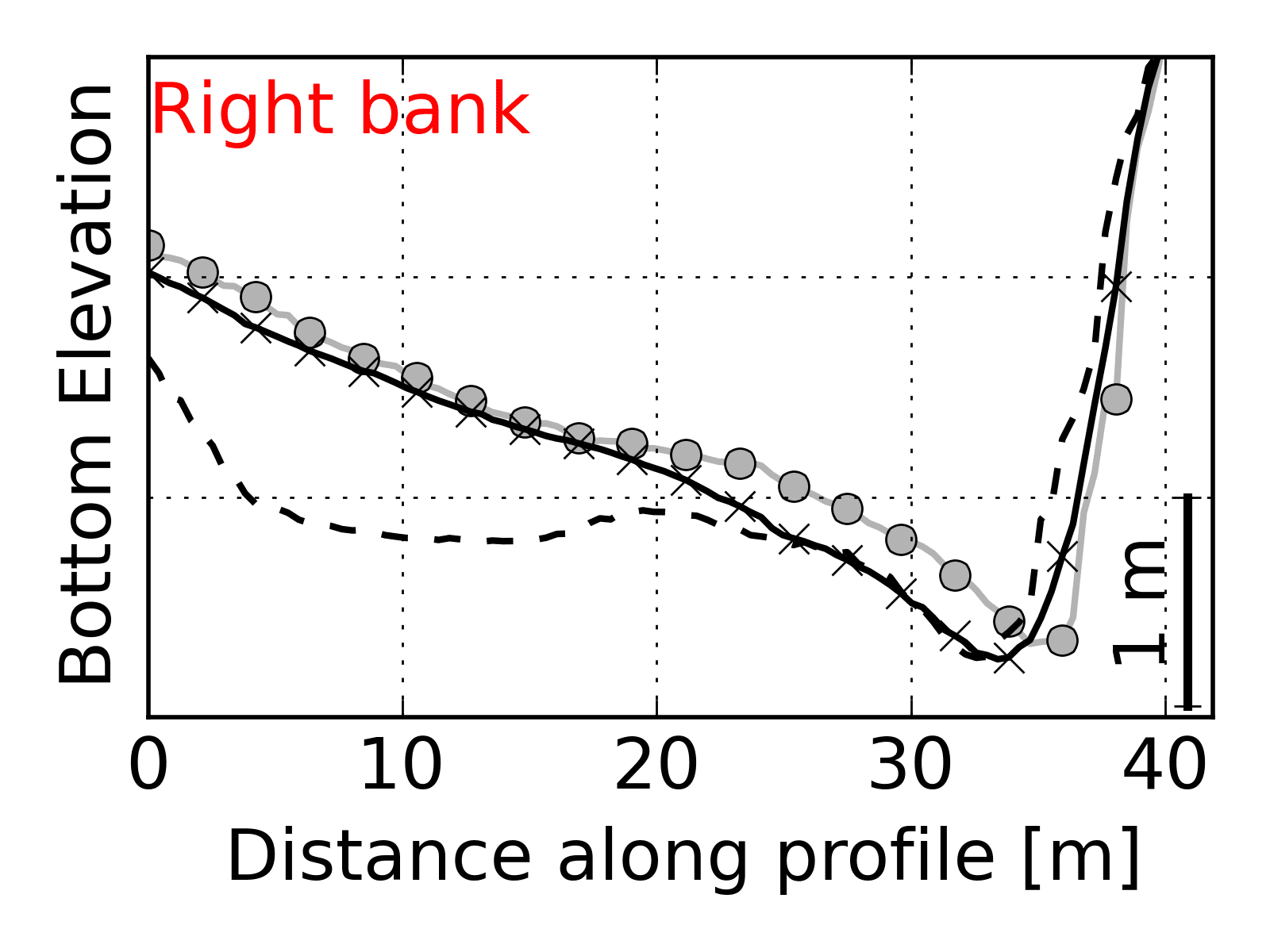}} \\
  \subfloat[][Middle - Date 5]{\includegraphics[trim={0cm 1cm 0cm 2.0cm},clip,scale=0.09]{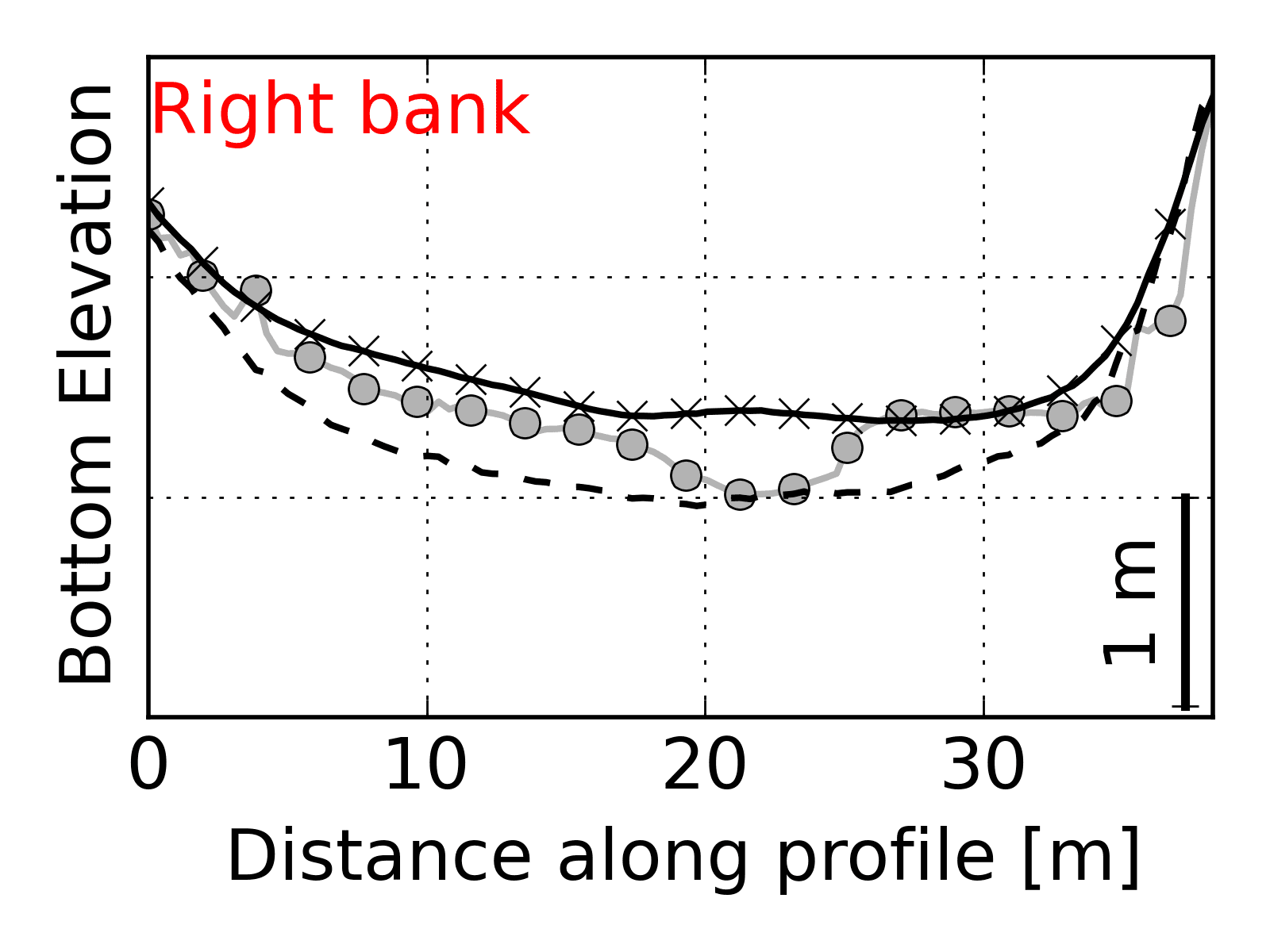}} 
  \subfloat[][Middle - Date 6]{\includegraphics[trim={0cm 1cm 0cm 2.0cm},clip,scale=0.09]{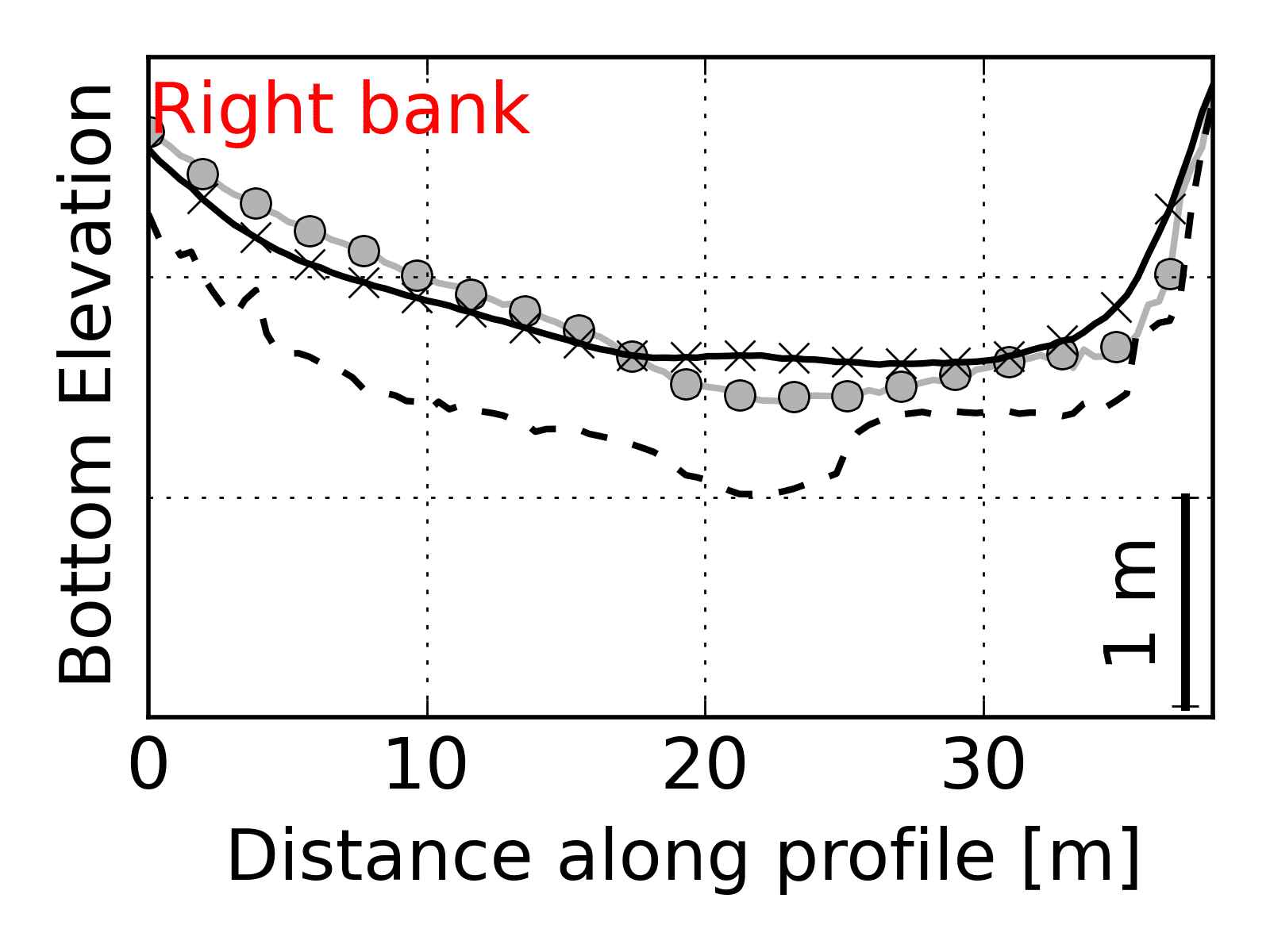}} 
  \subfloat[][Middle - Date 7]{\includegraphics[trim={0cm 1cm 0cm 2.0cm},clip,scale=0.09]{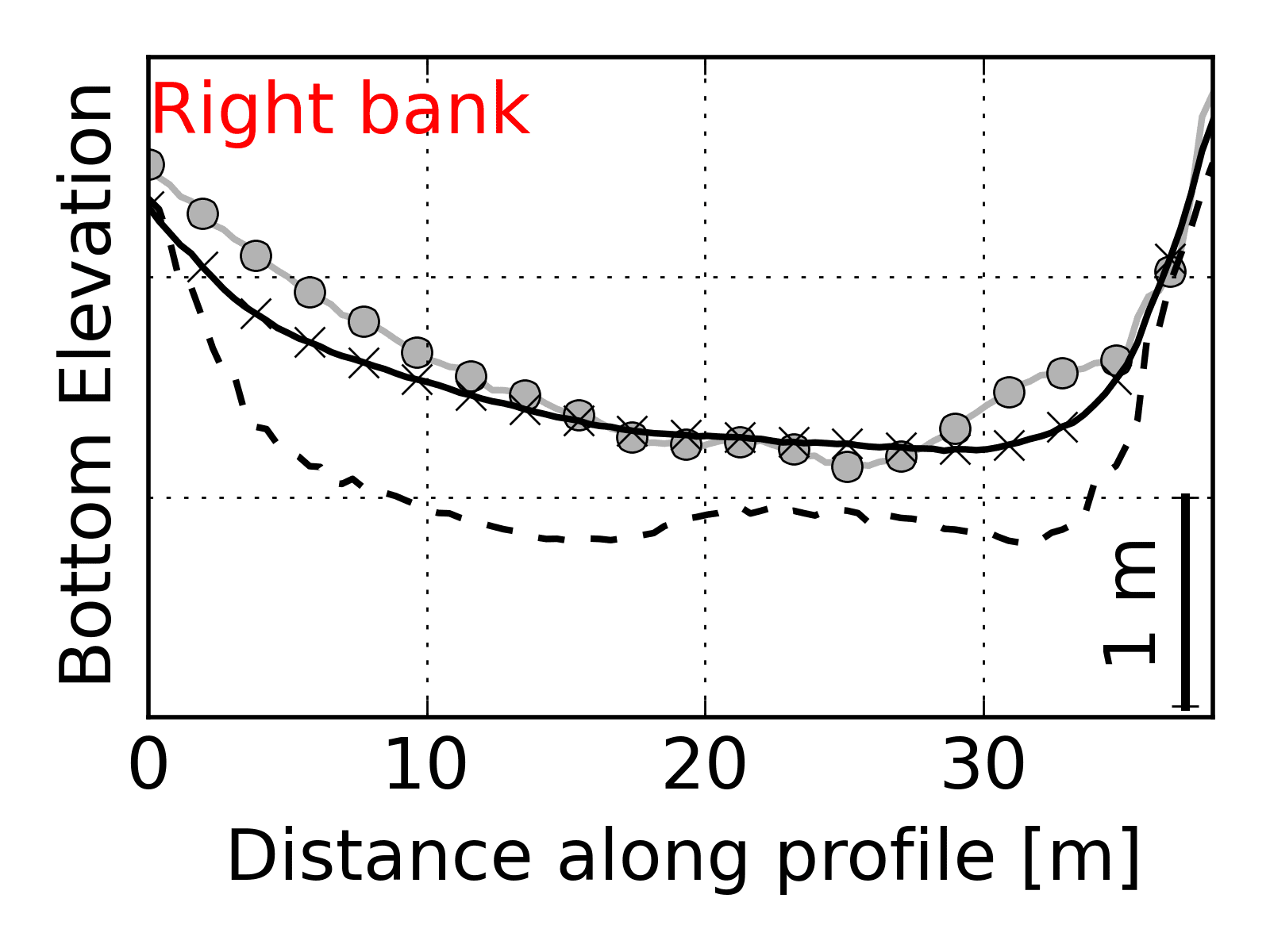}}
    \caption{POD-PCE prediction vs. reality on cross-sections at the entrance (a, b and c) and  middle of the first portion (d, e and f) of the intake.}
    \label{fig:Prediction:spatial:entrance_middle}
\end{figure}

The spatial details of the prediction were analyzed on cross-sections for specific prediction dates. First, sediment deposition was observed on a cross-section at the entrance of the intake (Figures \ref{fig:Prediction:spatial:entrance_middle}-a, b and c). 
In accordance with the previous conclusions from Figure \ref{fig:Prediction:sedimentationRates}, the POD-PCE prediction captured various sedimentation ranges, as shown with Dates 2 and 7. However, a slight artificial sedimentation was predicted whereas there was no dynamics in reality, for example for Date 1 (Figure \ref{fig:Prediction:spatial:entrance_middle}-a), due to the fact that the model is continuous, whereas threshold phenomena can occur in reality. Next, sediment deposition was observed on a cross-section at the middle of the first portion of the intake (Figures \ref{fig:Prediction:spatial:entrance_middle}-d, e and f). Mean sedimentation was well captured, but some details of the bathymetry were missing, such as formation of a new feature for Date 5 (distance 20 to 25 m) and Date 7 (distance 30 to 35 m). For Date 6, sediment deposition was slightly underestimated in the right bank and overestimated in the left bank. However, although the details of sediment deposition were not perfectly captured, the value of the sedimentation area corresponds well enough to reality. It can also be concluded that the way the RMSE and relative errors are averaged in space, for example in Figure \ref{fig:Prediction:errors}, actually penalizes the accuracy of the algorithm because it does not take account of the oscillation of the prediction around an accurate mean. \\

\begin{figure}[H]
  \centering
	\vspace{-0.2cm}
  \subfloat{\includegraphics[trim={0cm 6.3cm 0cm 0cm},clip,scale=0.5]{figs/profiles_legend.png}} \\
	\vspace{-0.1cm}
	\setcounter{subfigure}{0}
  \subfloat[][Bending - Date 4]{\includegraphics[trim={0cm 1cm 0cm 2.0cm},clip,scale=0.09]{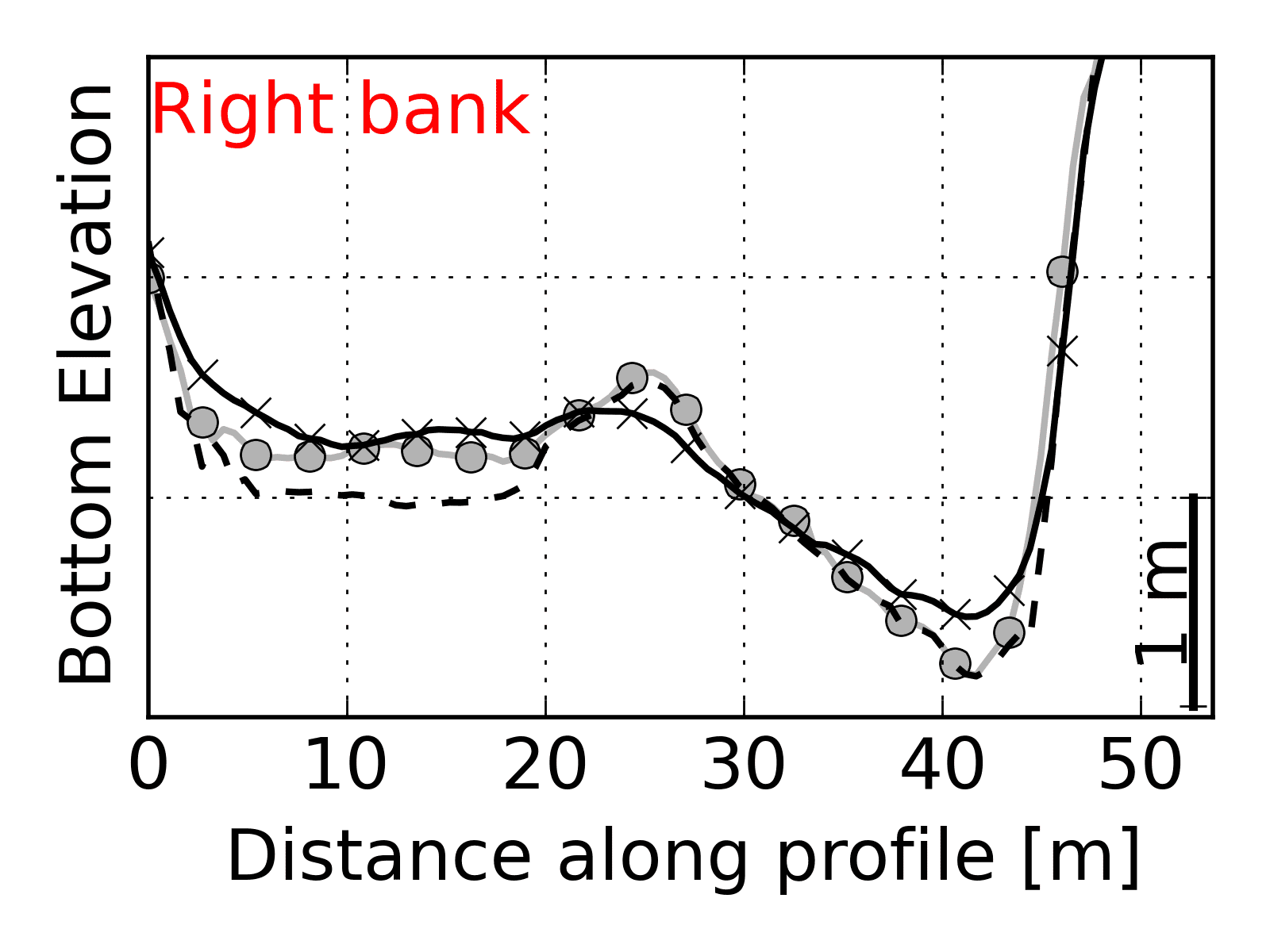}} 
  \subfloat[][Bending - Date 7]{\includegraphics[trim={0cm 1cm 0cm 2.0cm},clip,scale=0.09]{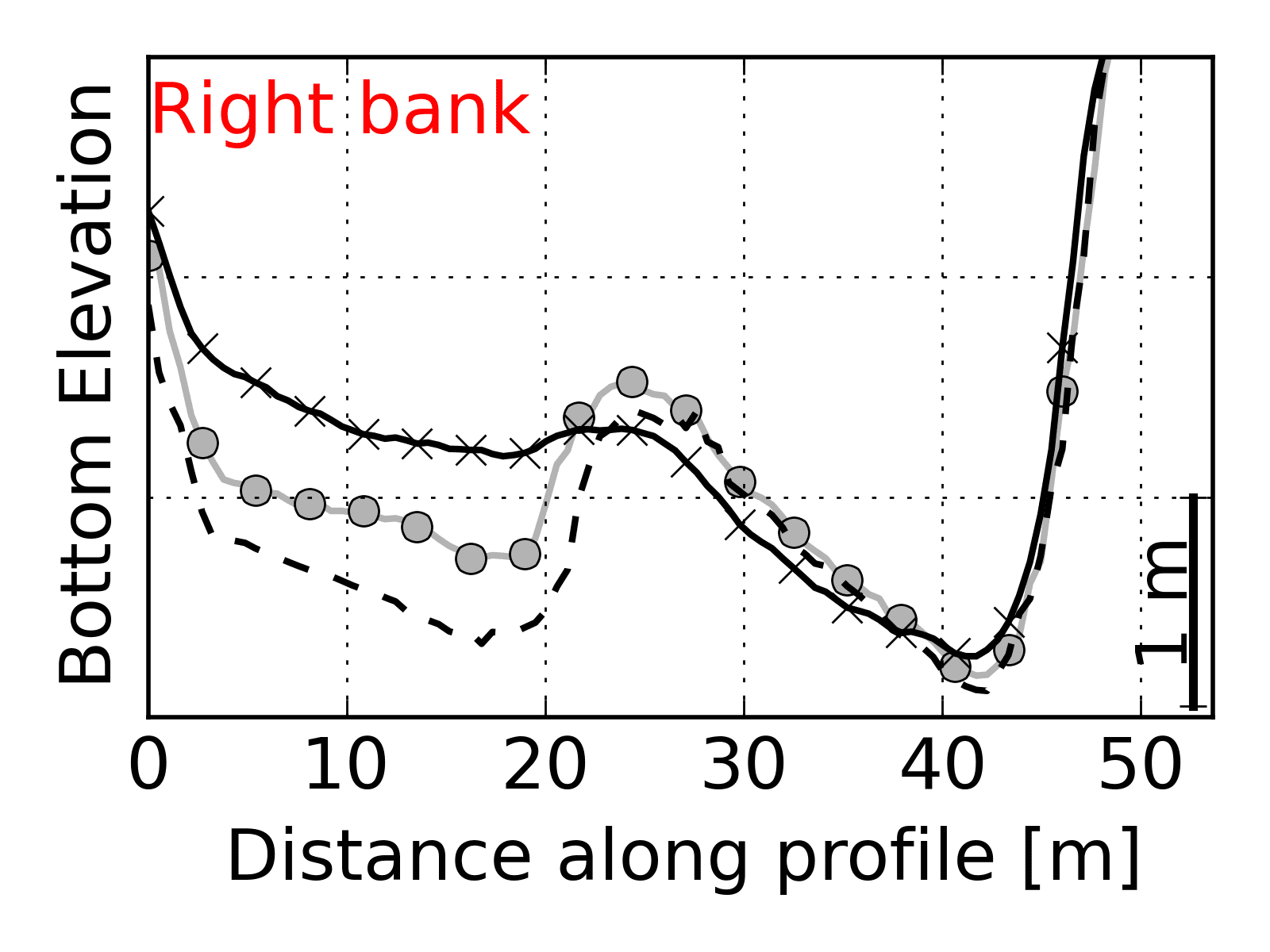}} 
  \subfloat[][Bending - Date 14]{\includegraphics[trim={0cm 1cm 0cm 2.0cm},clip,scale=0.09]{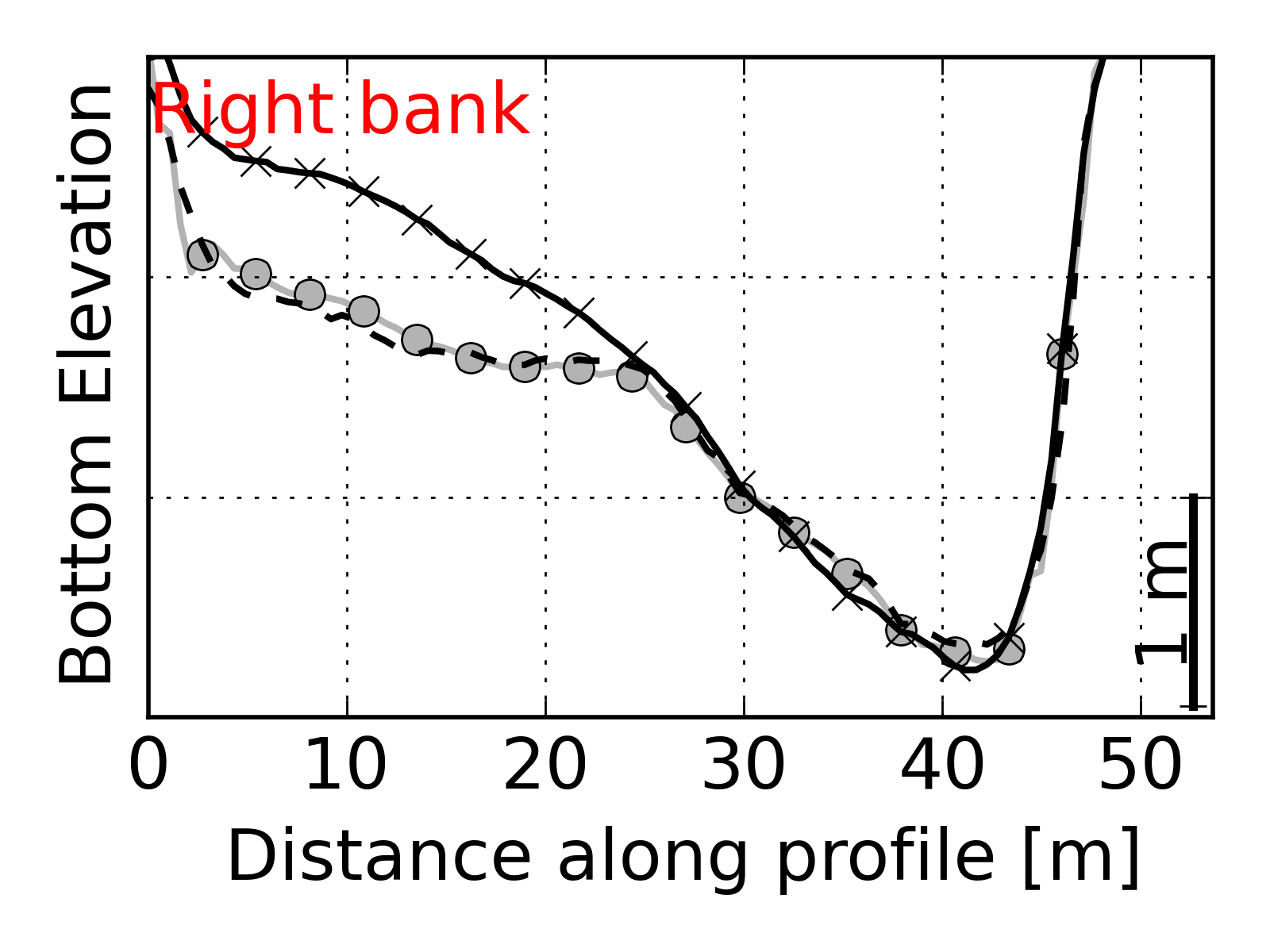}} \\
  \subfloat[][Last - Date 4]{\includegraphics[trim={0cm 1cm 0cm 2.0cm},clip,scale=0.09]{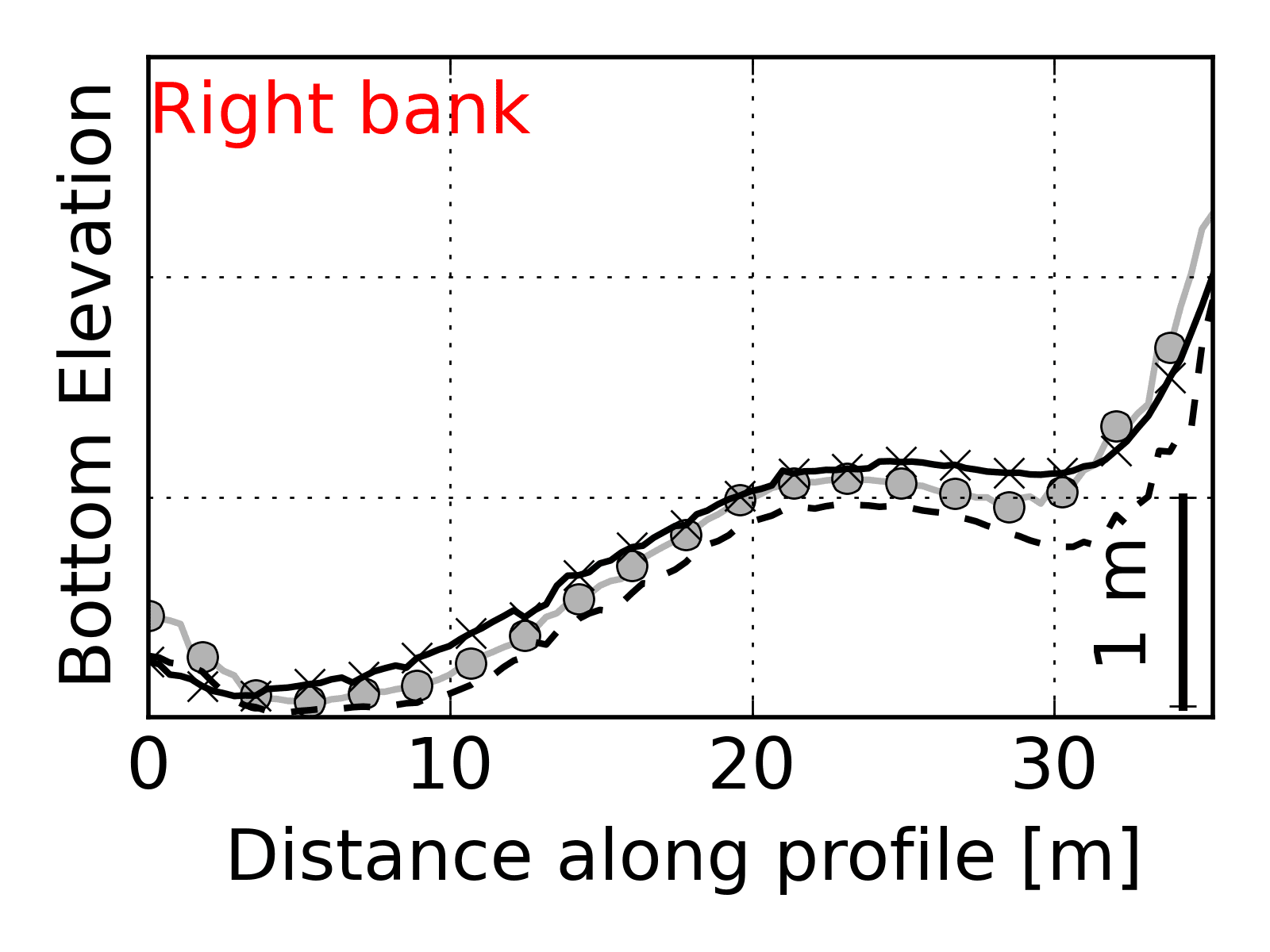}} 
  \subfloat[][Last - Date 5]{\includegraphics[trim={0cm 1cm 0cm 2.0cm},clip,scale=0.09]{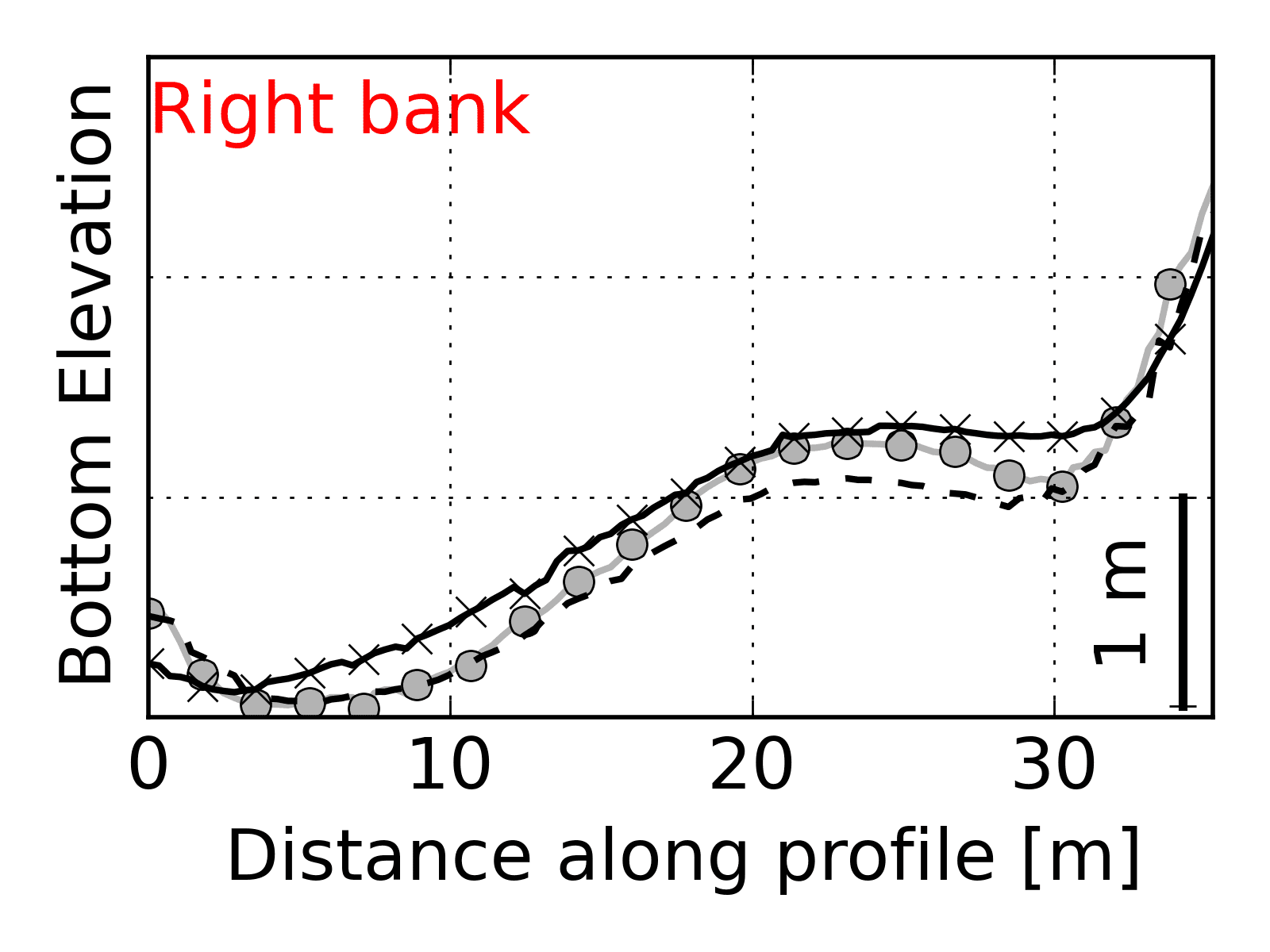}} 
  \subfloat[][Last - Date 11]{\includegraphics[trim={0cm 1cm 0cm 2.0cm},clip,scale=0.09]{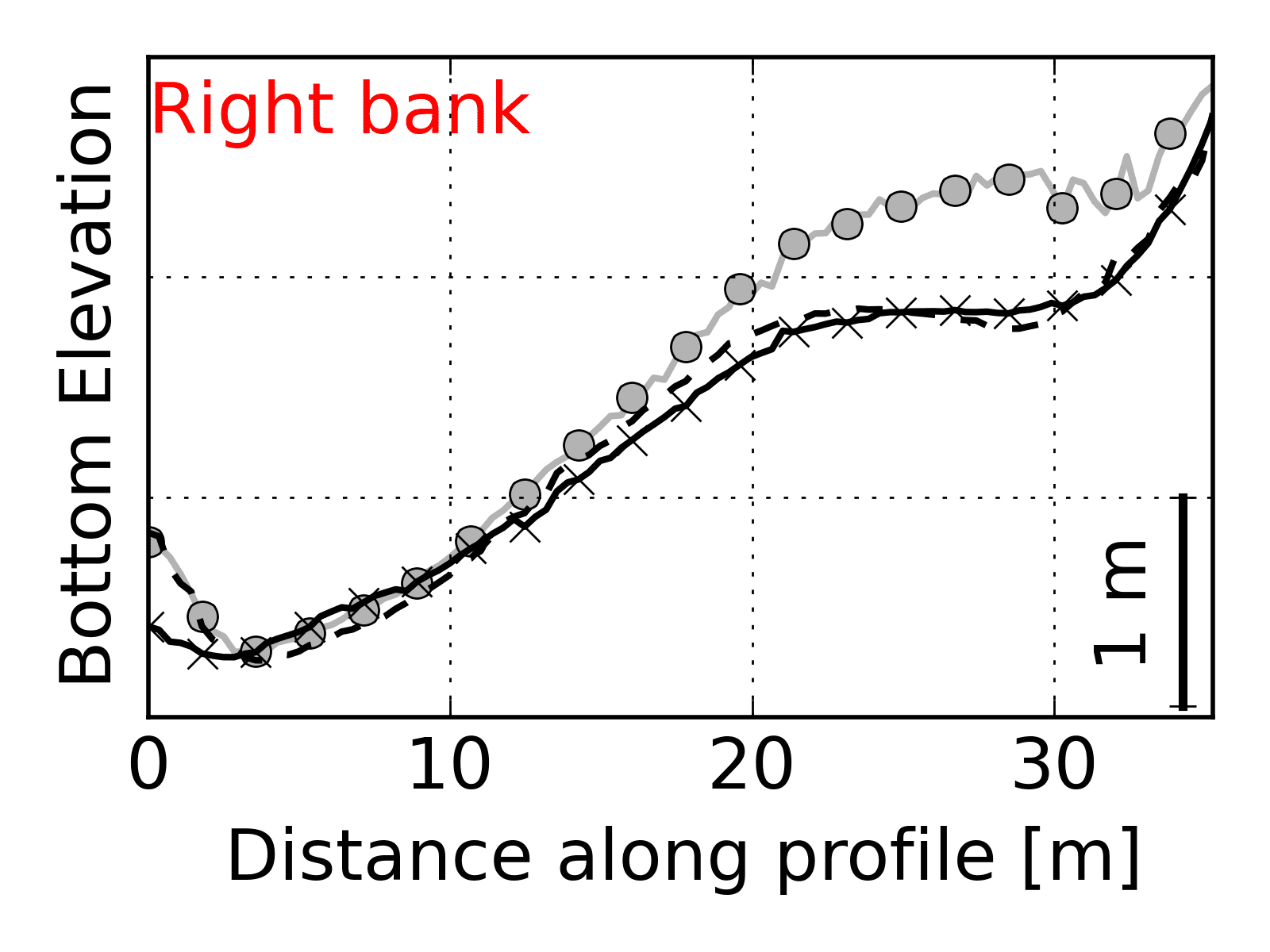}}
    \caption{POD-PCE prediction vs. reality on cross-sections at the bending (a, b and c) and last portion (d, e and f) of the intake.}
    \label{fig:Prediction:spatial:bend_downstream}
\end{figure}

 Then sediment deposition was observed on a cross-section at the bending part of the intake (Figures \ref{fig:Prediction:spatial:bend_downstream}-a, b and c). It shows that the prediction algorithm understood that the sediment deposition mainly occured in the right bank of the channel, for example for Date 7, even though it was overestimated. Furthermore, considering modes of higher rank from the previous measurement as an input, the algorithm captured the swing in the profile throughout its history, which is here observed from Date 4 to 14. Lastly, a cross-section of the last portion of the channel, in front of the downstream pumping station, is analyzed (Figures \ref{fig:Prediction:spatial:bend_downstream}-d, e and f). Once again, the prediction algorithm understood where the sediment deposition occurs, this time in the left bank of the channel, coherent with the pattern represented by Mode 2 (Figure \ref{fig:POD:SR:spatial}-b). However, it can be seen that unusual sediment deposition occurred for Date 11, which was not captured by the model, and may correspond to the arrival of less frequent fine sediment that appears in Mode 4 (Figure \ref{fig:POD:SR:spatial}-d). This also explains the sediment deposition error observed for Date 11 in Figure \ref{fig:Prediction:sedimentationRates}.

\subsubsection{Summary of POD-PCE algorithm performance}
\label{subsubsection:application:prediction:summary}

Overall, the proposed learning algorithm showed interesting prediction characteristics. The model complexity can be increased gradually, by increasing the number of POD modes when accurate. Plotting error against the number of modes shows a convergence that helps in selecting the optimal number of modes. The average RMSE of the predicted field remains reasonably low. It was around $6.9\%$ with the 11 modes selected in the present case. \\

As a comparison, additional learnings were performed using different NN set-ups in \hyperref[Appendix:B]{Appendix B}. It is shown that POD-PCE gives the best equilibrium between accuracy and fit-time. Among the tested algorithms, only one allowed a RMSE reduction of $0.2\%$. However, this is of little importance in view of the very large increase in fit-time. Indeed, the latter nearly equals two-hours whereas POD-PCE is fitted in seconds (Table \ref{table:PODPCE_NN_summary} in \hyperref[Appendix:B]{Appendix B}). In addition to fit-time, choosing an algorithmic set-up for NN may be time consuming, as many network architectures are possible, not to mention the numerous choices for the involved hyper-parameters (e.g. Activation Functions). \\

With the proposed POD-PCE algorithm, the trends are well captured for spatially averaged quantities (here sedimentation rate) and for detailed spatial representations of the field. Good spatial distribution of evolution is guaranteed by the POD modes, even when evolution amplitude is over- or under-estimated. \\

Nevertheless, some disadvantages should be noted. For example, less frequent events that are represented by modes of higher ranks can be overlooked. Furthermore, sudden changes in features were not sufficiently captured, due to the high temporal correlation between last state and future state that was incorporated in the learning. 



\shorthandoff{:}
\section{Summary and discussion}
\label{section:summary}
In this study, POD-PCE coupling for field-measurement based Machine Learning was proposed and assessed on a toy problem and a real case. The first one concerns the learning of an analytical solution to groundwater perturbations in an aquifer subject to tidal solicitations, and the second, in an industrial context in the field of geosciences, concerns bathymetry prediction. Both are complex physical phenomena involving non-linear dynamics and various forcings. \\

POD showed excellent performance on both applications, for dimensionality reduction and physical analysis. This is an important property of POD \citep{Kerschen2002}, where the mathematical basis ends up to be physically interpretable, because it efficiently expresses the dynamics. However, adding random perturbations to the toy problem data showed that noise may contaminate the POD patterns (temporal and/or spatial), although modes of high variance are robust. If the noise is significant enough, it may also take more important positions in the decomposition than physical patterns of smaller statistical occurrence. It can then be interesting, for real data, to eliminate modes showing completely stochastic behavior in favor of explainable modes of lower variances. Next, the investigation of POD coefficients is also physically informative: dependency to inputs can sometimes be directly noticed with appropriate plots, and the regularity of the modes can be related either to the representation of different space and time scale physics, or to less frequent events. The potential of POD for detecting biased and missing data was also assessed in the real case. POD was first applied to the whole set of measurements, but discontinuities emerged in the temporal signals of the decomposition. Such a procedure is important because, in most of cases, the data need to be filtered, which is a time-consuming task. The POD enabled fast recognition of elements that react differently from the average. However, many points of improvement are worth mentioning. Firstly, the choice of POD as a decomposition technique was here motivated by its simplicity and the possibility of interpretation when coupled to a linear learning formulation such as PCE. Other decomposition techniques exist, and many authors attempted comparisons, for example with Fourier \citep{Paul2017}, extensions of POD \citep{Hekmati2011} or other classes of decomposition \citep{Schmid2010}. For the real case application, other decomposition techniques such as Kernel Principal Component Analysis (KPCA) \citep{Mika1999} and Sparse PCA \citep{Johnstone2009} were analyzed, without significant improvements. Secondly, data filtering using POD consisted only in deleting the poor-quality measurements and extracting the spatial zones where data were always measured. POD can however be used to reconstruct missing data, by inverse projection on POD basis elements deduced from qualitative data \citep{Saini2016}. This could help to extend the statistical set for the learning. Lastly, a linear interpolation of the bathymetry was used to project all the measurements onto the same grid for POD application. The uncertainty that emerged from this interpolation process was not treated. This, added to the measurement errors, can shed light on model behavior. For example, comparison of mono-beam cross-sections with multi-beam measurements and uncertainty propagation of bathymetry errors through the learning could be attempted, especially because uncertainties in the bathymetric information may impact the flow field computation \citep{Legleiter2011}.  \\

PCE was used to learn the POD modes coefficients as 1D data. We showed the importance of the polynomial basis, and therefore of marginals choice, for the learning phase of the real case problem. Indeed, choosing for example uniform distributions, associated with Legendre family, might not be appropriate even though it is widely used when no input information is available \citep{Torre2019}. Moreover, the number of inputs can alter the learning. When using LARS, the presence of numerous variables can mislead the algorithm to overfitting. Hence, a good combination between polynomial basis and dimension choice could significantly improve convergence speed, centering of residuals and mean training and prediction errors. The proposed contribution analysis using the PCE coefficients has been successfully tested on the toy problem, resulting with physically coherent conclusions. On the measurements set, it showed that the last-state information is often the most influential input. A robustness test was conducted on the latter by varying the training set, and the observation was stable. Additionally, the noise tests performed on the toy problem showed that LARS selects physically significant polynomial patterns even when the noise contaminates the POD coefficients. PCE and in particular LARS are therefore robust to noise, that propagates from a two-dimensional field to its POD coefficients. This is coherent with the conclusions in \cite{Torre2019} about PCE robustness to noise in $1D$ data. In the bathymetry case, for the modes of small ranks associated with the largest variances, wave height was the most influential forcing, whatever the chosen learning set. This is consistent with physical knowledge of sediment mobilization in coastal configurations, where waves are known to be determining through the influence they have on bed shear stress \citep{VanRijn_2007_a}. For modes of higher rank, however, the only selected variable by LARS is the last-state information. Firstly, the forcings that were used as PCE inputs were simple statistical estimators deduced from the data (means, percentiles, etc.). This reduction was used instead of giving all the time series as an input, because the problem would become ultrahigh-dimensional. This unfortunately wastes the richness of the available data as tidal information that are measured on an hourly basis. A more accurate statistical reduction of the inputs could be used. For example, \citet{Lataniotis2018} used PCA and KPCA for surrogate modeling with PCE and Gaussian Processes on ultrahigh-dimensional problems. Secondly, dependencies were not specifically modeled. These can be incorporated using the mathematical setting for the construction of the polynomial basis established by \citet{SoizeGhanem2004}~. The dependencies, however, indirectly influenced the construction of the model via selection of basis elements by LARS, which avoids redundancy. Thirdly, the choice of tested input configurations for PCE was arbitrary. A more objective variable selection technique is necessary \citep{Noori2011}. For example, the information from previous times could also be used as inputs for temporal evolution problems. This may raise other technical questions, such as the number of previous times that should be accounted for (time lag estimation) \cite{Du2019}. Lastly, PCE was chosen for the interpretation possibilities that it allows when combined to POD, thanks to the direct computation of importance measures from the expansion coefficients. Other interesting properties can encourage the use of PCE, for example the developed theoretical frameworks for the treatment of discontinuity \cite{WanKarniadakis2005}. However, some limitations are noted. For example, PCE worked better for modes associated with high than low variances. Although it may be tempting to conclude that modeling of high rank modes is not necessary, it should be noted that their accurate prediction can make the difference between average forecasting and capturing of less frequent events and/or smaller scale features. Therefore, the present ML could be enhanced by improving the learning of high-rank modes. For example, the construction of marginals and the use of random draw with confidence intervals, or extreme statistics models \citep{Ghil2011}, instead of causal models like PCE, could be attempted. \\

Finally, the robustness and convergence properties added to the physical interpretability supported the choice of POD-PCE coupling as a ML prediction algorithm. It respects the PDR (Predictive, Descriptive, Relevant) framework defined in \citep{Murdoch2019}. It is characterized by both predictive and descriptive accuracy (simplicity) and is stable with respect to data disturbance. It offers the best equilibrium between accuracy and fit-time, compared to other NN configurations tested on the real-case problem. This is consistent with the conclusions by \citet{Torre2019}, where PCE errors are comparable to the best NN from literature, on classical ML cases, while being much faster. Additionally, POD-PCE is interpretable, as the sparsity, simulatability and modularity defined in \citep{Murdoch2019} are respected by construction. Finally, it is both interpretable at features level (POD components and their PCE) and at multidimensional output level (GW compared to the proposed GGW indicators). The POD-PCE ML was therefore implemented using maximum the first 4 modes for the toy problem, and using the first 11 modes for the real case problem, after sensitivity test to number of modes. Mean information (e.g. sediment deposition rate) was in general well reproduced. Profile-by-profile investigation and $2D$ maps comparisons also showed that POD-PCE coupling was promising, as the spatial distribution of the groundwater perturbation on one hand, and the sediment deposition patch locations and amplitudes on the other hand, were well represented. Some general limitations should be highlighted and could be good perspectives for improving the process. The small data-set was a clear handicap in the measurement based problem. Some events, such as sediment downstream the intake or variation in measurement intervals, were poorly represented. It would be interesting to test the methodology on an enriched data set in order to assess the real potential of POD-PCE Machine Learning. Due to the lack of such data, input distributions were certainly not well approximated. One way of improving POD-PCE coupling would be the development of hybrid measurement-based/process-based data learning \citep{SenentAparicio2019,Mosavi2019}. This could be used to enrich the data set, not only by increasing its size (emulated realistic scenarios) but also by adding new input parameters that are not measured but obtained from process-based modeling. Last, the used POD and PCE basis may not always be sufficient for fields whose dependence to conditioning parameters considerably varies over time. Namely, the PDFs of the inputs may evolve, and the number of the basis elements needed for a good representation of the output may increase in time (stochastic drift \cite{Gerritsma2010}). Solutions as the Time-Dependent generalized Polynomial Chaos (TD-gPC) \cite{Gerritsma2010} could be interesting to explore for long-time learning problems. In particular, an adaptive strategy is used to update the basis elements when needed. Alternatively, the Dynamically Orthogonal (DO) decomposition \cite{Musharbash2015}, where both the basis elements and expansion coefficients vary over time in a Karhunen-Loève form, offers a good perspective.



\section*{Acknowledgements}
This work is funded by the French National Association of Research and Technology (ANRT) through the Industrial Conventions for Training through REsearch (CIFRE) in agreement with EDF R\&D. The authors acknowledge their support, and are grateful for data collection and feedback from EDF operators. In particular, we would like to thank D. Roug\'e for providing the data-set used in this study and for his continuous availability. We also would like to thank Pr. L. Terray (CERFACS) and Dr. M. Rochoux (CERFACS) for constructive discussions on POD and PCE respectively, and Pr. B. Sudret (ETH Zurich) for providing key literature elements on the treatment of ultrahigh dimensional problems and functional inputs using PCE. The authors also gratefully acknowledge the OpenTURNS open source community  (An Open source initiative for the Treatment of Uncertainties, Risks'N Statistics). Finally, we would like to thank the anonymous reviewers, whose comments and suggestions helped improve the manuscript.

\shorthandoff{:}

\section*{Appendix A. Complementary results on the parametric toy problem}
\label{Appendix:A}
In addition to the content of Section \ref{section:toy}, supplementary materials are herein given for the parametric toy problem. Firstly, \textit{Garson Weights} (GW) and \textit{Generalized Garson Weights} (GGW), respectively presented in Sections \ref{subsection:theory:PCE} and \ref{subsection:theory:methodology}, were calculated for fitted PCE models of the first four POD modes, in the amplitude learning case, as in Table \ref{table:toy:sensitivity:generalizedWeights_amplitude}.  \\

\begin{figure}[H]
  \centering
  \vspace{-0.5cm}
  \includegraphics[trim={0cm 0.5cm 0cm 0.3cm},clip,scale=0.35]{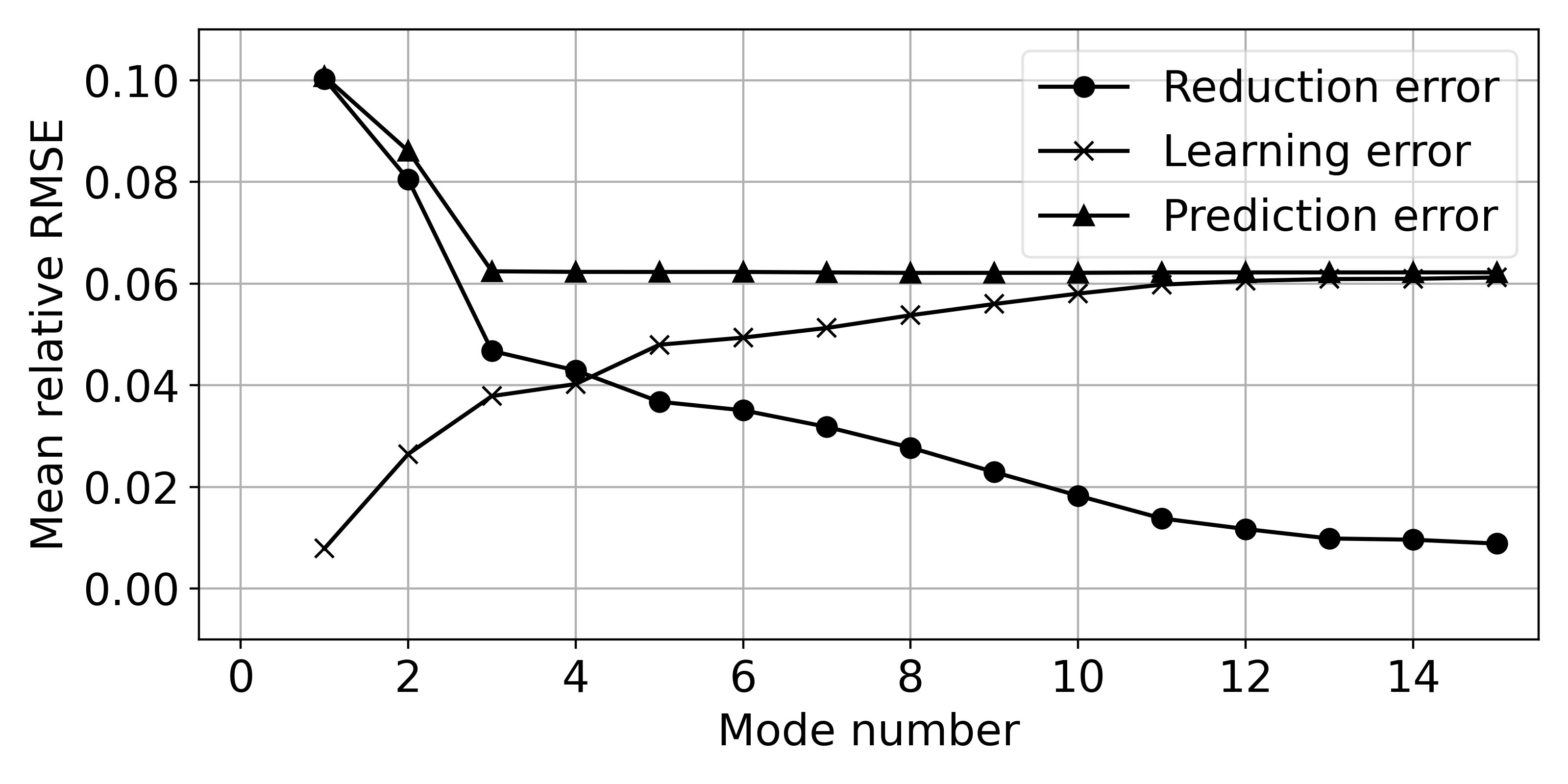}
    \caption{Mean relative RMSE generated at different steps of the POD-PCE ML applied to the time-lag case, with different approximation ranks.}
    \label{fig:toy:Prediction:errors_phase}
\end{figure}

\begin{figure}[H]
  \centering
  \vspace{-0.3cm}
  \subfloat[][Analytical]{\includegraphics[trim={0.7cm 1.5cm 1.2cm 3.2cm},clip,scale=0.35]{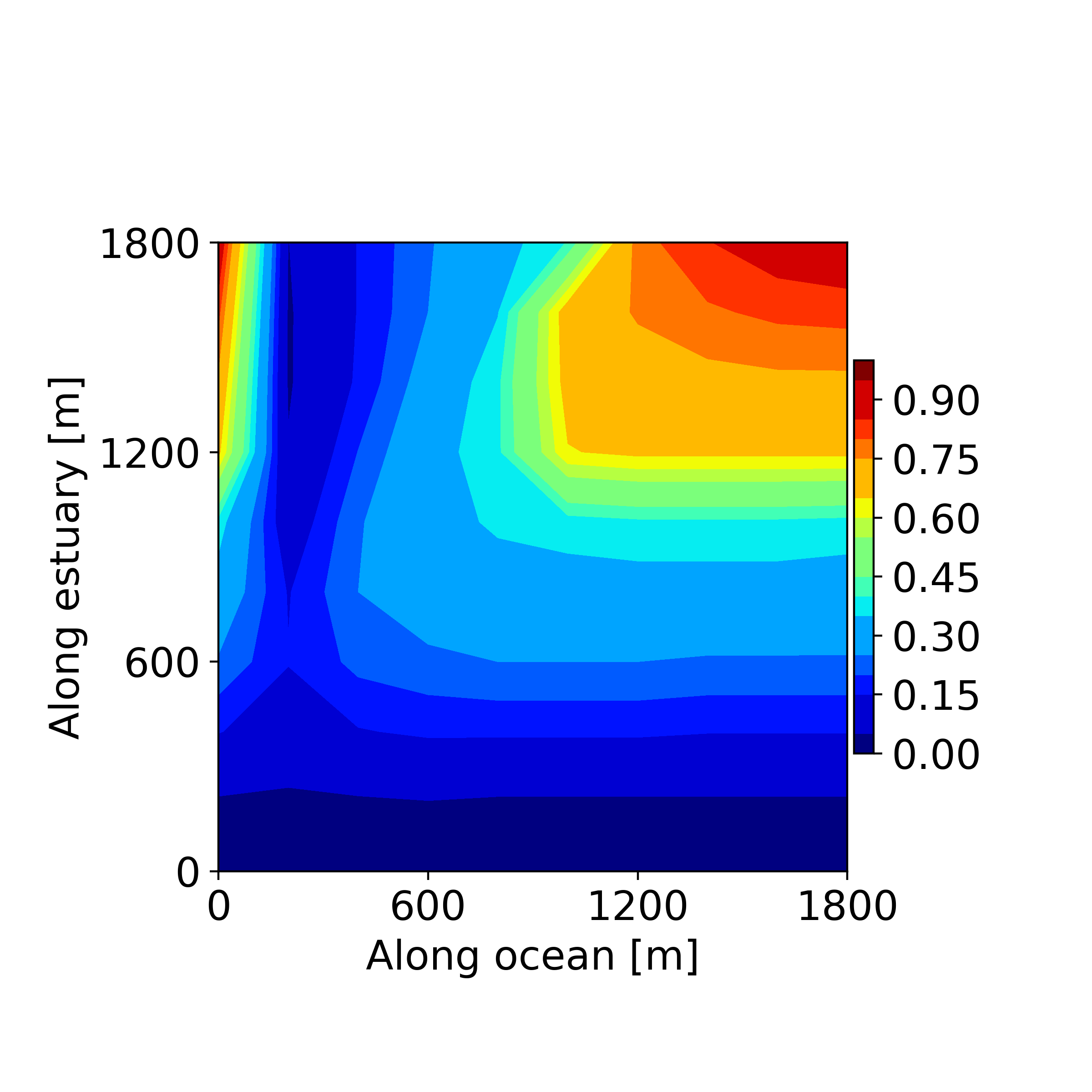}}
  \subfloat[][POD-PCE]{\includegraphics[trim={0.7cm 1.5cm 1.2cm 3.2cm},clip,scale=0.35]{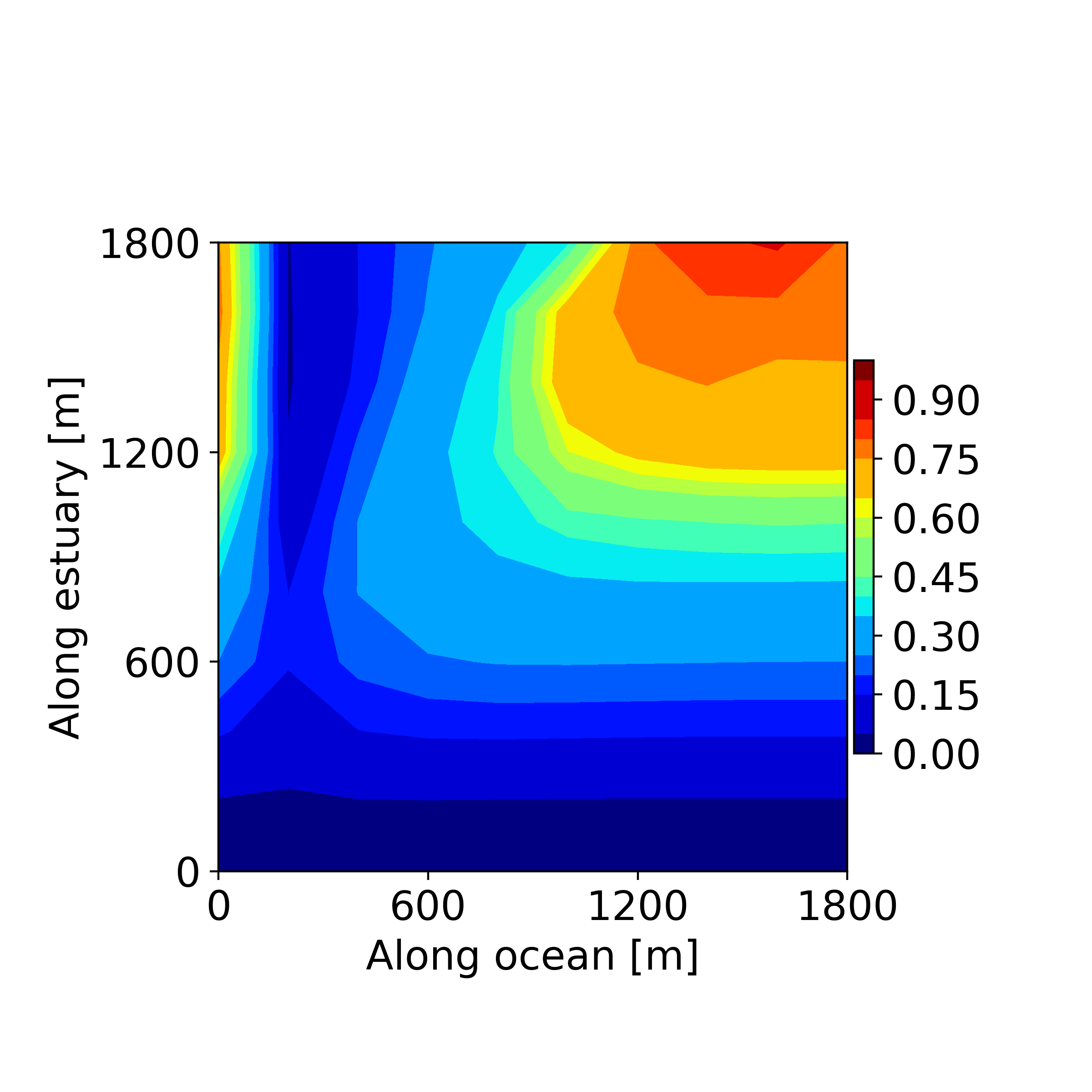}}
  \subfloat[][Absolute residual]{\includegraphics[trim={0.7cm 1.5cm 1.2cm 3.2cm},clip,scale=0.35]{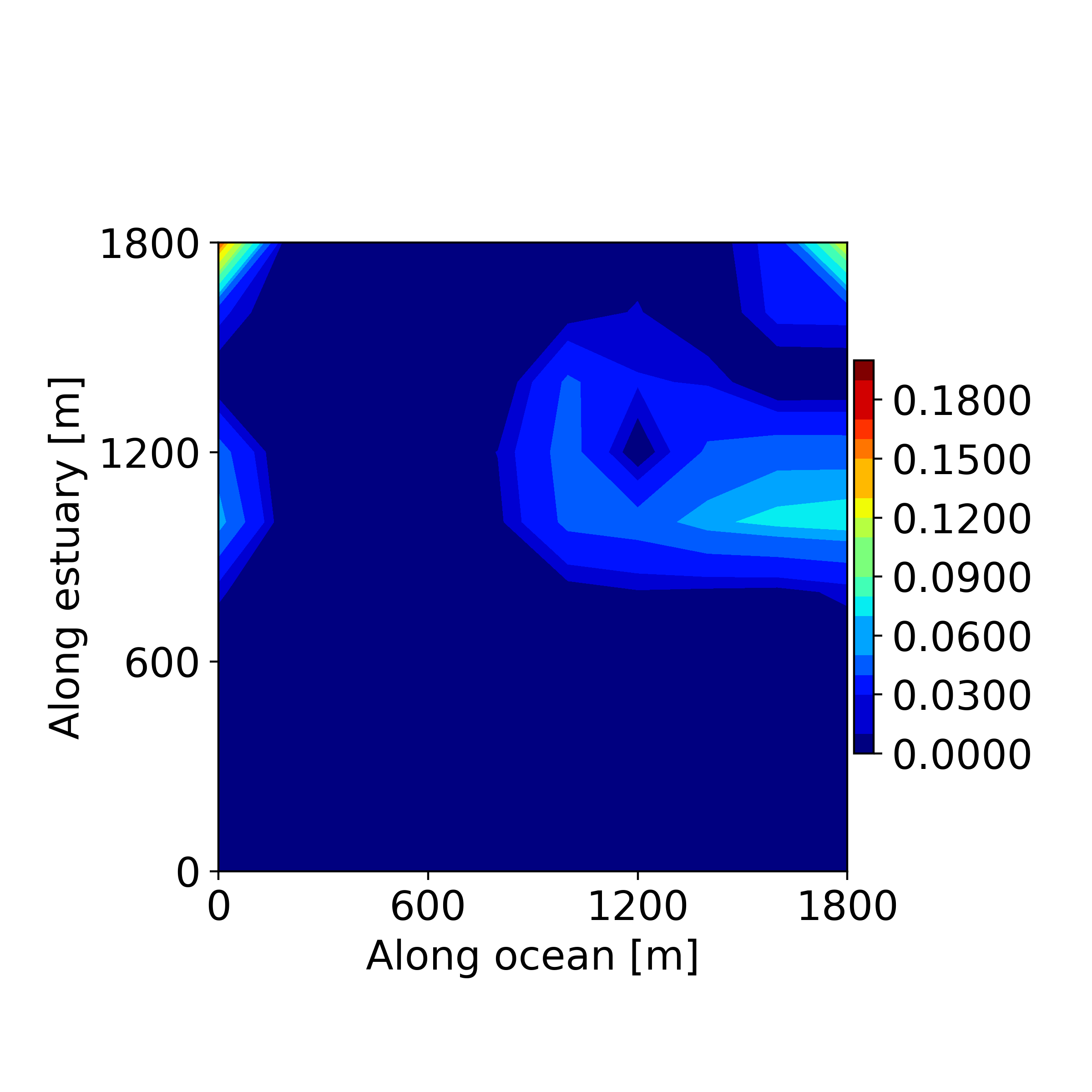}}
  \caption{Analytical solution vs. POD-PCE prediction of the time-lag, relative to the tidal period $T$ in the aquifer, and resulting absolute residual.} 
  \label{fig:toy:PODPCE:phase}
\end{figure}

Secondly, the POD-PCE strategy has also been deployed to learn the time lag between $f(x,y,t)$ and $f(0,0,t)$ relative to the period $T$, at each location $(x,y)$. POD was applied to the corresponding snapshot matrix. While $98\%$ of the variance is already represented by Mode 1, a total of 5 modes is needed to approach the $100\%$. This increase is slow compared to the EVR of the amplitude.  As a result, the POD-PCE performances, evaluated at each step of the algorithm, are different, as can be seen in Figure \ref{fig:toy:Prediction:errors_phase}.
\begin{table}[H]
  \begin{tabular}{|M{5.5cm}|M{3cm}|M{2cm}|M{0.8cm}|M{3cm}|}
    \hline
    Polynomial term & GGW & Total & Mode & GW\\
    \hline		
$\zeta_{\mathbf{\alpha}=(1)}(A)$ & 0.8016 & 0.80 & 1 & 0.84675 \\
\hline
$\zeta_{\mathbf{\alpha}=(1)}(D)$ & 0.10138 & 0.90 & 1 & 0.10709 \\
\hline
$\zeta_{\mathbf{\alpha}=(1)}(D)$ & 0.03199 & 0.93 & 2 & 0.71521 \\
\hline
$\zeta_{\mathbf{\alpha}=(1,1)}(A, D)$ & 0.02115 & 0.96 & 1 & 0.02234 \\
\hline
$\zeta_{\mathbf{\alpha}=(2)}(D)$ & 0.0145 & 0.97 & 1 & 0.01532 \\
\hline
$\zeta_{\mathbf{\alpha}=(1,1)}(A, D)$ & 0.00667 & 0.98 & 2 & 0.14921 \\
\hline
$\zeta_{\mathbf{\alpha}=(2)}(D)$ & 0.00375 & 0.98 & 2 & 0.08373 \\
\hline
$\zeta_{\mathbf{\alpha}=(1,2)}(A, D)$ & 0.00316 & 0.98 & 1 & 0.00334 \\
\hline
$\zeta_{\mathbf{\alpha}=(2)}(D)$ & 0.00238 & 0.99 & 3 & 0.47515 \\
\hline
$\zeta_{\mathbf{\alpha}=(1)}(\kappa_{er})$ & 0.002 & 0.99 & 1 & 0.00211 \\
\hline
$\zeta_{\mathbf{\alpha}=(1)}(\kappa_{er})$ & 0.00165 & 0.99 & 4 & 0.46206 \\
\hline
$\zeta_{\mathbf{\alpha}=(3)}(D)$ & 0.00118 & 0.99 & 1 & 0.00125 \\
\hline
$\zeta_{\mathbf{\alpha}=(1)}(\kappa_{er})$ & 0.00094 & 0.99 & 3 & 0.18655 \\
\hline
  \end{tabular}
  \caption{Polynomial terms of PCE models calibrated on the aquifer case, for the $4$ first modes ordered by their influence, using the GGWs in Equation \ref{eq:PODPCE:generalizedWeights}. Also shown are the GWs calculated as in Equation \ref{eq:PCE:weights}. The contributions are shown up to a total of $99\%$.}
  \label{table:toy:sensitivity:generalizedWeights_amplitude}
\end{table}
 The reduction error equals $10\%$ at rank 1 (compared to $3.8\%$ for the amplitude). It decreases following three slopes, the first one being from  $10\%$ at rank 1 to $5\%$ at rank 3. The learning error is here much higher, almost equal to $1\%$ for a 1-Mode approximation, and goes up to $4\%$ for 3-Mode approximation, where it keeps on increasing. The modes coefficients seem more difficult to learn for the phase. Consequently, the prediction error decreases from $10\%$ at rank 1 to $6\%$ rank 3, where it stabilizes. Indeed, even though adding more POD patterns is interesting, learning them with PCE becomes more and more difficult as the represented variance decreases. The gain in accuracy with POD modes is therefore compensated with the loss of precision in PCE learning. Hence, a 3-Modes POD-PCE model was selected for prediction.Examples of phase prediction are shown in Figure \ref{fig:toy:PODPCE:phase}. The model gives a good mapping of the two-dimensional time lag distribution along the estuary and through the aquifer. However, the residuals are more important compared to the amplitude prediction. For example, an absolute residual of $0.1~T$ time-lag is noticed in the middle of the aquifer in Figure \ref{fig:toy:PODPCE:phase}-c, where the analytical time lag (Figure \ref{fig:toy:PODPCE:phase}-a) is around $0.4~T$, representing a local error of $25\%$. The global performance of the model remains however satisfying.
The physical analysis of the latter are therefore performed using the GW and GGW indicators, reported in Table \ref{table:toy:sensitivity:generalizedWeights_phase}.  First, the most important polynomial pattern for the coupled POD-PCE model is the diffusivity $D$ at degree 1 (GGW $47\%$), whereas it was the tidal amplitude $A$ at degree 1 for the amplitude distribtion (GGW $80\%$). In particular, it barely represents half of the dynamics. It is completed by higher degree monomials of the same parameter $D$ up to $79\%$. The phase representation exhibits more non-linearities than the amplitude. The contribution of $D$ is followed by an interaction between $A$, $D$ and the wave number in the estuary $\kappa_{ei}$. As a reminder, the latter did not appear as an influencing parameter for the amplitude distribution. Globally, the phase problem involves higher polynomial degrees, and higher orders of interaction.  
\begin{table}[H]
  \begin{tabular}{|M{5.5cm}|M{3cm}|M{2cm}|M{0.8cm}|M{3cm}|}
    \hline
    Polynomial term & GGW & Total & Mode & GW\\
    \hline		
$\zeta_{\mathbf{\alpha}=(1)}(D)$ & 0.4684 & 0.47 & 1 & 0.54334 \\
\hline
$\zeta_{\mathbf{\alpha}=(3)}(D)$ & 0.14637 & 0.61 & 1 & 0.16979 \\
\hline
$\zeta_{\mathbf{\alpha}=(4)}(D)$ & 0.0912 & 0.71 & 1 & 0.10579 \\
\hline
$\zeta_{\mathbf{\alpha}=(2)}(D)$ & 0.08382 & 0.79 & 1 & 0.09723 \\
\hline
$\zeta_{\mathbf{\alpha}=(2,1,1)}(A, \kappa_{ei}, D)$ & 0.03942 & 0.83 & 1 & 0.04572 \\
\hline
$\zeta_{\mathbf{\alpha}=(1)}(D)$ & 0.02889 & 0.86 & 2 & 0.35127 \\
\hline
$\zeta_{\mathbf{\alpha}=(2)}(D)$ & 0.02066 & 0.88 & 3 & 0.37113 \\
\hline
$\zeta_{\mathbf{\alpha}=(1)}(D)$ & 0.01986 & 0.90 & 3 & 0.35668 \\
\hline
$\zeta_{\mathbf{\alpha}=(1,1)}(\kappa_{er}, D)$ & 0.01594 & 0.91 & 1 & 0.01849 \\
\hline
$\zeta_{\mathbf{\alpha}=(3)}(D)$ & 0.01515 & 0.93 & 3 & 0.27218 \\
\hline
$\zeta_{\mathbf{\alpha}=(2)}(D)$ & 0.01389 & 0.94 & 2 & 0.16892 \\
\hline
$\zeta_{\mathbf{\alpha}=(1,3)}(\kappa_{er}, D)$ & 0.01288 & 0.96 & 1 & 0.01494 \\
\hline
$\zeta_{\mathbf{\alpha}=(5)}(D)$ & 0.00907 & 0.97 & 2 & 0.11024 \\
\hline
$\zeta_{\mathbf{\alpha}=(2,2)}(A, D)$ & 0.00658 & 0.97 & 2 & 0.07994 \\
\hline
$\zeta_{\mathbf{\alpha}=(1,4)}(\kappa_{ei}, D)$ & 0.0059 & 0.98 & 2 & 0.07178 \\
\hline
$\zeta_{\mathbf{\alpha}=(2,1,1)}(A, \kappa_{ei}, D)$ & 0.00531 & 0.98 & 2 & 0.06453 \\
\hline
$\zeta_{\mathbf{\alpha}=(1,2)}(\kappa_{er}, D)$ & 0.00405 & 0.99 & 1 & 0.0047 \\
\hline
  \end{tabular}
  \caption{Polynomial terms of PCE models calibrated on the aquifer case, for the $3$ first modes ordered by their influence, using the GGWs in Equation \ref{eq:PODPCE:generalizedWeights}. Also shown are the GWs calculated as in Equation \ref{eq:PCE:weights}}
  \label{table:toy:sensitivity:generalizedWeights_phase}
\end{table}


\shorthandoff{:}

\section*{Appendix B. Confronting POD-PCE to NN}
\label{Appendix:B}

As an additional proof for the POD-PCE Machine Learning capacity, multiple NN configurations are tested on the measurement-based problem for confrontation. The latter is a small-data problem, considered as the most challenging case in the presented work. The python library Scikit-learn \citep{Pedregosa2011} (\url{https://scikit-learn.org}) was used for fitting. \\

A first NN set-up, aiming at learning the bathymetry fields $[z(x_i,t_j)]_{i,j} \in \mathbb{R}^{m \times n}$ directly from their previous values $[z(x_i,t_{j-1})]_{i,j} \in \mathbb{R}^{m \times n}$ and a set of parameters $\boldsymbol{\Theta}$, was attempted. A simple configuration was tested, where $z(x_i,t_j)$ is learned for each $x_i$ independently, from its own previous value $z(x_i,t_{j-1})$ and the seven physical parameters $\Delta t$, $TLmean$, $WvH$, $Wvper$, $Wvdir$, $Wv2m$ and $Wv2m\%$, as done in the most optimal POD-PCE configuration ($\mathcal{H}_i^P$ described in Subsection \ref{subsubsection:application:learning:PCE}). Therefore, $m$ independent learnings are performed (points number), each characterized with an input dimension of $V=8$, and an output dimension of $o=1$. \\

Two learning strategies are adopted. The first one consists in a single-layer NN, where only the \textit{Activation Function} (AF) and the number of \textit{neurons}, denoted $l$, are varied. The considered AFs are the ones available in Scikit-learn (identity, tanh, logistic and ReLu) \citep{Pedregosa2011}, and allow to be in the theoretical conditions for the "shallow and wide" \textit{Universal Approximation Theorem} \citep{Hornik1991}. The second alternative consists in a multi-layer NN using the ReLu AF, where the number of neurons $l$ is fix, and the number of layers, denoted $L$ varies. This allows to be in the framework of the "deep and narrow" version of the theorem \citep{Hanin2019}. \\

\begin{figure}[H]
  \centering
  \vspace{-0.5cm}
  \subfloat[][Comparison to NN]{\includegraphics[trim={0cm 0.5cm 1cm 0.5cm},clip,scale=0.4]{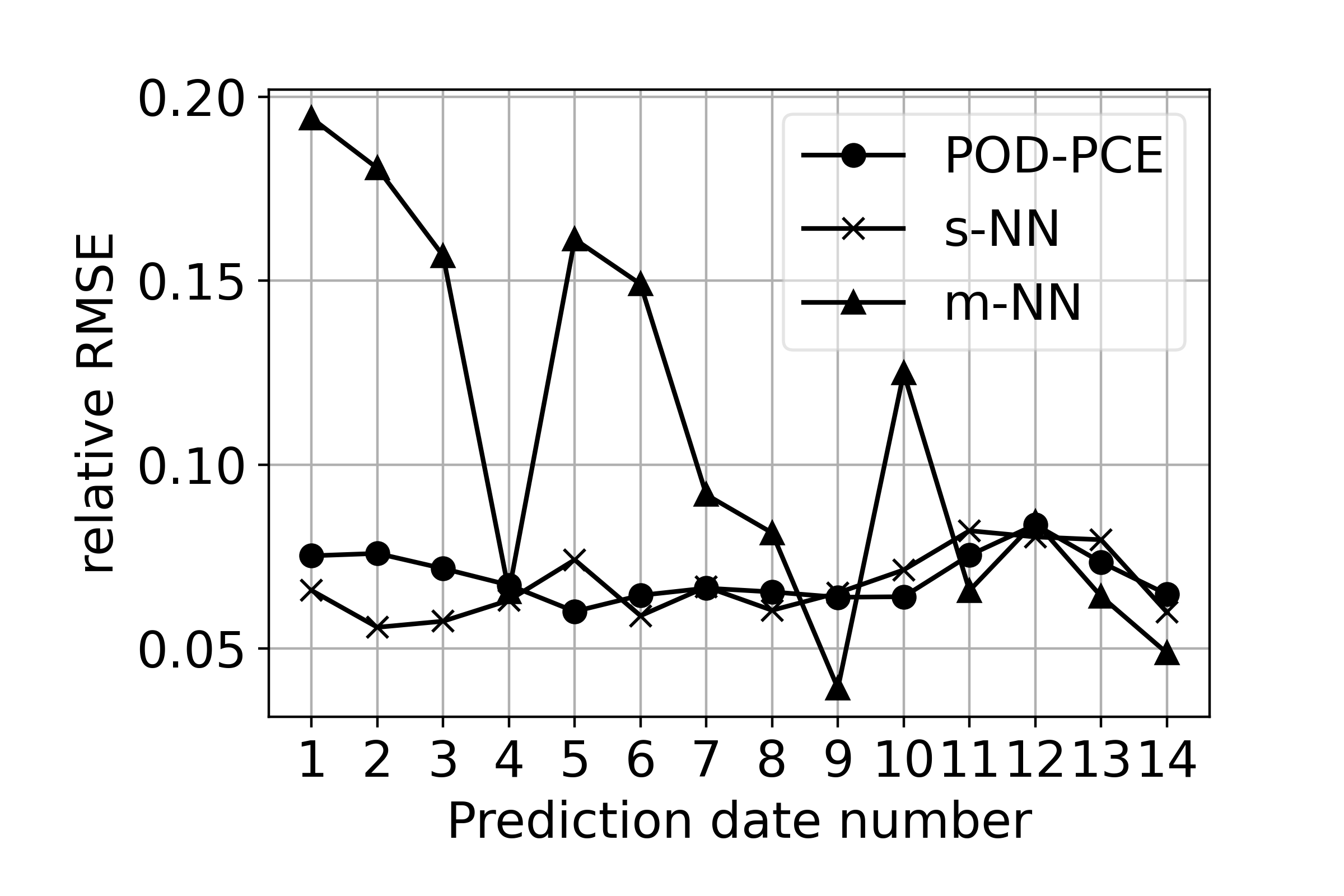}}
  \subfloat[][Comparison to POD-NN]{\includegraphics[trim={0cm 0.5cm 1cm 0.5cm},clip,scale=0.4]{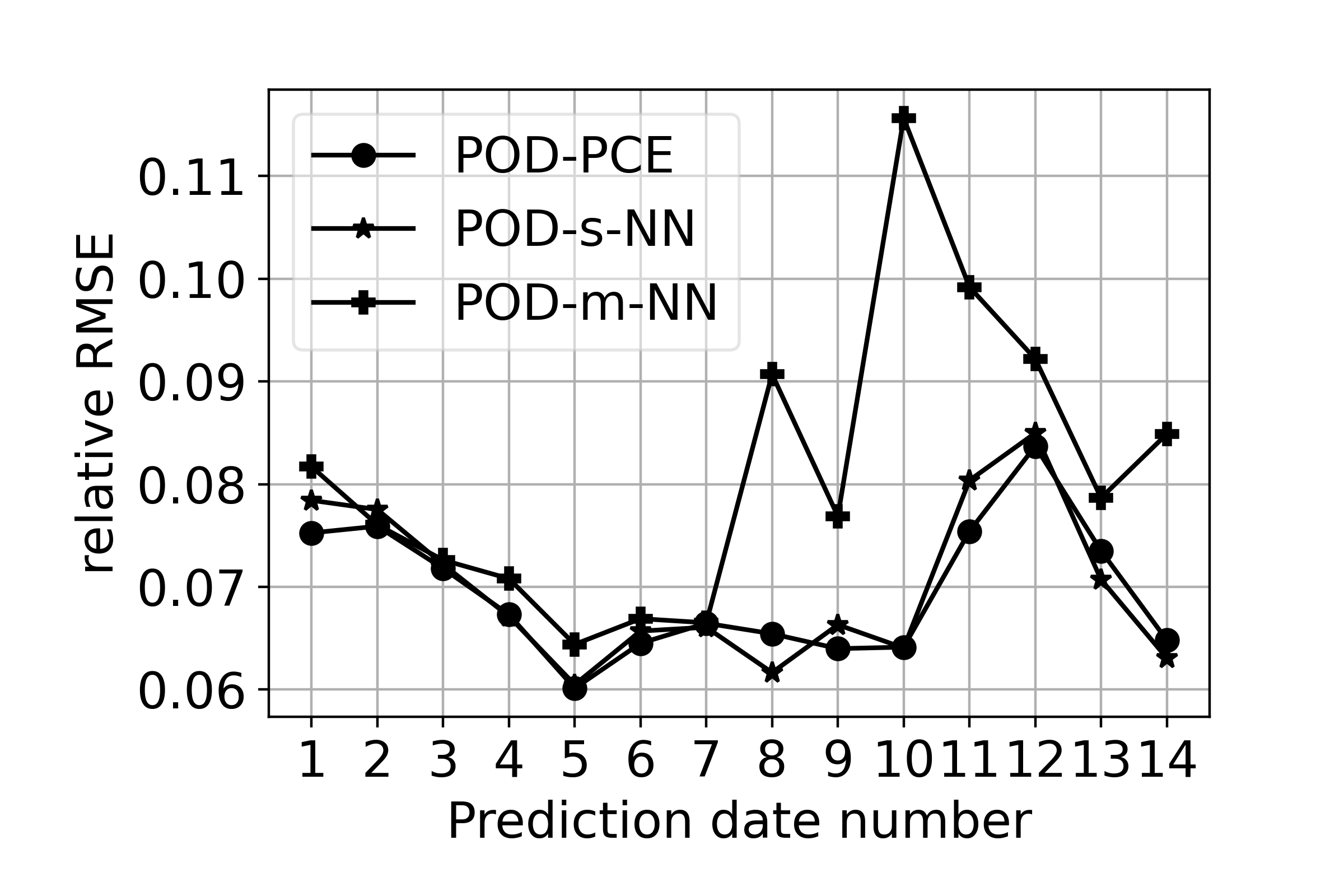}}
    \caption{Comparison of relative prediction RMSE between the POD-PCE algorithm and different NN set-ups.}
    \label{fig:PODPCE_NN_RMSE}
\end{figure}

For the single-layer NN, the maximal number of neurons is constrained to $l=5$. Indeed, the input-to-hidden connection matrix is of size $V \times l$, and the hidden-to-output matrix is of size $l \times o$. In this case, with $l=5$, a maximum number of $45$ matrix coefficients should be estimated from the training sample of size $50$. An additional neuron would result with an ill-posed problem. For the multi-layer ReLu NN, the number of neurons is set to $l=2$ and the maximum number of layers to $L=9$ (number of coefficients to estimate is $V\times l + (L-1)\times l^2 + l\times o = 50$). Using both configurations, the optimal choices (AF, neurons, layers) are selected for each coordinate $x_i$, based on the relative empirical error calculated on the test set, as in Equation \ref{eq:relativeEmpiricalError}~. The RMSE for each prediction date are then calculated with the whole field $z(.,t_j)$ (NN prediction vs. reality), and confronted to POD-PCE in Figure \ref{fig:PODPCE_NN_RMSE}-a. The single-layer NN is denoted s-NN and the multi-layer ReLu NN is denoted m-NN. \\

To account for spatial correlations, a supplementary set-up was tested, where POD is performed before NN. Similarly to the POD-PCE learning set-up, NN is used to learn the first $11$ POD coefficients, corresponding to the optimal POD-PCE learning in Section \ref{subsubsection:application:learning:PCE}, and a POD-NN coupling is performed. The learning configurations mentioned above (single-layer, and multi-layer ReLu) are tested, and the algorithmic choices corresponding to the minimal relative empirical error are selected for each POD mode independently. The RMSE results are shown in Figure \ref{fig:PODPCE_NN_RMSE}-b, where POD-s-NN and POD-m-NN denote the coupling of POD with s-NN and m-NN respectively. 

A last test is conducted, where an L2-penalty is used to fit sparse POD-NN. This is performed in scikit-learn \citep{Pedregosa2011} by adding a constraint to the learning minimization problem, consisting in a regularization term, controlled with an additional hyper-parameter. The values of hyper-parameters that were previously constrained can here be increased: the maximal number of neurons is set to $l=50$ for POD-s-NN, while the maximal number of layers is set to $L=10$ for POD-m-NN with a number of neurons fixed to $l=5$. The L2-penalty coefficient is varied from $10^{-4}$ (low sparsity) to $10^{4}$ (high sparsity). A comparison of all algorithms in terms of average RMSE and fit-time can be found in Table \ref{table:PODPCE_NN_summary}. \\

Firstly, it can be noticed in Figure \ref{fig:POD_PCE_NN} that the worst learning is performed with m-NN (average RMSE of $10.8\%$ in Table \ref{table:PODPCE_NN_summary}). It might be due to the fact that available data are not sufficient for a deep network fitting. This is followed in terms of worst performance by POD-m-NN (average RMSE of $8.1\%$), with the same interpretation. The other learning choices (POD-PCE, s-NN and POD-s-NN) have global similar behaviors. Among the last three, it is noted that s-NN performs the best for the three first dates, while it performs the worst for dates 5, 10, and 11 (Figure \ref{fig:POD_PCE_NN}). It scores the lowest average RMSE of $6.7\%$, but also by far the worst fit-time (Table \ref{table:PODPCE_NN_summary}). The performances of POD-PCE and POD-s-NN are very close, their average RMSE are $6.9\%$ and $7\%$, but POD-PCE is twice faster. Sparsity added to POD-s-NN and POD-m-NN helps reducing the errors by $0.1$ and $1.2\%$ respectively. The resulting RMSE are equivalent to POD-PCE using LARS, which takes much less fit-time.
\begin{table}[H]
  \centering
  \begin{tabular}{|M{3cm}|M{3cm}|M{3cm}|M{3cm}|}
    \hline
    Algorithm & Sparsity & Average RMSE & Fit-time \\
\hline
 POD-PCE & LARS \citep{Blatman2011} & $6.9 \%$ & $11s$ \\
\hline
 s-NN & None & $6.7\%$ & $1h47m14s$ \\
\hline
 m-NN & None & $10.8\%$ & $20m16s$ \\
\hline
 POD-s-NN & None & $7\%$ & $25s$ \\
\hline
 POD-m-NN & None & $8.1\%$ & $10s$ \\
\hline
 POD-s-NN & L2 penalty \citep{Pedregosa2011} & $6.9\%$ & $11m16s$ \\
\hline
 POD-m-NN & L2 penalty \citep{Pedregosa2011} & $6.9\%$ & $2m48s$ \\
\hline
  \end{tabular}
  \caption{Summary the performances for all tested learning algorithms.}
  \label{table:PODPCE_NN_summary}
\end{table}

The POD-PCE coupling methodology offers an interesting alternative to NN in terms of accuracy and fit-time balance. It competes with POD-s-NN which is slightly less accurate, but POD-PCE is here twice-faster. Additionally, the most optimal POD-s-NN is composed of different AFs for the different modes (combinations of logistic and ReLu), which makes the interpretation difficult compared to polynomial patterns, and results with a superiority of POD-PCE for physical analysis. However, these conclusions should be interpreted in light of the learning choices, which can be improved. For example, a combination of the best single-layer networks and best multi-layer ReLu networks can be attempted to optimize the previous set-ups. This can even be further improved by choosing PCE or NN when appropriate. Lastly, as was the case with PCE, limitations to the previous learnings can be noted, among which the physical parameters selection and time-lag choice for the previous field value. 


%
\bibliographystyle{abbrvnat}
\bibliography{refs}

\end{document}